\documentclass[aps,prx,reprint,floatfix,amsmath,amssymb,nofootinbib,bibnotes,longbibliography,superscriptaddress]{revtex4-2}

\usepackage[T1]{fontenc}
\usepackage{geometry}
\usepackage{color}
\usepackage{xcolor}
\usepackage{babel}
\usepackage{float}
\usepackage{amsmath}
\usepackage{amsfonts}
\usepackage{amssymb}
\usepackage{graphicx}
\usepackage[caption=false,position=top]{subfig}
\usepackage[normalem]{ulem} 

\usepackage{titlesec}
\usepackage[dont-mess-around]{fnpct}

\usepackage{siunitx}
\usepackage[version=4,textfontname=sffamily,mathfontname=mathsf]{mhchem}
\usepackage{nicefrac}

\usepackage{mathtools}

\usepackage{algorithm}
\usepackage{algpseudocode}

\usepackage{empheq}

\usepackage[braket, qm]{qcircuit}
\usepackage{ragged2e}
\usepackage{chngcntr}
\usepackage{etoolbox}

\makeatletter
\newcommand{\appendixtocext}{atoc}

\newcommand{\startappendixtoc}{%
  \let\orig@addcontentsline\addcontentsline
  \renewcommand{\addcontentsline}[3]{%
    \ifstrequal{##1}{toc}
      {\orig@addcontentsline{\appendixtocext}{##2}{##3}}
      {\orig@addcontentsline{##1}{##2}{##3}}%
  }%
}

\newcommand{\printappendixtoc}{%
  \@starttoc{\appendixtocext}%
}
\makeatother

\newcommand{\cupdot}{\mathbin{\mathaccent\cdot\cup}}

\usepackage[unicode=true,pdfusetitle,bookmarks=true,bookmarksnumbered=false,bookmarksopen=false,breaklinks=false,pdfborder={0 0 1},backref=false,colorlinks=true]{hyperref}

\newtheorem{definition}{Definition}

\newtheorem{theorem}{Theorem}

\newtheorem{example}{Example}
 \hypersetup{
     colorlinks   = true,
     linkcolor    = blue,
     urlcolor     = blue,
     citecolor    = blue
} 

 \algrenewcommand\algorithmicthen{}

\algrenewcommand\algorithmicrequire{\textbf{Input:}}
\algrenewcommand\algorithmicensure{\textbf{Output:}}
\algnewcommand\Input{\item[\algorithmicrequire]}
\algnewcommand\Output{\item[\algorithmicensure]}
\algrenewcommand\algorithmiccomment[1]{\hfill{\color{gray}\texttt{\textbackslash\textbackslash}~#1}}
\algnewcommand\LComment[1]{{\color{gray}\texttt{\textbackslash*}~#1~\texttt{*\textbackslash}}}

\global\long\def\ket#1{\left|#1\right\rangle }%
\global\long\def\bra#1{\left\langle #1\right|}%
\global\long\def\sket#1{\left.\left|#1\right\rangle \right\rangle }%
\global\long\def\sbra#1{\left\langle \left\langle #1\right|\right.}%
\global\long\def\norm#1{\left\Vert #1\right\Vert }%

\DeclarePairedDelimiterX\braketExp[1]{\langle}{\rangle}{#1}%

\makeatletter
\DeclareFontFamily{OMX}{MnSymbolE}{}
\DeclareSymbolFont{MnLargeSymbols}{OMX}{MnSymbolE}{m}{n}
\SetSymbolFont{MnLargeSymbols}{bold}{OMX}{MnSymbolE}{b}{n}
\DeclareFontShape{OMX}{MnSymbolE}{m}{n}{
    <-6>  MnSymbolE5
   <6-7>  MnSymbolE6
   <7-8>  MnSymbolE7
   <8-9>  MnSymbolE8
   <9-10> MnSymbolE9
  <10-12> MnSymbolE10
  <12->   MnSymbolE12
}{}
\DeclareFontShape{OMX}{MnSymbolE}{b}{n}{
    <-6>  MnSymbolE-Bold5
   <6-7>  MnSymbolE-Bold6
   <7-8>  MnSymbolE-Bold7
   <8-9>  MnSymbolE-Bold8
   <9-10> MnSymbolE-Bold9
  <10-12> MnSymbolE-Bold10
  <12->   MnSymbolE-Bold12
}{}
\let\llangle\@undefined
\let\rrangle\@undefined
\DeclareMathDelimiter{\llangle}{\mathopen}%
                     {MnLargeSymbols}{'164}{MnLargeSymbols}{'164}
\DeclareMathDelimiter{\rrangle}{\mathclose}%
                     {MnLargeSymbols}{'171}{MnLargeSymbols}{'171}
\DeclarePairedDelimiterX\kket[1]{\lvert}{\rrangle}{#1}
\DeclarePairedDelimiterX\bbra[1]{\llangle}{\rvert}{#1}
\DeclarePairedDelimiterX\bbrakket[2]{\llangle}{\rrangle}{#1\delimsize\vert\mathopen{}#2}%
\DeclarePairedDelimiterX\bbrakketExp[1]{\llangle}{\rrangle}{#1}
\DeclarePairedDelimiterX\bbrakketOP[3]{\llangle}{\rrangle}{#1\,\delimsize\vert\,\mathopen{}#2\,\delimsize\vert\,\mathopen{}#3}%
\newcommand{\ignore}[1]{}
\makeatother

\newcommand{\trp}{\ensuremath{T}}  

\DeclareMathOperator{\IF}{IF}

\DeclareMathOperator{\Frob}{Frob}
\DeclareMathOperator{\Cov}{\mathbb{C}\mathrm{ov}}

\makeatletter
\providecommand*{\cupdot}{%
  \mathbin{%
    \mathpalette\@cupdot{}%
  }%
}
\newcommand*{\@cupdot}[2]{%
  \ooalign{%
    $\m@th#1\cup$\cr
    \hidewidth$\m@th#1\cdot$\hidewidth
  }%
}
\makeatother

\makeatletter
\long\def\@makecaption#1#2{%
  \vskip\abovecaptionskip
  \begingroup
    \justifying
    \small
    \noindent
    \textbf{#1:} #2\par
  \endgroup
  \vskip\belowcaptionskip
}
\makeatother

\begin{document}

\title{
Reliable high-accuracy error mitigation for utility-scale quantum circuits}

\author{Dorit Aharonov}
\affiliation{Qedma Quantum Computing, Tel Aviv, Israel}
\affiliation{The Benin School of Computer Science and Engineering, Hebrew University, Jerusalem, Israel}

\author{Ori Alberton}
\affiliation{Qedma Quantum Computing, Tel Aviv, Israel}

\author{Itai Arad}
\affiliation{Qedma Quantum Computing, Tel Aviv, Israel}
\affiliation{Centre for Quantum Technologies, National University of Singapore, Singapore}

\author{Yosi Atia}
\author{Eyal Bairey}
\thanks{Correspondence to: Eyal.Bairey@Qedma.com, Ron.Melcer@Qedma.com}
\author{Matan Ben Dov}
\author{Asaf Berkovitch}
\affiliation{Qedma Quantum Computing, Tel Aviv, Israel}

\author{Zvika Brakerski}
\affiliation{Qedma Quantum Computing, Tel Aviv, Israel}
\affiliation{Faculty of Mathematics and Computer Science, Weizmann Institute of Science, Israel}

\author{Itsik Cohen}
\author{Eran Fuchs}
\author{Omri Golan}
\author{Or Golan}
\author{Barak D. Gur}
\author{Ilya Gurwich}
\author{Avieli Haber}
\author{Rotem Haber}
\author{Dorri Halbertal}
\author{Yaron Itkin}
\author{Barak A. Katzir}
\affiliation{Qedma Quantum Computing, Tel Aviv, Israel}

\author{Oded Kenneth}
\affiliation{Qedma Quantum Computing, Tel Aviv, Israel}
\affiliation{Department of Physics, Technion, Haifa, Israel}

\author{Shlomi Kotler}
\affiliation{Qedma Quantum Computing, Tel Aviv, Israel}
\affiliation{Racah Institute of Physics, The Hebrew University of Jerusalem, Jerusalem 91904, Israel}

\author{Roei Levi}
\author{Eyal Leviatan}
\author{Yotam Y. Lifshitz}
\author{Adi Ludmer}
\author{Shlomi Matityahu}
\author{Ron Aharon Melcer}
\thanks{Correspondence to: Eyal.Bairey@Qedma.com, Ron.Melcer@Qedma.com}
\author{Adiel Meyer}
\author{Omrie Ovdat}
\author{Aviad Panahi}
\author{Gil Ron}
\author{Ittai Rubinstein}
\author{Gili Schul}
\author{Tali Shnaider}
\author{Maor Shutman}
\author{Asif Sinay}
\author{Tasneem Watad}
\author{Assaf Zubida}
\affiliation{Qedma Quantum Computing, Tel Aviv, Israel}

\author{Netanel H. Lindner}
\affiliation{Qedma Quantum Computing, Tel Aviv, Israel}
\affiliation{Department of Physics, Technion, Haifa, Israel}

\date{\today{}}

\begin{abstract} 
Error mitigation is essential for unlocking the full potential of quantum algorithms and accelerating the timeline toward quantum advantage. As quantum hardware progresses to push the boundaries of classical simulation, efficient and robust error mitigation methods are becoming increasingly important for producing accurate and reliable outputs. However, existing error-mitigation approaches face a fundamental tradeoff between practical performance and reliability: heuristic methods such as zero-noise extrapolation (ZNE) enjoy faster runtime but lack accuracy guarantees, while rigorous techniques such as probabilistic error cancellation (PEC) provide unbiased estimates at prohibitive computational cost. We introduce a characterization-based, rigorously-grounded quantum error mitigation and error suppression framework (QESEM) that resolves this tradeoff by leveraging the accuracy guarantees of quasi-probabilistic mitigation with dramatically reduced overhead.  We explain the innovative methods underlying QESEM and demonstrate its capabilities in the largest utility-scale error mitigation experiment based on an unbiased method. This experiment simulates the kicked transverse field Ising model with far-from-Clifford parameters on an IBM Heron device. We further validate QESEM's versatility across arbitrary quantum circuits and devices through high-accuracy error-mitigated molecular VQE circuits executed on IBM Heron and IonQ trapped-ion devices. Compared with multiple variants of the widely used zero-noise extrapolation method, QESEM consistently achieves higher accuracy while avoiding the prohibitive runtime overhead associated with PEC. These results mark a significant step forward in accuracy and reliability for running quantum circuits on current devices across diverse applications. Finally, we provide projections of QESEM's performance on near-term devices toward quantum advantage.

\end{abstract}

\maketitle

\section{ Introduction \label{Sec:intro}}

Errors are the primary obstacle to realizing the potential of quantum computers today~\cite{Lanes2025FrameworkQuantumAdvantage, YEMpaper}. When left unaddressed, they dominate the outcomes of even modest-sized computations, limiting quantum circuit sizes to the order of the inverse hardware error rate, or significantly smaller when high accuracy is required.

To address errors, researchers rely on two complementary approaches: \emph{error correction} and \emph{error mitigation} (EM). Error correction encodes logical qubits into multiple physical ones, but requires substantial qubit and gate overheads, mid-circuit measurements, and high-fidelity operations~\cite{Shor1995, Steane1996, Aharonov2008}. These stringent demands render error correction impractical for executing useful quantum algorithms on near-term hardware. Error-detecting codes~\cite{Knill2004} ease these requirements slightly, but still incur significant qubit and gate overheads. In contrast, EM encompasses techniques that negate the impact of hardware errors through circuit recompilation and classical postprocessing. This increases quantum runtime while avoiding the substantial qubit overhead required for error correction.

In the long term, these error reduction methods will likely be combined to maximize achievable circuit volumes and accuracies for a given number of qubits and computational time~\cite{Suzuki2022, Toshio2024, Piveteau2021, Wahl2023, Zhang2025, myths2025, YEMpaper}. However, given the stringent hardware requirements of error correction, EM remains the only viable solution today. 

The use of EM significantly accelerates the timeline for achieving quantum advantage at increasing scales~\cite{The_future_of_superconducting_qubits}. The runtime overhead of EM scales exponentially with the total accumulated error in the circuit. This suggests that, on its own, EM cannot yield 
\textit{asymptotic exponential} quantum advantage~\cite{Takagi2022, Takagi2023universal, Tsubouchi2023universal, Quek2024, Zimboras2025}. Nevertheless, relative to unmitigated execution, EM boosts the circuit volumes executable to a desired accuracy by orders of magnitude~\cite{YEMpaper}. Therefore, EM is expected to play a central role in realizing \emph{finite} quantum advantages~\cite{YEMpaper}, where quantum computers outperform classical ones within practical runtimes for concrete use cases. 

Indeed, EM is widely used today in applications ranging from high-energy \cite{DiMeglio2024, Farrell2024, Angelides2025} and condensed-matter physics \cite{Shinjo2024} to quantum chemistry \cite{Kandala2019, Bauman2025}. As hardware improves and gate infidelities decrease, the runtime cost of EM decreases exponentially. This will enable high-accuracy execution of increasingly large quantum circuits, extending beyond those accessible to classical simulation. Looking ahead to the early fault-tolerant era, EM is expected to continue enhancing circuit volumes in combination with error correction and detection~\cite{Suzuki2022, Toshio2024, Piveteau2021, Wahl2023, Zhang2025, YEMpaper}, remaining crucial as long as high-quality qubits are a scarce resource. 

Accurate and efficient EM methods are thus becoming increasingly important. High output accuracies are necessary for deriving utility in many quantum applications; for example, in quantum chemistry \cite{McArdle2020}, where `chemical accuracy' is often considered necessary to reliably predict reaction energies and outperform classical approximation methods. Delivering such precision is challenging due to diverse error sources and hardware fluctuations, which make circuit errors hard to model and predict. The importance of accurate and reliable mitigation grows as cutting-edge devices push the boundaries of classical simulation, producing outputs that are increasingly hard to verify \cite{Arute2019, Kim2023utility, Farrell2024, Google2025, Quantinuum2025}. Yet, many EM methods rely on uncontrolled heuristics that can fail, with severity depending on hardware noise and the executed circuit. Often, these methods offer neither control nor guarantees over the output error, and can exhibit large mitigation biases \cite{Cai2021}. As problem sizes exceed classical simulability, it is essential to develop mitigation methods whose biases can be systematically controlled and bounded without access to the ideal circuit values. 

Beyond accuracy, practical EM also demands efficiency. Although mitigation times scale exponentially with the accumulated error in the circuit, and thus with its size, the exponent depends on the mitigation method. We measure circuit size in terms of \emph{active volume} \cite{YEMpaper}: the number of entangling gates in the circuit that affect the target observable. Active volume determines the complexity both for classical simulation and for EM, since any gate outside the `causal lightcone' of the observable can be omitted without loss of accuracy. It is crucial to reduce the exponent controlling the time overhead in order to unlock the largest active volumes and highest output accuracies achievable on a given noisy quantum device.

Here we present QESEM: a reliable, efficient, and accurate Quantum Error Suppression and Error Mitigation software. Built on an unbiased quasi-probabilistic (QP) mitigation framework~\cite{Temme2017, Endo2018}, QESEM preserves the reliability of QP-based mitigation methods while introducing key innovations that improve efficiency and scalability. These include the ability to trade minimal, controlled bias for substantial performance gains, and efficient \emph{multi-type} QP decompositions for both Clifford and fractional entangling gates, which reduce circuit depths by up to twice compared to standard mitigation methods that are incompatible with fractional gates. 

\begin{figure*}[t!]
\centering
\subfloat[
\label{fig:QESEM_workflow}]{
\includegraphics[width=0.95\linewidth]{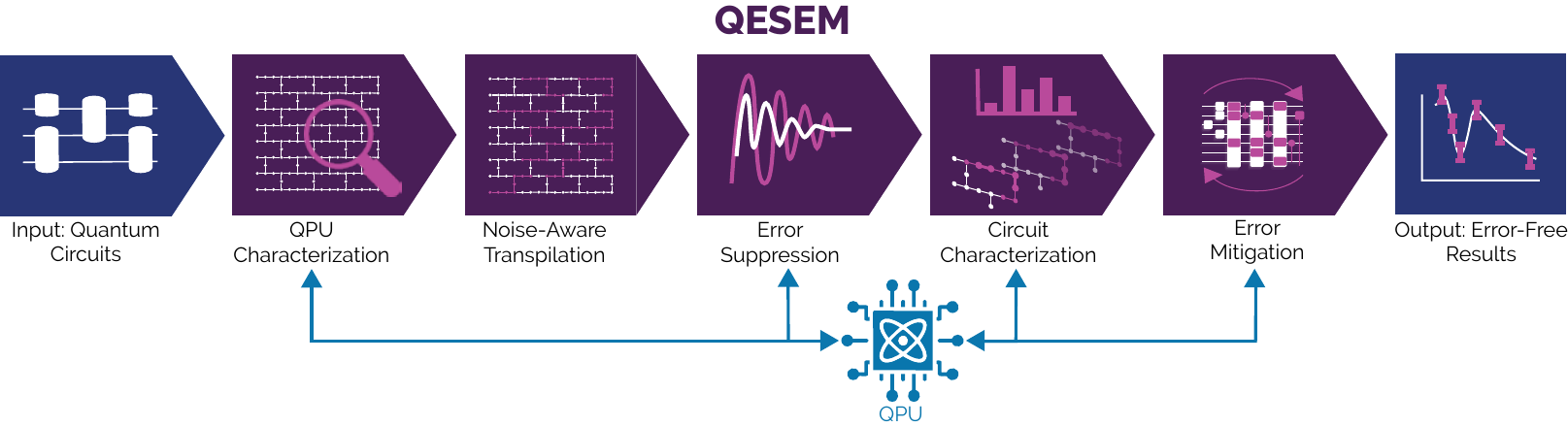}
}
\\
\subfloat[
\label{fig:QESEM_innovations}]{
\includegraphics[width=0.80\linewidth]{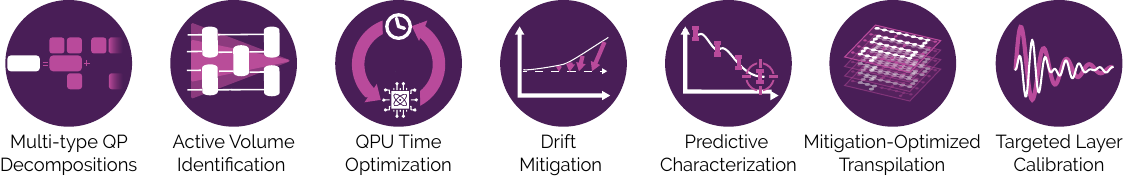}
}
\caption{QESEM overview.
\textbf{\protect\subref{fig:QESEM_workflow} QESEM workflow and stages.} 
\emph{Device characterization} maps gate fidelities and identifies coherent errors, providing real-time calibration data.
\emph{Noise-aware transpilation} generates and evaluates alternative qubit mappings, operation sets, and measurement bases, selecting the variant that minimizes estimated QPU runtime, with optional parallelization to accelerate data collection. 
\emph{Error suppression} redefines native gates, applies Pauli twirling, and optimizes pulse-level control (on supported platforms
) to improve fidelity. 
\emph{Circuit characterization} builds a tailored local error model and fits it to QPU measurements to quantify residual noise. 
\emph{Error mitigation} constructs multi-type quasi-probabilistic decompositions and samples from them in an adaptive process that minimizes mitigation QPU time and sensitivity to hardware fluctuations, achieving high accuracies at large circuit volumes.
\textbf{\protect\subref{fig:QESEM_innovations} Key innovations.} Core algorithmic and implementation features of QESEM that enable scalability, efficiency, and robustness, described in detail in the appendices.}
\end{figure*}

We begin with an overview of the QESEM workflow. As explained in Sec.~\ref{Sec:workflow} (see Fig.~\ref{fig:QESEM_workflow}), QESEM begins with a preliminary \emph{device characterization} that identifies the fidelities of available device operations and calibrates them, suppressing reversible coherent errors. It then transpiles the circuit based on the device characterization, choosing operations that minimize the required mitigation QPU time using procedures for \emph{active volume identification} and \emph{mitigation time estimation}. High accuracy is guaranteed by a \emph{predictive circuit characterization} protocol extracting an error model for the operations of the transpiled circuit, including Clifford and fractional 2-qubit gates, as well as state preparation and measurement errors. QESEM then constructs QP decompositions based on this error model, and samples from them efficiently in an adaptive \emph{drift-robust} flow to obtain an estimate and error bar for each requested observable. 

\begin{figure*}[tb]
\centering
\subfloat[
    \label{fig:ising_circ}]{   \includegraphics[width=0.26\textwidth]{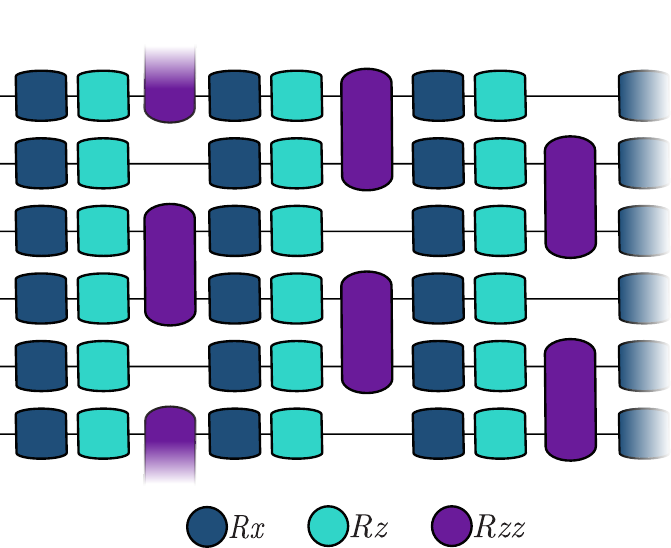}}
\hfill
\subfloat[
    \label{fig:ising_geometry}]{
    \includegraphics[width=0.26\textwidth]{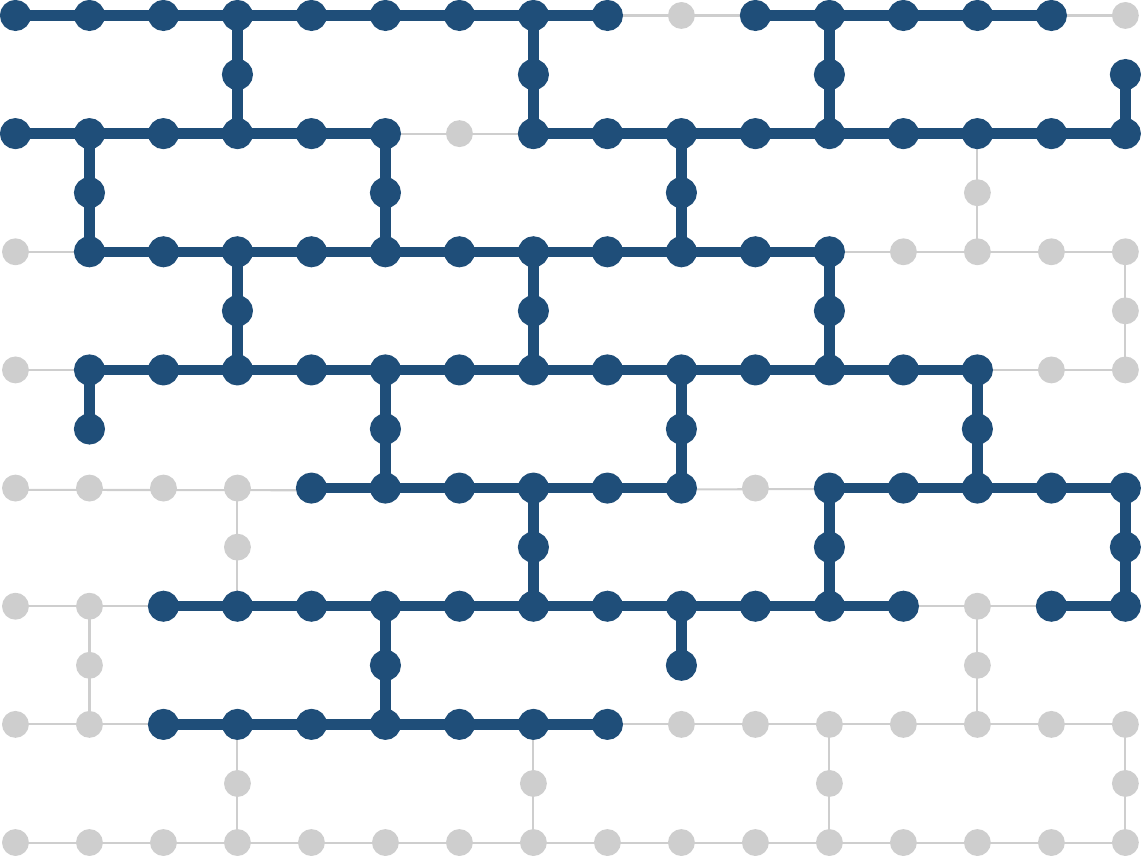}}
\hfill
\subfloat[
    \label{fig:ising_lc}]{
    \includegraphics[width=0.44\textwidth]{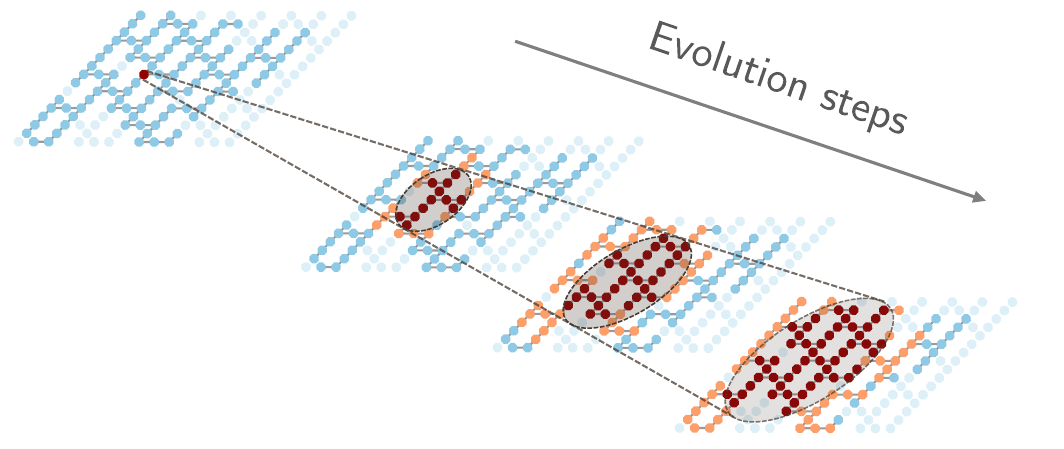}}
\\
\subfloat[
    \label{fig:ising_steps}]{
    \includegraphics[width=0.49\textwidth]{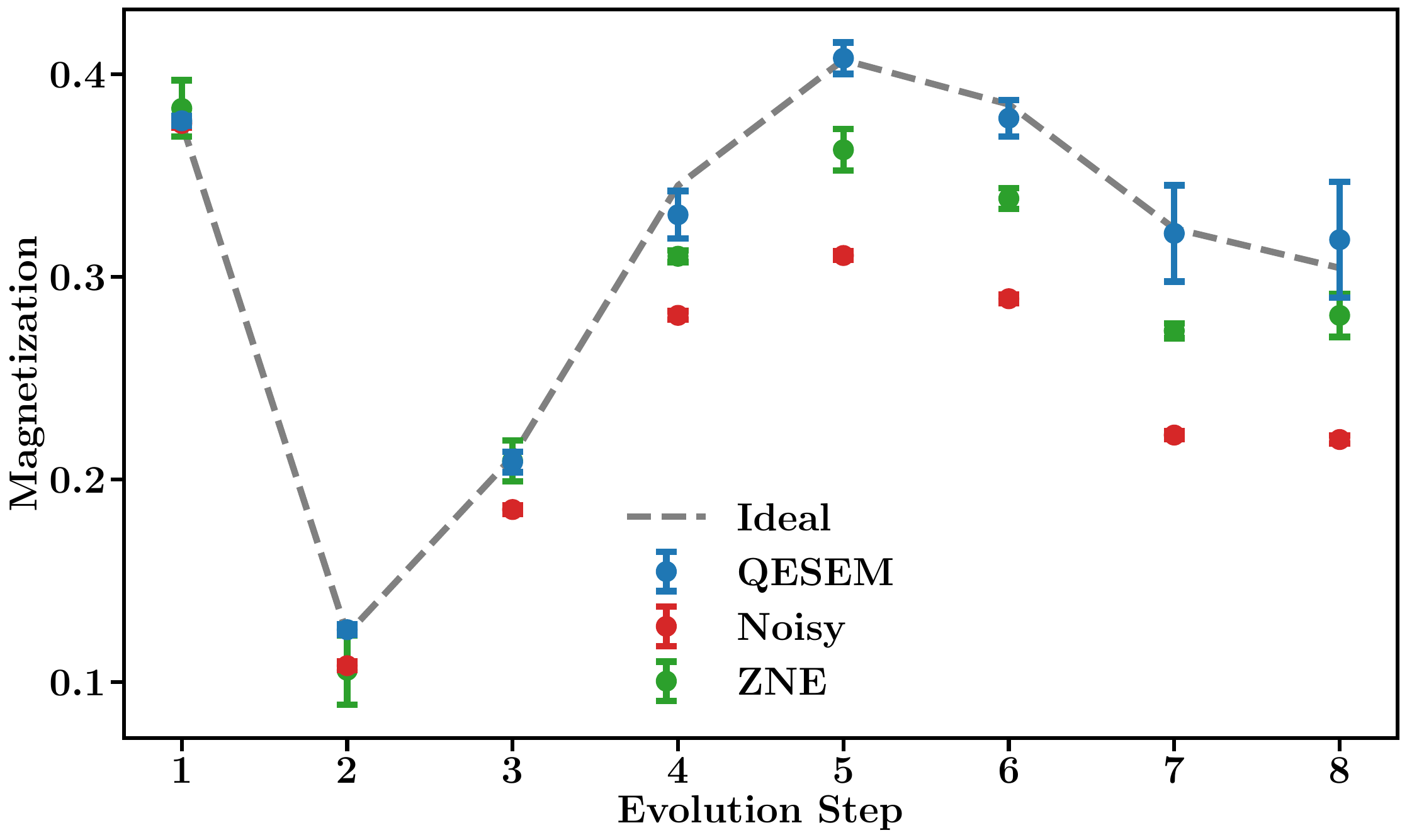}}
\hfill
\subfloat[
    \label{fig:ising_z_hist}]{
    \includegraphics[width=0.48\textwidth]{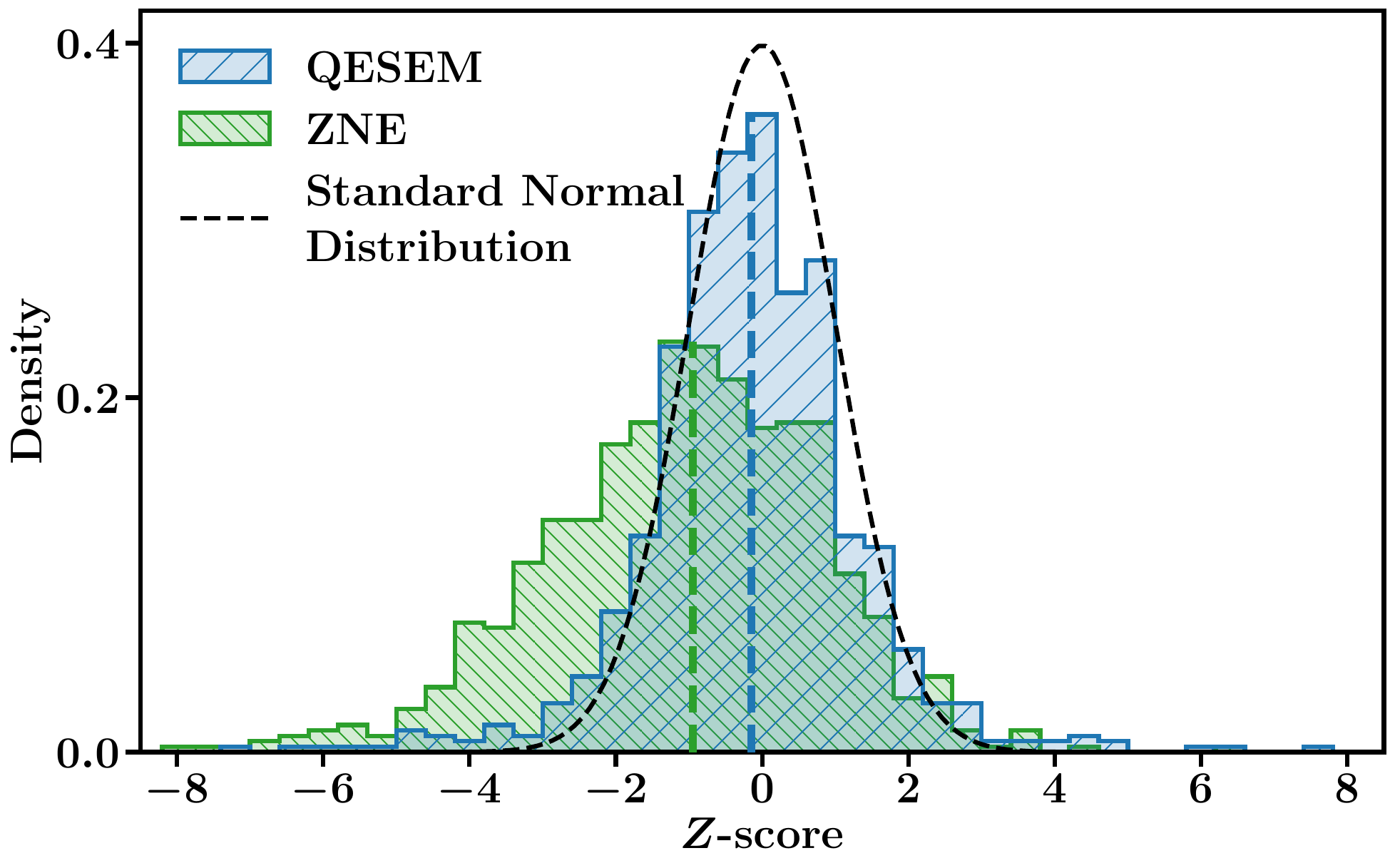}}
\caption{Unbiased mitigation of utility-scale Hamiltonian simulations with QESEM: 
\textbf{\protect\subref{fig:ising_circ} Kicked Ising circuit} -- The circuit includes three distinct layers of fractional $R_{ZZ}$ two-qubit gates, sandwiched between $R_X$ and $R_Z$ single-qubit gate layers.
\textbf{\protect\subref{fig:ising_geometry} Device geometry} -- The algorithm ran on a 103-qubit graph embedded in \texttt{ibm\_marrakesh}. The geometry was selected to have the minimum possible infidelity of the two-qubit gates.
\textbf{\protect\subref{fig:ising_lc} Active volume} -- The significance of different qubits for the mitigation of $\langle Z\rangle $ on qubit 47 at steps 0, 3, 5, and 7: In transparent blue -- qubits excluded from the experiment geometry. In solid blue -- qubits outside the observable's lightcone; their errors do not affect its expectation value. In orange -- qubits that are naively inside the lightcone, but their errors have a negligible effect on the observable (inside the commutativity-lightcone and outside the active-lightcone, see App. \ref{appendix:active-volume}). These are not mitigated by QESEM. In red -- qubits inside the lightcone, whose errors affect the expectation value of the target observable and are thus mitigated by QESEM. The distinction QESEM offers between red and orange qubits facilitates a significant reduction in mitigation overhead, leading to a reduction in the QPU time required to reach the desired statistical uncertainty. 
\textbf{\protect\subref{fig:ising_steps} Mitigation results} -- QESEM provides unbiased mitigation, recovering the ideal magnetization, Eq. \eqref{eq:def_magnetization} (computed via the PEPS-BP method). In contrast, ZNE fails to recover the ideal, showing a clear bias (see App. \ref{sec:zne} for comparison of different ZNE methods and further discussion). 
\textbf{\protect\subref{fig:ising_z_hist} Statistical consistency} -- Densities of $Z$-scores (see Eq. \eqref{eq:zscore}) across all single-qubit $\langle Z \rangle$ (822 total) for QESEM and ZNE. The vertical dashed lines denote the median $Z$-score. QESEM results follow the standard normal distribution as depicted by the black dashed line, validating unbiased mitigation. On the other hand, ZNE results are too wide and left-centered, consistent with systematic under-mitigation. One outlier for ZNE and one outlier for QESEM were removed for clarity.
\label{fig:ising_results}}
\end{figure*}

We demonstrate QESEM in several experimental settings, achieving high-accuracy mitigation for circuits where standard methods fail to produce reliable results.  We apply QESEM to a 100-qubit circuit that is challenging for both classical simulation and heuristic mitigation methods. In particular, we simulate a kicked Ising model on an IBM Heron device, reaching a total of 832 fractional entangling gates (or 1664 CZ gates in a standard Clifford compilation). We tune the circuit parameters to generic single-qubit and two-qubit gate angles featuring high lightcone velocity, reaching an active volume of 301 fractional entangling gates (see Fig.~\ref{fig:ising_results} and Sec.~\ref{main_demo}). 

QESEM maintains its high accuracy for all quantum circuits. To demonstrate its generality, we further benchmark it on a molecular variational quantum eigensolver (VQE) circuit with a generic, non-periodic circuit structure and a target observable requiring mitigation across 65 distinct measurement bases. Leveraging QESEM's automatic patch parallelization, we distribute the workload across 9 QPU regions and achieve a $\times 5$ speedup compared to mitigating a single patch (see Sec.~\ref{vqe_demo} and Fig.~\ref{fig:vqe_res}). 

QESEM can be applied to a variety of hardware platforms. The adaptations required to implement QESEM on new devices are mostly confined to the characterization modules and involve optimizing the corresponding error models. In this work, we benchmark QESEM on IBM quantum computers of the Heron (Secs.~\ref{main_demo} and~\ref{vqe_demo}) and Eagle generations, as well as on the IonQ Aria trapped-ion device (App.~\ref{other_hardware}).

We compare our results with the most widely used EM method: zero-noise extrapolation (ZNE) \cite{Temme2017, Endo_Benjamin_Ying, Kandala2019, LaRose2022, Kim2023utility}. ZNE estimates the ideal observable value by executing circuit variants with amplified noise and extrapolating the result to the zero-noise limit. Together with Clifford data regression \cite{Czarnik2020, LaRose2022}, designed for near-Clifford circuits, it belongs to a family of heuristic mitigation methods, which generally lack accuracy guarantees. Many ZNE variants exist, differing in their noise amplification and extrapolation strategies. Using experiments on IBM Heron devices and simulations, we show that QESEM consistently achieves significantly higher accuracy for the benchmarks presented here across a broad range of ZNE approaches (see Figs.~\ref{fig:ising_results} and~\ref{fig:vqe_res}). We expect this advantage over existing ZNE approaches to persist across a variety of important use cases.

In contrast to ZNE, QP-based approaches such as QESEM can produce unbiased expectation values: in the limit of infinitely many shots, the results converge to the exact value. The first and most widely studied unbiased QP-based method 
is Probabilistic (or Pauli) error cancellation (PEC) \cite{Temme2017, Endo2018}. However, standard PEC incurs a prohibitive runtime overhead, and has thus been demonstrated experimentally only on relatively small-scale circuits \cite{VandenBerg2023, Ferracin2024}. QESEM incorporates key elements that boost the efficiency of QP-based unbiased mitigation, enabling it to scale to large circuits while maintaining reliability and universal applicability. It also offers tunable parameters that can be set to guarantee unbiased results, or relaxed to reduce QPU time, introducing controlled biases that have been tested experimentally across a wide range of circuits to be negligible compared to statistical errors. 
Together with its advanced characterization protocols, QESEM provides a full EM solution yielding high accuracy at scale. 

We demonstrate several methods to accurately estimate QESEM's runtime in advance. Obtaining such estimates is crucial for correctly planning executions with EM, and for designing quantum algorithms that employ EM. As we explain in Sec.~\ref{sec:QA}, the runtime of QESEM depends on the properties of the quantum circuit, the observables whose expectation value is computed, the required accuracy, and the underlying hardware. We show that QESEM exhibits predictable runtime scaling that can be estimated either by analytical methods (using effective parameters for the specific circuit properties) or by using dedicated executions of QESEM. These dedicated time-estimation executions require only short QPU time, and can provide highly accurate estimates of QESEM's runtime on specific circuits, observables, and hardware.

Accurately estimating QESEM's runtime allows us to make predictions for the active volumes 
enabled using QESEM as a function of the hardware parameters and the quantum circuit properties. 
It is interesting to consider those projections in the context of the major goal of achieving verifiable quantum algorithmic advantage on near-term devices. As recently discussed in Refs.~\cite{Zimboras2025, YEMpaper, Lanes2025FrameworkQuantumAdvantage, Kechedzhi2024EffectiveQuantumVolume}, EM techniques are expected to be the first to achieve this goal, with one of the likely candidates being quantum simulation, such as in the 100-qubit experiment we report on here (Fig.~\ref{fig:ising_results}). As the scale of the active volumes approaches that of quantum advantage and becomes challenging for classical simulations, verifiability of the results becomes a crucial component. Control of the bias in the EM, as is done in QESEM, is required to produce high-accuracy verifiable results.
We note that our estimates of QESEM runtime can be used
as a guide for the hardware specifications (in particular, 2-qubit gate fidelity) and the properties of the quantum circuits (specifically, light cone velocity), which are best suited for demonstrating verifiable quantum advantage.

\section{QESEM Overview \label{Sec:workflow}}

QESEM implements a workflow designed for high-accuracy mitigation of generic quantum circuits. It has been validated across diverse hardware platforms, including utility-scale experiments on IBM Eagle and Heron devices and the largest mitigated VQE circuit executed on an IonQ device (see App.~\ref{other_hardware}). This workflow is composed of multiple stages (Fig.~\ref{fig:QESEM_workflow}) incorporating a set of algorithmic innovations (Fig.~\ref{fig:QESEM_innovations}). The stages are as follows:

\begin{enumerate}
    \item \textbf{Device characterization:} QESEM begins by executing on the QPU a protocol for rapid characterization of the whole device, intended for collecting data which would enable minimization of the error mitigation runtime. The protocol identifies suppressible coherent errors and maps the fidelities of all available device operations. By supplying real-time characterization data to the error suppression and noise-aware transpilation modules, this stage ensures that mitigation leverages the highest-fidelity available operations. 
    
    \item \textbf{Noise-aware transpilation:} QESEM constructs a family of optional transpilations of the input circuit. Each of these different transpilations specifies a physical qubit mapping and a sequence of operations chosen from the available operations in the QPU. For instance, on Heron devices, transpilations may construct Clifford-only, fractional-only, or mixed Clifford-fractional circuits. For each transpilation, QESEM estimates the total characterization and mitigation QPU time using data from the device characterization together with a procedure for \emph{active volume identification} (see App.~\ref{appendix:active-volume}). These are used to estimate the amount of noise affecting the target observable and predict the required QPU time. QESEM then selects the transpilation that minimizes the estimated QPU runtime. QESEM's noise-aware transpilation can yield dramatic savings in QPU time of up to x1000 (see App.~\ref{app:transpilation}).
    
    QESEM's transpilation also performs automatic measurement basis optimization and parallelized execution for mitigation. When target observables contain non-commuting terms, it generates candidate measurement basis sets using several heuristic algorithms and selects the set that minimizes the number of distinct bases. For 'narrow' circuits utilizing a small subset of the device's qubits, the transpilation variants constructed by QESEM include multiple parallelization options, where the circuit is mapped to different qubit subsets and executed on them simultaneously. Parallelization increases the data collection rate but may lower gate fidelities; QESEM chooses the number of parallel executions that best balances these effects. Parallelization can yield significant speedups in the mitigation; for instance, in Sec.~\ref{vqe_demo}, we demonstrate a x5 speedup in a typical VQE circuit.

    \item \textbf{Error suppression:} QESEM reconfigures the native device operations to optimize for fidelity and predictability in error-mitigated execution. We redefine native gates to reverse coherent errors identified in the preliminary characterization, such as coherent errors in entangling gates. For example, these calibrations reduce the over-rotation of \(R_{ZZ}\) gates on Heron devices by a typical factor of $\times10$ (see App.~\ref{app:calibration}). QESEM applies Pauli twirling to any remaining errors \cite{Ware2021}, and structures the gates in the circuit in layers consisting of alternating parallel applications of entangling and single-qubit gates, merging consecutive single-qubit operations. Depending on the hardware platform, this stage may also include dynamical decoupling of idle qubits and pulse-level calibration and optimization of gates, including operations that are not natively available (see Appendix \ref{other_hardware}). 
    
    \item \textbf{Circuit characterization:} 
    At this stage, QESEM comprehensively characterizes the remaining errors affecting the operations of the chosen transpilation for the circuit. Based on prior diagnostic experiments performed on the QPU, we construct a local Pauli error model for the entangling layers and state preparation and measurement (SPAM) operations; on Heron devices, the model contains up to two-local Pauli error rates. We construct and execute a set of characterization circuits tailored to the transpiled circuit, sensitive to all model parameters, while minimizing the QPU time required to achieve the desired mitigation accuracy. We then self-consistently optimize all model parameters to fit local observables in the characterization circuits, minimizing the difference between measurements and model predictions (see App.~\ref{Appendix: characterization}).
    
    \item \textbf{Error mitigation:} This stage constitutes the bulk of the QESEM execution. To eliminate remaining gate and SPAM errors in the transpiled circuit, QESEM constructs quasi-probabilistic (QP) decompositions for the noisy operations comprising the circuit, based on the data from the circuit characterization stage. QESEM utilizes ``multi-type'' QP decompositions that efficiently mitigate both Clifford and non-Clifford noisy operations. 
    
    QESEM then executes a sequence of interleaved characterization and mitigation circuits, while post-processing the resulting measurement data continuously. The mitigation circuits are sampled from the multi-type QP distribution, and their measurement data is used to compute the expectation values and empirical variances of the target observables. The processing of the data collected from the mitigation circuits employs \emph{active volume identification} to avoid sampling overhead due to unnecessary mitigation of noisy operations outside the ``causal lightcone'' of each observable. The interleaved characterization circuits yield updates to the current noise model. QESEM uses the collected data to update the QP distribution and to introduce ``retroactive'' corrections to mitigation data. The adaptive execution process outlined above both minimizes mitigation QPU time and ensures robustness to hardware drifts. QESEM's efficient use of reliable QP-based error mitigation enables it to reach high accuracies, at which error mitigation enhances accessible circuit volumes by 1000x or more \cite{YEMpaper}.
    
\end{enumerate}
    
 Within the error mitigation stage, QESEM performs an \textbf{empirical time estimation} to accurately determine the resources required to reach the requested statistical accuracy. This estimate relies on data from an initial batch of mitigation circuits, and is used to optimize resource allocation throughout the execution. The empirical time estimation can also be used as a standalone feature to obtain accurate time estimates for full error mitigation executions using a minimal amount of QPU time. This ability to perform accurate time estimation for error mitigation executions is important for planning circuit executions and guiding quantum algorithm design.

With the workflow and its methods established, we now demonstrate QESEM on two leading near-term quantum applications that require high-accuracy results.

\section{Hamiltonian Simulation Benchmark} \label{main_demo}

We demonstrate QESEM at scale by applying it to simulate the dynamics of a canonical quantum spin model: the tilted-field Ising (TFI) model,
\begin{align}
\mathcal{H} = J \sum_{\langle i,j \rangle} Z_i Z_j + g_x \sum_i X_i + g_z \sum_i Z_i,
\end{align}
where $\langle i,j \rangle$ denotes nearest neighbors on a lattice. Simulating the time evolution of many-body quantum systems is widely regarded as a computationally hard task for classical computers~\cite{feynman1982simulating,lloyd1996universal}. Quantum computers, by contrast, are naturally designed to perform Hamiltonian time evolution efficiently~\cite{lloyd1996universal,berry2007efficient}. Indeed, such Hamiltonian simulation is considered a promising candidate for early demonstrations of quantum advantage~\cite{babbush2015chemical,preskill2018quantum}. The TFI model, in particular, has become a popular benchmark on quantum hardware due to its rich physical behavior and hardware-friendly implementation~\cite{mi2022time, mi2022noise, chen2022error, Quantinuum2025, Kim2023utility}.

Rather than simulating continuous-time dynamics, we adopt the closely related \emph{kicked Ising} model~\cite{Prosen2000}, which is directly expressed as a quantum circuit. Specifically, we consider a spin system on the heavy-hex lattice under time-periodic dynamics corresponding to the sequential application of the TFI Hamiltonian terms. This hardware-native setting provides a natural platform for investigating diverse quantum many-body non-equilibrium phenomena such as Floquet many-body localization~\cite{Zhang2016}, Floquet-engineered topological phases~\cite{Google_FloquetKitaev_2025, Quantinuum_Floquet_2023}, entanglement spreading and many-body quantum chaos~\cite{Bertini2019}, and discrete time-crystalline phases~\cite{Shinjo2024}. The dynamics can be expressed \emph{exactly} as a periodic quantum circuit, where each evolution step consists of three layers of fractional two-qubit gates \(R_{ZZ}\left(\alpha_{ZZ}\right)\), interleaved with layers of single-qubit gates \(R_X\left(\alpha_X/3\right)\) and \(R_Z\left(\alpha_Z/3\right)\) (see Fig.~\ref{fig:ising_circ}).

We tuned the parameters of this circuit family to generic angles that are challenging for both classical simulation and error mitigation. Specifically, we chose \(\alpha_{ZZ}=1.0\), \(\alpha_{X}=1.6\), and \(\alpha_{Z}=0.3\), placing the model far from any integrable point. Large $\alpha_{ZZ}$ angles, interleaved layer-wise with non-commuting $R_X$ gates, drive rapid entanglement growth in each evolution step. An important property of a quantum circuit is the lightcone of an \emph{observable}, which is the set of gates in the circuit that affect it. The fast growth of entanglement results in a broad lightcone base already at low circuit depths, enlarging the number of qubits that determine local expectation values.

Using efficient device characterization (see App.~\ref{sec:qubit-selection}), we selected a 103-qubit, high-connectivity subgraph of the native heavy-hex geometry of \texttt{ibm\_marrakesh} (see Fig.~\ref{fig:ising_geometry}), on which we performed the simulation. The geometry was selected such that the constituting bonds have the highest available two-qubit gate fidelity, averaging approximately $99.4\%$ per $R_{ZZ}$ gate. On the selected subgraph, we simulated the evolution of a “ferromagnetic” state $\left|0\right\rangle^{\otimes N}$ under the kicked-Ising dynamics to steps 1 through 8.\footnote{At step 8, we had to remove three qubits from the algorithm due to hardware deterioration.} At the final step, the circuits have a maximal \emph{active volume}—the number of gates in the lightcone that have a significant effect on the observable (see App.~\ref{appendix:active-volume} and Fig.~\ref{fig:ising_lc})—of 301 gates for a local $\langle Z\rangle$ observable, with 54 qubits at the lightcone base.

Such a large number of active qubits places these circuits beyond the limitations of state-vector simulations and in a regime where controlled tensor network (TN) simulation methods become challenging~\cite{YEMpaper}. Furthermore, the far-from-Clifford angles chosen for all gates increase the difficulty for simulation with Pauli propagation methods~\cite{Chan2023, Aharonov2023, Holmes2024}. Finally, this far-from-Clifford regime is also challenging for error-mitigation methods such as Clifford data regression, which works well for near-Clifford circuits~\cite{Czarnik2020, LaRose2022}.

While we aim for a parameter regime that is beyond the reach of brute-force simulation and as challenging as possible for controlled approximate simulation methods, we are still able to simulate the ideal values of local observables in our circuits using recently introduced methods for approximate contraction and compression of tensor networks using the belief-propagation (BP) method~\cite{Begusic2024, Tindall2024, Arad2021}. The BP-based approaches are heuristic and have no rigorous error bounds, but they have been shown to work effectively for Hamiltonian simulation circuits on the heavy-hex lattice in certain instances. Specifically, we simulated the magnetization via the BP approach for evolving a PEPS tensor network introduced in Ref.~\cite{Begusic2024} (which we refer to as PEPS-BP), using the tensor-network Python package \textit{quimb}~\cite{gray2018quimb} (see App.~\ref{app: simulation} for details).

As the kicked-Ising circuits are very challenging for error mitigation, QESEM offers a significant advantage over other QP-based error-mitigation methods such as PEC. A core advantage of QESEM comes from two mechanisms that dramatically reduce the mitigation volume. First, QESEM applies an active-volume identification algorithm (described in App.~\ref{appendix:active-volume}) to determine which gates affect the measured observable and avoid unnecessary mitigation of errors that have a negligible effect. For example, at step 8, naively mitigating all the errors in the \emph{commutativity lightcone} (see App.~\ref{appendix:active-volume} and Fig.~\ref{fig:ising_lc}) would increase the mitigation volume from 301 to 590 two-qubit gates. Second, QESEM uses calibrated (see App.~\ref{app:calibration}) \emph{non-Clifford} two-qubit gates in characterization and mitigation (see App.~\ref{sec:char_frac} and App.~\ref{app:QP}). This allows for the natural compilation of the circuits to fractional $R_{ZZ}$ gates, further halving the mitigation volume. Standard implementations of PEC would require to compile the circuit to Clifford entangling gates (such as $CZ$), doubling the volume requiring mitigation from 590 to 1180 two-qubit gates at step 8. We estimated (see Appendix ~\ref{app:pec}) the time required for PEC to mitigate the noise of step 8 to be larger than the age of the universe.

The results for the average magnetization,
\begin{equation}
\label{eq:def_magnetization}
M=\frac{1}{N}\sum_{i=1}^N Z_i,
\end{equation}
are presented in Fig.~\ref{fig:ising_steps}. The unmitigated (noisy) results decay significantly at high depths and no longer indicate the value of the ideal magnetization. Nonetheless, QESEM successfully mitigates the errors, reproducing the ideal values within statistical uncertainty using 6.2 hours of QPU time (see Sec.~\ref{sec:QA} for further discussion on QPU time usage). In comparison, ZNE generally achieves smaller statistical uncertainty but fails to reproduce the ideal magnetization, providing a biased estimator of the ideal result (for details of the ZNE experiment, see App.~\ref{sec:zne}).

We demonstrate the statistical consistency of QESEM via the distribution of $Z$-scores for all single-qubit local $\langle Z\rangle$ observables across all time steps (822 results in total). For a given observable, the $Z$-score is defined as
\begin{align}
\label{eq:zscore}
\textrm{$Z$-score}=\frac{\textrm{mitigated}-\textrm{ideal}}{\textrm{standard deviation}}~.
\end{align}
In the absence of systematic errors, the $Z$-score should follow a standard normal distribution. Indeed, Fig.~\ref{fig:ising_z_hist} shows that the $Z$-scores of QESEM results are mostly consistent with the expected distribution, depicted by a black dashed line. The ZNE $Z$-scores, however, do not follow a normal distribution, indicating the presence of bias.

Finally, we considered averages of higher-weight observables, up to local four-point correlators. The corresponding mitigation results are presented in App.~\ref{app:heavy_weight}. They demonstrate accurate mitigation with QESEM for all observables considered, reaching a maximal active volume of 338 \(R_{ZZ}\) gates.

\section{Variational Quantum Eigensolver Benchmark} \label{vqe_demo}

The \emph{Variational Quantum Eigensolver} (VQE) is an algorithmic framework for preparing approximate ground states of quantum systems on a quantum computer, particularly those arising in quantum chemistry and condensed matter physics. VQE operates by preparing a parameterized quantum state, known as a variational ansatz, on quantum hardware using a parameterized quantum circuit. The ansatz parameters are iteratively optimized using classical routines to minimize the expectation value of the system's Hamiltonian as estimated from measurements taken from the QPU at each iteration. This process yields a circuit that prepares an approximate variational ground state, enabling the measurement of its energy as well as other properties, such as linear response functions~\cite{Reinholdt2025} and hyperfine coupling constants~\cite{Jensen2025}. 

To estimate the energy for a given set of parameters, VQE measures the Pauli operators that constitute the system’s Hamiltonian, expressed as a weighted sum of such terms. Since many of these operators are not diagonal in the computational basis, a final basis-rotation layer is typically appended to the circuit to align each Pauli operator -- or groups of jointly measurable Pauli operators -- with the measurement axis. At each optimization step, the circuit is executed multiple times under different measurement settings to accumulate the statistics required for all terms. Once optimized, the circuit can be used to probe additional properties of the approximate ground state by measuring further observables.

For our benchmark, we apply VQE to the water molecule (\ce{H_2O}) at a stretched geometry (see App.~\ref{app:vqe} for details). The electronic structure is modeled at the (4,4)/6-31G level of theory, where four electrons in four spatial orbitals define the active space. This active space is mapped to an 8-qubit register using the Jordan–Wigner transformation. A parameterized VQE circuit is constructed using the \textsf{SlowQuant} software package~\cite{slowquant2025}, implementing an orthogonally optimized tiled Unitary Product State (oo-tUPS) ansatz in the one-layer configuration. We bypass the classical optimization loop by directly assigning a set of pre-optimized parameters to the circuit, yielding the optimized circuit used as input to QESEM.

\begin{figure}[!ht]
\centering
\subfloat[\label{fig:vqe_patches}]{
    \includegraphics[width=0.56\columnwidth]{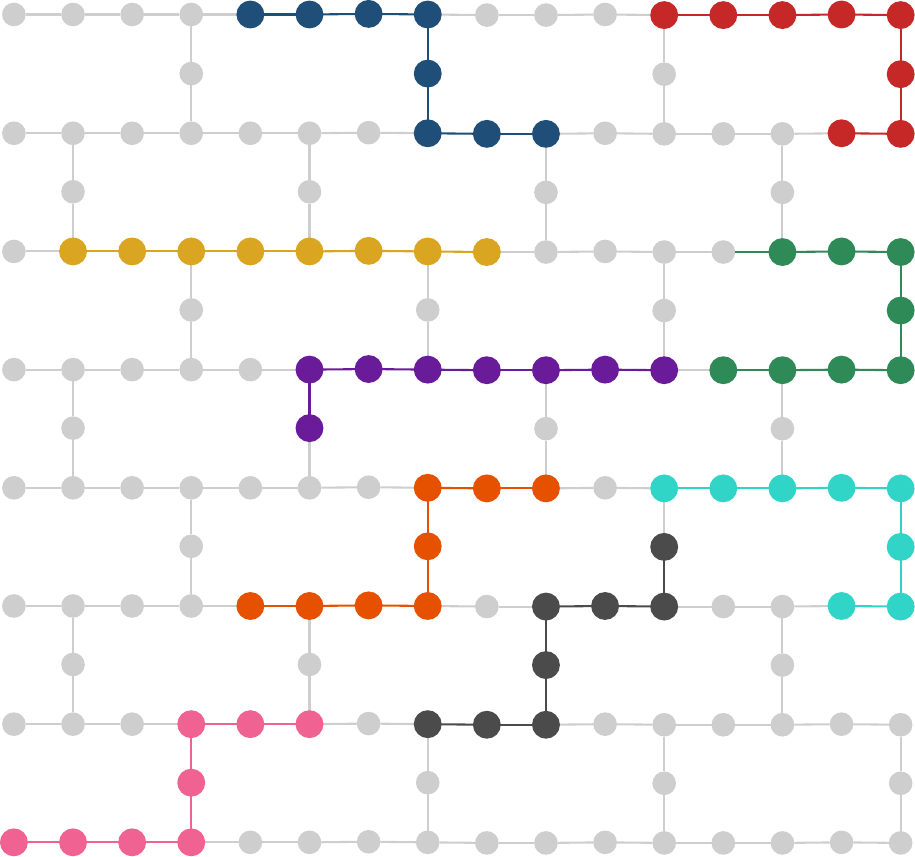}
    }
\\
\subfloat[\label{fig:vqe_res1}]{
    \includegraphics[width=0.95\columnwidth]{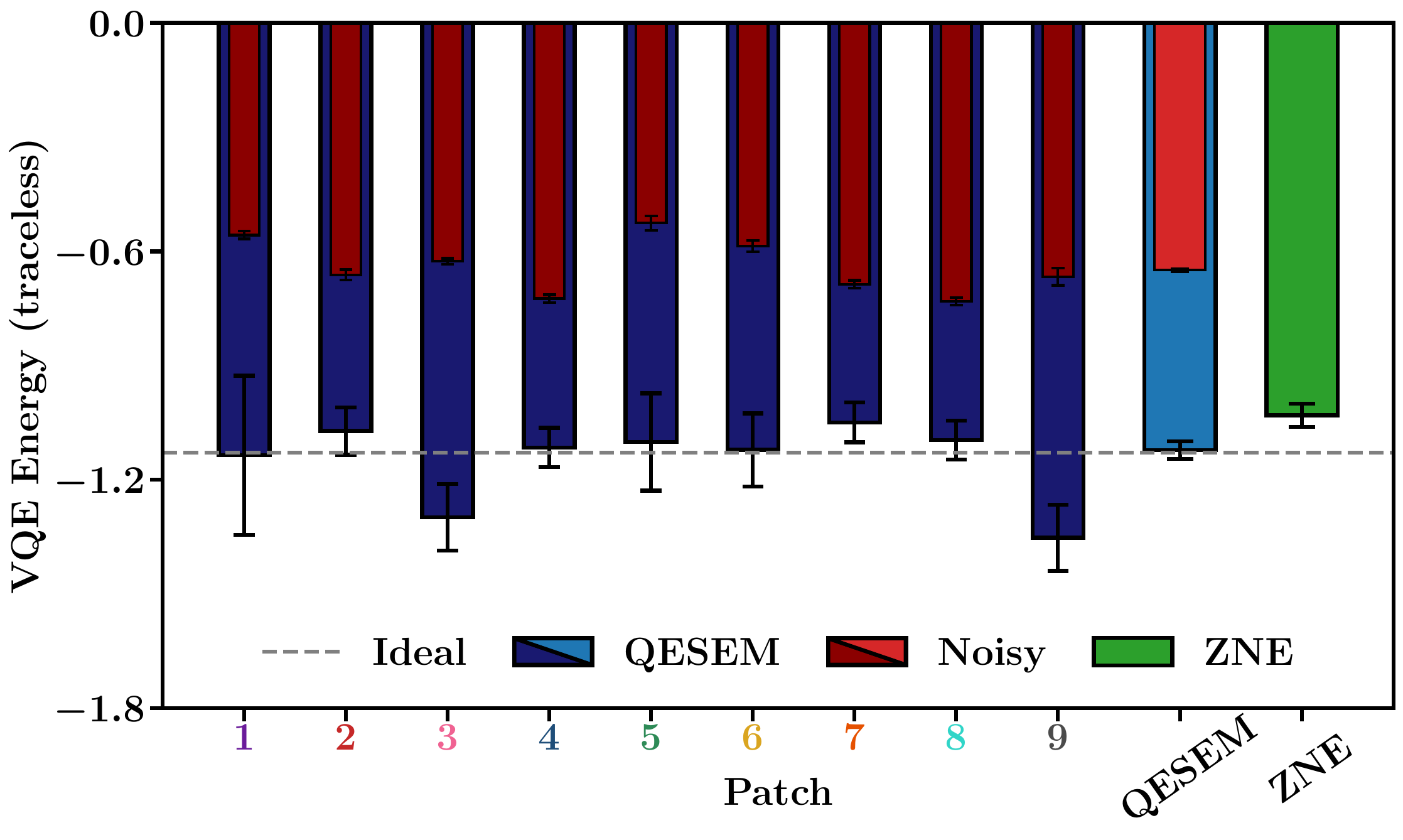}
}
\\
\subfloat[\label{fig:vqe_res2}]{
    \includegraphics[width=0.95\columnwidth]{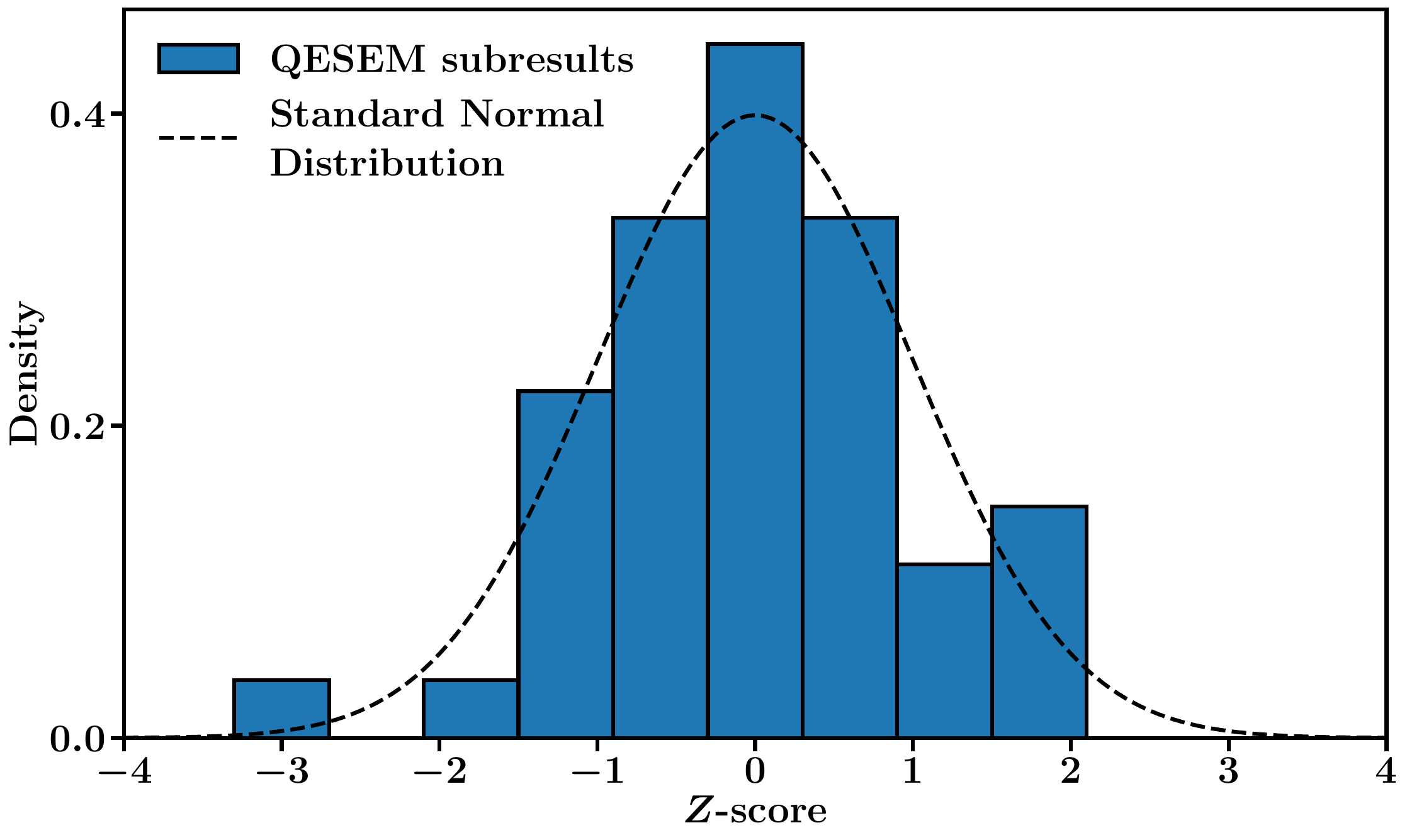}
}
\caption{The VQE benchmark results: 
\textbf{\protect\subref{fig:vqe_patches}} The nine patches used in the VQE benchmark, in different colors, on Marrakesh's connectivity graph.
\textbf{\protect\subref{fig:vqe_res1}} The variational ground state energy of \ce{H_2O} as obtained from the VQE circuit, for each qubit-patch and the final QESEM estimator given by the inverse-variance-weighted average. QESEM-mitigated values appear in blue, and noisy values in red. The ideal result is shown as a dashed gray line. The result of a ZNE experiment using a similar QPU time ($\sim$30 mins) is shown in green for comparison.
\textbf{\protect\subref{fig:vqe_res2}} Histogram of $Z$-scores [see Eq.~\eqref{eq:zscore}] of the 45 independent mitigation results building up the QESEM result. A black dashed curve shows the standard normal distribution.
\label{fig:vqe_res}
}
\end{figure}

We executed the QESEM workflow on IBM's \texttt{ibm\_marrakesh} device, using the optimized circuit and the system's Hamiltonian, $H$, as input. 
Rather than specifying a target precision, we constrained the total QPU usage time. In this operation mode, the QPU-time optimization algorithm (see App.~\ref{app:QPU time minimization}) distributes resources to minimize the statistical error bar on the mitigation result within the allotted QPU time.

QESEM transpilation arranged the input circuit into 63 layers of two-qubit gates, interleaved with single-qubit gate layers. Among these, 12 are unique, including layers consisting solely of CZ gates, layers containing only fractional-angle $R_{ZZ}$ gates, and hybrid layers combining fractional-angle $R_{ZZ}$ gates with ones at the Clifford angle ($\pi/2)$. The full list of unique layers is provided in App.~\ref{app:vqe}. 

The Hamiltonian's active volume in the transpiled circuit consists of all 120 two-qubit gates in the circuit, as well as 264 instances of single qubits idling during two-qubit gate layers. To extract the full set of Pauli observables in the Hamiltonian, QESEM constructed 65 measurement bases by grouping terms into sets of jointly measurable Pauli operators.

Because the input circuit has a narrow layout, i.e., utilizes relatively few qubits, we enabled QESEM's parallel execution feature (see App. \ref{app:transpilation} for details). QESEM created nine copies of the transpiled circuit, each mapped to a distinct 8-qubit patch, shown in Fig.~\ref{fig:vqe_patches} overlaid on Marrakesh's connectivity graph. The nine qubit-patches are also listed explicitly in App.~\ref{app:vqe}.

QESEM executed 7901 mitigation circuits across 5 mitigation batches (see App.~\ref{app:vqe} for details) and all measurement bases, totaling 121632 mitigation shots. The distribution of circuits per basis is shown in Fig.~\ref{fig:vqe_nc_ns_bases}. Overall, the entire QESEM workflow used 33 minutes of QPU time, of which approximately 10 minutes were devoted to characterizations.

Figure \ref{fig:vqe_res1} shows the QESEM-mitigated Hamiltonian expectation values for all nine qubit-patches compared to the ideal value, 
\begin{align}
    E_{0} - h_I =-1.129~,
\end{align}
depicted as a horizontal gray dashed line. Here, $h_I = 2^{-N_q}\text{tr}(H) = 80.728 \ldots$ is a constant contribution to the energy that does not depend on the state and thus requires no computation. The figure also displays the non-mitigated, or noisy, values for each qubit patch. The full QESEM-mitigated Hamiltonian expectation value for the experiment 
\begin{align}
\label{eq:vqe_mitigated}
    E_{\text{QESEM}} - h_I = -1.122 \pm 0.023~.
\end{align}
is obtained as the inverse-variance weighting of all patches. It also appears on the figure along with the inverse-variance weighted noisy (unmitigated) value, $-0.649 \pm 0.004$.

Finally, Fig.~\ref{fig:vqe_res2} shows the distribution of $Z$-scores for all 45 independent mitigation results (5 mitigation batches $\times$~9 qubit-patches). These closely follow a standard normal distribution, illustrated by the black dashed line, confirming the expected statistical behavior of the mitigation error bars.

The use of parallel execution improved the precision by more than a factor of two relative to the best-performing individual patch. Indeed, the QESEM-mitigated result of the \emph{best} patch [teal in Fig.~\ref{fig:vqe_patches}; index 8 in Fig.~\ref{fig:vqe_res1}] has an error bar of $\sim0.051$. Achieving the final QESEM estimator error bar of $0.023$ without parallelization would have required $\left(\frac{0.051}{0.023}\right)^2\approx 4.95$ times as many mitigation circuits, and therefore approximately five times as much QPU time.

\section{QESEM Runtime and Prospects for Quantum Advantage}
\label{sec:QA}

\begin{figure}[b]
\centering
\includegraphics[width=0.95\columnwidth]{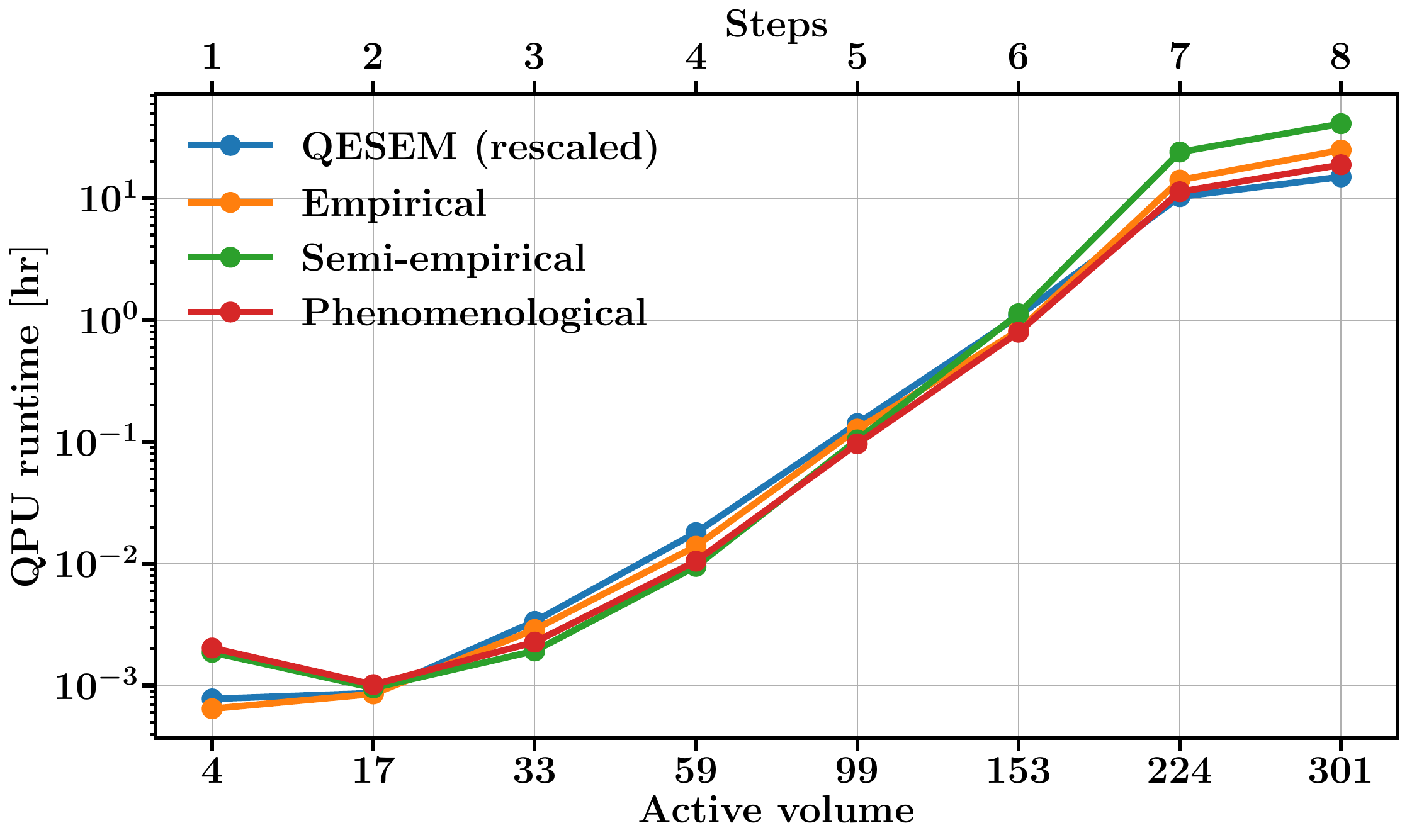}
\caption{
 The QPU runtime of QESEM in the generic kicked Ising benchmark, compared to various estimations, with active volume determined by Algorithm \ref{comm_lc_algo} with $\epsilon_{\text{LC}}=0.03$:
 QESEM's QPU runtime (blue) is extracted from IBM's reported workload usage~\cite{IBMJobRuntimeEstimation} and is rescaled to a precision $\epsilon=0.01$ in each step via $T_\text{QPU}\propto\epsilon^{-2}$, for ease of comparison to estimations;
 The empirical time estimate (orange) is based on a small sample of mitigation circuits, which run as part of every QESEM job (40 QPU seconds per step) as part of QESEM's QPU runtime optimization.
 The semi-empirical estimate (green) relies on execution of the circuit on the QPU without QESEM, and requires even less QPU time. 
 Estimate based on the phenomenological model with fitted parameter $r = V_\text{eff}/V_{\text{A}} \approx 1/4$ and ideal Pauli expectation values $\braketExp{P}^{(\text{ideal})}_{\mathcal{C}_0}$ obtained from PEPS-BP simulation.
}
\label{fig:QA}
\end{figure}

Error mitigation incurs a runtime overhead that is generally exponential in the accumulated circuit error. Hence, in order to plan experiments and quantum algorithms utilizing error mitigation, it is important to accurately estimate in advance the required QPU time $T_{\text{QPU}}$ for a given hardware, circuit, and set of observables. 

In this section, we present the different methods for estimating the QPU runtime of QESEM. To obtain the most quantitatively accurate estimates, we use methods that utilize data collected using short QPU access (requiring significantly less QPU time than full mitigation executions). We show that these methods are able to accurately predict the required QPU time for QESEM experiments, enabling in-advance resource estimation and planning of executions using QESEM. We also provide a phenomenological model for QESEM's QPU runtime where this empirical data is replaced by a small number of effective parameters, and show that it yields excellent predictions. This model allows us to provide reliable projections for accessible circuit volumes with QESEM for future hardware with improved gate fidelities and control electronics.

The discussion here is kept at an informal level; for full details, see App.~\ref{app:QPU time minimization}. We start by explaining the source of the time overhead in QESEM, which arises from a quasi-probability (QP) decomposition~\cite{Temme2017,Endo2018} of the ideal circuit into a linear combination of circuits executable on the noisy device: 
\begin{equation}
\mathcal{C}_0^{(\text{ideal})} = \sum_j c_j \mathcal{C}_j ,
\label{eq:QP decomp}
\end{equation}
where $c_j\in\mathbb{R}$ satisfy $\sum_j c_j = 1$, but may take negative values (hence the term quasi-probability). 
Given a QP decomposition \eqref{eq:QP decomp} and observable $O$ we can write $\langle O \rangle_{\mathcal{C}_0}^{(\text{ideal})} = \sum_j c_j \langle O \rangle_{\mathcal{C}_j}$, where throughout this section $\langle O \rangle_{\mathcal{C}_0}^{(\text{ideal})}$ is the expectation value of $O$ with respect to the original circuit without noise and $\langle O \rangle_{\mathcal{C}_j}$ is the expectation value of the observable in the circuit $\mathcal{C}_j$ executed on the actual device (with infinite shots and without error mitigation).
We estimate the ideal expectation value by sampling from the ensemble of QP circuits according to the probability $p_j = |c_j| / W$, where we define the \emph{QP norm} $W = \sum_j |c_j| \geq 1$, and our estimate for $\langle O\rangle_{\mathcal{C}_0}^{(\text{ideal})}$ is given by
\begin{equation} 
\hat{O}= \frac{W}{N_{\text{c}}} \sum_{i=1}^{N_c} \mathrm{sign}(c_i) \frac{1}{N_{\text{s}}}\sum_{s=1}^{N_{\text{s}}} o_{is},
\label{eq:QPestimator}
\end{equation}
where $o_{is}$ is the sampled eigenvalue\footnote{
    For simplicity, we consider measurement in the eigenvalue basis. For a generic observable, a decomposition into measurement bases would be more practical, as is used in the VQE benchmark in Sec.~\ref{vqe_demo}.
} of $O$ in shot $s$ of circuit $\mathcal{C}_i$, and the circuits $\mathcal{C}_i$ are sampled independently according to the QP distribution defined above. The multiplication by the normalization factor $W$ is what causes the increase of the variance of the estimator $\hat{O}$ and is the source of the runtime overhead (compared to sampling from an ideal QPU) in a QP-based EM such as the one used in QESEM. 

\begin{figure}[ht]
\centering
\subfloat[\label{subfig:time-est-theory-sc}]{
 \includegraphics[width=\linewidth]{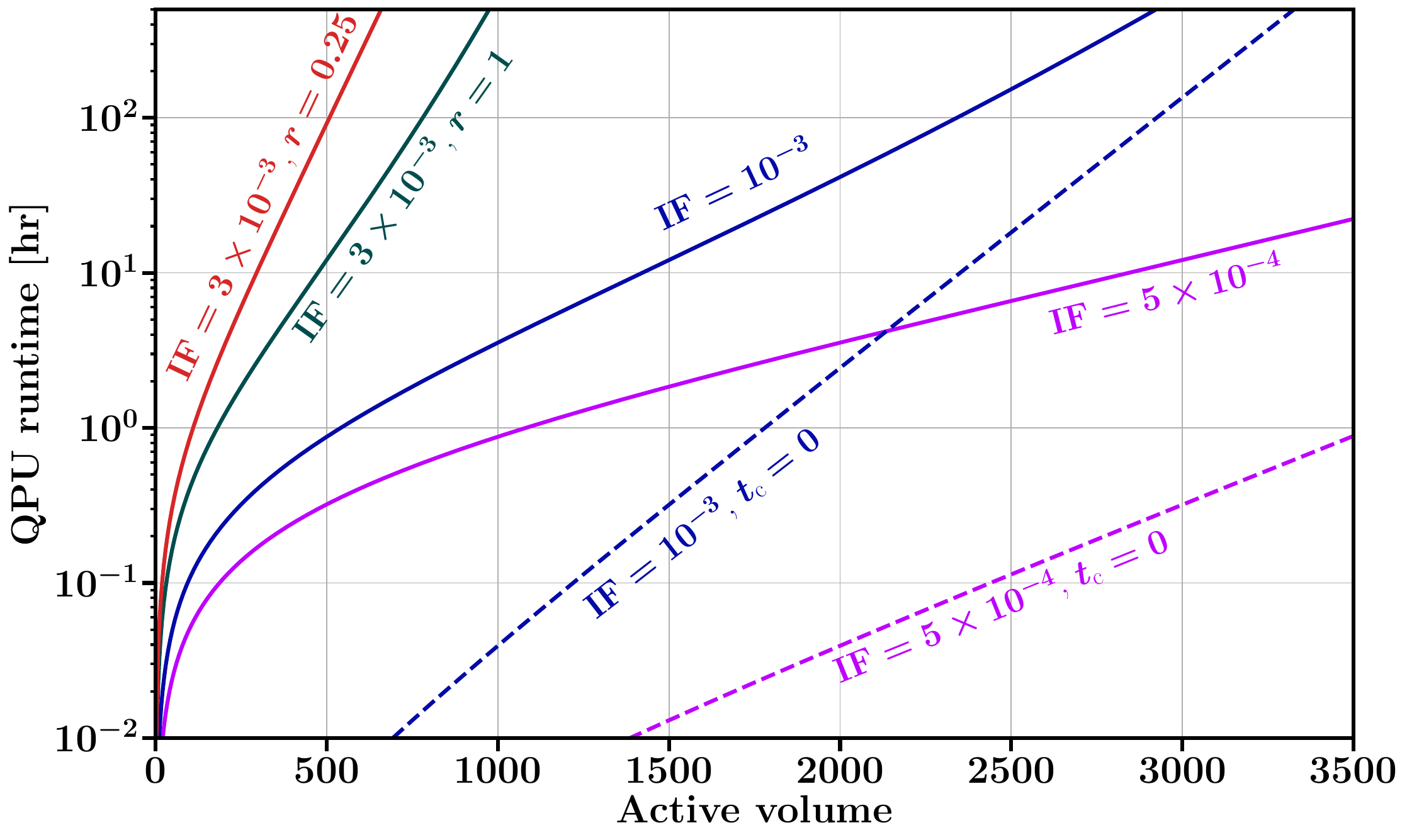}}
\\
\subfloat[\label{subfig:time-est-theory-ions}]{
 \includegraphics[width=\linewidth]{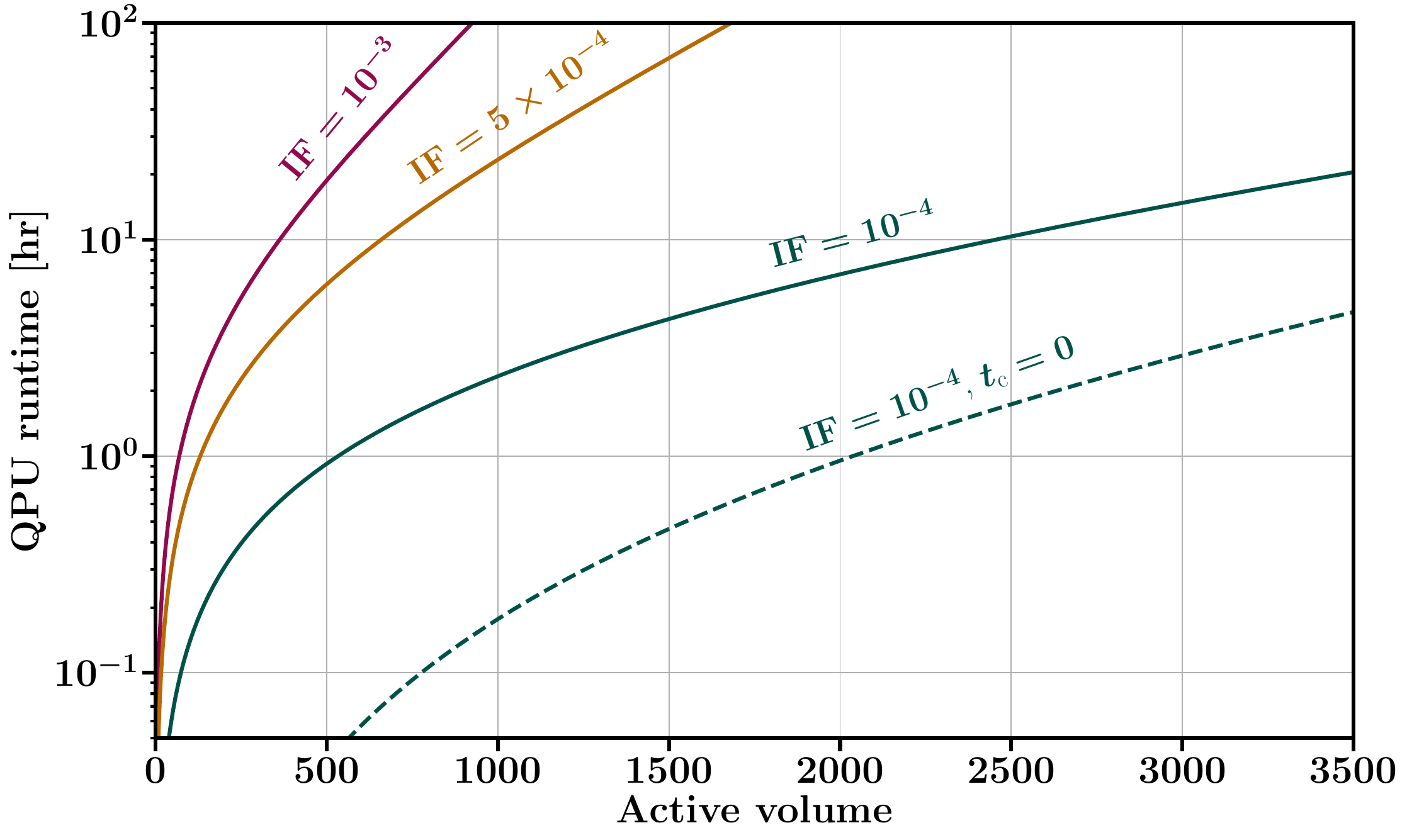}}
\caption{
 Analytical extrapolation of QPU runtimes of QESEM for larger active volumes and with varying hardware properties, using the phenomenological model \eqref{eq:phenomenological-vc-vs-pauli}. In the two plots, we use $\langle O\rangle ^{(\text{ideal})}_{\mathcal{C}_0}=1$, required accuracy $\epsilon=0.01$, and $r=1$ unless specified otherwise. The $t_\text{c}=0$ lines correspond to QPUs without any controller delays.
 \textbf{\protect\subref{subfig:time-est-theory-sc}} Projections for superconducting qubits, using timescales close to those found in the benchmarks presented in Secs.~\ref{main_demo} and \ref{vqe_demo} of $t_{\text{c}}=\qty{0.16}{sec}$ and $t_{\text{s}}=\qty{300}{\micro s}$. The infidelity $\IF=\num{3e-3}$ roughly matches the values found in the benchmarks presented in Secs.~\ref{main_demo} and \ref{vqe_demo}.
 \textbf{\protect\subref{subfig:time-est-theory-ions}} Projections for ion-based QPUs, with time scale of $t_\text{c}=\qty{2}{sec}$ and $t_\text{s}\approx \qty{3}{msec}+\qty{0.2}{msec}\times V_{\text{A}}$.
}
\label{fig:QA-future}
\end{figure}
Qualitatively, the QP norm in the QP decomposition used in QESEM behaves as $W \approx \exp(2 \IF \times V_{\text{A}})$ where $\IF=1-\operatorname{F}$ with $\operatorname{F}$ the gate fidelity, and the \emph{active volume} $V_{\text{A}}$ is the volume (number of gates) of the part of the circuit which affects the value $\langle O \rangle_{\mathcal{C}_0}^{(\text{ideal})}$ (see App.~\ref{appendix:active-volume} for details). 
More precisely, we note that $e^{2\IF\times V_\text{A}}$ is a lower bound of the QP norm of a generic QP decomposition and the QP norm of the EM used in QESEM is close to saturating this bound (see App.~\ref{app:QP} for details).
In quantum circuits with local connectivity, the active volume can be substantially smaller than the total circuit volume, and the EM overhead depends only on $V_{\text{A}}$ since we only have to mitigate noise on gates inside the active volume of $O$. Note that in practice, in most QPUs, the two-qubit gates are the dominant source of error, and therefore, the infidelity and the active volume refer to two-qubit gates.\footnote{The infidelity carried by idle gates and single-qubit gates can be accounted for by a finer approach, such as that discussed in App.~\ref{Appendix: characterization}.}

A naive upper bound on the overhead with QESEM can be seen directly from Eq.~\eqref{eq:QPestimator} and is given by $W^2$. This upper bound corresponds to a scaling of the QPU runtime
with the active volume as $T_{\text{QPU}} \sim \exp(4 \IF \times V_{\text{A}})$. This form for the scaling of $T_{\text{QPU}}$ turns out to be too pessimistic. Moreover, it overlooks the fact that in actual QPUs, two separate timescales contribute to $T_{\text{QPU}}$: (i) the ``circuit time'' $t_{\text{c}}$, which is the time needed to compile a new circuit into pulse instructions by the control electronics; and (ii) the ``shot time'' $t_{\text{s}}$, the time required to execute a given circuit once, including state preparation, application of the quantum gates, and measurement. Due to these timescales, we use a two-step sampling procedure in which we sample $N_{\text{c}}$ circuits and from each circuit we sample $N_{\text{s}}$ shots.
The total QPU time given by $T_{\text{QPU}} = N_{\text{c}} (t_{\text{c}} + N_{\text{s}} t_{\text{s}})$ is the time required to collect $N_{\text{s}}$ shots from each of the $N_{\text{c}}$ circuits. To make use of this behavior, it is useful to express the variance in the estimator $\hat{O}$ with $N_{\text{c}}$ circuits and $N_{\text{s}}$ shots each as
\begin{equation}
 \mathbb{V}[\hat{O}] = \frac{1}{N_{\text{c}}} \left(\mathbb{V}_{\text{c}} + \frac{\mathbb{V}_{\text{s}}}{N_{\text{s}}}\right), \label{eq:variance decomp}
\end{equation} 
where we defined the circuit-to-circuit variance 
\begin{equation}
\mathbb{V}_{\text{c}} = W^2 \operatorname*{\mathbb{V}}_{C \sim \text{QP}} [\text{sign}(C) \langle O \rangle_C]
\label{eq:circuittocircuit}
\end{equation}
and the shot-to-shot variance
\begin{equation}
\mathbb{V}_{\text{s}} = W^2 \operatorname*{\mathbb{E}}_{C \sim \text{QP}} [\langle O^2 \rangle_C - \langle O \rangle^2_C],
\label{eq:shottoshot}
\end{equation}
The variance in 
Eq.~(\ref{eq:circuittocircuit}) and the expectation value in Eq.~\eqref{eq:shottoshot} are taken with respect to the QP distribution of circuits.
The decomposition of the variance in Eq.~(\ref{eq:variance decomp}) originates from the two step sampling procedure, where $\mathbb{V}_{\text{c}}$ is the variance which originates from sampling different circuits while $\mathbb{V}_{\text{s}}$ originates from sampling shots for each circuit. Using this decomposition of the variance, we find (App.~\ref{app:QPU time minimization}) that the optimal QPU time for a given statistical precision $\epsilon$ satisfies
\begin{equation}
\epsilon^2 T_{\text{QPU}} = \left(\sqrt{\mathbb{V}_{\text{c}}t_{\text{c}}} + \sqrt{\mathbb{V}_{\text{s}}t_{\text{s}}}\right)^2. \label{eq:opt_t_qpu}
\end{equation}

From Eq.~\eqref{eq:opt_t_qpu} we see that knowledge of $\mathbb{V}_{\text{c}}, \mathbb{V}_{\text{s}}$ for a given circuit and observable will allow us to determine the required QPU time for mitigation. In general, it is not possible to compute those quantities exactly in advance without doing a full classical simulation of multiple circuits from the QP ensemble in the presence of noise. In the following, we present several methods for estimating these variances.

The most accurate estimate for $T_\text{QPU}$ is obtained by estimating the variances $\mathbb{V}_{\text{c}}$ and $\mathbb{V}_{\text{s}}$ empirically from a small sample of circuits sampled from the QP 
distribution and executed on the QPU.
This is the approach taken when running QESEM's \emph{empirical time estimation}, and it is also
used during QESEM's mitigation workflow to set the optimal number of circuits and shots for achieving the statistical
accuracy requested by the user (for the full algorithm see App.~\ref{app:QPU time minimization}).
This empirical estimation relies on a small sample size, relative to the sample size required for full mitigation of large-scale circuits to high precision.
We then estimate $\mathbb{V}_{\text{s}}$ and $\mathbb{V}[\hat{O}]$ directly from the sample and extract
$\mathbb{V}_{\text{c}}$ according to Eq.~\eqref{eq:variance decomp}.

While the empirical time estimation procedure requires only a modest amount of QPU time, we now present an approximate estimate that requires even less QPU time and gives insight into the behavior of $T_\text{QPU}$.
The method relies on the approximation $\mathbb{E}_{C\sim \text{QP}}\braketExp{O}_C^2 \lesssim \braketExp{O}_{\mathcal{C}_0}^2$, which essentially assumes that the sampled circuit perturbations reduce the expectation value relative to the unmitigated circuit.
For the case when the observable $O$ is a single Pauli string $P$ this leads to the approximation
\begin{equation}\label{eq:semi-empirical-vc-vs-pauli-main}
 \mathbb{V}_\text{c} \approx W^2 \langle P\rangle_{\mathcal{C}_0}^2 -(\langle P\rangle_{\mathcal{C}_0}^{(\text{ideal})})^2, \quad \mathbb{V}_\text{s} \approx W^2(1-\langle P\rangle_{\mathcal{C}_0}^2).
\end{equation}
Thus, for large circuit volumes where $|\langle P\rangle_{\mathcal{C}_0}^{(\text{ideal})} |\ll W |\langle P \rangle_{\mathcal{C}_0}|$, we can neglect the ideal value in the approximation for $\mathbb{V}_\text{c}$ in Eq.~\eqref{eq:semi-empirical-vc-vs-pauli-main}, and obtain an estimate for the optimal QPU runtime that is based on $W$ and $\langle P\rangle_{\mathcal{C}_0}$ alone, which we call \emph{semi-empirical estimation}. This suggests a protocol for EM resource estimation by measuring the expectation value $\langle P \rangle_{\mathcal{C}_0}$ on the QPU without error mitigation, which requires a very short amount of QPU time.

The above semi-empirical estimate allows us to write a phenomenological model for $T_\text{QPU}$. We relate $\langle P \rangle_{\mathcal{C}_0}$ to the ideal value as $\braketExp{P}_{\mathcal{C}_0}=e^{-\IF \times V_{\text{eff}}} \langle P \rangle_{\mathcal{C}_0}^{(\text{ideal})}$ where $V_\text{eff}$ is termed the \emph{effective volume} of observable $O$ on circuit $\mathcal{C}_0$, quantifying its sensitivity to circuit noise. 
Writing $V_{\text{eff}} = r V_{\text{A}}$ defines $r$ as a factor capturing sensitivity to errors in the active volume; in many practical regimes $r<1$.
We can thus express the variances as
\begin{equation}\label{eq:phenomenological-vc-vs-pauli}
\begin{aligned}
 \mathbb{V}_\text{c} &\approx e^{2(2-r) \IF V_{\text{A}}} (\langle P \rangle_{\mathcal{C}_0}^{(\text{ideal})})^2, \\ 
 \mathbb{V}_\text{s} &\approx e^{4 \IF V_{\text{A}}} - e^{2(2-r) \IF V_{\text{A}}} (\langle P \rangle_{\mathcal{C}_0}^{(\text{ideal})})^2.
\end{aligned}
\end{equation}
These expressions for $\mathbb{V}_{\text{c}}$ and $\mathbb{V}_{\text{s}}$ explain a mechanism which leads to a smaller runtime overhead 
compared to the naive pessimistic estimate, exhibiting an exponential growth of $\mathbb{V}_{\text{c}}$ with an exponent coefficient of $2(2-r)$, which is smaller than the naive value of $4$.
In many cases $r$ can be estimated (for example by extrapolation from a small number of Trotter steps in Hamiltonian simulation circuits), and Eq.~\eqref{eq:phenomenological-vc-vs-pauli} can be used for time estimation, with a bound $(\langle P \rangle^{(\text{ideal})}_{\mathcal{C}_0})^2 \sim \mathcal{O}(1)$.

In Fig.~\ref{fig:QA} we demonstrate how the different time estimation methods work for 
the Hamiltonian simulation experiment described in Sec.~\ref{main_demo}, where we measure the average magnetization $M=\sum_{i=1}^{N_{\text{qubits}}} Z_i / N_{\text{qubits}}$. We compare the estimations to the QPU time used
in practice in the experiment where we normalize the actual runtime for each number of Trotter steps according to $T_\text{QPU}^{\text{normalized}} = T_\text{QPU} \times \epsilon^2(n_{\text{steps}}) / 0.01^2$, where $\epsilon(n_{\text{steps}})$ is the empirically estimated error-bar of the mitigated expectation value at the Trotter step $n_\text{steps}$. That is, we normalize the QPU runtime to that required to achieve statistical precision of $0.01$ at each step.
We can see that the empirical time estimate provides accurate predictions for the experiment QPU runtime.

To use the semi-empirical estimation and the phenomenological model for the non-Pauli observable $M$, we use the approximation $\mathbb{V}_{\text{c}/\text{s}}[M] \approx \alpha \sum_i \mathbb{V}_{\text{c}/\text{s}}[Z_i] / N_{\text{qubits}}^2$. The scale factor $\alpha=\mathbb{V}_{\mathcal{C}_0}[M]/(N_{\text{qubits}}^{-2}\sum_i \mathbb{V}_{\mathcal{C}_0}[Z_i])$ is chosen to take into account some of the effect of the statistical dependence between $Z_i$ on different qubits, with $\mathbb{V}_{\mathcal{C}_0}[O] = \braketExp{O^2}_{\mathcal{C}_0} - \braketExp{O}^2_{\mathcal{C}_0}$  estimated on the QPU without error mitigation. This allows us to estimate $\mathbb{V}_{\text{c}/\text{s}}[Z_i]$ via the semi-empirical estimator or phenomenological model. In the phenomenological model we obtained $r$ as fitting parameter to the simulated ideal values of $\langle Z_i \rangle_{\mathcal{C}_0}^{(\text{ideal})}$, yielding a fitted value of $r\approx 1/4$, which provides good agreement with the actual QPU time. 
A detailed treatment of the semi-empirical and phenomenological estimations for a general observable given as a linear combination of Pauli operators is given in App.~\ref{app:QPU time minimization}.

The phenomenological model allows us to obtain predictions for the QPU runtime of QESEM on future QPUs with improved specifications. These are shown in Fig.~\ref{fig:QA-future} in which we use the phenomenological model Eq.~(\ref{eq:phenomenological-vc-vs-pauli}) assuming a single Pauli observable with $\langle P\rangle ^{\text{(ideal)}}_{\mathcal{C}_0}=1$ to provide resource estimation 
projections for different values of 
$r$ and $\IF$. We also show a projection of the QPU runtime of QESEM for a QPU with an ideal controller, for which $t_{\text{c}}=0$, showing the significant runtime gain that can be obtained from using improved control electronics.

The predictions for $T_{\text{QPU}}$ enable making projections for demonstrating verifiable quantum advantage using QESEM \cite{YEMpaper}. First, we note that circuits exhibiting fast lightcone growth, such as the circuit used in Sec.~\ref{main_demo}, are good candidates for demonstrating quantum advantage. In these circuits, the large number of qubits at the base of the light-cone pose a challenge for many simulation methods, see Fig.~2 of Ref.~\cite{YEMpaper}. In our experiment, the active lightcone base already involves 54 qubits, beyond the capacity of standard statevector simulation. 

Moreover, recent studies of Hamiltonian simulation circuits on 2D square lattices suggest that state-of-the-art tensor-network (TN) simulation methods begin to fail on circuits with an active volume of $\sim 1000$ two-qubit gates \cite{Quantinuum2025,rudolph2025}, which, according to our projections in Fig.~\ref{fig:QA-future}, should be within reach of QESEM with near-term hardware fidelities. In particular, Ref.~\cite{Quantinuum2025} shows a discrepancy between the classical TN simulation results and results obtained on a trapped-ion QPU using ZNE for a circuit simulating the transverse-field Ising model, with $V_{\text{A}} \sim 1000$ fractional-angle $R_{ZZ}$ gates. Notably, since ZNE is an uncontrolled heuristic EM method, it is not clear whether the estimates produced by classical simulation, ZNE, or neither accurately reproduce the ideal values. This exemplifies the necessity of verifiable EM methods such as QESEM, in which the bias can be controlled.

\section{Conclusions \label{Sec:conclusions}}

We have introduced QESEM: an efficient, reliable, and high-accuracy quantum error mitigation workflow that bridges a critical gap in the current mitigation landscape. On one end of the spectrum lie widely used heuristic methods that lack accuracy guarantees, while on the other, standard quasi-probabilistic approaches offer theoretical rigor but are limited by prohibitive runtime overheads. QESEM balances reliability and efficiency by introducing algorithmic innovations that boost the efficiency of QP-based mitigation; offering tunable parameters that allow trading minimal, controlled bias for substantial efficiency gains; and incorporating error suppression protocols and noise-aware transpilation to reduce the physical errors before mitigation is applied. Backed by high-precision characterization protocols and a drift-resilient workflow, QESEM enables reliably high-accuracy mitigation for high-volume circuits.

Beyond the results shown here on IBM Heron processors, QESEM has been demonstrated on diverse hardware platforms, including fixed-frequency superconducting devices as well as trapped ions. These demonstrations span various use cases, including Hamiltonian simulation and VQE, underscoring its utility as a general-purpose error mitigation tool.

By boosting accessible circuit volumes and delivering verifiable high-accuracy results at predictable runtimes, error mitigation methods such as QESEM chart a reliable timeline towards quantum advantage. This timeline can be further accelerated by combining error-mitigated quantum hardware with classical high-performance computing. For example, observable backpropagation through circuit fragments has been shown to accelerate mitigation and boost accessible circuit volumes, as demonstrated experimentally using both Pauli Propagation techniques ~\cite{fuller2025improvedquantumcomputationusing} and tensor network methods ~\cite{HPCQESEM} for the classical backpropagation.

As quantum hardware progresses towards the era of early fault tolerance, it will continue to play a key role by addressing the logical errors that remain after the application of error correction. The building blocks and workflow of QESEM, developed here for mitigating physical errors, can be naturally extended to logical errors. In fact, novel combinations of error mitigation and error correction can exploit the structures of error-correcting codes and leverage information gained by syndrome measurements to further enhance QESEM's performance at the logical level \cite{YEMpaper, SALEMpaper, qedma_logical_errors_2024}.

\acknowledgments We thank Abhinav Kandala, Alireza Seif, Minh Tran, Ewout van den Berg, and Oles Shtanko for fruitful discussions on classically-challenging circuits for heavy-hex devices; Jay Gambetta, Blake Johnson, Stefan Elrington, Shesha Shayee Raghunathan, and the rest of the IBM team for their ongoing support; Christopher Monroe, Jungsang Kim, John Gamble, Matthew Keesan, Richard Moulds, Eric Kessler, Stefan Natu, Daniela Becker, and the rest of the IonQ and Amazon Braket teams for their support during the IonQ demonstrations; Nicc Lewis for assistance with visual design; and Karl Jansen's research group, Seiji Yunoki's research group, Masahito Yamazaki, Muqing Zheng, Hiroshi Yamauchi, and Erik Kjellgren for their insights and early feedback on QESEM. We acknowledge the use of IBM Quantum Credits via the IBM Quantum Startups Program for this work. The views expressed are those of the authors and do not reflect the official policy or position of IBM or the IBM Quantum Platform team.

\medskip
\noindent\textbf{Competing Interests}~--- This work describes methods that are the subject of U.S. and international patent applications filed by Qedma Quantum Computing Ltd., including Refs. \cite{qedma_qp_sampling_patent_app,qedma_multi_qp_patent,qedma_multi_qp_patent,qedma_characterization_calibration_2023,qedma_logical_errors_2024,qedma_efficient_lightcones_2025,qedma_drift_robust_2025,qedma_pauli_fractional_2025,qedma_clifford_char_2025}.

\clearpage

\clearpage
\appendix
\counterwithin{figure}{section}
\renewcommand{\thefigure}{\thesection.\arabic{figure}}

\startappendixtoc

\section*{Appendix}

This appendix details the core components and key innovations of QESEM, serving as a reference for users. Much of this work was completed earlier and documented in a series of patent applications \cite{qedma_qp_sampling_patent_app,qedma_multi_qp_patent,qedma_multi_qp_patent,qedma_characterization_calibration_2023,qedma_logical_errors_2024,qedma_efficient_lightcones_2025,qedma_drift_robust_2025,qedma_pauli_fractional_2025,qedma_clifford_char_2025}. 

\section*{Contents}
\setcounter{tocdepth}{2}
\renewcommand{\theequation}{\thesection\arabic{equation}}
\renewcommand{\thefigure}{\thesection\arabic{figure}}
\setcounter{equation}{0}
\setcounter{figure}{0}
\printappendixtoc

\section{Construction of QP distributions}\label{app:QP}

 Quasi-probability (QP) error mitigation \citep{IBM_PEC_ZNE,Endo_Benjamin_Ying} is based on the representation\footnote{We avoid the terminology `PEC', for `probabilistic (or Pauli) error cancellation', as it has often been used to refer specifically to the `Clifford mitigation' case described below \cite{berg2022probabilistic, Ferracin2024}.}
  \begin{align}
  G_0=\sum_i c_i B_i\label{Eq: QP rep}
\end{align}
of an ideal quantum operation $G_0$ as a linear combination of noisy quantum operations $\mathcal{B}=\{B_i\}_i\ni G$, including the noisy version $G$ of $G_0$, with coefficients $\{c_i\}_i$. Under standard assumptions,\footnote{Namely, both $G_0$ and all basis elements $B_i$ are assumed to be hermiticity- and trace-preserving.} the coefficients $c_i$ are real and normalized, $\sum_i c_i =1$, and therefore define a QP distribution. Assuming the errors in the operations $\mathcal{B}$ are known, one can solve for the optimal coefficients $c_i$, as discussed in Appendix \ref{Sec: Multi-type QP algorithm}. The un-available ideal operation $G_0$ can then be implemented on a given QPU by replacing the outcome of an ideal quantum circuit containing $G_0$ by an appropriate average over the outcomes of noisy quantum circuits obtained by randomly replacing $G_0$ with an operation $B_i$, sampled according to the QPs $c_i$ (see Appendix \ref{app:QPU time minimization}). 

We refer to the set $\mathcal{B}$ as a \emph{QP basis}, or simply \emph{basis}, though it may not be linearly independent or span a predefined sub-space of quantum operations. A fundamental challenge in applying the QP method is the construction of a basis $\mathcal{B}$ which is expressive enough to efficiently mitigate all significant errors, and yet comprised of operations which are natively available on the QPU, and preferably have a short duration and high fidelity. An additional useful feature in case the operations $B_i$ act on a subset of qubits is that they can be applied simultaneously with each other and with other gates in the circuit. 
  
QP bases which are expressive enough to allow for the mitigation of any quantum operation on $n$ qubits, and involve only the noisy operation $G$ and layers of single-qubit operations, have been proposed \citep{Endo_Benjamin_Ying} and even fused in a 2-qubit experiment \citep{QPexperimentWithReset}. These are given by $\mathcal{B}=\{G\}\cup \mathcal{S}_M$, and $\mathcal{B}= G\mathcal{S}_M:=\{GS|\ S\in \mathcal{S}_M\}$, where $\mathcal{S}_M$ denotes the set of layers (ideally tensor products) of 16 linearly-independent single-qubit operations. A practical difficulty with these bases is that 6 of the above 16 linearly independent single-qubit operations involve non-unitary operations, such as $\Pi_0:\rho\mapsto \ket{0}\bra{0}\rho\ket{0}\bra{0}$, corresponding to measuring a qubit mid-circuit in the computational basis, and replacing the result of any end-of-circuit measurement with 0 unless the mid-circuit measurement gives 0. Though mid-circuit measurements are becoming increasingly available in QPUs, they often suffer an unfavorable trade-off between their fidelity and duration, in comparison to unitary single-qubit and even two-qubit gates.\footnote{As an example, in order to obtain a high fidelity implementation of $\Pi_0$, the mid-circuit measurement is sometimes followed by a reset operation, which includes another mid-circuit measurement followed by an $X$ gate conditioned on the measurement outcome \cite{riste_feedback_reset, ibm_mid_circuit_reset}. Alternatively, the reset operation can be implemented using an ancilla qubit and a SWAP gate, as in Ref.~\citep{QPexperimentWithReset}. In both cases, significant resources are used to implement a high-fidelity $\Pi_0$.}

Accordingly, larger experiments utilizing the QP method did not make use of non-unitary operations as part of the QP basis they employed \citep{berg2022probabilistic, Ferracin2024}. Instead, these experiments used the following mitigation strategy, based on 2-qubit Clifford gates, such as $C\!X$. Every quantum circuit can be compiled onto alternating layers of $C\!X$ gates and single-qubit gates. By ``absorbing'' single qubit gate errors into the adjacent two-qubit gate $C\!X$ layers, the method focuses on the $C\!X$ layers (see Appendix \ref{Appendix: characterization} for details on noise characterization of the two-qubit gates). These can be Pauli twirled, resulting in a simple Pauli error channel before (or equivalently, after) the ideal gate, which can be mitigated with the basis $\mathcal{B}=G\mathcal{P}:=\{GP|\ P\in \mathcal{P}\}$, where $\mathcal{P}$ denotes the group of Pauli layers. In particular, no non-unitary basis elements are used. We refer to this mitigation strategy as `Clifford mitigation'. 

QESEM utilizes \emph{multi-type QP bases} in order efficiently mitigate errors in quantum circuits compiled using 2-qubit \emph{non-Clifford} gates, without relying on non-unitary operations, or on a significantly noisier compilation onto 2-qubit Clifford gates (see Patent \cite{qedma_multi_qp_patent}).

\subsection{Multi-type QP bases\label{Sec: general description}} 

A multi-type QP basis $\mathcal{B}$ may be defined by a noisy gate $G$ (with ideal version $G_0$), and a user-specified set of `mitigation operations' $\mathcal{S}$, corresponding to a set of `simple' quantum operations which are available on the relevant QPU. The elements of $\mathcal{S}$ act on the same number of qubits as $G$. As a simple yet already useful example, we may take $\mathcal{S}$ to be a set of layers of single-qubit unitary gates, and in particular, we may take $\mathcal{S}$ to be the group of Pauli layers $\mathcal{P}$. Another useful example is the case where $G=G_\alpha$ is an element of a parameterized family of gates (where the parameter $\alpha$ may e.g. be a rotation angle), and $\mathcal{S}$ contains additional elements $G_{\alpha'}$ from the family. A multi-type basis $\mathcal{B}$ then consists of basis elements $B_i$ which are sub-circuits constructed from $G$ and the elements of $\mathcal{S}$, where each `type' corresponds to a sub-circuit structure, e.g. 
\begin{align}
  \mathcal{B}=\{G\}\cup \mathcal{S}\cup G\mathcal{S} \cup \mathcal{S}G \cup \mathcal{S}^2 \cup \mathcal{S}G\mathcal{S},\label{Eq: no-reset basis elements}
\end{align}
where, e.g., $\mathcal{S}G\mathcal{S}:=\{SG\tilde{S}|\ S,\tilde{S}\in\mathcal{S}\}$. Equation \eqref{Eq: no-reset basis elements} includes the necessary trivial type given by the noisy gate $G$, as well as five non-trivial types, involving up to a single instance of $G$ and up to two instances of elements from $\mathcal{S}$. These numbers may be enlarged to include additional types, such as $\mathcal{S}G\mathcal{S}G$ etc. Adding types gives a basis that potentially allows for the mitigation of additional kinds of errors, or the more efficient mitigation of errors that can be mitigated with a simpler basis. Mathematically, the basis constructed from all types (including any number of instances of $G$ and $\mathcal{S}$) spans the algebra generated by $\{G\}\cup \mathcal{S}$, and taking a larger number of types gives access to larger vector sub-spaces of this algebra, or allows to span an already attained sub-space more efficiently. 

The operation $G$ may be a single gate (acting on e.g., a pair of qubits), or a sub-circuit involving a number of gates acting in parallel and/or sequentially, $G=G_K \cdots G_1$. The latter case allows for more general multi-type bases, where basis elements $B_i$ correspond to sub-circuits constructed from the set of gates $\mathcal{G}=\{G_1,\dots,G_K\}$ and the set of mitigation operations $\mathcal{S}$. Each type corresponds to a sub-circuit structure, of the form 
\begin{align}
  \mathcal{G}^{a_1}\mathcal{S}^{b_1}\cdots \mathcal{G}^{a_M}\mathcal{S}^{b_{M}},\label{Eq: most general multi-type basis}
\end{align}
defined by $a_j,b_j,M\in \mathbb{N}_0$. To clarify the notation, $\mathcal{G}^2\mathcal{S}=\{G_i G_jS|\ G_i,G_j\in\mathcal{L},\ S\in\mathcal{S}\}$. 

Given a noisy circuit to be mitigated, there is freedom in decomposing it into sub-circuits $G$ for which QP distributions are constructed. Larger sub-circuits admit more types of basis elements, allowing for the mitigation of errors that may not be mitigated with smaller sub-circuits. On the other hand, multi-type QP distributions for larger sub-circuits are computationally harder to construct -- either the depth or width (and preferably both) of sub-circuits must be bounded to allow for efficient computation. In particular, constructing a QP distribution for the entire circuit is computationally harder than simulating the (ideal) circuit, and therefore intractable for useful quantum circuits.

\subsection{QP norm and accuracy \label{Appendix: QP norm and accuracy}}

As discussed in Appendix \ref{app:QPU time minimization}, the QP norm $W=\|c\|_1=\sum_i |c_i|\geq 1$ is an important factor in determining the QPU time overhead for EM based on the QPs $c_i$. It can be shown that $W$ obeys a lower bound 
\begin{align}
  W\geq \frac{1+\IF(G)}{1-\IF(G)} \geq e^{2 \IF(G)}\geq 1+2\IF(G),\label{Eq: OIB 1}
\end{align}
where $\IF(G)$ is the (entanglement) infidelity of the available noisy version $G$ of $G_0$, which must be included in the basis $B_i$.\footnote{In Eq.~\eqref{Eq: OIB 1}, the operation $G$ is mathematically defined as the basis element $B_i$ with the highest fidelity relative to $G_0$.} In many cases, this bound is approximately tight, in the sense that QP distributions with $W=1+2 \IF(G)+\mathcal{O}(\IF(G)^2)$ can be constructed. 

Of course, an exact implementation of the ideal gate $G_0$ is too strict a requirement, and approximate QP distributions
\begin{align}
  G_0\approx G_{\text{QP}}=\sum_i c_i B_i\label{QP rep 1.1}
\end{align}
may be considered. The distance\footnote{Measured in, e.g., fidelity, diamond distance, or Frobenius distance.} of $G$ from $G_{\text{QP}}$ may be referred to as the \emph{inaccuracy of the QP distribution}, and controls mitigation biases. For an approximate QP distribution, the infidelity in the RHS of Eq.~\eqref{Eq: OIB 1} is reduced to the infidelity eliminated by mitigation,
 \begin{align}
  W\geq e^{2 [\IF(G)-\IF(G_{GP})]}\geq 1+2[\IF(G)-\IF(G_{GP})],\label{Eq: OIB 2}
\end{align}
demonstrating the trade-off between inaccuracy and sampling overhead, corresponding respectively to systematic and statistical errors, involved in constructing (approximate) QP distributions. 

 A main challenge involved in error mitigation with the QP method is the construction of QP distributions that minimize both inaccuracy and sampling overhead as much as possible, and prioritize these according to a given specification when minimizing both isn't possible. 

Generally speaking, a good accuracy requires that the basis $\mathcal{B}$ spans a large enough subspace in the space of super-operators. To identify the required subspace, it is useful to factor out the ideal gate and write $G=G_0e^{L_b}$ in terms of a before-error generator $L_b$,\footnote{An after-error generator can equally be used.} which is essentially a Lindbladian \citep{Taxonomy_Sandia}. Equation \eqref{Eq: QP rep} then takes the form
\begin{align}
   e^{-L_b}\approx\sum_i c_i B_i',\label{Eq: QP rep 2}
\end{align}
where $B_i'=G^{-1}B_i$.\footnote{Note that as opposed to $B_i$ the $B_i'$s involve the inverse of a noisy gate and are therefore not completely positive -- they are not physical quantum operations that can be applied on the QPU.} We see that the basis $\mathcal{B}'=\{B_i'\}$ should span a subspace that includes, or is as close as possible to, the inverse noise channel $e^{-L_b}$, and we can interpret $\operatorname{Span}(\mathcal{B}')$ as the space of `mitigatable error channels'. The condition number of the basis $\mathcal{B}'$ (when viewed as a matrix whose columns are the vectorized basis elements) determines the efficiency with which $\mathcal{B}'$ spans this space of mitigatable errors, and controls the QP norm $W$. Generally speaking, a large condition number implies a large $W$.
  
To get a feeling for the basis $\mathcal{B}'$ in the multi-type case, consider the example in Eq.~\eqref{Eq: no-reset basis elements}, where 
\begin{align}
   \mathcal{B}'=\{I\}\cup G^{-1}\mathcal{S} \cup \mathcal{S} \cup G^{-1}\mathcal{S}G\cup G^{-1}\mathcal{S}G\mathcal{S},\label{Eq: no-reset basis elements '}
\end{align}
and $\mathcal{S}$ corresponds to a large enough set of ideal layers of single-qubit unitary gates.\footnote{The type $\mathcal{S}^2$ is redundant in this case.} Single-qubit unitary gates span a 10-dimensional vector subspace containing all single-qubit unital and trace-preserving super-operators, of the form
\begin{align}
  \left(
    \begin{array}{cccc}
   * & 0 & 0 & 0 \\
   0 & * & * & * \\
   0 & * & * & * \\
   0 & * & * & * \\
  \end{array}
  \right)
\end{align}
in the Pauli basis.\footnote{
 $G_{P,P'}=\sbra{P}G\sket{P'}=\operatorname{Tr}(PG[P'])/2^n$, where $P,P'\in \mathcal{P}$, with the first basis element being the identity.
} An example for a basis for this subspace is given by $I$, $X_{\pi}$, $Y_{\pi}$, $Z_{\pi}$, $X_{\pi/2}$, $Y_{\pi/2}$, $Z_{\pi/2}$, $Z_{\pi/2}X_{\pi}$, $Y_{-\pi/2}X_{\pi}$, $X_{\pi/2}Y_{\pi}$, and we may take $\mathcal{S}$ to be the size-$10^n$ set of layers ($n$-fold tensor products) of these operations, spanning the corresponding $10^n$-dimensional tensor-product subspace. Note that this subspace is much smaller than the $16^n$-dimensional space of all $n$-qubit super-operators. Moreover, for $n>1$, the above $10^n$-dimensional subspace is contained but is much smaller than the $(4^n-1)^2+1$-dimensional subspace containing all trace-preserving and unital $n$-qubit super-operators. It follows that each of the types in Eq.~\eqref{Eq: no-reset basis elements '} which contains a single $\mathcal{S}$, namely $G^{-1}\mathcal{S},\mathcal{S}, G^{-1}\mathcal{S}G$, spans a $10^n$-dimensional subspace, which is insufficient to mitigate a generic error channel $e^{L_b}$. However, these $10^n$ dimensional subspaces are rotated w.r.t to each other by the conjugation and multiplication by $G^{-1}$, so the multi-type basis $G^{-1}\mathcal{S}\cup \mathcal{S}\cup G^{-1}\mathcal{S}G$ generically spans a much larger space, as demonstrated in Appendix \ref{Sec: numerical simulations}. Note that, since in this example all basis elements $B_i$ are ideally unitary, they cannot efficiently span non-unital or non-trace-preserving operations, leaving the subspace of efficiently mitigated errors with a maximal dimension $1+(4^n-1)^2$. Note that the elements in a multi-type basis are often linearly-dependent, covering their span redundantly. This is a feature rather than a bug, allowing for the optimization of the QP norm over coefficients $\{c_i\}$ satisfying Eq.~\eqref{Eq: QP rep 2}, as we now discuss.

\subsection{Constructing QP distributions for a given basis\label{Sec: Multi-type QP algorithm}}

In this section we describe an algorithm for choosing the coefficients $c_i$ in Eq.~\eqref{Eq: QP rep}, given a \emph{fixed} multi-type basis $\mathcal{B}$, and the characterization of the noisy operation $G$, and of the mitigation operations $\mathcal{S}$ used to construct $\mathcal{B}$. The noisy gate $G$ will usually be twirled (see Appendix \ref{Appendix: characterization}), and the corresponding before-error $e^{L_b}$ will then be the twirled error. In the simplest case the mitigation operations $\mathcal{S}$ can be assumed ideal, but this is not a requirement. 

Loosely speaking, our goal is to solve the linear equation Eq.~\eqref{Eq: QP rep 2} for the coefficients $c_i$, given a characterization of the before-error $L_b$, and of the noisy basis operations in $\mathcal{B}'$ (obtained from the characterizations of $G$ and $\mathcal{S}$). As discussed above, certain types of errors may not admit mitigation with the chosen basis, so a more robust approach is to minimize an appropriate distance measure between the two sides of Eq.~\eqref{Eq: QP rep 2}. Taking this distance measure to be the Frobenius distance leads to a least squares problem 
\begin{align}
  \min_{\{c_i\}}\|\sum_i c_i B_i' - e^{-L_b}\|_{F}.\label{Eq: full least squares}
\end{align}
We use the Frobenius distance as opposed to the diamond distance (used in Ref.~\cite{Piveteau2022Quasiprobability}) for two reasons. First, the Frobenius distance leads to simpler optimization problems, as described below. Second, the Frobenius distance is an average-case distance measure, as opposed to the worst-case diamond distance. Since the mitigation bias on a measured observable is due to the accumulated inaccuracy of many local QP distributions, the bias is `self-averaging', such that an average-case distance measure better predicts the bias than a worst-case measure. 

It is useful to consider expansions of Eq.~\eqref{Eq: full least squares} in $L_b$. As an example, consider the first order expansion $e^{-L_b}=I-L_b+\mathcal{O}(L_b^2)$. For $L_b=0$, we have $c_i=\delta_{i,0}$, where $c_0$ is the coefficient corresponding to $B_0'=I$. It follows that $c_i=\mathcal{O}(L_b)$ for $i\neq 0$, so we can set $L_b=0$ in $B_i'$ and obtain a consistent first order expansion of $\sum_i c_i B_i' - e^{-L_b}$. We can similarly ignore the errors on $B_i'$ and the elements of $\mathcal{S}$. This leads to the least squares problem 
\begin{align}
   \min_{\{x_i\}}\|\sum_i x_i B_{i,0}' + L_b\|_{F},\label{Eq: first order least squares}
\end{align}
where $B_{i,0}'$ is the ideal version of $B_{i}'$, and $x_i=c_i-\delta_{i,0}$. Such expansions are useful as starting points for analytic solutions as a function of gate angles and error parameters, which are indeed utilized within QESEM. Moreover, assuming locality of $L_b$, such expansions allow to reduce the least-squares problem from the $16^n$-dimensional space of super-operators to a subspace with dimension $d=\operatorname{poly}(n)$. The latter may be spanned by a set of local Choi basis elements $\{P_{i_k}\cdot P_{j_k}\}_{k=1}^d$, where $P_{i_k},P_{j_k}\in \mathcal{P}$ and the combined support of $P_{i_k}$ and $P_{j_k}$ is bounded according to the locality of $L_b$. With this basis, we obtain a standard least squares problem,
\begin{align}
\min_{x}\|Ax-y\|_2,\label{Eq: standard least squares}
\end{align}
with
\begin{align}
   A_{j,i}&=\left<(P_{i_j}\cdot P_{k_j}),B_{i,0}'\right>,\label{Eq: least squares data}\\
   y_{j}&=-\left<(P_{i_j}\cdot P_{k_j}),L_b\right>,\nonumber
\end{align}
which can be solved via the SVD-based pseudo-inverse, $x^{(-1)}:=A^{-1}y$. 

An important feature of the least-squares problems obtained from multi-type bases is that the matrix $A$ has a non-trivial kernel,\footnote{The kernel is defined numerically by identifying singular values below a threshold as "0" in defining the pseudo inverse $A^{-1}$. The choice of threshold corresponds to the tradeoff between accuracy and QP norm discussed in Appendix \ref{Appendix: QP norm and accuracy}. \label{Foot: "0"}} and as a result there is in fact an affine space of solutions $x^{(-1)}+\operatorname{ker}(A)$. The pseudo-inverse solution $x^{(-1)}$ minimizes the $\ell_2$-norm over this space. Denoting $y_\parallel=Ax^{(-1)}$, which is the orthogonal projection of $y$ onto $\operatorname{Image}(A)$, this means that $x^{(-1)}$ is the solution to 
\begin{align}
  \min_{x:\ Ax=y_\parallel}\|x\|_2.
\end{align}
However, our goal is to minimize the QP norm $\|c\|_1=|1+x_0|+\sum_{i\neq0}|x_i|$, rather than $\|x\|_2$. The former can be identified as a linear program, and therefore solved efficiently, as follows. 

\begin{figure}[b]
\begin{algorithm}[H]
  \caption{Multi-type QP distributions (first order)}\label{Algo: 1st order QP rep}
  \begin{flushleft}
    \Input A local before-error generator $L_b$ for the ideal gate $G_0$; 
    an ideal multi-type QP basis $\{B_{i,0}'\}$ (e.g., of the form Eq.~\eqref{Eq: no-reset basis elements '}), with $B_{0,0}'=I$.
    \Output Coefficients $\{c_i^*\}$ which provide an optimal first-order mitigation in the sense
    \begin{align}
      \|\sum_i c_i^* B_i' - e^{-L_b}\|_{F}
      =\min_{\{c_i\}}\|\sum_i c_i B_i' - e^{-L_b}\|_{F}+\mathcal{O}(L_b^2),\nonumber
    \end{align}
    and further minimize the QP norm $\|c\|_1$ among all coefficient vectors $c$ satisfying the above condition.
  \end{flushleft}
  \algblock{PerformanceMetric}{EndMetrics}
  \begin{algorithmic}[1]
    \Procedure{constructQPdecomp}{}
      \State Construct a local Choi basis $\{P_{i_j}\cdot P_{k_j}\}_{j=0}^{16^{n}-1}$ with $P_{i_0}=P_{k_0}=I$.
      \State Construct the least squares data $A$ and $y$ according to Eq.~\eqref{Eq: least squares data}.
      \State Compute the Moore-Penrose pseudo-inverse $A^{-1}$. 
      \State $x^{(-1)}\gets A^{-1}y$ \Comment{Minimal $\ell_2$-norm solution, Eq.~\eqref{Eq: standard least squares}} 
      \State $y_\parallel \gets Ax^{(-1)}$\Comment{
        Orthogonal projection of $b$ onto $\operatorname{Image}(A)$. This is the part of $L_b$ that can be mitigated with the chosen basis
      }
      \State Obtain a solution $(x^+,x^-)^*$ to the linear program Eq.~\eqref{Eq: linear program 2}
      \State $c^*_i \gets x_i^{+}-x_i^-+\delta_{i,0}(1-\max(x_0^+,x_0^-))$ \Comment{Minimal $\ell_1$-norm solution to Eq.~\eqref{Eq: first order least squares}}
      \State \Return $c^*$
    \EndProcedure
    \Statex
    \PerformanceMetric
      \State Relative residual Frobenius distance:
      \[\frac{\Frob_{\text{QP}}}{\Frob}=\frac{\|\sum_i c_i^*B_i-G_0\|_F}{\|G-G_0\|_F}\]
      \State QP norm relative to lower bound: $\|c^*\|_1/(1+2\IF)$.
    \EndMetrics
  \end{algorithmic}
\end{algorithm}
\end{figure}

In terms of $c$ the constraint $Ax=y_\parallel$ can be written as $Ac=b_{\parallel}$, where $b_{\parallel, j}=y_{\parallel,j}+\delta_{j,0}$. We then have a standard `basis pursuit' problem,
\begin{align}
  \min_{c:\ Ac=b_\parallel}\|c\|_1.
\end{align}
The latter can be mapped to a linear program by using the positive and negative parts $c^\pm_i=\Theta(\pm c_i)|c_i|$,
where $\Theta$ is the Heaviside step function. The linear program reads\footnote{
  Note that by definition, for each $i$ either $c^+_i=0$ or $c_i^-=0$, and the linear program is missing this non-linear constraint. However, any solution $c^{\pm}_i$ to the linear program must obey this constraint, since the transformation $c_i^{\pm}\mapsto c_i^{\pm}-\min(c_i^+,c^-_i)$ maintains the constraints of the linear program, but lowers its cost function if $\min(c_i^+,c^-_i)>0$ for some $i$.
}
\begin{align}
  \min_{c^+,c^-}&\ \sum_i(c_i^++c_i^-)\label{Eq: linear prog}\\
  \text{subject to }&\ A(c^+_i-c^-_i)=b_\parallel, \ c^+\succcurlyeq \mathbf{0},\ c^-\succcurlyeq\mathbf{0}.\nonumber
\end{align}
Here $\mathbf{0}=(0,\dots,0)$ and $\succcurlyeq$ denotes the element-wise inequality. In terms of $x$, the linear program takes an almost identical form,\footnote{Here the problem is completely invariant under $x_0^\pm \mapsto x_0^\pm-\min(x_0^+,x_0^-)$, so we need to apply this transformation to any minimizer in order to obtain a minimizer satisfying $x_0^+=0$ or $x_0^-=0$.}
\begin{align}
  \min_{x^+,x^-}&\ x_0^+-x_0^-+\sum_{i\neq0}(x_i^++x_i^-)\label{Eq: linear program 2}\\
  \text{subject to }&\ A(x^+_i-x^-_i)=y_\parallel, \ x^+\succcurlyeq \mathbf{0},\ x^-\succcurlyeq\mathbf{0}.\nonumber
\end{align}
The problem for $x$ may be more stable numerically, since $y_\parallel$ has only $\mathcal{O}(L_b)$ entries, while $b_{\parallel,j}=\delta_{j,0}+\mathcal{O}(L_b)$ contains two scales. We summarize the full procedure in Algorithm \ref{Algo: 1st order QP rep}.

\subsection{Numerical simulations\label{Sec: numerical simulations}}

\begin{figure}[tb]
\begin{centering}
\includegraphics[width=0.95\columnwidth]{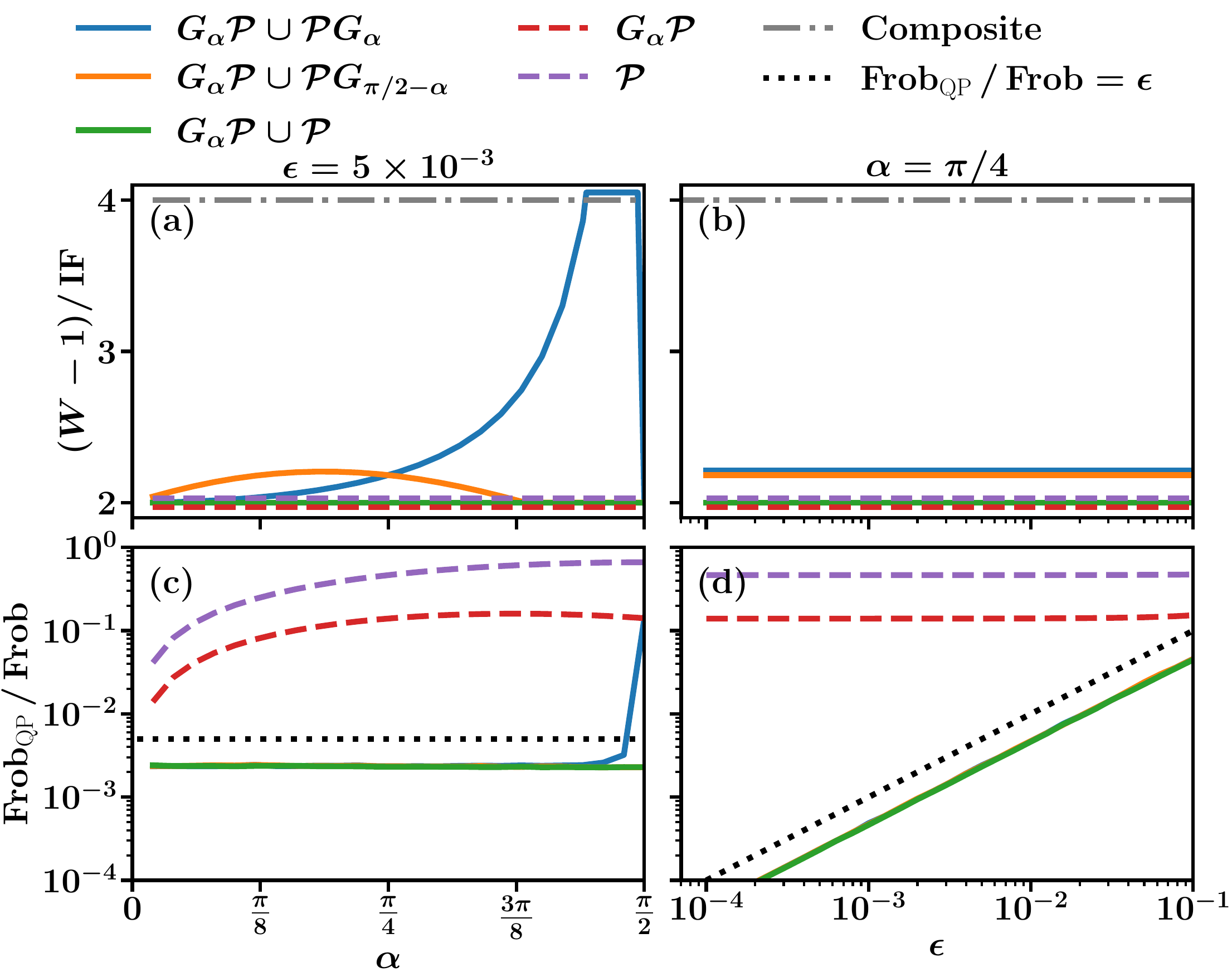}
\par\end{centering}
\caption{Performance of 2-qubit multi-type QP distributions for a family of non-Clifford gates $G_\alpha=R_{ZZ}(\alpha)$. Multi-type distributions (solid lines) can attain a near-optimal and even optimal sampling overhead, quantified by the QP norm $W\approx 1+2\IF$ (apart from the distribution for the basis $\mathcal{P}G_\alpha \cup G_\alpha \mathcal{P}$ at $\alpha>\pi/4$), as well as an optimal accuracy, quantified by the relative Frobenius distance $\Frob_{\text{QP}}/\Frob<\epsilon$. In comparison, single-type strategies (dashed lines) can only attain optimality in one of these performance metrics.  \label{fig: POC 1}
}
\end{figure}

Figure \ref{fig: POC 1} demonstrates the performance of multi-type QP distributions in the mitigation of the family of 2-qubit non-Clifford $ZZ$-rotation gates $G_\alpha=R_{ZZ}(\alpha)$, with $\alpha\in(0,\pi/2)$.

We exclude the Clifford points $\alpha=0,\pi/2$, which can be twirled with the full Pauli group and handled with `Clifford mitigation' \citep{berg2022probabilistic, Ferracin2024}. An existing strategy for mitigating non-Clifford two-qubit gates is to compile them using two-qubit Clifford gates and single-qubit gates, e.g., 
\begin{align}
G_{\alpha}^{\text{comp}}=C\!X\,R_{IZ}(\alpha)\,C\!X,
\end{align}
and applying Clifford mitigation to the two-qubit Clifford gates ($C\!X$). The downside of this `composite mitigation' is that $G_{\alpha}^{\text{comp}}$ will generically have a significantly larger infidelity than a directly implemented $G_{\alpha}$. The reason is that $C\!X$ is equal to $G_{\pi/4}$ up to single-qubit gates. Assuming the infidelity $\IF$ of $G_\alpha$ is independent of $\alpha$ then leads to $\IF^{\text{comp}}\approx 2 \IF$, where $\IF^{\text{comp}}$ is the infidelity of $G_{\alpha}^{\text{comp}}$. This is a very conservative assumption - it is in fact expected and has been experimentally reported that the infidelity of $G_\alpha$ can be made proportional to the rotation angle, $\IF\propto \alpha$ \cite{Quantinuum_data_sheet, PhysRevResearch.3.033171}. This leads to $\IF^{\text{comp}}\approx (\pi/\alpha)\IF$, which reduces to $ 2 \IF$ as $\alpha\rightarrow\pi/2$, but gives a much stronger advantage to the direct implementation at smaller angles. 

The noise model we take for $G_\alpha$ in Fig.~\ref{fig: POC 1} is given by the standard amplitude damping ($T_1$) and pure dephasing ($T_\phi$) dissipative processes occurring on each qubit \emph{during} the gate operation, with rates $t_g/2T_1=t_g/T_\phi=\epsilon/3$, where $t_g$ is the gate duration. The rate of total dephasing ($T_2$) is given by $t_g/T_2=t_g/2T_1+t_g/T_\phi=2\epsilon/3$. The parameter $\epsilon$ is essentially the 2-qubit gate (entanglement-) infidelity, 
\begin{align}
  \IF=\epsilon+\mathcal{O}(\epsilon^2).\label{Eq: epsilon}
\end{align}
We note that qualitatively similar results are obtained for any single-qubit dissipative errors during the gate operation. The noisy gates $G_\alpha$ are then twirled with the sub-group of Pauli layers commuting with $ZZ$, as described in Sec.~\ref{Appendix: characterization}. For the composite implementation $G_\alpha^{\text{comp}}$ we make the conservative assumption $\IF^{\text{comp}}=2\times \IF$. As a function of the parameters $\alpha$ and $\epsilon$ defining the above twirled noisy gates, we numerically test the performance of first-order 2-qubit QP distributions constructed using Algorithm \ref{Algo: 1st order QP rep}, and utilizing several single-type and multi-type bases. In the Left Panels in Fig.~\ref{fig: POC 1} we fix $\epsilon=5\times 10^{-3}$ and vary $\alpha\in(0,\pi/2)$, while in the Right Panels we fix $\alpha=\pi/4$ and vary $\epsilon\in(10^{-4},10^{-1})$. 

The basis operations $B_i$ in all bases $\mathcal{B}$ considered in Fig.~\ref{fig: POC 1} (see legend in Panel (a)) are constructed from the gate $G_\alpha$ to be mitigated (not indicated explicitly), and an \emph{ideally-unitary} set of mitigation operations. The latter is given by the set of 16 2-qubit Pauli layers $\mathcal{P}$ (with similar results for any set of single-qubit layers), and in one case an additional 2-qubit gate $G_{\alpha'}$ with $\alpha'\neq \alpha$. From these mitigation operations we construct several multi-type bases (solid lines), as well as single-type bases for reference (dashed lines). To clarify the notation, the basis $\mathcal{B}=G_\alpha \mathcal{P} \cup \mathcal{P}G_{\pi/2-\alpha} $ includes basis elements where any Pauli layer $P\in \mathcal{P}$ is applied before the mitigated gate $G_\alpha$, or after the additional 2-qubit gate $G_{\pi/2-\alpha}$. The legend entry `Composite' refers to the `composite mitigation' of $G_\alpha^{\text{comp}}$.

As a first performance metric (Upper Panels in Fig.~\ref{fig: POC 1}) for the QP distributions constructed with the above bases, we use the `blowup rate' $(W-1)/\IF$, quantifying the exponential dependence of the QP-norm $W^V$ of a volume-$V$ circuit on the `total infidelity' $\IF\times V$. As discussed in Sec.~\ref{Appendix: QP norm and accuracy}, the blow-up rate has an optimal value of $2$, which sets the lower limit for the $y$-axis in the Upper Panels. Note that in both Upper Panels we slightly separate vertically overlapping lines for visibility. Panel (b) merely demonstrates that, as expected, the blowup rate is independent of $\epsilon\approx \IF$, and will not be discussed further. 

As a second performance metric (Lower Panels in Fig.~\ref{fig: POC 1}) we use the relative residual Frobenius distance $\Frob_{\text{QP}}/\Frob$, as a measure for the inaccuracy of QP distributions.
This is $\mathcal{O}(\epsilon)$ for a QP distribution which mitigates \emph{all} errors to first order. The dotted lines in the Lower Panels correspond to $\Frob_{\text{QP}}/\Frob=\epsilon$. 

To demonstrate the advantages of multi-type mitigation, we first consider a number of single-type mitigation strategies, indicated by dashed lines in Fig.~\ref{fig: POC 1}. The first single-type mitigation we consider is `composite mitigation', based $G^{\text{comp}}_\alpha$ (gray dashed). This method can easily be applied non-linearly, and therefore gives a vanishing inaccuracy (which is why the gray dashed line is absent in the lower panels). Additionally, the method gives an optimal blowup rate $(W-1)/\IF^{\text{comp}}=2$ in terms of the infidelity $\IF^{\text{comp}}$ of $G_\alpha^{\text{comp}}$. However, even under the conservative assumption $\IF^{\text{comp}}=2\IF$, this translates to $(W-1)/\IF=4$ in terms of the infidelity $\IF$ of the direct implementation (Panel (a)), implying the possibility of an exponential advantage for a mitigation strategy based on the direct implementation $G_\alpha$, which may, in principle, have a blowup rate as low as $(W-1)/\IF=2$. Indeed, both single-type bases $G_\alpha \mathcal{P}$ (red dashed) and $\mathcal{P}$ (purple dashed) produce this optimal blowup rate, as seen in Panel (a). However, both bases cannot mitigate all errors to first order, with $\Frob_{\text{QP}}/\Frob\gg \epsilon$ for small $\epsilon$, as seen in Panels (c) and (d). [The relative inaccuracy slightly decreases at small $\alpha$, where $G_\alpha$ approaches a noisy idle gate.]

We now describe in detail the performance of multi-type QP distributions in Fig.~\ref{fig: POC 1}. The simplest multi-type QP basis we consider is $\mathcal{B}_1=G_\alpha \mathcal{P} \cup \mathcal{P}G_\alpha$, inserting Pauli layers before or after the noisy gate $G_\alpha$. As seen in Panel (a), the blowup rate of $\mathcal{B}_1$ is close to the optimal value of 2 for small $\alpha$, with $(W-1)/\IF<2.18$ for $\alpha<\pi/4$. In this range, one also obtains a full first-order mitigation, with relative error $\Frob_{\text{QP}}/\Frob<\epsilon$, as seen in Panels (c) and (d). The multi-type basis $\mathcal{B}_1$ therefore gives a near-optimal first-order mitigation for $\alpha<\pi/4$. However, as $\alpha\rightarrow \pi/2$ the blowup rate itself blows up, giving an increasingly non-optimal mitigation, until plummeting back to the optimum very near $\alpha=\pi/2$. As seen in Panel (c), this plummeting is accompanied by a sharp spike in the inaccuracy. The correlated plummeting of the blowup rate and spike in the inaccuracy are due to the trade-off between QP norm and inaccuracy incorporated into Algorithm \ref{Algo: 1st order QP rep} (see Footnote \ref{Foot: "0"}). 

The excessive QP norm obtained for $\mathcal{B}_1=G_\alpha \mathcal{P} \cup \mathcal{P}G_\alpha$ as $\alpha\rightarrow \pi/2$ stems from the fact that $G_{\pi/2}$ is (ideally) Clifford, mapping the Pauli group to itself via conjugation. As a result $G_{\pi/2} \mathcal{P} = \mathcal{P}G_{\pi/2}$, effectively reducing the multi-type basis $\mathcal{B}_1$ to a single-type basis at $\alpha=\pi/2$. A generic solution to this issue is to include additional 2-qubit gates $G_{\alpha'}$ as mitigation operations. As an example, we consider the basis $\mathcal{B}_2=G_\alpha \mathcal{P} \cup \mathcal{P}G_{\pi/2-\alpha}$. As seen in Fig.~\ref{fig: POC 1}, this basis produces a near-optimal first-order mitigation for any angle $\alpha\in (0,\pi/2)$, and in particular for angles $\alpha>\pi/4$. For $\alpha<\pi/4$ the blowup rate of $\mathcal{B}_2$ is slightly higher than that of $\mathcal{B}_1$, so it's clear that the two bases can be combined to further improve the blowup rate.\footnote{The basis $\mathcal{B}_2$ requires the calibration of an additional gate $G_{\pi/2-\alpha}$, that may not be needed for the ideal quantum circuit(s) one is interested in. Nevertheless, for the purpose of generating first-order QP distributions, the characterization of this additional gate isn't needed, as discussed in Sec.~\ref{Sec: Multi-type QP algorithm}.}

\begin{figure}[tb]
\begin{centering}
\includegraphics[width=\columnwidth]{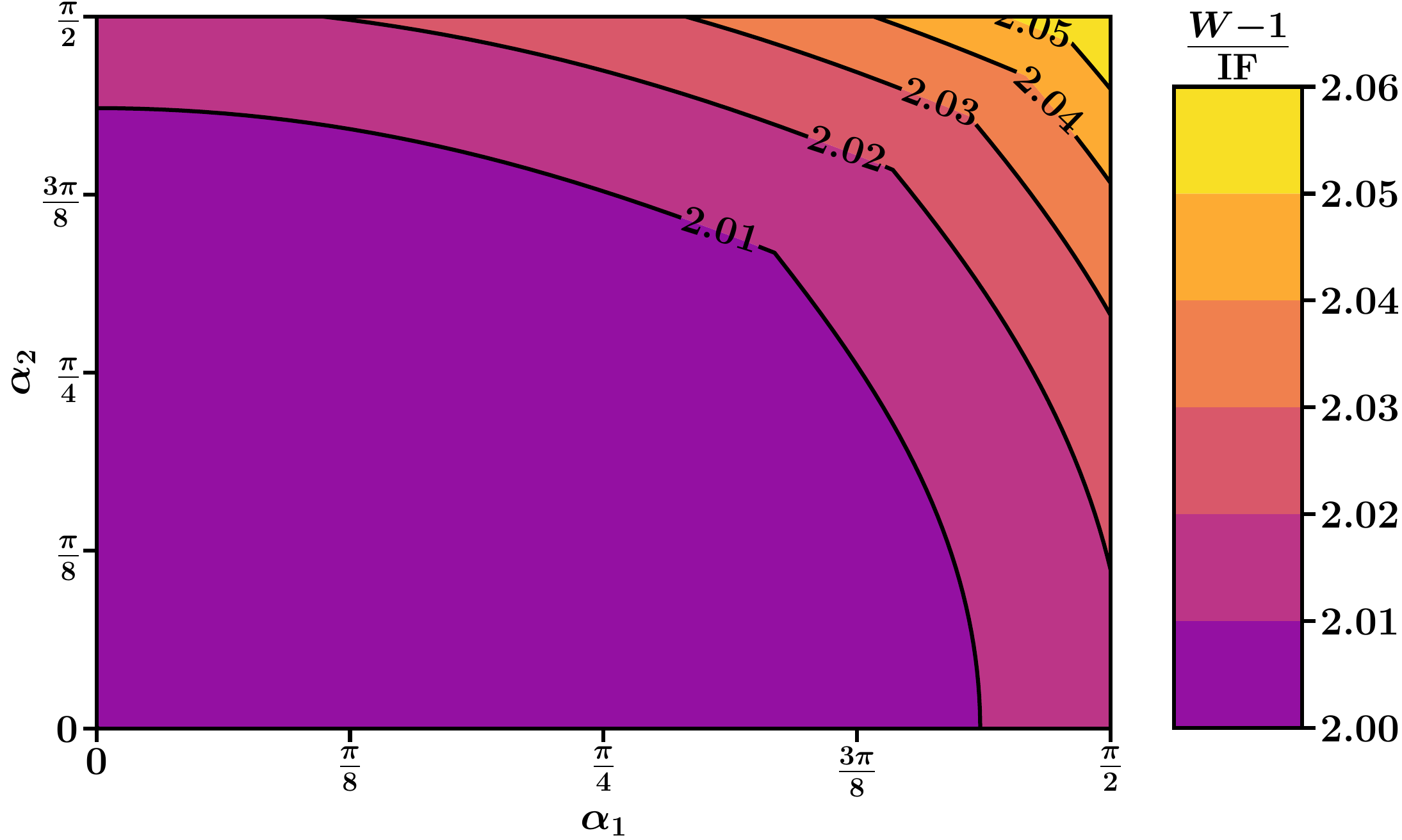}
\par\end{centering}
\caption{Near-optimal multi-type QP distributions for Pauli cross-talk errors during the ideal 4-qubit layer $G=G_{1}G_{2}=R_{Z_1Z_2}(\alpha_1)R_{Z_3Z_4}(\alpha_2)$, with angles $\alpha_1,\alpha_2\in [0,\pi/2]$. The infidelity $\IF$ is evenly split between the 9 cross-talk Pauli terms. The set of mitigation operations is given by the 4-qubit Pauli group $\mathcal{P}$. The multi-type QP basis demonstrates the general form in Eq.~\eqref{Eq: most general multi-type basis}, and includes 7 non-trivial types: $\mathcal{P}\cup G\mathcal{P}\cup \mathcal{P}G \cup G_1 \mathcal{P} \cup G_2 \mathcal{P} \cup \mathcal{P} G_1 \cup \mathcal{P} G_2$. QP coefficients $c_i$ were obtained by solving analytically Eq.~\eqref{Eq: standard least squares} and \eqref{Eq: linear program 2}, and are first-order-accurate. We plot the blow-up rate as a function of $\alpha_1,\alpha_2$, showing it is approximately equal to the lower bound 2, reaching a maximal value $\approx 2.05$ at $\alpha_1=\alpha_2=\pi/2$. \label{fig: POC 1.5}
}
\end{figure}

Next, we consider the multi-type basis $\mathcal{B}_3=G_\alpha \mathcal{P} \cup \mathcal{P}$, where we treat the second type as a special case of $\mathcal{P}G_{\alpha'}$ with $\alpha'=0$, corresponding to a Pauli layer $P\in\mathcal{P}$ followed by a noisy idle operation. As seen in Fig.~\ref{fig: POC 1}, this multi-type basis produces an optimal first-order mitigation, achieving both the lower bound $(W-1)/\IF=2$ on the blowup rate and an inaccuracy $\Frob_{\text{QP}}/\Frob<\epsilon$, for any angle $\alpha\in (0,\pi/2)$. This basis is used in QESEM to mitigate (non-cross-talk) `during' Pauli errors. Cross-talk errors are similarly mitigated by QESEM using multi-type bases, see Fig.~\ref{fig: POC 1.5}.

\begin{figure}[tb]
\begin{centering}
\includegraphics[width=\columnwidth]{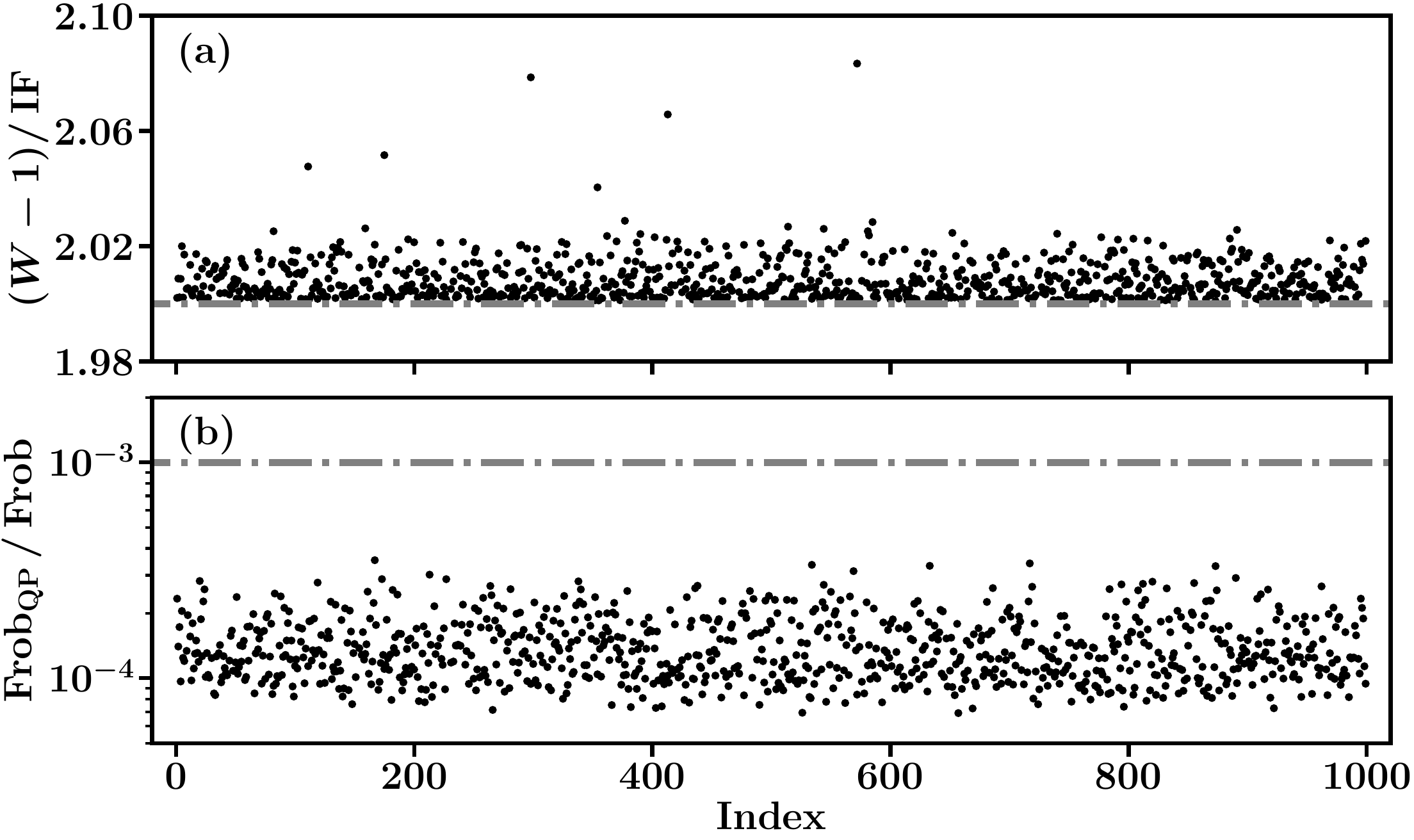}
\par\end{centering}
\caption{Optimal multi-type QP distributions for the family $G_{\alpha,\beta,\gamma}$ of non-Clifford KAK gates. Distributions based on the multi-type basis $\mathcal{B}=G_{\alpha,\beta,\gamma}\mathcal{P}\cup \mathcal{P}G_{\alpha,\beta,\gamma}\cup \mathcal{P}$ are able to attain both an optimal blow-up rate $(W-1)/\IF \approx 2$, and a better-than-first-order accuracy, quantified by the relative Frobenius distance $\Frob_{\text{QP}}/\Frob\ll\epsilon=10^{-3}$. The $x$-axis indexes one-thousand points $\{(\alpha_i,\beta_i,\gamma_i)\}_{i=1}^{1000}$, uniformly sampled from the cube $[0,\pi/2]^3$. \label{fig: POC 2}
}

\end{figure}

The results in Fig.~\ref{fig: POC 1} qualitatively generalize to the larger family of gates $G_{\alpha,\beta,\gamma}=e^{h_{\alpha,\beta,\gamma}+L}$, where the ideal generator is $h_{\alpha,\beta,\gamma}=(i/2)[\alpha XX+\beta YY+\gamma ZZ,\cdot ]$, and the during-error generator $L$ corresponds to single-qubit dissipation. The parameters $\alpha,\beta,\gamma$ are assumed to take values in $[0,\pi/2]$. The fractional iSWAP gate $G_{\alpha,\alpha,0}$ is an important example which has been natively implemented in super-conducting qubit platforms \citep{Rigetti_XY_Gate}. The significance of the family $G_{\alpha,\beta,\gamma}$ stems from the KAK decomposition \citep{KAK}, stating that every two-qubit gate $G$ is ideally equal to some $G_{\alpha,\beta,\gamma}$, up to multiplication by single-qubit layers from the left and right, $G=S_1 G_{\alpha,\beta,\gamma}S_2$. All gates in the family $G_{\alpha,\beta,\gamma}$ can be twirled with the subgroup $\{II,XX,YY,ZZ\}\subset \mathcal{P}$. As shown in Fig.~\ref{fig: POC 2}, for this family of twirled gates, the multi-type basis $\mathcal{B}=G_{\alpha,\beta,\gamma}\mathcal{P}\cup \mathcal{P}G_{\alpha,\beta,\gamma}\cup \mathcal{P}$, along with the non-linearized version of Algorithm \ref{Algo: 1st order QP rep}, produces QP distributions with approximately-optimal blow-up rates, and better-than-first-order accuracy. 

\subsection{Comparison to recent work}

After completing our work on QP distributions and incorporating it into QESEM, Ref.~\cite{Layden2024ErrorMitigationNonClifford} appeared on the arXiv, addressing the same problem discussed in this appendix, the construction of QP distributions for non-Clifford gates. Ref.~\cite{Layden2024ErrorMitigationNonClifford} focused on the gate $G_\alpha=R_{ZZ}(\alpha)$, and proposed the QP basis $\mathcal{P}G_\alpha\mathcal{P}$, a special case of the last type in Eq.~\eqref{Eq: no-reset basis elements}. For $G_\alpha$ with the error model considered in Fig.~\ref{fig: POC 1}, this basis can be shown to be equivalent to a particular multi-type basis, $\mathcal{P}G\cup \mathcal{P}G$,\footnote{To see this, split $\mathcal{P}=\mathcal{P}_c \cup\mathcal{P}_{ac}$, where $\mathcal{P}_c$ ($\mathcal{P}_{ac}$) corresponds to Pauli operators commuting (anti-commuting) with $ZZ$, and note that (i) $\mathcal{P}G_\alpha \mathcal{P}_{c}=\mathcal{P}G_\alpha$, and (ii) that $\mathcal{P}_{ac}G_\alpha \mathcal{P}_{ac}$ is irrelevant for error models we considered.} corresponding to the blue line in Fig.~\ref{fig: POC 1}. As shown in Fig.~\ref{fig: POC 1}(a), this basis gives a near-optimal blowup rate for small $\alpha$, but a highly non-optimal blowup rate for $\alpha$ close to $\pi/2$. Such non-optimality was indeed noted in Ref.~\cite{Layden2024ErrorMitigationNonClifford} (for different error models). As discussed above, the green line in Fig.~\ref{fig: POC 1} demonstrates that an optimal mitigation can be achieved using more general multi-type bases.
For the examples considered in Fig.~\ref{fig: POC 1.5} and \ref{fig: POC 2}, the basis proposed in Ref.~\cite{Layden2024ErrorMitigationNonClifford} does not allow for a first-order-accurate mitigation, which, as shown above, can be achieved with more general multi-type bases.

\section{QPU time estimation and minimization}\label{app:QPU time minimization}

Quantum error mitigation techniques reduce errors in the output of quantum computations with a significant overhead in time complexity. 
QESEM is no different, but to alleviate the overhead as much as possible it utilizes several QPU runtime estimation and optimization techniques, which are covered in this appendix, and are tailored for QESEM's structure and workflow. These estimation and optimization techniques are covered under patent application \cite{qedma_qp_sampling_patent_app}.

The core ideas for runtime estimation are detailed in Sec.~\ref{subsec:circuit-sampling-opt-est}.
The first runtime optimization relies on the fact that QESEM's mitigation involves sampling quantum circuits from a circuit QP distribution. This allows for optimizing over the number of sampled circuits and the number of sampled shots separately, and is also detailed in Sec.~\ref{subsec:cost-opt}.
Another optimization is that of assigning optimal target precision per measurement basis and is detailed in Sec.~\ref{subsec:opt-prec-per-mb}. Lastly, we detail optimization of the estimator in the case of several measurement bases in Sec.~\ref{subsec:obs-choice}.

While these estimation and optimization ideas are utilized by QESEM's error mitigation, they are applicable to other error mitigation algorithms, such as ZNE \cite{IBM_PEC_ZNE} and TEM \cite{Filippov2023}, and other quantum-algorithm-related protocols such as Pauli twirling, sampling of measurement bases, and VQE.

\subsection{Quantum circuit sampling programs}\label{subsec:circuit-sampling-opt-est}

We start by recalling the QP decomposition of the channel $G_\text{QP}$ to channels $G_{\mathcal{C}}$ implemented by (possibly noisy) quantum circuits $\mathcal{C}$
\begin{equation}
    G_{\text{QP}} = \sum_j c_j G^{\mathcal{C}_j}.
\end{equation}
In the following discussion, we ignore the difference between the superoperator $G_\text{QP}$ and the channel we want to emulate via the QP decomposition (see Appendix \ref{app:QP} for discussion on biases in QESEM's QP decomposition).
For an observable $O$ measurable in the standard basis and some initial state $\rho_\text{in}$, we have the relation $O_{\text{target}} = \bbrakketOP{O}{G_{\text{QP}}}{\rho_{\text{in}}} = \sum_{j} c_{j} \braketExp{O}_{\mathcal{C}_j}$, where $\braketExp{O}_{\mathcal{C}}$ is the expectation value of $O$ on the output state of circuit $\mathcal{C}$ as run on the QPU (with infinite shots).
We consider a circuit random variable $C$ with a probability distribution given by $p_j = \operatorname{Prob}(C = \mathcal{C}_j)$.
Then, an unbiased expectation-value estimator for $O_{\text{target}}$ is given by $f_C \braketExp{O}_C$ where $f_{\mathcal{C}_j} = c_j / p_j$.
Commonly, one chooses\footnote{
    This choice is motivated by Monte-Carlo sampling techniques, see Ref.~\cite{IBM_PEC_ZNE}, and is proved to be optimal in some cases \cite{QP_estimation}. This probability distribution can be improved by choosing different probabilities, or by splitting the distributions into complementary sub-distributions and sampling circuits from each, see patent application \cite{qedma_qp_sampling_patent_app} and the more recent Ref.~\cite{chen2025fasterprobabilisticerrorcancellation}.
} 
\begin{equation}\label{eq:canonical-qp-probs}
    p_j = |c_j| / W
    \text{ and }
    f_{\mathcal{C}_j}=W\operatorname{sign}(c_j)~,
\end{equation} 
where $W=\sum_k |c_k|$ is the \emph{QP norm}, or for Pauli twirling one has $p_j=c_j=1/(\text{number of circuits in ensemble})$ and $f_{\mathcal{C}}=1$.

Two variances of importance are the \emph{circuit-to-circuit variance} and \emph{shot-to-shot variance} given by
\begin{align}
    \mathbb{V}_{\text{c}}
    &=
    \mathop{\mathbb{V}}_{C\sim p}[f_C \braketExp{O}_{C}]~, \label{eq:circuit-var-def}
    \\
    \mathbb{V}_{\text{s}}
    &= \mathop{\mathbb{E}}_{C\sim p} \left[
        f_C^2 \bigl(\braketExp{O^2}_{C} - \braketExp{O}_{C}^2\bigr)
    \right]~,  \label{eq:shot-var-def}
\end{align}
respectively. The first relates to the step of sampling circuits, and is unaware of shot noise, and the second to the sampling of shots. 

In a circuit sampling program, we sample $N_{\text{c}}$ quantum circuits $C_j$ for $j=1,\ldots N_{\text{c}}$ and sample $N_{\text{s}}$ shots of each yielding a measurement output per shot $o_{jk}$ for $k=1,\ldots N_{\text{s}}$.
Then, the canonical unbiased estimator for the expectation value $O_{\text{target}}$ is given by
\begin{equation}
    \hat{O}_{\text{tot}} = \frac{1}{N_{\text{c}}} \sum_{j=1}^{N_{\text{c}}} \frac{f_{C_j}}{N_{\text{s}}} \sum_{k=1}^{N_{\text{s}}} o_{jk}~.
\end{equation}
The variance of this estimator is related to $\mathbb{V}_{\text{c}}$ and $\mathbb{V}_{\text{s}}$ by the law of total variance
\begin{equation}
    \mathbb{V}\hat{O}_{\text{tot}} = N_{\text{c}}^{-1} (\mathbb{V}_{\text{c}} + N_{\text{s}}^{-1}\mathbb{V}_{\text{s}})~.
\end{equation}

From the dataset $o_{jk}$, we can also estimate the different variances.
First, we can estimate the expectation value per circuit $\hat{O}_j = N_{\text{s}}^{-1} f_{C_j} \sum_{k=1}^{N_s} o_{jk}$ unbiasedly. Then, the following are  unbiased estimators for $\mathbb{V}\hat{O}_{\text{tot}}$, $\mathbb{V}_{\text{s}}$ and $\mathbb{V}_{\text{c}}$,
\begin{align}
    \hat{S}_{\text{tot}}^2 &= \frac{1}{N_{\text{c}} ( N_{\text{c}} - 1 )}\sum_{j=1}^{N_{\text{c}}} (\hat{O}_j - \hat{O}_{\text{tot}})^2~,
    \label{eq:est-var-tot}\\
    \hat{S}_{\text{s}}^2 &= \frac{1}{N_{\text{c}}} \sum_{j=1}^{N_{\text{c}}} \frac{f_{C_j}^2}{N_{\text{s}} - 1} \sum_{k=1}^{N_{\text{s}}} (o_{jk} - \hat{O}_j)^2,
    \label{eq:est-var-shot}\\
    \hat{S}_{\text{c}}^2 &= N_{\text{c}} \hat{S}_{\text{tot}}^2 - \hat{S}_{\text{s}}^2 / N_{\text{s}}~,
    \label{eq:est-var-circ}
\end{align}
respectively.\footnote{
    Note that the $\hat{S}_{\text{c}}^2$ estimator can take negative values, which is usually unwanted for a variance estimate.
    In such cases, one can replace the value with 0. This also motivates replacing $\hat{S}_{\text{tot}}^2$ with $\max\{\hat{S}_{\text{tot}}^2, \hat{S}_{\text{s}}^2 / N_{\text{s}}\}$ for the variance estimator of $\hat{O}_{\text{tot}}$.
}

\subsubsection{Variance and resource estimation}\label{subsec:variance-estimation-is-practical}

For a QP-based EM, it is common to bound the number of circuit samples required to attain a precision $\epsilon$ within some probability, using Hoeffding's inequality \cite{Hoeffding01031963}, e.g.~Ref.~\cite{IBM_PEC_ZNE}. 
Assuming $\norm{O}_{\text{op}}\leq 1$, we have the bound $|\hat{O}_j| < W$ with $W$ being the QP norm, then the Hoeffding bounds guarantee that if
\begin{equation}\label{eq:standard-hoeffding-bound}
    N_{\text{c}} \geq 2 \log\left(\frac{2}{\delta}\right)\frac{W^2}{\epsilon^2}~,
\end{equation}
then we have an absolute estimation error of $\epsilon$ with probability at least $1-\delta$.
However, a tighter bound, also due to Hoeffding \cite[Eq.~(2.9)]{Hoeffding01031963}, can be used.
The bound involves the common variance of each $\hat{O}_j$, which we denote $v = \mathbb{V}\hat{O}_{j} =  N_{\text{c}}\mathbb{V}\hat{O}_{\text{tot}}$.
Assuming $\epsilon < W+1$, the tighter bound asserts that it suffices to have
\begin{multline}\label{eq:hoeffding-variance-bound}
    N_{\text{c}} 
    \geq \log\left(\frac{2}{\delta}\right) 
    \frac{W+1}{\epsilon}\left[\frac{1+\xi}{\xi}\ln(1+\xi)-1\right]^{-1}~,
\end{multline}
where $\xi = (W+1)\epsilon/v$.
Since this bound is monotonically increasing with $v$, we consider the limit of large variance $\xi\ll 1$. At this regime, the bound is 
\begin{equation}
    N_{\text{c}}^{\text{min}} \simeq 2\log\left(\frac{2}{\delta}\right) \frac{v}{\epsilon^2}.
\end{equation}

Assuming that the variance saturates Popoviciu's inequality $v \leq W^2$, we regain the naive Bound \eqref{eq:standard-hoeffding-bound}.
However, for a smaller variance $v$, we can get a significantly better bound. 
For example, consider mitigating the expectation value of a traceless observable under a global depolarizing model, using QP distributions. In this case it can be shown that for $N_{\text{s}}\simeq W$
\begin{equation}\label{eq:qp-var-glob-depol-bound}
    v \leq 2 W + 1,
\end{equation}
giving a significant improvement for large QP norms.
Note that the regime $\xi\ll 1$ considered in Eq.~\eqref{eq:hoeffding-variance-bound} with the assumption $v=\Theta(W)$ is simply $\epsilon \ll W$ which is a reasonable, since $W\geq 1$.

To use the improved sample bounds, we require the variance or an estimate of it, for which the task of obtaining it might be just as hard as the expectation-value estimation.
We now argue that, in the context of EM, variance estimation is an easier problem than expectation-value estimation when considering a desired relative precision.

We denote by $\hat{S}^2 = N_\text{c}\hat{S}^2_{\text{tot}}$ the variance $v$ estimator. 
The standard deviation of the estimators $\hat{S}^2$ and $\hat{O}_\text{tot}$ vanish as $1/\sqrt{N_\text{c}}$ for large $N_\text{c}$, but their decay coefficients may differ.
On one hand, the relative error of the expectation-value estimator is $\sqrt{\mathbb{V}\hat{O}_\text{tot}}/{|\mathbb{E}\hat{O}_\text{tot}|} \geq \sqrt{v/N_\text{c}}$.
On the other hand, the relative error of the variance estimator $\hat{S}^2$ is $\sqrt{\mathbb{V}\hat{S}^2}/\mathbb{E}\hat{S}^2$.
It is known that $\mathbb{V}\hat{S}^2 = \frac{1}{N_{\text{c}}}\left(\mu_4 - \frac{N_\text{c}-3}{N_\text{c}-1}v^2\right)$ with $\mu_4$ the fourth central moment of $\hat{O}_j$.
We can compare the relative errors
\begin{equation}\label{eq:var-est-EV-est}
    \zeta
    =\frac{
        \sqrt{\mathbb{V}\hat{S}^2}/\mathbb{E}\hat{S}^2
    }{
        \sqrt{\mathbb{V}\hat{O}_\text{tot}}/|\mathbb{E}\hat{O}_\text{tot}|
    }~,
\end{equation}
in two simple scenarios: global depolarizing noise and Gaussian random variables.
In the first scenario, we revisit the mitigation of an expectation value of a traceless observable under a global depolarizing model, using QP distributions and $N_\text{s}\simeq W$. For such a scenario, one can show that $\mu_4 = \mathcal{O}(W^2)$. Together with $v\propto \Omega(W)$ as per Eq.~\eqref{eq:qp-var-glob-depol-bound}. In the second scenario, we consider $\hat{O}_j$ to be random variables with mean in $[-1,1]$, and variance $aW$ for some constant $a$. Thus, we again find $\mu_4 = 3 a^2 W^2$. In both scenarios, we find that the relative-error ratio $\zeta=\mathcal{O}(W^{-1/2})$ as $W\to\infty$, which translates to a significant improvement in variance estimation relative error over the expectation value estimation given the same number of circuit samples.
Moreover, in these two scenarios, the variance estimation requires a fixed number of samples for a given relative error
\begin{equation}
    \sqrt{\mathbb{V}\hat{S}^2} / \mathbb{E}\hat{S}^2 = \mathcal{O}(N_\text{c}^{-1/2})\quad\text{as }W\to \infty~.
\end{equation}

Beyond these two scenarios, $\zeta$ is bound
\begin{equation}\label{eq:var-est-better-than-EV-est}
    \zeta
    \leq \frac{1}{v}\sqrt{W^2 + W^2 / v + 1}~,
\end{equation}
which can be shown using the fourth-central-moment bound given in Theorem~2.1 of Ref.~\cite{sharma2019boundsspreadsmatricesrelated}. This bound assures, in the worst case, variance scaling of $v=\Omega(W^2)$ as $W\to\infty$, this ratio is $\zeta=\mathcal{O}(W^{-1})$, meaning the variance requires fewer samples for a given fixed relative error.
Even for the common scaling of $v=\Omega(W)$ as $W\to\infty$, the ratio is $\zeta=\mathcal{O}(1)$, meaning that for a fixed relative error, we require at most the same number of samples for variance estimation when compared to expectation-value estimation.
For smaller variances, the problem of estimating the expectation value to some relative error might require fewer samples, in which case one probably does not need many samples to directly estimate the expectation value without estimating the variance first.

The key important feature for the ease of variance estimation over the expectation-value estimation is our prior knowledge that the estimator $\hat{O}_j$ lies in a segment of size several orders of magnitude smaller than that of its variance estimator (i.e., $[-1, 1]$ versus $[-W, W]$ for $W\gg 1$).
For this reason, the variance estimation is beneficial also for other quantum EM techniques such as ZNE, where the expectation value has a large decay factor due to noise $\Lambda\gg 1$ and the mitigation effectively ``stretches'' the estimate by this factor, leading to large variances with a small expected ideal value.

\subsubsection{Cost optimization}\label{subsec:cost-opt}

While the bound \eqref{eq:hoeffding-variance-bound} yields a concrete promise of a target precision within some specified probability, we now explore a more practical scheme that estimates the required $N_{\text{c}}$ to obtain a desired precision.
Using estimations of $\mathbb{V}_{\text{c}}$ and $\mathbb{V}_{\text{s}}$, we can plan an experiment with desired properties of the precision and cost $T(N_{\text{c}}, N_{\text{s}})$,
where $T$ is a cost function.
The cost function $T$ can be whatever cost one associates with the program, e.g., the sampling complexity $T=N_{\text{c}}N_{\text{s}}$,
the QPU runtime, or the monetary cost of running the planned program. 
We consider two optimization problems. Minimizing the cost given the target precision $\epsilon$,
\begin{equation}
    \begin{aligned}
        \text{min}\quad 
        &
        T(N_{\text{c}},\, N_{\text{s}})
        \\
        \text{subject to}\quad 
        & \epsilon^2 \geq
        N_{\text{c}}^{-1} (
	       \mathbb{V}_{\text{c}}
	       +
	       N_{\text{s}}^{-1} \mathbb{V}_{\text{s}}
        ),
    \end{aligned}
    \label{eq:plan-min-time-set-prec}
\end{equation}
and minimizing the precision given a cost $t$
\begin{equation}
    \begin{aligned}
	\text{min}\quad 
	&
	N_{\text{c}}^{-1} \left(
	    \mathbb{V}_{\text{c}}
	    +
	    N_{\text{s}}^{-1} \mathbb{V}_{\text{s}}
	\right)
	\\
	\text{subject to}\quad 
	& t \geq T(N_{\text{c}},\, N_{\text{s}})~.
    \end{aligned}
    \label{eq:plan-min-prec-set-time}
\end{equation}
We will consider the cost function 
\begin{equation}\label{eq:simple-cost-function}
    T=N_{\text{c}}(t_{\text{c}} + N_{\text{s}}t_{\text{s}})
\end{equation}
for simplicity.
This cost function can be thought of as the QPU runtime of the program, where $t_{\text{s}}$ is the time to run a single shot on average and $t_{\text{c}}$ signifies controller delays in loading a new circuit to run on the QPU.

In both cases, the minimum is that of $\epsilon^2 t$ for $\epsilon^2 = N_{\text{c}}^{-1} (\mathbb{V}_{\text{c}} + N_{\text{s}}^{-1} \mathbb{V}_{\text{s}})$ and $t=T$. When $\epsilon$ and $t$ are substituted, the expression for $\epsilon^2 t$ depends on $N_\text{s}$, but not on $N_\text{c}$. We are left with an unconstrained single-variable minimization problem. Allowing for non-integer solutions, the optimal number of shots and corresponding minimal $\epsilon^2 t$ are 
\begin{align}
    N_{\text{s}} &= \sqrt{
	   \frac{
        \mathbb{V}_{\text{s}} t_{\text{c}}
      }{
        \mathbb{V}_{\text{c}} t_{\text{s}}
      }
    }~, \\
    \epsilon^2 t &= \left(\sqrt{\mathbb{V}_{\text{c}}t_{\text{c}}} + \sqrt{\mathbb{V}_{\text{s}}t_{\text{s}}}\right)^2~,\label{eq:opt_t_nc_ns}
\end{align}
and the corresponding optimal number of circuits is
\begin{equation}
    N_{\text{c}} \sqrt{\frac{t_{\text{c}}}{\mathbb{V}_{\text{c}}}}
    = \begin{cases}
        \left(\sqrt{\mathbb{V}_{\text{c}} t_{\text{c}}} + \sqrt{\mathbb{V}_{\text{s}} t_{\text{s}}}\right) / \epsilon^2~,
        & \text{given }\epsilon~, \\
        t / \left(\sqrt{\mathbb{V}_{\text{c}} t_{\text{c}}} + \sqrt{\mathbb{V}_{\text{s}} t_{\text{s}}}\right)~,
        & \text{given }t~.
    \end{cases}
\end{equation}
Equation~\eqref{eq:opt_t_nc_ns} shows the optimal cost has two regimes depending on $\eta=\frac{\mathbb{V}_\text{c}t_\text{c}}{\mathbb{V}_{\text{s}}t_\text{s}}$. At $\eta\to\infty$ we are limited by the controller timescale and the circuit-to-circuit variance, and at $\eta\to0$ we are limited by the shot time of the shot-to-shot variance. Both limits are shown in Fig.~\ref{fig:nc_ns_comp}.
To obtain integer values, we can round up to get a good approximate solution to the two minimization problems\footnote{
    In the special case of $\mathbb{V}_{\text{c}} = 0$ there is no reason to sample more than a single circuit,
    so a solution to problem \eqref{eq:plan-min-time-set-prec} is
    $N_{\text{c}}=1$ and $N_{\text{s}}=\left\lceil \mathbb{V}_{\text{s}} / \epsilon^{2} \right\rceil$.
    In the special case that $t_{\text{c}}=0$ or negligible, a solution is one-shot per circuit $N_{\text{s}}=1$ and $N_{\text{c}}=\left\lceil \mathbb{V}_{\text{s}} / \epsilon^{2} \right\rceil$.
}.

These solutions suggest the following optimized runtime procedure, where a small variance estimation batch is executed before a main batch of circuits. The basic procedure corresponding to Problem~\eqref{eq:plan-min-time-set-prec} is detailed in Algorithm~\ref{algo:runtime-opt} [a similar one can be applied to Problem~\eqref{eq:plan-min-prec-set-time}].
This procedure can be further refined for example by: aggregating the estimation from both batches and not just the main batch (this also reduces the required precision in the main batch); running more than a single main batch to mitigate drift errors (see App.~\ref{app:drift}); use relative precision instead of absolute precision; and adding more batches with each batch correcting the estimations of remaining required number of circuits and number of shots to run.
The principle running a variance estimation batch is applied to every job executed by QESEM.

\begin{figure}[b]
\begin{algorithm}[H]
\caption{Optimal estimation for circuit sampling programs}\label{algo:runtime-opt}
    \begin{flushleft}
        \Input Quantum circuit $\mathcal{C}$, observable $O$, precision $\epsilon > 0$, cost function $T$, and a circuit sampling function \textsc{getSample} 
        \Comment{
            \textsc{getSample} is part of the circuit sampling program,
            and takes as input a circuit and an observable and outputs a circuit and an observable.
            The sampled observable is used in a few use cases: for Pauli twirling, it is the unaltered observable; for a QP distribution, it is the observable with the QP-related weights $f_\mathcal{C}$ included; and for measurement bases sampling, it is an observable measurable in a sampled measurement basis.
        }
        \Output Estimate of $\langle O\rangle_C$ with precision $\epsilon$ obtained with minimal QPU time
    \end{flushleft}

    \begin{algorithmic}[1]
        \State Constants $N_\textrm{c}^{\textrm{init}}$, $N_\textrm{s}^{\textrm{init}}$ are set independent of the sampling procedure
        \State \LComment{Variance estimation batch}
        \State Sample quantum circuits by invoking \textsc{getSample} $N_\textrm{c}^{\textrm{init}}$ times.
        The samples are a stack of pairs per sample (\texttt{sampled\_circuit}, \texttt{sampled\_observable}).
        \State Execute $N_\textrm{s}^{\textrm{init}}$ shots per of each sampled circuit and obtain matching counts per \texttt{sampled\_circuit}.
        \State \LComment{Plan optimal batch}
        \State Estimate the \emph{circuit-to-circuit variance} and \emph{shot-to-shot variance} from returns counts via Eqs.~\eqref{eq:est-var-circ} and \eqref{eq:est-var-shot}
        \State Solve Problem~\eqref{eq:plan-min-time-set-prec} with precision $\epsilon$ and obtain optimal sampling numbers $N_\textrm{c}$ and $N_\textrm{s}$ so as to minimize $T$
        \State \LComment{Run optimal batch}
        \State Sample $N_\textrm{c}$ quantum circuits from circuit ensemble based on $C$ and $O$.
        \State Execute $N_\textrm{s}$ shots per of each sampled circuit and obtain matching counts
        \State Evaluate the expectation value from the counts
    \end{algorithmic}
\end{algorithm}
\end{figure}

\subsubsection{Cost estimation and bounds}

For the qualitative understanding of the behavior of $\mathbb{V}_\text{c}$ and $\mathbb{V}_\text{s}$ and the resulting optimal cost $T$ as per Eq.~\eqref{eq:simple-cost-function}, we derive several estimations and bounds.
We start by giving a re-derivation of the semi-empirical and phenomenological models that appear in the main text [Eqs.~\eqref{eq:semi-empirical-vc-vs-pauli-main} and \eqref{eq:phenomenological-vc-vs-pauli}]. 
In line with an error mitigation application, $O_\text{target}$ is the expectation value of the observable $O$ on some circuit $\mathcal{C}_0$ ran on a fault-free quantum computer, $O_\text{target} = \braketExp{O}_\mathcal{C_0}^{(\text{ideal})}$. Correspondingly, we also have an expectation value of $O$ on the output state when the circuit $\mathcal{C}_0$ in the presence of QPU errors (in the infinite shots limit),
\begin{equation}
    O_\text{noisy} = \braketExp{O}_{\mathcal{C}_0}.
\end{equation}
These two quantities play key roles in the following approximation and bounds.

We can use them to approximate $\mathbb{V}_\text{c}$ and $\mathbb{V}_\text{s}$, which by definitions Eqs.~\eqref{eq:circuit-var-def} and \eqref{eq:shot-var-def}, satisfy $\mathbb{V}_{\text{c}}=W^2\mathbb{E}_{C\sim p}\braketExp{O}_C^2 - O_{\text{target}}^2$ and $\mathbb{V}_{\text{s}} = W^2 \mathbb{E}_{C\sim p}[\braketExp{O^2}_C - \braketExp{O}_C^2]$.
A good rule of thumb is to approximate $\mathbb{E}_{C\sim p}\braketExp{O}_C^2 \approx O_\text{noisy}^2$. This approximation is not trivial, but can be roughly justified as follows. In an EM application, the expectation value of interest is usually large; otherwise, error mitigation is usually not practical. The errors cannot be too large as well, so $O_\text{noisy}$ is not too small. The QP-related ensemble of circuits is composed of the target circuit $\mathcal{C}_0$ with several local modifications as per the local QP decompositions (see App.~\ref{app:QP}) that roughly behave as inserting errors in the circuit $\mathcal{C}_0$ on average. Under these conditions we expect that $\mathbb{E}_{C\sim p}\braketExp{O}_C^2 \lesssim O_\text{noisy}^2$. This leads us to the first approximation for a Pauli operator $O$
\begin{equation}\label{eq:almost-semi-empirical-vc-vs-pauli}
    \mathbb{V}_\text{c} \approx W^2 O_{\text{noisy}}^2 -O_{\text{target}}^2, \quad \mathbb{V}_\text{s} \approx W^2 -W^2 O_{\text{noisy}}^2.
\end{equation}
For large volume circuits, we expect the circuit-to-circuit variance to be dominated by the QP norm, and we are led to the semi-empirical approximation
\begin{equation}\label{eq:semi-empirical-approx}
    \text{(semi-empirical)}\quad\begin{aligned}
        \mathbb{V}_\text{c} &\approx W^2 O_{\text{noisy}}^2,\\
        \mathbb{V}_\text{s} &\approx W^2 -W^2 O_{\text{noisy}}^2.
    \end{aligned}
\end{equation}

The reason for the term semi-empirical, is because the $O_\text{noisy}$ can be replaced by an expectation-value estimate obtained by running the circuit $\mathcal{C}_0$ on the actual QPU with a finite number of shots (and with Pauli twirling if the noise model that the QP decomposition is based on requires it).
Eq.~\eqref{eq:semi-empirical-approx} yields an estimate that requires even less QPU time than an empirical estimation via the estimators \eqref{eq:est-var-circ} and \eqref{eq:est-var-shot}, and although it is less reliable, it often succeeds in predicting the empirically estimated values.

We can avoid the requirement of empirically collected data and obtain a phenomenological model (i.e., with some small number of free parameters that can be fitted to small problem sizes) by replacing $O_\text{noisy}$ with an expression with $O_\text{target}$. To this end we relate $O_\text{noisy}=e^{-\IF \times V_{\text{eff}}} O_{\text{target}}$ where $\IF$ is the average (over the circuit frequency) of gate entanglement infidelity and $V_\text{eff}$ is the effective volume.
The effective volume is commonly replaced in QP-based mitigations by using the active volume $V_\text{ac}$, or some approximation thereof (see Def.~\ref{def:active-vol} in App.~\ref{appendix:active-volume}) to reduce the mitigated volume and its related (exponential) sampling overhead.
On the one hand, the size of active volume (the number of gates in the region enclosed by the active volume) of the mitigated circuit and observable expectation value is bounded from below by the effective volume
\begin{equation}\label{eq:veff-over-vactive}
    V_{\text{eff}} = r |V_{\text{ac}}| \leq |V_{\text{ac}}|~.
\end{equation}
On the other hand, the active volume and the average infidelity can be used to lower bound the QP norm as per Eq.~\eqref{Eq: OIB 1}.
Since QESEM's QP decomposition is close to saturating this lower bound, 
\begin{equation}\label{eq:phenom-I}
    \text{(phenomenological I)}\quad W = e^{2 \IF |V_\text{ac}|}~,
\end{equation}
we are led to 
\begin{equation}\label{eq:phenom-II}
    \text{(phenomenological II)}
    \quad
    \begin{aligned}
        \mathbb{V}_\text{c} &\approx (W^{2-r} -1) O_\text{target}^2~, \\
        \mathbb{V}_\text{s} &\approx W^2 - W^{2-r} O_\text{target}^2~.
    \end{aligned}
\end{equation}
Expressions \eqref{eq:phenom-I} and \eqref{eq:phenom-II} together with the optimized cost \eqref{eq:opt_t_nc_ns} are those used in Fig.~\ref{fig:QA-future}.
These are also the equations used in Ref.~\cite{YEMpaper} to derive projections of QESEM's QPU runtime on future hardware.

The phenomenological model is useful in the analysis of a mitigation application of a problem with varying problem size.
By first obtaining or estimating the QPU properties $t_\text{c}$, $t_\text{s}$ and $\IF$, one can assume some functional dependence of $|V_\text{ac}|$, $O_\text{target}$ and $r$ on the problem size and fit these three according to this assumption and empirical data on small problem sizes. 

For a composite observable $O=\sum_P \alpha_P P$, one often assigns a different QP norm and assigns the factors $f_C$ based also on the active volume associated with each Pauli.
Thus, the sampling overhead is reduced, not by the size of the active volume of the full observable (which for a non-adaptive circuit with magnetization is the whole circuit volume), but by the size of the active volume of each of its local constituents.
We denote the QP norm associated with the Pauli operator $P$ by $W_P$.
Upon re-deriving Eq.~\eqref{eq:almost-semi-empirical-vc-vs-pauli}, one can then neglect correlations between Pauli operators to find
\begin{equation}\label{eq:semi-empirical-vc-vs-comp}
\begin{aligned}
    \mathbb{V}_{\text{c}} &\approx \sum_P\alpha_P^2 W_P^2 P_\text{noisy}^2 - O_{\text{target}}^2~,
    \\
    \mathbb{V}_{\text{s}} &\approx \sum_P\alpha_P^2 W_P^2\left(1 - P_\text{noisy}^2\right)~.
\end{aligned}
\end{equation}
From the last pair of approximations, we can deduce a semi-empirical estimate
\begin{equation}
\begin{aligned}
    \mathbb{V}_{\text{c}} &\approx \sum_P\alpha_P^2 W_P^2 P_\text{noisy}^2~,
    \\
    \mathbb{V}_{\text{s}} &\approx \sum_P\alpha_P^2 W_P^2\left(1 - P_\text{noisy}^2\right)~,
\end{aligned}
\end{equation}
and phenomenological model
\begin{equation}\label{eq:phenom-no-cov}
\begin{aligned}
    \mathbb{V}_{\text{c}} &\approx \sum_P\alpha_P^2 W_P^{2-r} P_\text{target}^2 - O_\text{target}^2~,
    \\
    \mathbb{V}_{\text{s}} &\approx \sum_P\alpha_P^2 \left(W_P^2 - W_P^{2-r}P_\text{target}^2\right)~.
\end{aligned}
\end{equation}
The phenomenological model can be refined by allowing different $r$ ratios for different Pauli observables.

The variance estimates given by the phenomenological model for a general observable, Eq.~\eqref{eq:phenom-no-cov}, can be modified to include the effect of covariance between Pauli observables expectation-value estimators.
To that end, we define the \emph{circuit-to-circuit covariance} and \emph{shot-to-shot covariance} of the observables $O$ and $O'$ as
\begin{equation}
    \Cov_\text{c}(O,O') = \operatorname*{\Cov}_{C\sim p}[f_{C;O}\braketExp{O}_C, f_{C;O'}\braketExp{O'}_C],
\end{equation}
and
\begin{equation}
    \Cov_\text{s}(O,O') = \operatorname*{\mathbb{E}}_{C\sim p}[f_{C;O} f_{C;O'}(\braketExp{O O'}_C - \braketExp{O}_C\braketExp{O'}_C)],
\end{equation}
respectively, where the factors $f_{C;O}$ and $f_{C;O'}$ are $f_{C}$ factors but each only includes contributions from local QP decompositions that lie in the active volumes used for the two observables $O$ and $O'$, respectively.
We can then write
$\mathbb{V}_{\text{c}/\text{s}}=\sum_{P,Q} \alpha_P \alpha_Q \Cov_{\text{c}/\text{s}}(P,Q)$.
Next, note that for the probabilities \eqref{eq:canonical-qp-probs} we have $f_{C;P} f_{C;Q} = \frac{W_P W_Q}{W_{P\triangle Q}}f_{C;P\triangle Q}$ where $W_{P\triangle Q}$ and $f_{C;P\triangle Q}$ are the QP norm and $f_C$-factor accounting only for local QP decompositions that lie in the symmetric difference of the active volumes of $P$ and of $Q$.
Following the approximations leading to Eq.~\eqref{eq:almost-semi-empirical-vc-vs-pauli}, we find
\begin{equation}
\begin{aligned}
    \operatorname*{\mathbb{E}}_{C\sim p} f_{C;P} f_{C;Q}\braketExp{P}_C\braketExp{Q}_C 
    &\approx P_\text{noisy} Q_\text{noisy} \operatorname*{\mathbb{E}}_{C\sim p} f_{C;P} f_{C;Q} \\
    &= P_\text{noisy} Q_\text{noisy} \frac{W_P W_Q}{W_{P\triangle Q}}~.
\end{aligned}
\end{equation}
Moreover, we can write
\begin{equation}
    \operatorname*{\mathbb{E}}_{C\sim p} f_{C;P} f_{C;Q}\braketExp{P P'}_C = \frac{W_P W_Q}{W_{P\triangle Q}} \braketExp{PQ}_{\mathcal{C}_0}^{(\text{noise $P\cap Q$})}
\end{equation}
where $\braketExp{O}_{\mathcal{C}}^{(\text{noise $P\cap Q$})}$ is the expectation value of $O$ evaluated on the output of circuit $C$ in the presence of QPU noise only on the intersection of the active volumes of $P$ and $Q$. Then, we can extend the phenomenological model via
$\braketExp{PQ}_{\mathcal{C}_0}^{(\text{noise $P\cap Q$})} = e^{-\IF \times V_\text{eff}(P\cap Q)} (PQ)_\text{target}$ where $V_\text{eff}(P\cap Q)$ is the effective volume of $\braketExp{PQ}_{\mathcal{C}_0}^{(\text{ideal})}$ in presence of noise limited to the intersection of the active volumes of $P$ and $Q$. Finally, the phenomenological model with the inclusion of covariances between Pauli expectation-value estimators
\begin{equation}
\begin{aligned}
    \mathbb{V}_\text{c} &\approx \sum_{P,Q} \alpha_P \alpha_Q \frac{ W_P^{1-r/2} W_Q^{1-r/2}}{W_{P\triangle Q}} (PQ)_\text{target} - O_\text{target}^2~, \\
    \mathbb{V}_\text{s} &\approx \sum_{P,Q} \alpha_P \alpha_Q \left(\frac{ W_P W_Q}{W_{P\triangle Q}^{1+r/2}} - \frac{ (W_P W_Q)^{1-r/2}}{W_{P\triangle Q}}\right) (PQ)_\text{target}~.
\end{aligned}
\end{equation}

\begin{figure}[tb]
\centering
    \includegraphics[width=\columnwidth]{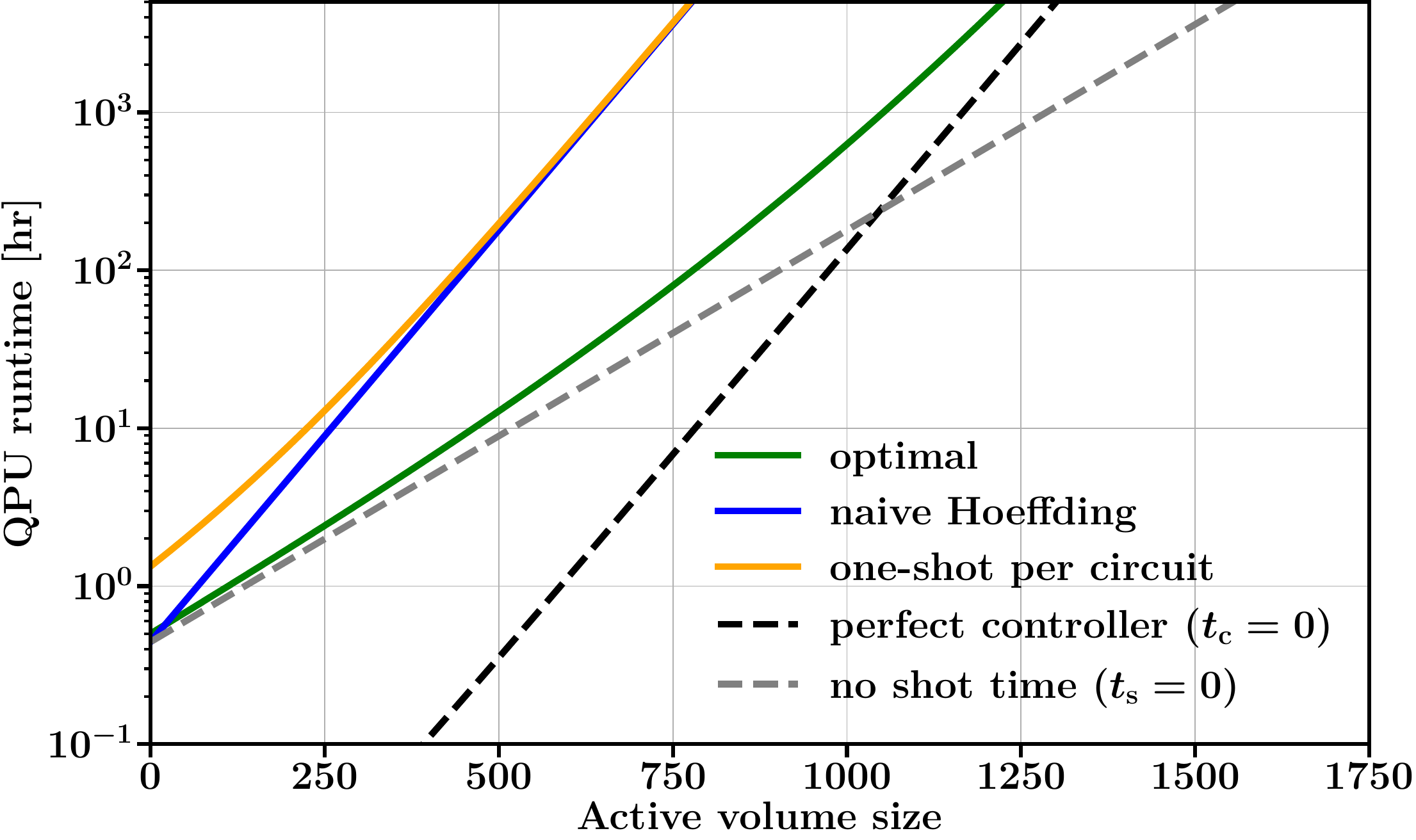}
    \caption{
        Simulated comparison of QPU runtime for $t_{\text{c}}=\qty{0.16}{sec}$, $t_{\text{s}}=\qty{300}{\micro s}$, $\epsilon=0.01$, gate entanglement infidelity of $\IF=\num{3e-3}$, which are values close to those encountered in the benchmarks in Secs.~\ref{main_demo} and \ref{vqe_demo}.
        We use expression $\mathbb{V}_{\text{c}}=W$ and $\mathbb{V}_{\text{s}}=W^2 + W$ found by the bounds \eqref{eq:vc_vs_glob_depol_bounds} and $W=e^{2 \IF \times \text{volume}}$ taken from \eqref{Eq: OIB 1}, where the volume here is the active volume size (see Def.~\ref{def:active-vol}).
        The optimal solution (green) as per \eqref{eq:opt_t_nc_ns} interpolates between being limited by the controller (dashed-gray) and being limited by the shot-time (dashed-black). The runtime of one shot per circuit (orange) and the naive Hoeffding bound of $N_\text{c}=W^2/\epsilon^2$ and $N_\text{s}=1$ are significantly larger than the optimized QPU time.
        \label{fig:nc_ns_comp}
    }
\end{figure}

Next, we discuss bounds on $\mathbb{V}_\text{c}$ and $\mathbb{V}_\text{s}$ for the case of a QP-based mitigation of global depolarizing error channel applied to an $n$ qubit circuit with $L$ layers after each layer.
As discussed above, we consider the mitigation in the context of expectation-value estimation of an observable on the output state of the circuit $\mathcal{C}_0$ (i.e., $O_\text{target}$). We assume that the operator norm is bounded $\norm{O}_\text{op}\leq 1$ in the following.
In such a case, expectation values evaluated on the QPU will be reduced by some factor $\lambda$ per layer, yielding
\begin{equation}
    \braketExp{O}_\mathcal{C} = \lambda^L \braketExp{O}^{(\text{ideal})}_\mathcal{C} + (1-\lambda^L) O_\text{MM}~,
\end{equation}
where $O_\text{MM}=\operatorname{tr}O/2^n$ is the expectation value of $O$ at the maximally-mixed state.
The corresponding QP distribution, obtained via ``Clifford mitigation'' (see Appendix \ref{app:QP}), has a QP norm of
\begin{equation}
    W = \lambda^{-2L(1-4^{-n})}~,
\end{equation}
and the circuit distribution mimics the noise exactly
\begin{equation}
    \mathbb{E}_{C\sim p}\braketExp{O}^{(\text{ideal})}_{C} = \lambda^L \braketExp{O}^{(\text{ideal})}_{\mathcal{C}_0} + (1-\lambda^L) O_\text{MM}~.
\end{equation}
Using these, it is straightforward to show that 
\begin{equation}\label{eq:global-depol-vc-vs-bound}
\begin{aligned}
    \mathbb{V}_\text{c} &\leq W + W|O_\text{MM}|(W |O_\text{MM}| +4)~,\\
    \mathbb{V}_\text{s} &\leq W^2 \Delta^2(O)_\text{MM} + W(1 + 4|O_\text{MM}|)~, 
\end{aligned}
\end{equation}
where $\Delta^2(O)_\text{MM}=\operatorname{tr}(O^2)/2^n - O_\text{MM}^2$ is the quantum variance of $O$ at the maximally-mixed state.

For a traceless observable, we find 
\begin{equation}\label{eq:vc_vs_glob_depol_bounds}
    \mathbb{V}_{\text{c}} \leq W,\quad \mathbb{V}_{\text{s}} \leq W^2 + W
\end{equation}
which can be used to derive Inequality~\eqref{eq:qp-var-glob-depol-bound}. In Fig.~\ref{fig:nc_ns_comp}, we use Bounds \eqref{eq:vc_vs_glob_depol_bounds}, assuming $\mathbb{V}_\text{c}$ and $\mathbb{V}_\text{s}$ saturate them, to compare the optimal cost to an unoptimized cost and the naive Hoeffding bound, Eq.~\eqref{eq:standard-hoeffding-bound}. The naive Hoeffding bound is used with a single shot per circuit, similarly to the analysis in Ref.~\cite{IBM_PEC_ZNE}, and we neglect the $2\log(2/\delta)$ term, which can only increase the Hoeffding bound.
Figure \ref{fig:nc_ns_comp} shows that while the large QP norm limit (or equivalently, the large active volume in the figure) scales the same in the optimal and the unoptimized costs, there are a few orders of magnitude separating them.

The large QP norm is dominated by the shot-to-shot variance, which scales as $W^2$, so the optimized and naive Hoeffding result coincide on a device with a perfect controller, i.e., $t_\text{c}=0$. In other words, the optimization is seemingly unneeded on devices with a perfect controller. While, as discussed in Sec.~\ref{subsec:variance-estimation-is-practical}, Algorithm \ref{algo:runtime-opt}, for optimizing the cost, does not increase the runtime of a mitigation program; it is worthwhile to note that even on devices with optimal control, it can still reduce the runtime significantly. 
To see this, we note that the scaling behavior for a device with a perfect controller is dominated by $\mathbb{V}_\text{s}$ and in the toy model we are considering by the quantum variance $W^2 \Delta^2(O)_\text{MM}$. This latter variance can be a significantly smaller one, yielding a better bound on the optimal cost. For example, if $O$ is a bit-string projector we find $\mathbb{V}_\text{s}= \mathcal{O}(W^2 / 2^n)$. This factor of $2^n$ matches an exponential improvement in the optimal cost over the naive Hoeffding bound.

\subsubsection{Multistep sampling}

As an aside, note that the two variances $\mathbb{V}_\text{c}$ and $\mathbb{V}_\text{s}$ correspond to the two steps of sampling in the program: the first sampling of quantum circuits and the second the sampling of shots of each circuit. We can generalize this to multistepped sampling. Such multistep programs occur in many use cases in quantum computations and EM applications, for example:
\begin{enumerate}
    \item The case where we want to sample expectation values of observables from an ensemble of ideal quantum circuits with EM. Such an ensemble of ideal quantum circuits is encountered in disorder realizations or other variational quantum circuits.
    \item If we want to apply both circuit sampling from measurement-bases-based ensemble and EM-based ensemble.
    \item While sampling for EM usually comes hand in hand with sampling from a twirled circuit, these two samplings can be separated to two distinct steps. This can be useful if the timescale for running different twirled versions of an error-mitigated circuit and running different EM samples differ. The difference in timescale can be significant as some quantum hardware vendors allow for parallelization of compilation to controller of Pauli-twirl circuit re-compilations.
\end{enumerate}
For example, in a three-stepped sampling, where we sample $N_1$ samples, for each we sample $N_2$ samples in the second step, and for each we sample $N_3$ in the third step, we find an estimator $\hat{X} = (N_1 N_2 N_3)^{-1} \sum_{i=1}^{N_1} \sum_{j=1}^{N_2} \sum_{k=1}^{N_3} x_{ijk}$ with corresponding variances $\mathbb{V}_a$ for $a=1,2,3$ satisfying
\begin{equation}\label{eq:three-step-var-rel}
    \mathbb{V}[\hat{O}] = \frac{\mathbb{V}_1}{N_1} + \frac{\mathbb{V}_2}{N_1 N_2} + \frac{\mathbb{V}_3}{N_1 N_2 N_3}~,
\end{equation}
and $\mathbb{V}_a$-s can be estimated from a given dataset.

\subsection{Measurement bases optimization}\label{subsec:meas-bases-opt}

We now describe two more optimization techniques for the case where a few measurement bases are used. The first is to optimize the resources expended on each measurement basis, and the second is to choose efficient estimators for each measurement basis. We do not discuss here the optimization done when choosing which measurement bases to run on the QPU.
For the case of several measurement bases, the estimand is
\begin{equation}\label{eq:sum-observables}
    D = \sum_{i} \braketExp{O_{i}}_{\mathcal{C}_i},
\end{equation}
with $i$ going over the different measurement bases, $\mathcal{C}_i$ circuit implementing the different measurement bases as a final gate layer, and $O_i$ the corresponding observables measurable in each measurement basis.
Using independent estimators $\hat{O}_i$ of $\braketExp{O_{i}}_{\mathcal{C}_i}$ we have the $D$-estimator $\hat{D} = \sum_i \hat{O}_i$, with variance $\mathbb{V}\hat{D} = \sum_{i} \mathbb{V}\hat{O}_{i}$.

\subsubsection{Optimal precision per basis}\label{subsec:opt-prec-per-mb}

We relate the $i$-th estimator with a precision-dependent cost $T_i(\epsilon_i) = \chi_i / \epsilon_i^2$ for some $\chi_i$, a total cost of $T=\sum_i T_i$, and $\epsilon_i$ is the target precision for the $i$-th expectation value estimate, i.e., $\epsilon_i^2=\mathbb{V}\hat{O}_i$.

The values $\chi_i$ depend on the timescale and variance of the $\hat{O}_i$ estimator. If $\hat{O}_i$ is a single circuit sampled some number of shots we have $\chi_i = t_{\text{s}} (\braketExp{O_i^2}_{\mathcal{C}_i} - \braketExp{O_i}_{\mathcal{C}_i}^2)$.
Alternatively, if $\hat{O}_i$ is obtained from a QP-based EM (or some other circuit sampling program) with the simple cost function \eqref{eq:simple-cost-function}, then we can use Eq.~\eqref{eq:opt_t_nc_ns} to find 
\begin{equation}\label{eq:chi_i-opt-circuit-sampling}
    \chi_{i}=(\sqrt{\mathbb{V}_{i,\text{c}}t_{\text{c}}}+\sqrt{\mathbb{V}_{i,\text{s}}t_{\text{s}}})^2
\end{equation}
with $\mathbb{V}_{i,\text{c}}$ and $\mathbb{V}_{i,\text{s}}$ the circuit-to-circuit and shot-to-shot variance associated with $\hat{O}_i$, respectively.

For the problem of minimizing the total cost given a total precision $\epsilon$,
\begin{equation}\label{eq:tagged_sum_opt_runtime}
    \begin{aligned}
        \text{min}\quad & \sum_{i} T_{i}(\epsilon_{i})
        \\
        \text{subject to}\quad & \epsilon^2 =\sum_{i} \epsilon_{i}^2~,
    \end{aligned}
\end{equation}
we have the solution
\begin{equation}
    \epsilon_{i}^2 = \epsilon^2 \frac{\sqrt{\chi_{i}}}{\sum_{j} \sqrt{\chi_{j}}}.
\end{equation}
Moreover, for the problem of minimizing the precision given a total cost of $t$,
\begin{equation}\label{eq:tagged_sum_opt_precision}
    \begin{aligned}
        \text{min}\quad & \sum_{i} \epsilon_{i}^2
        \\
        \text{subject to}\quad & t = \sum_{i} T_{i}(\epsilon_{i}),
    \end{aligned}
\end{equation}
the solution is
\begin{equation}
    \epsilon_{i}^2 = \frac{\sqrt{\chi_{i}}}{t} \sum_{j}\sqrt{\chi_{j}}~.
\end{equation}

These two solutions mesh well with the scheme of using an initial variance estimation batch as in Algorithm~\ref{algo:runtime-opt}, matching the optimizations of Sec.~\ref{subsec:cost-opt}.
The initial variance estimation batch can be used to estimate the different $\chi_i$, and then the optimizations \eqref{eq:tagged_sum_opt_precision} and \eqref{eq:tagged_sum_opt_runtime} can be solved to plan a main batch of circuits to run.
This is exemplified in the Algorithm~\ref{algo:runtime-opt-meas-bases}, which can be refined similarly to Algorithm~\ref{algo:runtime-opt} (see discussion in Sec.~\ref{subsec:cost-opt}).
The principle of this Algorithm, of using a batch for $\chi$ estimation and optimizing for the precision in each measurement basis, is applied by QESEM for each job requiring multiple measurement bases in conjunction with Algorithm~\ref{algo:runtime-opt}.

\begin{figure}[!th]
\begin{algorithm}[H]
\caption{Optimal estimation of multiple circuit}\label{algo:runtime-opt-meas-bases}
    \begin{flushleft}
        \Input Quantum circuits $C_1,\ldots C_M$, and matching observables $O_1,\ldots O_M$, precision $\epsilon > 0$, and cost function $T$
        \Output Estimate of $\sum_i\langle O_i\rangle_{C_i}$ with precision $\epsilon$ obtained with minimal QPU time
    \end{flushleft}
        
    \begin{algorithmic}[1]
        \State \LComment{$\chi$ estimation batch}
        \State Execute quantum programs on QPU to estimate each $\langle O_i \rangle_{C_i}$ with equal cost allotted to each.
        \State \LComment{Plan optimal batch}
        \State Estimate all $\chi_i$ based on returned counts from QPU execution
        \Comment{
            This is based on variance estimation, see, e.g., Eq.~\eqref{eq:chi_i-opt-circuit-sampling}
        }
        \State Solve Problem~\eqref{eq:tagged_sum_opt_precision} with precision $\epsilon$ and obtain optimal precision per circuit $\epsilon_i$
        \State Calculate, based on $\epsilon_i$, the required number of samples per $i$ 
        \State \LComment{Run optimal batch}
        \State Execute quantum programs on QPU to estimate each $\langle O_i \rangle_{C_i}$ with the optimal number of samples.
        \State Evaluate $\sum_i\langle O_i\rangle_{C_i}$ based on estimated in previous step.
    \end{algorithmic}
\end{algorithm}
\end{figure}

As an aside, we note that the expression \eqref{eq:sum-observables} can be used for more than measurement bases, for example in the estimation of the super-operator $-i[ZZ,\cdot]$ using parameter shift rules \cite[Eq.~(7)]{VQE-rev-2021}, and even beyond that we can use a functional expression $D = f(\braketExp{O_1}_{\mathcal{C}_1}, \ldots, \braketExp{O_M}_{\mathcal{C}_M})$ using the estimator $\hat{D} = f(\hat{O}_1,\ldots, \hat{O}_M)$. For a function $f$ one can repeat the optimization problems \eqref{eq:tagged_sum_opt_precision} and \eqref{eq:tagged_sum_opt_runtime} with an approximate variance $\mathbb{V}\hat{D} \approx (\partial_j f)^2 \mathbb{V}\hat{O}_i$.\footnote{
    An estimate of $(\partial_i f)^2$ might require an expectation-value estimate, nullifying the benefit of estimating the variance alone [recall Eq.~\eqref{eq:var-est-better-than-EV-est}]. In such cases, one should validate the estimate of $(\partial_i f)^2$ in later batches.
}\footnote{Using the variance approximation of $\mathbb{V}\hat{D}$ can be used as a first step in the optimization to find an initial point before a finer optimizer using a more complex expression for $\mathbb{V}\hat{D}$.}
As an example for a function, consider two-point exponential ZNE with the unmitigated, noisy, expectation-value and an expectation at some point with the noise-amplified by $s$, $D=\braketExp{O}_\text{target} = \braketExp{O}_1\left(\braketExp{O}_1 / \braketExp{O}_s\right)^{1/(s-1)}$. Then we can find 
\begin{equation}
    \begin{aligned}
    \epsilon_1^2 &= \frac{\sqrt{\chi_1}}{\frac{s}{s-1}e^{\Lambda} } \cdot \frac{\epsilon^2}{\frac{s}{s-1} e^{\Lambda} \sqrt{\chi_1} + \frac{1}{s-1} e^{s\Lambda} \sqrt{\chi_s}},
    \\
    \epsilon_s^2 &= \frac{\sqrt{\chi_s}}{\frac{1}{s-1}e^{s\Lambda} } \cdot \frac{\epsilon^2}{\frac{s}{s-1} e^{\Lambda} \sqrt{\chi_1} + \frac{1}{s-1} e^{s\Lambda} \sqrt{\chi_s}},
    \end{aligned}
\end{equation}
where $\Lambda$ is the decay factor $\log(\braketExp{O}_s / \braketExp{O}_{\text{target}}) = -s\Lambda $. One might also optimize over the amplification $s$ before running another batch, but that requires some further assumptions on the dependence of $\chi_s$ on $s$.

\subsubsection{Observable choice}\label{subsec:obs-choice}

In many cases, the observable choice for each circuit $\mathcal{C}_i$ in Eq.~\eqref{eq:sum-observables} is not unique.
As a motivating example, consider the expectation value of $O = Z_1 + X_1 + Z_2$ where the indices correspond to the measured qubits, for some circuit $\mathcal{C}_1$. 
To estimate $D=\braketExp{O}_{\mathcal{C}_1}$ we use two circuits. The original $\mathcal{C}_1$ to estimate $\braketExp{Z_1}_{\mathcal{C}_1}$ and $\braketExp{Z_2}_{\mathcal{C}_1}$, and the circuit $\mathcal{C}_2$ which is $\mathcal{C}_1$ with a Hadamard gate appended on qubit 1 to estimate $\braketExp{Z_1}_{\mathcal{C}_2} = \braketExp{X_1}_{\mathcal{C}_1}$ and $\braketExp{Z_2}_{\mathcal{C}_2}=\braketExp{Z_2}_{\mathcal{C}_1}$.\footnote{
    This simple example can also be resolved by using a single circuit in which we append an $R_Y(-\pi/4)$ gate on qubit 1. However, such circuit optimizations are not always as trivial and may increase the volume significantly, and our focus in this section is on the optimization over the choice of observables per circuit.
}
Overall,
\begin{equation}
    \braketExp{O}_{\mathcal{C}_1} = \braketExp{Z_1 + \lambda Z_2}_{\mathcal{C}_1} + \braketExp{Z_1 + (1-\lambda)Z_2}_{\mathcal{C}_2}
\end{equation}
with any scalar $\lambda$.
The freedom in choosing $\lambda$ allows us to choose an optimal estimator in the sense that its variance is minimized.

While the choice of observables is continuous, as in the example above, there is usually a finite number of decompositions under consideration that span the space of candidate decompositions. 
We label each decomposition in the spanning set by some label set $\mathcal{A}$, this way each labeled decomposition is specified by its observables $O_i^{I}$ so that $D = \sum_{i} \braketExp{O_{i}^{I}}_{\mathcal{C}_i}$ for $I\in \mathcal{A}$.
A candidate estimator for $D$ is 
\begin{equation}\label{eq:tagged_sum_optimal}
    \hat{D} = \sum_{i} \sum_{I\in\mathcal{A}} \lambda_{I} \hat{O}^{I}_{i},
\end{equation}
where $\sum_{I\in \mathcal{A}} \lambda_{I} = 1$.
However, since estimators $\hat{O}^{I}_{i}$ sharing the same $i$, i.e., are based on the same measurements, are in general not independent, and their analysis will require consideration of the covariance between them. One can employ precision weighting which uses a given covariance matrix $\mathbb{C}^{I,J} = \sum_{i,j} \Cov[\mathsf{O}_{i}^{I},\mathsf{O}_{j}^{J}]$, and assures that the optimal choice of $\lambda_I$, 
minimizing $\mathbb{V}\hat{D}$ is 
$\lambda_{I} \propto (\mathbb{C}^{-1} \mathbf{1})_I$,
where $\mathbf{1} = (1,\,\ldots, 1)^\trp$ is a vector of ones of length $|\mathcal{A}|$ and $\mathbb{C}^{-1}$ is the Moore-Penrose pseudo-inverse of the covariance matrix $\mathbb{C}$.
If $\mathbb{C}$ is not known but only estimated, then the estimation must be a symmetric and non-negative matrix.

Since estimating reliably or calculating the covariance can be a hard problem, we are led to neglect the covariances in the optimization.
To that end, consider a decomposition of the observables $O_i$ to some measurable observables $E \in \mathcal{M}$ such that $\mathcal{M}$ spans all $O_i^I$. We can then write 
\begin{equation}
    D = \sum_{E\in \mathcal{M}} \sum_{\substack{i \text{ such that}\\\mathcal{C}_i\text{ measures }E}} \lambda_{E,i} \braketExp{E}_{\mathcal{C}_i}
\end{equation}
where $\lambda_{E,i}$ over the circuits that measure $E$ sum to one.  
We can then optimize $\lambda_{E,i}$ for a fixed $E$. For the case of measurement bases, we have the benefit that $\braketExp{E}_{\mathcal{C}_i}$ are independent of $i$.\footnote{
    This is true for ideal expectation values. For noisy values with either cross-talk noise in the final gate layer implementing the measurement basis change, or measurement correlations in the measurement, the expectation value can be circuit implementation dependent.
}
Then, we can aggregate the estimators $\hat{O}_{E,i}$ for $\braketExp{E}_{\mathcal{C}_i}$, using well-known statistical aggregations such as inverse-variance weighting $\lambda_{E,i} \propto (\mathbb{V}\hat{O}_{E,i})^{-1}$ with $\mathbb{V}\hat{O}_{E,i}$ either given or estimated in a previously run batch of quantum circuits.
Another option is to aggregate by the number of samples $\lambda_{E,i} \propto N_i$ where $N_i$ is shots taken of circuit $\mathcal{C}_i$. 

As presented, the choice of $\lambda$-s directly improves the precision obtained from a given dataset (relative to some naive choice of $\lambda$-s).
To further use the optimization of $\lambda$ in conjunction with the runtime estimation, one needs to solve Problems~\eqref{eq:tagged_sum_opt_precision} and \eqref{eq:tagged_sum_opt_runtime} where they are optimized both over $\epsilon_i$-s and the $\lambda$ parameters.

The VQE benchmark uses the latter algorithm, of weighting by the number of samples, since it is more robust when many measurement bases are applied.\footnote{
    For many measurement bases, the covariance and variance-based weights require reliable estimates of the covariance and variance, respectively, and without a sufficient number of samples, would add biases.
}

\section{Active volume identification}\label{appendix:active-volume}
Here, we describe how we exploit locality in quantum computation to trade off the variance and bias of a QP estimator.
We will show how the variance can be reduced dramatically while introducing only a small bias, in a controllable way.

\subsection{Lightcones and active volume}
An ideal quantum circuit has the property that the expectation value $\langle O \rangle$ of a local observable $O$ generally depends only on part of the gates in the circuit, so that gates outside a ``lightcone'' associated with $O$ can be removed from the circuit without introducing a ``bias''. The simplest such lightcone, which we will refer to as the \textit{connectivity lightcone (ConnLC)}, is easily obtained by tracing two-qubit gates starting from the support of $O$.

This observation leads to the following general question. Consider transforming the ideal circuit by inserting quantum channels after gates and idle times. By inserting quantum channels, we can remove gates, add local unitary errors, add Pauli error channels, etc. Which transformations of the ideal circuit introduce a bias smaller than a given $\delta > 0$? 
Formally, we define (this is a refinement of the definition given in \cite{YEMpaper}):

\begin{definition}[Theoretical active volume]\label{def:active-vol}
Let $C$ be a layered ideal quantum circuit, $O$ an observable, and $\delta > 0$.
Consider transforming $C$ by letting $\mathcal{T}^{(l)}_{m_l} \dots \mathcal{T}^{(l)}_{1}$ act after layer $l$ and before layer $l+1$, where $\mathcal{T}^{(l)}_{1}, \dots, \mathcal{T}^{(l)}_{m_l}$ are quantum channels. 
Consider all subsets $S$ of layer-qubit pairs in $C$ such that if any subset of $\{\mathcal{T}^{(l)}_{j} \text{ without support in $S$} \}$ is allowed to act in $C$ the bias of $\langle O \rangle$ is at most $\delta$.  
Let the minimal cardinality of these sets be $M$.
Then each of the sets with $|S|=M$ is an active volume for the tuple $(C,O, \{\mathcal{T}^{(l)}_j\}, \delta)$.
\end{definition}

In the context of QP error mitigation, given a noisy circuit with its associated noise channels, we need to find the smallest set of gates and idle times on which the noise must be mitigated such that the bias is at most $\delta$.
In a circuit with approximately uniform noise, as typically happens in QPUs, this will yield close to the smallest possible estimation variance given $\delta$.
Generally, however, finding the active volume is intractable---in particular, estimation of the effect of quantum channels on an expectation value is as hard as the ideal quantum computation, and the number of possible subsets of gates and idle times is prohibitively large.
Therefore, we focus on finding an approximation to the theoretical active volume by considering options with the shape of an \textit{expanding lightcone}:
\begin{definition}[Expanding lightcone]\label{def:lc}
Let $O$ be an observable and $C$ a circuit with $L$ layers. An expanding lightcone of $O$ is a list $(S_1, S_2, \dots, S_{L})$, where $S_j$ is a subset of qubits corresponding to layer $j$, such that $S_{1} \supseteq \dots \supseteq S_{L} \supseteq support(O)$.
\end{definition}

We obtain a relatively small number of candidate expanding lightcones using the \textit{Commutativity Lightcone Algorithm}, which we describe in Sec.~\ref{sub:comm_lc_alg}.  Then we give analytic bounds on the bias and heuristic estimates from QP experimental data (Sec.~\ref{heur_est_from_qp_data}).
In Sec.~\ref{subsub:nested_lc} we present the \textit{Nested Lightcones Algorithm}, which we use to choose from the available candidates for mitigation.

\subsection{Related work}
In \cite{tran2023locality}, the ConnLC was used to reduce the sampling overhead of QP error mitigation
and get a tighter error bound for ZNE. 
In the case of QP, the experiment is performed as done in the case where all the error channels in the circuit are mitigated.
Then, in post-processing, the error channels outside the ConnLC are ignored (i.e., not mitigated) when constructing the estimator for the expectation value. 
As ConnLC is used, the QP estimator is unbiased. However, in many cases, ConnLC is much larger than the theoretical active volume.

In \cite{eddins2024lightcone}, the lightcone approach of \cite{tran2023locality} is improved. 
An algorithm for bounding the effect of Hermitian (e.g., Pauli) errors is given, which enables application of QP with a controllable trade-off of bias and variance.
The algorithm computes a bound for every possible local Hermitian error.
If possible, the error is propagated to the beginning or the end of the circuit to yield a bound. 
When brute-force propagation of an error $E$ is too hard, an efficient recursive algorithm is used for bounding the operator norm $\|[E, V^{\dagger} A V] \|$,
where $ V^{\dagger} A V$ is the observable propagated backwards to the time of the error. 
However, the bounds of \cite{eddins2024lightcone} generically overestimate the bias, and we propose alternative bounds. In Sec.~\ref{comp_to_ibm_bound} we present an analytic example where our bound can be significantly better, due to a ``destructive interference'' phenomenon that is not captured by the bounds of \cite{eddins2024lightcone}.

\subsection{Post-processing QP EM data with a lightcone}\label{qp_post_proc_with_lc}
We give more details on how a lightcone is used for post-processing QP experimental data.
It is convenient to express the noise model as follows. Suppose that the error channels in the noisy circuit can be expressed using an `after' error channel $\Lambda_{l}$ for the $l$th layer, such that the noisy circuit is 
\[
\Lambda_{L}\mathcal{U}_{L}\dots\Lambda_{2}\mathcal{U}_{2}\Lambda_{1}\mathcal{U}_{1},
\]
where each channel has the form $\Lambda_{l}=\Lambda_{l,m_{l}}\dots\Lambda_{l,2}\Lambda_{l,1}$, and each $\Lambda_{l,j}$
is acting on at most some small number of qubits and has the form
\begin{align}\label{noise_model_lj}
\Lambda_{l,j}=(1-\sum_{k}p_{k}^{(l,j)})\mathcal{I}+\sum_{k}p_{k}^{(l,j)}\mathcal{E}_{k}^{(l,j)}
\end{align}
where $\mathcal{E}_{k}^{(l,j)}$ is a unitary error, occurring with probability $p_{k}^{(l,j)} > 0$
(`before' and `during' operations can be represented in this form by applying the inverse of a gate followed by the required operation).

Given a candidate lightcone $LC_{\text{cand}}$, we can, in post-processing, mitigate only those $\Lambda_{l,j}$ whose support intersects $LC_{\text{cand}}[l+1]$, at the cost of introducing an estimation bias. 
Let $\sigma^{(cand)}$ map a sampled circuit $C$ to its associated sign, corresponding to $LC_{\text{cand}}$. 
That is, $\sigma^{(cand)}(C)=1$ if the number of sampled QP operations with negative probability is even, and $-1$ otherwise. 
Let the function $w^{(cand)}$ map the noise layers to their QP norm, associated with $LC_{\text{cand}}$:
\begin{align}
w^{(cand)}(\Lambda_{l}) & =\prod_{1\leqslant j\leqslant m_{l}:\ support(\Lambda_{l,j})\cap LC[l+1]\ne\emptyset}w(\Lambda_{l,j}),
\end{align}
where $w$ maps a channel to its corresponding QP norm. The total QP norm associated with $LC_{\text{cand}}$ is
\[
W_{cand}=\prod_{l=1}^{L}w^{(cand)}(\Lambda_{l}).
\]

A key observation here is that the channels $\Lambda_{l,j}$ that are not mitigated are approximately doubled in the sense that $p_{k}^{(l,j)} \mapsto 2p_{k}^{(l,j)}$ for all $k$ with second-order corrections in the $p^{(l,j)}_{k}$s. Example~\ref{pauli_example_doubling} shows this explicitly for a Pauli channel.
This physical and injected noise acting outside $LC_{\text{cand}}$ is the source of the mitigation bias. 

\begin{example}[Pauli channel]\label{pauli_example_doubling}
Consider the case where the $\Lambda_{l,j}$ are of the form $\mathcal{N} = (1-p)\mathcal{I} + p\mathcal{P},$ where $0 < p < \frac{1}{2}$ and $\mathcal{P}(\cdot) = P\,\cdot\,P$ with $P$ a Pauli operator.
Then $\mathcal{N}^{-1}=(1-2p)^{-1}\left[(1-p)\mathcal{I}-p\mathcal{P}\right]$ and $W=(1-2p)^{-1}$. 
In post-processing we implement $\mathcal{N}^{-1}$ as
\[
W\left[W^{-1}\frac{1-p}{1-2p}\mathcal{I}+(-1)W^{-1}\frac{|-p|}{1-2p}\mathcal{P}\right].
\]
If we choose not to mitigate $\mathcal{N}$, we do not multiply by $W$ and do not multiply by $-1$ when we sample $\mathcal{P}$, so that we obtain
\[
\left[W^{-1}\frac{1-p}{1-2p}\mathcal{I}+W^{-1}\frac{|-p|}{1-2p}\mathcal{P}\right]=(1-p)\mathcal{I}+p\mathcal{P}=\mathcal{N}.
\]
Therefore, the effect of not mitigating $\mathcal{N}$ is that $\mathcal{N} \mapsto \mathcal{N}^2$, that is $p \mapsto 2p-2p^2$.
\end{example}

\subsection{The commutativity lightcone algorithm}\label{sub:comm_lc_alg}
Here, we give an algorithm for computing expanding lightcones in a meaningful way. This algorithm scales to any number of qubits and operates only on the ideal circuit.

\begin{figure*}[tb]
\centering
\subfloat[\label{fig:comm_lc_brute_force}]{
    \includegraphics[width=0.43\textwidth]{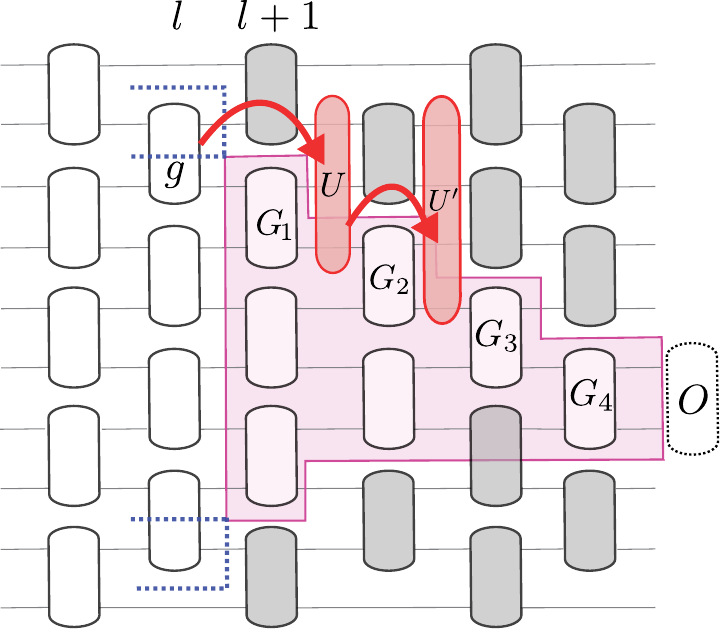}
}
\hfill
\subfloat[\label{fig:nested_diff_vol}]{
    \includegraphics[width=0.53\textwidth]{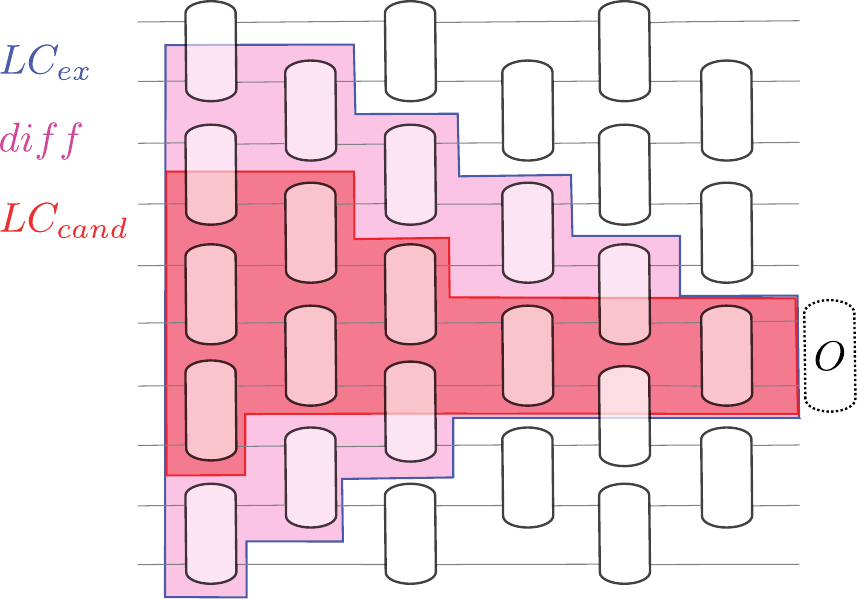}
}
\caption{
\textbf{\protect\subref{fig:comm_lc_brute_force}} Illustration of Alg.~\ref{comm_lc_algo}.
The purple area shows the expanding lightcone found up to layer $l+1$. The gate $g$ is propagated through $G_1$ to give $U = G_1 g G^\dagger_1$. The gray gates are not contained in the lightcone and are ignored in propagation. \textbf{\protect\subref{fig:nested_diff_vol}} $diff$ refers to the ``difference volume'' between $LC_{\text{ex}}$ and $LC_{\text{cand}}$. 
}
\label{fig:comm_lc}
\end{figure*}

Algorithm~\ref{comm_lc_algo} shows the pseudo-code of the main procedure \textproc{CommutativityLC}, which we use to get an approximation to the active volume in the form of an expanding lightcone. \textproc{CommutativityLC} receives an ideal circuit, observable $O$, and threshold $\epsilon > 0$. 
\textproc{CommutativityLC} is based on propagation of gates forward in time towards $O$, unlike \cite{eddins2024lightcone} which propagates errors.
The lightcone is represented by a list $LC$ of length $L+1$ initialized with $LC[L] = support(O)$ and $LC[j]=\emptyset$ for $0 \leq j \leq L-1$. 
Then the algorithm processes layer by layer, from $L-1$ to $0$, and for each decides which qubits are to be added to the lightcone (it can only expand).
When the algorithm processes layer $l$, it checks which gates in that layer are not contained in, but do intersect $LC[l+1]$. 
Figure \ref{fig:comm_lc_brute_force} illustrates the routine at layer $l$.
For each such gate $g$, the procedure \textproc{ApproxExtractable} checks if $g$ can be ``extracted'' from the (previously constructed) lightcone up to $l+1$, by propagating it as far as possible forward in time. This can be done in a brute-force manner, as shown in the pseudo-code \ref{approx_extract_algo}, or in a sampling-based way, as we discuss in Sec.~\ref{sec:samp_extr_check}.
If the propagated gate is $U$, it then checks if $\|U - I\otimes B^{*}\|_F < \epsilon$ or $\|[U,O] \|_F < \epsilon$, depending on whether $g$ could be propagated to the end of the circuit or not. Here $\|\cdot\|_F$ denotes a normalized Frobenius norm $|A|_F = 2^{-\lvert\mathrm{supp}(A)\rvert/2}\sqrt{\mathrm{tr}(A^\dagger A)}$ and $I\otimes B^{*}$ is the best approximation of $U$ as the tensor product where $I$ acts only inside and $B^{*}$ (not necessarily unitary) acts only outside the lightcone. That is, $B^{*}$ is obtained by a partial trace of $U$ over the qubits inside the lightcone.

The idea of the routine is clear for the case of $\epsilon = 0$. In this case, the resulting expanding lightcone has the property that any unitary error outside the lightcone does not affect the expectation value. We refer to circuits corresponding to $\epsilon= 0$ as \textit{exact}. Generically, this lightcone is smaller than the corresponding ConnLC, sometimes even significantly. Using $\epsilon >0$ results in a smaller lightcone for which errors outside introduce a bias. This allows us to trade bias and variance.

\begin{figure}[tb]
\begin{algorithm}[H]
\caption{Commutativity Lightcone Algorithm}\label{comm_lc_algo}
\begin{flushleft}
\Input Quantum circuit $C$ with $L$ layers, observable $O$, threshold $\epsilon > 0$.\\
\Output Expanding lightcone for $O$.
\end{flushleft}
\begin{algorithmic}[1]
\Procedure{CommutativityLC}{$C$, $O$, $\epsilon$}
\State $LC \gets [\emptyset, \dots, \emptyset, \textproc{Supp}(O)]$
\For{$l = L-1$ to $0$}
    \State $LC[l] \gets LC[l+1]$
    \For{each gate $g$ in layer $l$}
        \If{$\textproc{Supp}(g)\cap LC[l+1]\neq\emptyset \,\land\, \textproc{Supp}(g)\not\subset LC[l+1]$}
            \State $q\gets$ qubit of $g$ not in $LC[l+1]$ 
            \If{not \textproc{ApproxExtractable}($C$, $g$, $LC$, $l$, $\epsilon$)}
                \State $LC[l]\gets LC[l]\cup\{q\}$
            \EndIf
        \EndIf
    \EndFor
\EndFor
\State \Return $LC[0:L]$
\EndProcedure
\end{algorithmic}
\end{algorithm}
\end{figure}

\begin{figure}[tb]
\begin{algorithm}[H]
\caption{Brute-force extractability check}\label{approx_extract_algo}
\begin{flushleft}
\Input Quantum circuit $C$, gate $g$, expanding lightcone $LC$, layer index $l$, threshold $\epsilon > 0$.\\
\Output Boolean indicating whether $g$ can be approximately extracted.
\end{flushleft}
\begin{algorithmic}[1]
\Procedure{ApproxExtractable}{$C$, $g$, $LC$, $l$, $\epsilon$}
\State $U\gets$ unitary of $g$
\For{$k=l+1$ to $L$}
    \State $gates \gets$ gates in layer $k$ satisfying:
    \Statex \hspace{\algorithmicindent}$\textproc{Supp}(G)\subset LC[k]$ and $\textproc{Supp}(G)\cap\textproc{Supp}(U)\neq\emptyset$
    \For{$G$ in $gates$}
        \If{$|\textproc{Supp}(U)\cup\textproc{Supp}(G)|>maxsupp$}
            \State \textbf{break}
        \EndIf
        \State $U\gets G U G^\dagger$
    \EndFor
\EndFor
\If{$U$ did not reach end}
    \State $B^*\gets\min_B \|U - I\otimes B\|_F$
    \If{$\|U - I\otimes B^*\|_F  \geq \epsilon$}
        \State \Return False
    \EndIf
\Else
    \If{$\|[U,O]\|_F \geq \epsilon$}
        \State \Return False
    \EndIf
\EndIf
\State \Return True
\EndProcedure
\end{algorithmic}
\end{algorithm}
\end{figure}

\begin{figure}[tb]
\begin{algorithm}[H]
\caption{Tracing-based extractability check}\label{non_samp_alg_1}
\begin{flushleft}
\textbf{Input:} Expanding lightcone $LC$, $g \in \partial LC$ a gate in the circuit, and a threshold $\epsilon > 0$.

\textbf{Output:} Boolean indicating whether $g$ can be approximately extracted.
\end{flushleft}
\begin{algorithmic}[1]
    \State Let $G$ be the superoperator associated with $g$ (i.e., $G = g \otimes g^\dag$). Let $l$ be the layer index of $g$.
    \While{$l > L$}
        \State{$Q \gets $ the set of qubits on which $G$ acts, written as a disjoint union $Q = Q_{\text{in}} \cupdot Q_{\text{out}}$, where $Q_{\text{in}} = Q \cap LC[l+1]$.}
        \State Trace out the part of $G$ acting on $Q_{\text{out}}$ and divide by $2^{|Q_{\text{out}}|}$ to maintain normalization. The resulting $G$ acts only on $Q_{\text{in}}$.
        \If{$\|G - I\| < \epsilon$}
            \State \Return True
        \Else
            \State Conjugate $G$ by gates of layer $l+1$ acting on $Q_{\text{in}}$; 
            \State $l \gets l + 1$
        \EndIf
    \EndWhile
    \State \Return False
\end{algorithmic}
\end{algorithm}
\end{figure}

\subsection{Sampling-based extractability check}\label{sec:samp_extr_check}

As explained above, given some expanding lightcone $LC$, we wish to check whether a gate $g$ on the ``boundary'' of $LC$ is extractable. We will denote the set of all gates that have support inside and outside $LC$ by $\partial LC$.
When $g \in \partial LC$, some of the qubit lines exiting $g$ are known to be outside $LC$.
We would like to take advantage of this fact to reduce the complexity of our calculation. We do that by tracing over indices related
to irrelevant qubits. Algorithm~\ref{non_samp_alg_1} describes this process more explicitly.

The fact that Alg.~\ref{non_samp_alg_1} uses superoperators increases the dimension of the variable $G$ from $4^{|Q_{in}|}$ to $16^{|Q_{in}|}$. This may be costly when $|Q_{in}|$ gets large. 
This is reminiscent of the difficulty one often encounters when required to do computations using density matrices instead of pure states.
A standard remedy for that is the use of the ``quantum trajectory method''.
Inspired by this idea, we may similarly decide to continue working with operators $g$ instead of superoperators.
The irrelevant indices of $g$ corresponding to $Q_{out}$ will then be contracted with randomly chosen single-qubit operators
(e.g., one of $I,X,Y,Z$ or alternatively one of $\{|i\rangle\langle j|\}$).
Averaging over a large enough ensemble of such random choices will then give the same information as in the above method. This sampling method is detailed in \cite{qedma_efficient_lightcones_2025}.

\subsection{Comparison to recent work}\label{comp_to_ibm_bound}
In \cite{eddins2024lightcone}, an iterative bound on the effect of a given gate on the measurement of observable $O$ was found.
In the most generic case, this bound grows, however, as $4^l$ with the number of layers $l$, making it not very useful.
In the special case where the gates considered are close to the lightcone edge, this bound can be much better.
For a Trotterized Ising model (and a gate $g$ near the boundary of $LC$), the bound of  \cite{eddins2024lightcone} and the estimate
from our method both scale with the number of layers $l$ as $\xi^l$ with the same constant $\xi<1$.
Specifically $\xi=\sin(2\alpha)\sin(2\beta)$ where we took each layer to consist 
of the gates $e^{i\beta\sum_i X_i}e^{i\alpha\sum_i Z_{2i+i_0}\otimes Z_{2i+i_0+1}}$  with $i_0=0,1$.
It may be noted, however, that taking the direction of the magnetic field to be $\hat{x}\cos\theta+\hat{y}\sin\theta$ with arbitrary $\theta$
has no effect on $\xi$ in our method, but increases the value of $\xi$ derived from \cite{eddins2024lightcone} thus making it less tight.

\subsection{Analytic estimation of QP mitigation bias}
Consider the noise model defined in Eq.~\ref{noise_model_lj}, where errors act only outside an expanding lightcone $LC$—that is, $\mathrm{supp}(\mathcal{E}_{k}^{(l,j)}) \cap LC[l] = \emptyset$ for all $l, j, k$. Then the following bound on the bias holds:

\begin{theorem}\label{thm_main_analytic_bound}
\begin{align}|\langle\Lambda_{L}\mathcal{U}_{L}\dots\Lambda_{1}\mathcal{U}_{1}(\rho_{0}),O\rangle-\langle\mathcal{U}_{1:L}(\rho_{0}),O\rangle|\le 2\sum_{l,j,k}p_{k}^{(l,j)}\sum_{g}\eta_{g}\end{align}

where $g$ runs over all gates in $\partial LC$ that are also contained in the forward lightcone of
$\mathcal{E}_{k}^{(l,j)}$ and $\eta_{g}$ is the norm corresponding to $g$ in Alg.~\ref{comm_lc_algo}, that is $\eta_{g} = \|\left[U,O\right]\|$ if $g$ was propagated to the end of the circuit, and $\eta_{g} = 2\|U-I\otimes B^{*}\| + \|(B^{*})^\dagger B^{*} - I \|$ otherwise, where $\|\|$ is the operator norm.
\end{theorem}

We note that Alg.~\ref{comm_lc_algo} employs the normalized Frobenius norm, despite the theoretical bound being derived for the operator norm. In practice, the Frobenius norm is expected to yield a more accurate bound for the bias, for reasons analogous to those discussed following Eq.~\eqref{Eq: full least squares}. Theorem~\ref{thm_main_analytic_bound} implies that when $\epsilon = 0$ is used in Alg.~\ref{approx_extract_algo} the bias is exactly 0. When $\epsilon$ is increased to get a smaller lightcone, the bound can become grossly overestimated depending on the circuit and noise model.

\subsection{Estimation of QP mitigation bias from experimental data}\label{heur_est_from_qp_data}
Thus far, we have considered only analytic bounds. While they can be useful in some cases, they are fundamentally limited in that they do not simultaneously account for both the state and the observable. In what follows, we construct significantly improved estimates of the bias based on experimental data. Although heuristic, these estimators work very well in practice.

Consider the setup described in Sec.~\ref{qp_post_proc_with_lc}. We would like to estimate, or at least bound, the bias induced to $\hat{O}_{cand}$ by not mitigating noise outside $LC_{\text{cand}}$ which is contained in $LC_{\text{ex}}$, as illustrated in Fig.~\ref{fig:nested_diff_vol}. We refer to the volume between $LC_{\text{cand}}$ and $LC_{\text{ex}}$ as the \emph{difference volume}. As explained in Sec.~\ref{qp_post_proc_with_lc}, mitigation of noise only inside $LC_{\text{cand}}$ in post-processing effectively doubles the noise outside $LC_{\text{cand}}$. 
In the following, we estimate the resulting bias.

To motivate the heuristic method we use, we first consider the case of a Clifford circuit with Pauli error channels as in Example \ref{pauli_example_doubling} and a Pauli observable.
In this setting, the effect of each Pauli channel is to multiply the ideal value $O_{id}$ by a factor, independently of the other channels. 
Let $\gamma_{cand}$ and $\gamma_{diff}$ be scaling factors associated with all channels in $LC_{\text{cand}}$ and in the difference volume, respectively. Note that error channels outside $LC_{\text{ex}}$ do not affect the expectation value.
The noisy expectation value is
\begin{align}
O_{n} & = O_{id}\gamma_{cand}\gamma_{diff}
\end{align}
and if errors outside of $LC_{\text{cand}}$ are not mitigated
\begin{align}
\mathbb{E}\hat{O}_{cand} & = O_{id}\gamma_{diff}^2,
\end{align}
so that the mitigation bias is
\begin{align}
B_{cand} = O_{id}\left(\gamma_{diff}^2-1\right).
\end{align}

Consider the mean expectation value over the subset of circuits in which there is no QP insertion in $LC_{\text{cand}}$. The expectation for this set is 
\begin{align}
O_{diff} = O_{id}\gamma_{cand}\gamma_{diff}^2
\end{align}
such that 
\begin{align}
\gamma_{diff} = \frac{O_{diff}}{O_n}.
\end{align}
These observations lead to the heuristic Alg.~\ref{heur_bias_estim} that we use on general circuits. $\gamma_{diff}$ is estimated using a form of ``importance sampling''. We compute two bounds---$\hat{B}$ which uses for $O_{id}$ the unbiased estimator $\hat{O}_{ex}$, and $\hat{B}_{triv}$ which uses $|O_{id}| \leq 1$. We then return the estimator that gives a tighter bound. A generalization of this heuristic is discussed in \cite{qedma_efficient_lightcones_2025}.

\begin{figure}[t!]
\begin{algorithm}[H]
\caption{Heuristic bias estimate}\label{heur_bias_estim}
\begin{flushleft}
\Input QP experiment data, Pauli observable $O$, noisy measurement $\hat{O}_n$, a candidate lightcone $LC_{\text{cand}}$ for $O$. 

\Output An estimate of $B_{cand}$ with its statistical error.
\end{flushleft}

\begin{algorithmic}[1]
\State Let $S_{diff}$ be the subset of sampled circuits without an insertion in $LC_{\text{cand}}$ and with at least one insertion in the difference volume.

\State Let $\hat{O}_{diff}$ and $\hat{\sigma}_{diff}^{2}$ be the empirical mean and variance of the measured expectation values of the circuits in $S_{diff}$.

\State Compute from the data $\hat{\gamma}$ and its variance $\hat{\sigma}^{2}_{\gamma}$:
\[
\hat {\gamma} \gets \mathbb{P}(diff) \frac{\hat{O}_{diff}}{\hat{O}_{n}}+ 1 - \mathbb{P}(diff),
\]
where $\mathbb{P}(diff)$ is the probability that there is an insertion in the difference volume (computed from the noise model). If $\hat{O}_n$ and $\hat{O}_{diff}$ are independent (otherwise add a covariance term),
\[
\hat{\sigma}_{\gamma}^{2}\gets\mathbb{P}(diff)^{2}\left[\frac{\hat{\sigma}_{diff}^{2}}{\hat{O}_{n}^{2}}+\left(\frac{\hat{O}_{diff}}{\hat{O}_{n}^{2}}\right)^{2}\hat{\sigma}_{n}^{2}\right].
\]

\State Compute estimate of $B_{cand}$ using $\hat{O}_{ex}$:
\[
\hat{B}\gets\hat{O}_{ex}(\hat{\gamma}^{2} - 1);
\]

\State Compute the statistical error on $\hat{B}$:
\[
D_{1}\gets \hat{\gamma}^{2} - 1
\]
\[
D_{2}\gets 2 \hat{O}_{ex}\hat{\gamma}
\]
\[
\hat{E} \gets \sqrt{D_{1}^{2}\hat{\sigma}_{ex}^{2}+D_{2}^{2}\hat{\sigma}_{\gamma}^{2}+2 |D_{1}D_{2}|\hat{\sigma}_{ex}\hat{\sigma}_{\gamma}}
\]

\State Compute estimate of $B_{cand}$ using the trivial bound $|O_{id}| \leq 1$:
\[
\hat{B}_{triv} \gets  \textrm{sign}(O_n) (\hat{\gamma}^{2} - 1);
\]

\State Compute the statistical error on $\hat{B}_{triv}$:
\[
\hat{E}_{triv} \gets  2|\hat{\gamma}|\hat{\sigma}_{\gamma},
\]
where we assume $\textrm{sign}(O_n)$ is constant.

\State Return $(\hat{B}, \hat{E})$ if
\begin{multline*}
    \max( |\hat{B} + \hat{E}|, |\hat{B} - \hat{E}|) 
    \\
    < \max( |\hat{B}_{triv} + \hat{E}_{triv}|, |\hat{B}_{triv} - \hat{E}_{triv}|),
\end{multline*}
and $(\hat{B}_{triv}, \hat{E}_{triv})$ otherwise.
\end{algorithmic}
\end{algorithm}
\end{figure}

\subsection{The Nested Lightcones Algorithm}\label{subsub:nested_lc}
Here, we address the problem of finding optimal lightcones for post-processing the measurements from a QP error mitigation experiment.
Given a local observable, each lightcone defines a distinct QP estimator derived from the same experimental data, which has an associated bias and variance. We wish to minimize some metric that is a function of these two quantities.

Algorithm~\ref{meta_nested_alg} presents the pseudocode for the \textit{Nested Lightcone Algorithm}. Let $O = \sum_{b,j} O_{b,j}$, where $b$ denotes a measurement basis and each $O_{b,j}$ is a local Pauli operator measurable in basis $b$. As depicted in Fig.~\ref{High-level-nested}, Step~1 computes a sequence of candidate lightcones for each measurement basis and $Z$-string using a sequence of increasing $\epsilon$ values starting from $\epsilon = 0$---the name of the algorithm arises from the nesting of these lightcones.

While the variance $\hat{V}$ can be estimated directly from experimental data, estimating the bias $\hat{B}$ remains the principal challenge. In Step~2, we estimate the bias using the most accurate method available—either analytically or from data—via Alg.~\ref{heur_bias_estim}. Figure~\ref{test_Z73_nc2536} shows a representative example of this bias estimation procedure applied to $Z_{73}$ in the step-6 circuit of Sec.~\ref{main_demo}.
Finally, in Step~4, we select the set of lightcones that minimize the variance subject to a constraint on the bias: $\hat{B}^2 < \tau^2 \hat{V}$, where $\tau$ is a hyperparameter that governs the bias–variance tradeoff.

We demonstrate in simulation the robustness of Alg.~\ref{meta_nested_alg} in the case where $O$ is a single Pauli observable. Figure~\ref{fig:ising_simuls_nested} presents results for one- and two-qubit Pauli operators, $\tau = \frac{1}{2}$, evaluated on 20 Ising-type circuits spanning a variety of geometries as well as a range of two-qubit gate infidelities and circuit depths.

\begin{figure}[t!]
\begin{algorithm}[H]
\caption{Nested Lightcones Algorithm} \label{meta_nested_alg}

\begin{flushleft}
\Input Ideal quantum circuit $C$, observable $O = \sum_{b,j} O_{b,j}$, where $b$ is a measurement basis and $O_{b,j}$ is a local Pauli operator measurable in basis $b$,  data from a QP experiment for the estimation of $O$, threshold $\tau > 0$.

\Output Estimate and variance corresponding to the found lightcones per $O_{b,j}$.
\end{flushleft}

\begin{algorithmic}[1]

\State For each $O_{b,j}$ compute a sequences of nested candidate lightcones $LC^{(b,j)}_{1}, \dots , LC^{(b,j)}_{n_{b,j}}$, where $LC^{(b,j)}_{1}$ is the smallest available exact lightcone.

\State For each $O_{b,j}$ estimate or bound the mitigation bias for each $LC^{(b,j)}_{i}$. For each candidate, this can be done either analytically or based on the experimental data.

\State Let $\hat{V}$ map a choice of $LC^{(b,j)}_i$s to the empirical mitigation variance. Let $\hat{B}^2$ map a choice of $LC^{(b,j)}_i$s to an estimate of the upper bound of the mitigation bias using the result in 2.

\State Solve (exactly or approximately) the combinatorial optimization problem 
\begin{align*}
\min_{ \{{LC^{(j)}_i}\}} \quad & \hat{V}\\
\textrm{subject to} \quad & \hat{B}^{2} < \tau^2 \hat{V}.
\end{align*}

\State Return QP estimate and variance for the solution found in step 4.
\end{algorithmic}
\end{algorithm}
\end{figure}

\begin{figure*}[tb]
\centering
\subfloat[\label{High-level-nested}]{
    \includegraphics[width=0.38\textwidth]{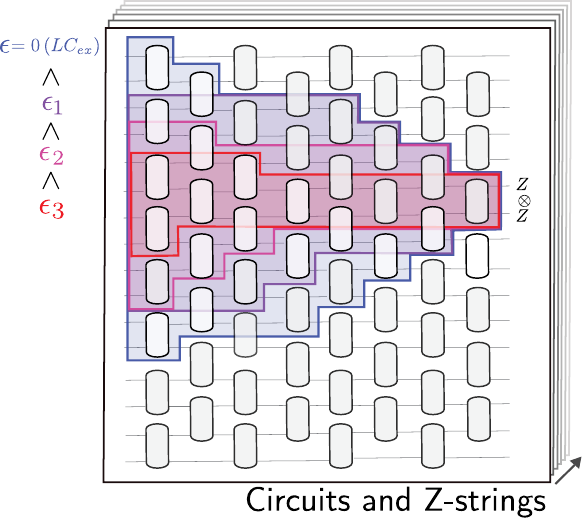}
    }
\hfill
\subfloat[\label{test_Z73_nc2536}]{
    \includegraphics[width=0.6\textwidth, trim={0 0 0 2cm},clip]{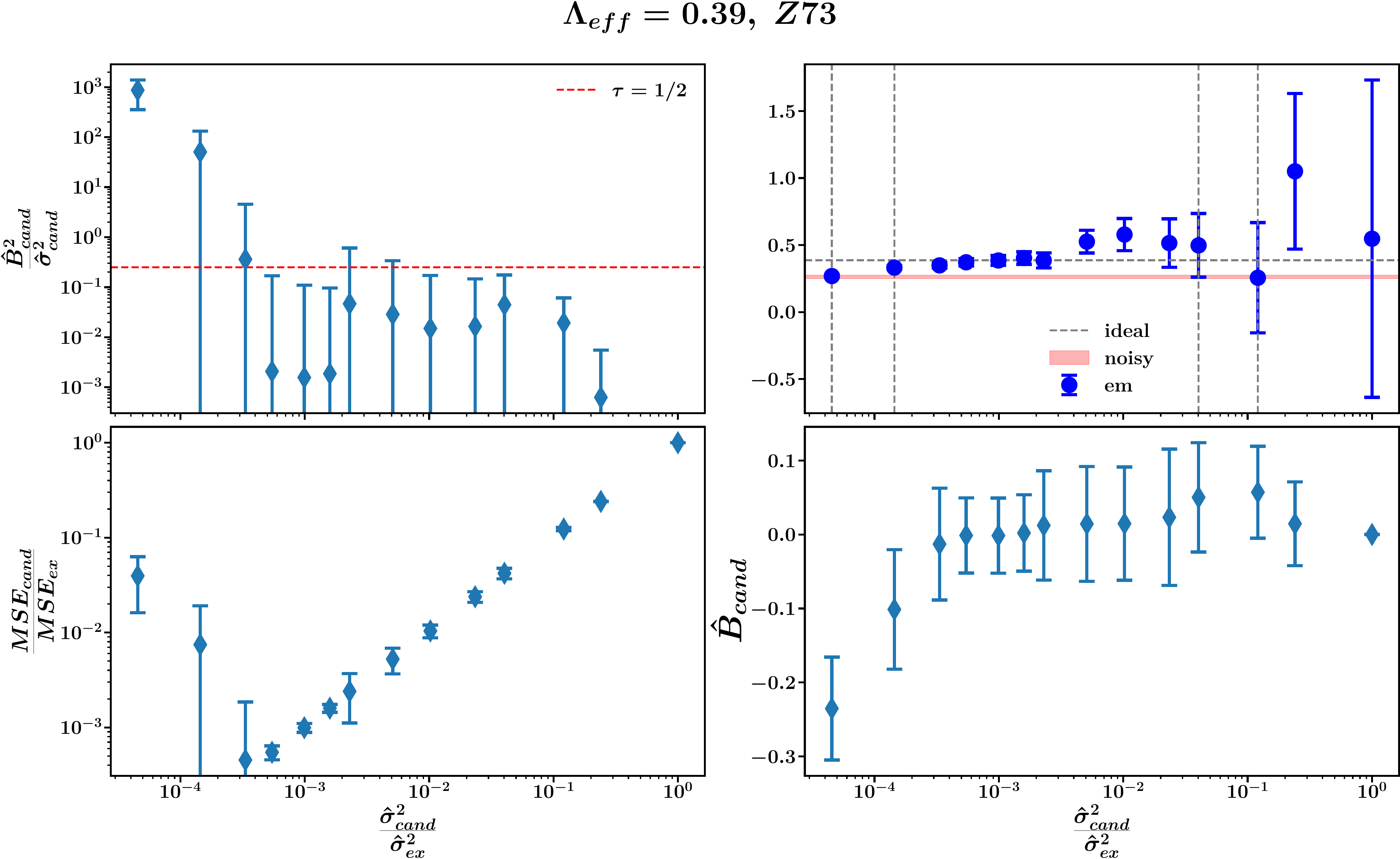}}
\caption{
\textbf{\protect\subref{High-level-nested}} For each Pauli $Z$-string observable that requires estimation, we compute a sequence of expanding lightcones with increasing values of $\epsilon$, starting from $\epsilon = 0$. This initial lightcone yields an unbiased estimator as described in~\ref{sub:comm_lc_alg}. For each lightcone candidate, we estimate the corresponding mitigation bias as illustrated in \protect\subref{test_Z73_nc2536}.
\textbf{\protect\subref{test_Z73_nc2536})} Example of the heuristic bound computed by Alg.~\ref{heur_bias_estim} for $Z_{73}$ in the 6 steps circuit of Sec.~\ref{main_demo}. The top-right plot shows the mitigation results for a list of candidate lightcones, where the rightmost one corresponds to $\epsilon = 0$. The vertical dashed lines show the candidates for which $\hat{B}_{triv}$ was chosen. The bottom-right plot shows the estimates of Alg.~\ref{heur_bias_estim} for each candidate.
}
\label{fig:nested_high_level}
\end{figure*}

\begin{figure*}[t!]
    \includegraphics[width=0.98\linewidth,trim={0cm 0cm 0cm 0cm},clip]{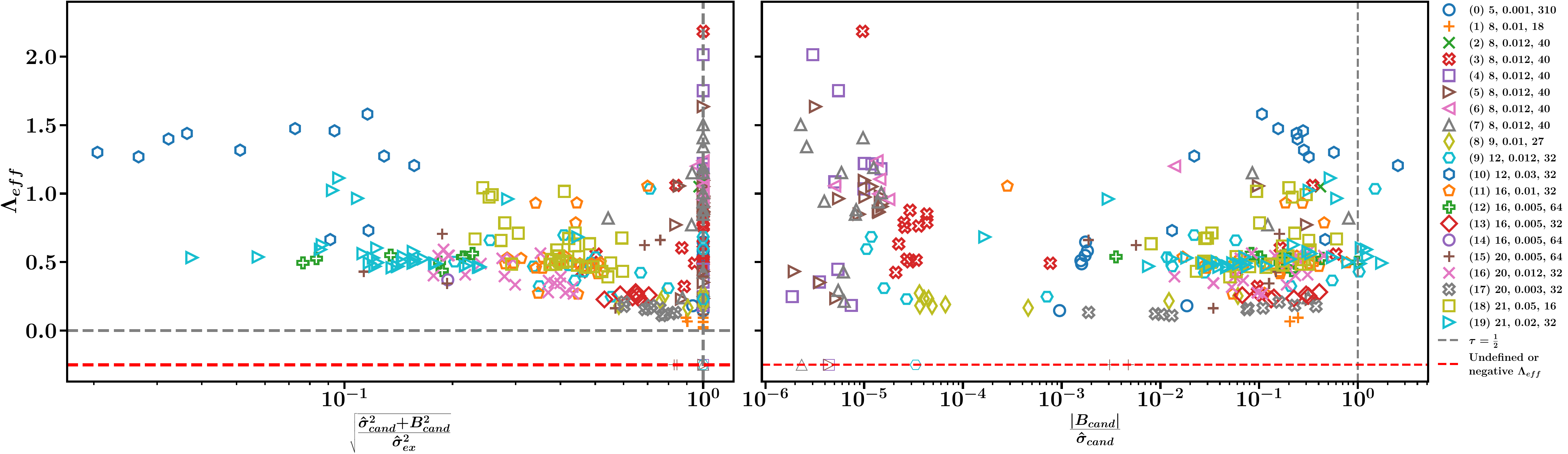}
    \label{subfig:a}   
    \label{fig:lightcones}
    \caption{
    Results of Alg.~\ref{meta_nested_alg} with Alg.~\ref{heur_bias_estim} for Step~2 are shown for 20 Ising Trotter circuits across various geometries (linear, square grid, heavy-hex, and ladder), two-qubit gate infidelities, and circuit depths. The legend indicates the circuit index along with the tuple \textit{(number of qubits, two-qubit gate infidelity, number of two-qubit gate layers)}. Each point in the plot corresponds to either a one- or two-qubit $Z$-string.
    Here, $B_{\mathrm{cand}}$ denotes the simulated bias corresponding to the candidate lightcone selected for each Pauli operator by Alg.~\ref{meta_nested_alg}, with threshold parameter $\tau = \frac{1}{2}$. As seen in the right plot, for the vast majority of tested observables, the ratio $|B_{\mathrm{cand}}| / \hat{\sigma}_{\mathrm{cand}} < \tau$, as intended. Furthermore, in all cases, the mean squared error (MSE) did not increase as shown in the left plot, where here $\Lambda_{eff} = \log(O_{id}/O_{n})$.}
    \label{fig:ising_simuls_nested}
\end{figure*}

\subsection{Experimental results}
\begin{figure*}[tb]
\centering
\begin{minipage}[t]{0.48\textwidth}
\centering
\subfloat[\label{fig:eps_distr_hists}]{
\includegraphics[width=0.86\columnwidth]{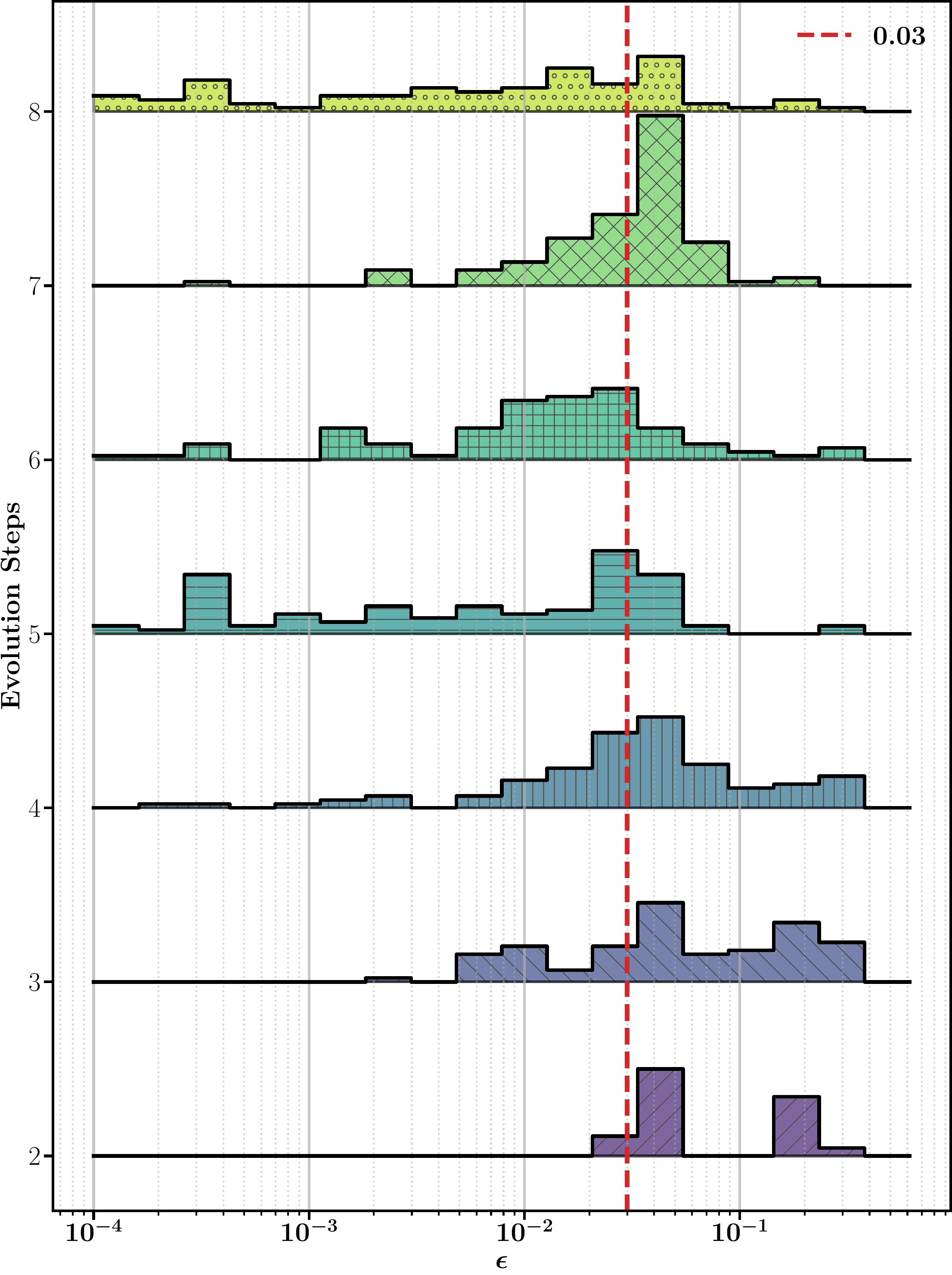}
}
\end{minipage}
\hfill
\begin{minipage}[t]{0.48\textwidth}
\subfloat[\label{fig:eps_vs_nested}]{       \includegraphics[width=0.9\columnwidth]{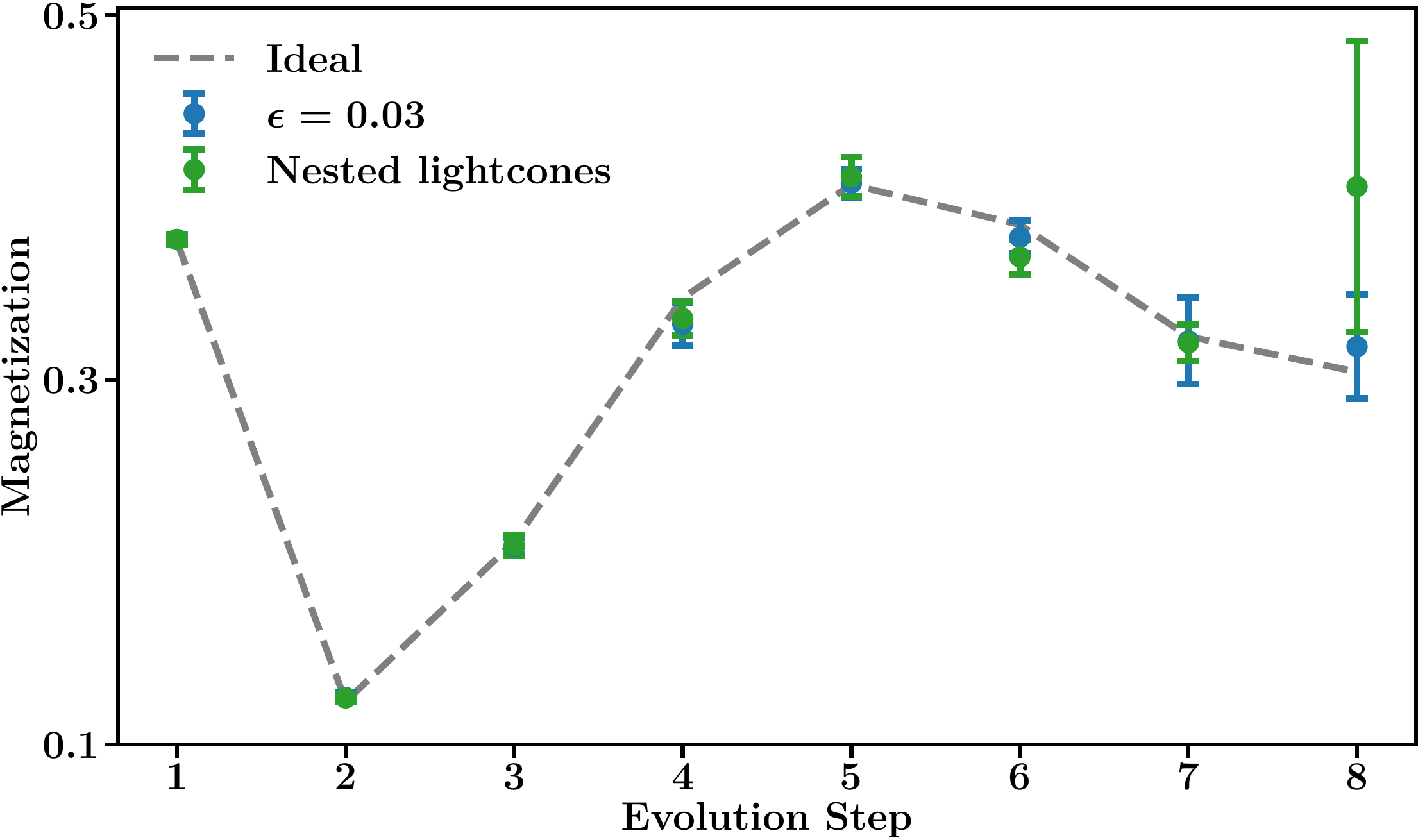}}
\\ 
\subfloat[\label{fig:nested_vs_eps_histogram}]{        \includegraphics[width=0.9\columnwidth]{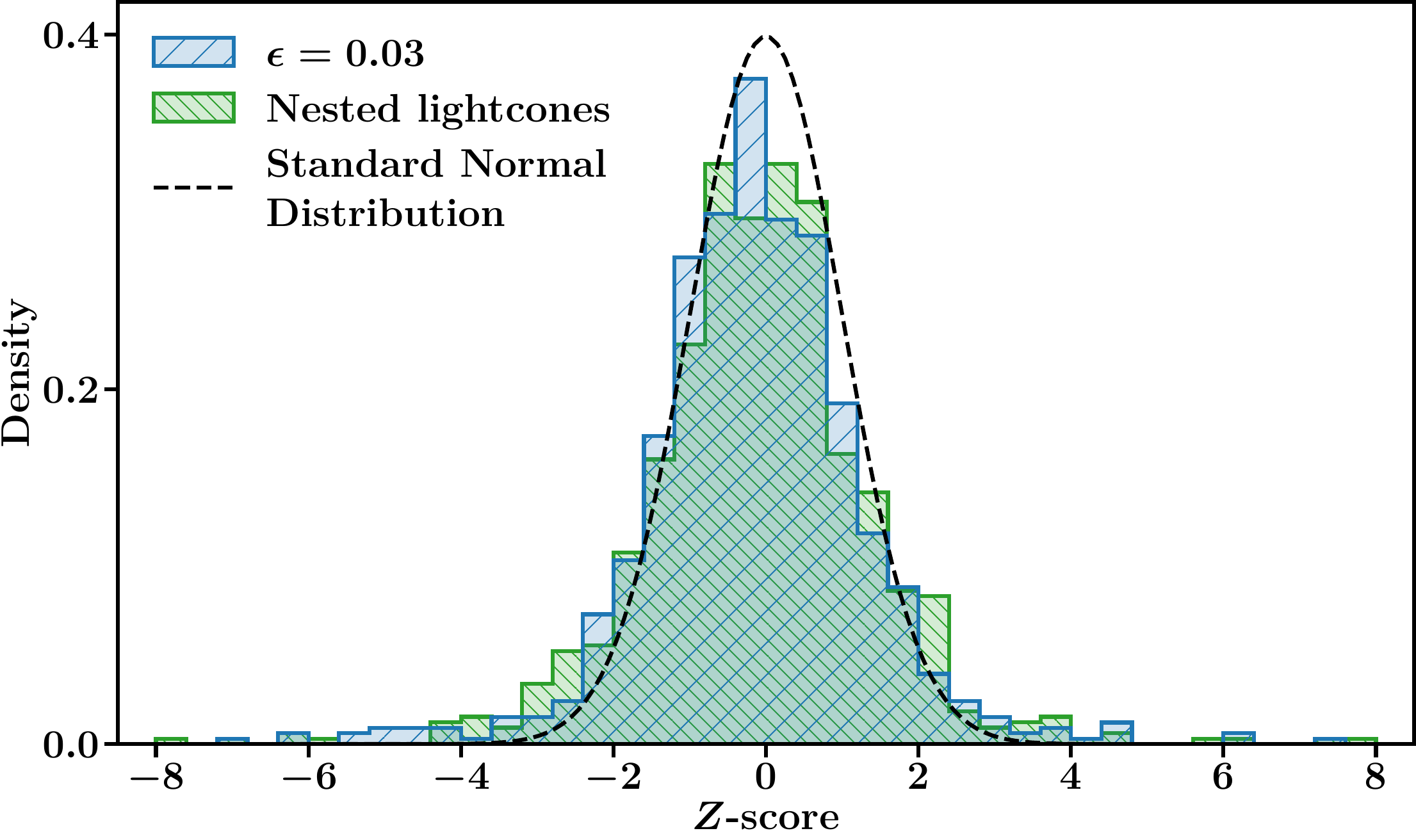}}
        \end{minipage}
\caption{
\textbf{\protect\subref{fig:eps_distr_hists}} Distributions of $\epsilon$ values found by Alg.~\ref{meta_nested_alg} for each evolution step. For step 1, the algorithm chose $\epsilon = 0$ for all Paulis.
\textbf{\protect\subref{fig:eps_vs_nested}} Comparison of the constant epsilon results from the main text and the Nested  Lightcones results.
\textbf{\protect\subref{fig:nested_vs_eps_histogram}} Densities of $Z$-scores across all single-qubit $\langle Z \rangle$ (822 total) for QESEM with constant $\epsilon=0.03$ and for the set of $\epsilon$ values obtained using the Nested Lightcones Algorithm.
\label{fig:nested_summary_results}}
\end{figure*}

We applied the Nested Lightcones Algorithm to each Trotter step of the experiment shown in Fig.~\ref{fig:ising_results}, obtaining the $\epsilon$ distributions shown in Fig.~\ref{fig:nested_summary_results}. The $\epsilon$ values found for step~8 were deemed suboptimal due to drifts (Fig.~\ref{fig:drift-step-8}) and bias overestimation.
Because the circuit is uniform in both time and space, we selected a single $\epsilon$ value, that is $\epsilon= 0.03$, which is at the mode of the distribution or bellow it for steps 2--8, which was used for the analysis presented in the main text. The resulting performance closely matches that obtained using the full set of $\epsilon$ values from the Nested Lightcones Algorithm, except for step~8 (Fig.~\ref{fig:eps_vs_nested}). We next detail the application of Alg.~\ref{meta_nested_alg} in our setting.

We applied Alg.~\ref{meta_nested_alg} for each number of Trotter steps as follows. For each Pauli observable $Z_j$, we construct a sequence of $\epsilon$ values to be used in Alg.~\ref{comm_lc_algo} and Alg.~\ref{approx_extract_algo}: the first is set to $\epsilon = 0$, and the remaining values are selected from 50 points uniformly spaced on a logarithmic scale between $10^{-4}$ and $0.6$. We ensure that the lightcones corresponding to any two consecutive $\epsilon$ values differ by at least two two-qubit gates. In Alg.~\ref{approx_extract_algo} we use $maxsupp=8$.
Steps~2–4 are executed only on the first mitigation batch, and the resulting lightcones are reused across all remaining batches.
We find an approximate solution to the combinatorial optimization problem in Step~4 by mapping it to a continuous one by treating the logarithm of the variance for each Pauli as a continuous variable. Given $N_b$ mitigation batches, we minimize the objective function
\[
\frac{\hat{V}}{N_b} + \hat{B}^2
\]
using the limited-memory BFGS optimization algorithm, and then perform a stochastic local search on the solution. 

Figure~\ref{fig:eps_distr_hists} shows the distributions of the $\epsilon$s for all $Z_j$ for evolution steps 2--8. For 1 step, the algorithm chose $\epsilon =0$ for all Paulis. Figure~\ref{fig:eps_vs_nested} shows the mitigation results for a constant $\epsilon=0.03$ for all $Z_j$ (main text result) and the results obtained from the Nested Lightcones Algorithm.

For Steps~1--8, we obtained the following values of $ |\hat{B}| / \sqrt{\hat{V}/N_b}$, which quantifies our bound for the bias of the result: $(0, 0.26, 0.27, 0.34, 0.46, 1.7, 0.44, 2.1)$. The value for 6,7, and 8 steps is expected to be overestimated as the trivial bound is used in Alg.~\ref{heur_bias_estim} for a large fraction of the Paulis due to the high variance of $\hat{O}_{ex}$. For steps 1 through 7, the results are comparable. For 8 steps, two main effects can degrade performance. First, the 8 step experiment exhibited strong fluctuations, as shown in Fig.~\ref{fig:drift-step-8}, which, in particular, displays a jump in infidelity between the first and the second batch. As we used only the first batch, we are expected to choose larger lightcones than necessary for the other batches. In addition, as mentioned above, the bias is overestimated, which also causes larger lightcones (smaller $\epsilon$s) to be chosen.

\section{Drift mitigation}
\label{app:drift}

\subsection{The problem of drift}

As discussed previously, many error mitigation protocols, such as those based on quasi-probability (QP) distributions, probabilistic error amplification, and tensor error mitigation \cite{Endo2018, filippov2024scalabilityquantumerrormitigation, Ferracin2024} are based on a characterization of the noise in the device. 
In these error mitigation methods, the characterization defines an ensemble of circuits $\mathcal{C}$ and a QP distribution $\left\{ q_{C}\right\} _{C\in\mathcal{C}}$ such that
\begin{equation}
\left\langle O\right\rangle _{C_{0},\mathrm{ideal}}=\sum_{C\in\mathcal{C}}q_{C}\cdot\left\langle O\right\rangle _{C,\mathrm{noisy}}~,
\label{eq:QP}
\end{equation}
where $C_{0}$ is the target circuit. Instead of equality, there can be an approximation, with the approximation being either heuristic or rigorous. 
Moreover, instead of the ideal expectation value $\left\langle O\right\rangle _{C_{0},\mathrm{ideal}}$, we can aim, more generally, at a different target $\left\langle O\right\rangle _{C_{0},\mathrm{target}}$, that can be the noise reduced or amplified circuit to use in methods like zero noise extrapolation (ZNE) or probabilistic error amplification (PEA).

Since the ensemble $\mathcal{C}$ is typically too large to evaluate exhaustively, it is not feasible to run all circuits in $\mathcal{C}$ on the device. Instead, the QP distribution is implemented by mapping it to a sampling distribution over $\mathcal{C}$.
The ideal expectation value $\left\langle O\right\rangle _{C_{0},\mathrm{ideal}}$ of the target circuit $C_{0}$ is the weighted average of the expectation values of the different circuits $C\in\mathcal{C}$:
\[
\left\langle O\right\rangle _{C_{0},\mathrm{ideal}}=\sum_{C\in\mathcal{C}}\mathrm{Prob}\left(C\right)\cdot\mathrm{weight}\left(C\right)\cdot\left\langle O\right\rangle _{C,\mathrm{noisy}}~,
\]
where $\mathrm{weight}\left(C\right)=\frac{q_{C}}{\mathrm{Prob}\left(C\right)}$.

We sample $N_{C}$ circuits $C_{1},\ldots,C_{N_{C}}$ according to the distribution $\left\{ \mathrm{Prob}\left(C\right)\right\} _{C\in\mathcal{C}}$ (independently), run these circuits on the device to obtain expectation values $\left\langle O\right\rangle _{C_{1},\mathrm{noisy}},\ldots,\left\langle O\right\rangle _{C_{N_{C}},\mathrm{noisy}}$, and aggregate the results to produce the estimator:
\[
\mathrm{EST}=\frac{1}{N_{C}}\sum_{i=1}^{N_{C}}\mathrm{weight}\left(C_{i}\right)\cdot\left\langle O\right\rangle _{C_{i},\mathrm{noisy}}~.
\]
Equation (\ref{eq:QP}) implies that this estimator is (approximately) unbiased, 
\[
\mathbb{E}\left[\mathrm{EST}\right]=\left\langle O\right\rangle _{C_{0},\mathrm{ideal}}~.
\]

When a large number of circuits must be run for the mitigation procedure (for instance, to achieve high statistical precision), there is a risk of drift in the device’s behavior during the execution. 
Methods for addressing drift in characterization-based error mitigation protocols have received little attention in the literature. 
We note Ref.~\cite{Kim2023utility}, which performed a large ZNE experiment, based on PEA, and included a method for dealing with drifts. We compare the approach of this reference to our approach in App.~\ref{sec:Batches-of-characterization}. 

Our drift-mitigation strategies can be applied to any error mitigation method consisting of two stages: 
\begin{enumerate}
    \item \textbf{Characterization step} — Characterize the device noise or the noisy behavior of the circuit.
    \item \textbf{Estimation step} — Execute a set of circuits on the device and combine their expectation values to estimate the ideal expectation value.
\end{enumerate}
Such techniques are not limited to QP-based approaches; they also include learning-based methods, such as Clifford data regression and free-fermion modeling \cite{EM_review_Babbush}. The concepts presented in this section are also covered under patent application~\cite{qedma_drift_robust_2025}.

\subsection{Batches of characterization and mitigation}
\label{sec:Batches-of-characterization}

\begin{figure}[tb]
\centering
\includegraphics[width=0.95\columnwidth]{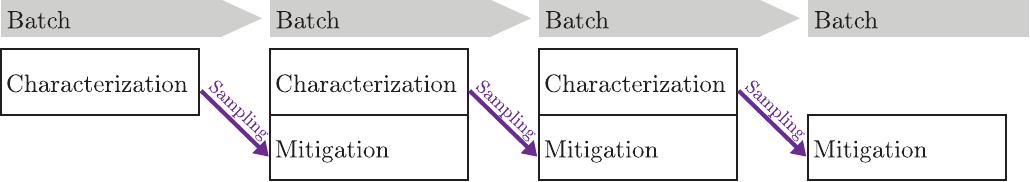}
\caption{QESEM's batches protocol. Instead of running one characterization and then all the mitigation circuits, we run a few batches. In each batch, we use the characterization that was run in the previous batch to randomly choose circuits for the mitigation circuits of the current batch.}
\label{fig:batches-protocol}
\end{figure}

We combat drift by running a few batches of characterization and mitigation. We use the word ``batch'' do describe circuits that are run on the quantum processor, without a pause for classical computation. This way, we run the full number of circuits we wanted to run to get the desired precision, but the characterization is being updated. In each batch, we use the characterization that was run in the previous batch to randomly choose circuits for the mitigation circuits of the current batch. See Fig.~\ref{fig:batches-protocol} for a visualization of this process.

For each batch $i$ we get the estimated expectation value $\mu_{i}$ with variance $\sigma_{i}^{2}$. We aggregate these values with Inverse-Variance Weighting (IVW). IVW is a method for aggregating random variables to minimize the variance of the weighted average. Each random variable is weighted in inverse proportion to its variance. The output estimation for the expectation value is:
\[
\hat{O}=\frac{\sum_{i}\mu_{i}/\sigma_{i}^{2}}{\sum_{i}1/\sigma_{i}^{2}}~.
\]
The variance is
\[
\mathrm{Var}\left(\hat{O}\right)=\frac{1}{\sum_{i}1/\sigma_{i}^{2}}~.
\]
A useful insight is that batched characterizations can be leaner, i.e., characterization runtime can be scaled down with the size of the mitigation batch, so batching does not cause total characterization time to increase.

In Ref.~\cite{Kim2023utility}, a large ZNE experiment was performed. They used batches to run the circuits. 
They describe their protocol in the supplementary material in IIB.
They use the batches to identify outliers among the batches, and they exclude these batches from all further processing. 
They obtain the final observable estimate by averaging the $\mu_{i}$ values of the remaining batches. 
Notice our different approach: First, we use all data from the experiment, and do not exclude data. Second, instead of averaging the $\mu_{i}$ with a plain average, we use the IVW as described before. 

Another difference is that we run multiple characterizations, while they run only one. 
The approach in \cite{Kim2023utility} can only be useful when noise in the hardware is generally stable, apart from short times during which it jumps. In more general and common circumstances, the noise slowly drifts, and there is no single noise model that accurately represents the device during the majority of circuit runs. 
Also, even if the hardware is generally stable, simply discarding samples showing large fluctuations during mitigation can lead to biases. One has to be careful to treat characterization and mitigation consistently, but fluctuations affect characterization circuits and mitigation circuits differently.

\subsection{Retroactive resampling}
\label{sec:Retroactive-resampling}

\begin{figure}
\includegraphics[width=0.5\columnwidth]{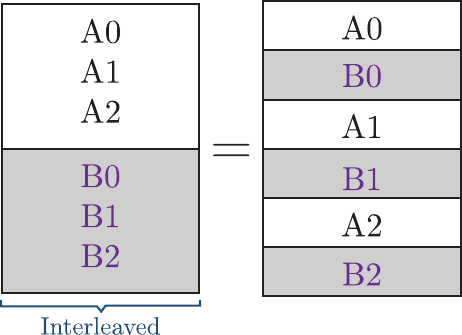}
\caption{By \textquotedblleft interleaving\textquotedblright{} circuits A and circuits B, we mean running a circuit from A and then a circuit from B, and then a circuit from A and so on (or a few circuits instead of one each time). If we interleave characterization circuits with mitigation circuits in the same batch, this characterization represents the state of the device at the time frame that the mitigation circuits ran.}
\label{fig:interleaving}
\end{figure}

In App.~\ref{sec:Batches-of-characterization} we described a method to avoid drifts, while here we will describe a method to correct the effects of drifts, if they occur. The method in this Section is called retroactive resampling. We run characterization circuits interleaved with the error mitigation circuits to get a characterization that better represents the state of the device at the time when the error mitigation circuits are run. See Fig.~\ref{fig:interleaving} to understand what we mean by interleaving. 

We can calculate the new weights $\mathrm{weight}_{new}\left(C_{i}\right)$ for every circuit, but the circuits $C_{1},\ldots,C_{m}$ were sampled from the wrong distribution. In retroactive resampling, we correct the probability with which the circuits were randomly generated. Each circuit $C'$ is multiplied by 
\[
\frac{\mathrm{Prob}_{new}\left(C'\right)}{\mathrm{Prob}_{old}\left(C'\right)}~.
\]
So in total, the output of retroactive resampling is 
\begin{multline*}
    \mathrm{EST}_{retro} = \\
    \frac{1}{N_{C}}\sum_{i=1}^{N_{C}}\frac{\mathrm{Prob}_{new}\left(C_{i}\right)}{\mathrm{Prob}_{old}\left(C_{i}\right)} \cdot \mathrm{weight}_{new}\left(C_{i}\right) \cdot \left\langle O\right\rangle _{C_{i},\mathrm{noisy}}~.
\end{multline*}
This kind of resampling is popular in various applications. To the best of our knowledge, resampling was never applied in the context of error mitigation.

While QP error mitigation is unbiased, the retroactive resampling may cause bias. If $\mathrm{Prob}_{old}\left(C_{i}\right)=0$ but $\mathrm{Prob}_{new}\left(C_{i}\right)>0$, we will never have $\left\langle O\right\rangle _{C_{i},\mathrm{noisy}}$, because the circuits $C_{i}$ where sampled according to the probability distribution $\mathrm{Prob}_{old}$, and therefore $C_{i}$ could not have been sampled. The consequence of this discrepancy between these two probabilities is a biased result. The bias's value is 
\[
\sum_{C|\mathrm{Prob}_{old}\left(C\right)=0}\mathrm{Prob}_{new}\left(C\right)\cdot\mathrm{weight}_{new}\left(C\right)\cdot\left\langle O\right\rangle _{C,\mathrm{noisy}}~,
\]
and the estimated value $\mathrm{EST}_{retro}$ has the following expectation value:
\begin{multline*}
\mathbb{E}\left[\mathrm{EST}_{retro}\right]=\\
\sum_{C|\mathrm{Prob}_{old}\left(C\right)>0}\mathrm{Prob}_{new}\left(C\right)\cdot\mathrm{weight}_{new}\left(C\right)\cdot\left\langle O\right\rangle _{C,\mathrm{noisy}}~.
\end{multline*}
This can be used to bound the bias due to retroactive resampling, which is generally expected to be small, if the drift (i.e., change
in model parameters) is small. 

Retroactive resampling can inflate the variance, and therefore inflate the QPU time overhead (the number of circuits and shots that we need to run in order to get an estimation with a desired precision). Most quasi-probability error mitigation protocols define the probability that is associated with the quasi-probability in the following way: if $\left\{ q_{C}\right\} _{C\in\mathcal{C}}$ is such that 
\[
\left\langle O\right\rangle _{C_{0},\mathrm{ideal}}=\sum_{C\in\mathcal{C}}q_{C}\cdot\left\langle O\right\rangle _{C,\mathrm{noisy}}~,
\]
then define $\mathrm{Prob}\left(C\right)\propto\left|q_{C}\right|$,
and more explicitly define 
\[
\mathrm{Prob}\left(C\right)=\frac{\left|q_{C}\right|}{\sum_{C\in\mathcal{C}}\left|q_{C}\right|}.
\]
 This widespread practice has a justification in \cite{QP_estimation}.

The variance of the estimator in regular quasi-probability error mitigation (without retroactive resampling) is
\begin{multline*}
\mathbb{V}\left[\mathrm{EST}\right]=\\
\frac{1}{N_{s}N_{c}}\sum_{c}\mathrm{Prob}\left(C\right)\cdot\mathrm{weight}\left(C\right)^{2}\left[\left\langle O^{2}\right\rangle _{C,\mathrm{noisy}}-\left\langle O\right\rangle _{C,\mathrm{noisy}}^{2}\right]+\\
\frac{1}{N_{c}}\left[\sum_{c}\mathrm{Prob}\left(C\right)\cdot\mathrm{weight}\left(C\right)^{2}\left\langle O\right\rangle _{C,\mathrm{noisy}}^{2}-\left\langle O\right\rangle _{C_{0},\mathrm{ideal}}^{2}\right]~,
\end{multline*}
where $N_{C}$ is the number of circuits run for the estimation, $N_{S}$ is the number of shots for each circuit, $\mathrm{EST}$ is our estimator for $\left\langle O\right\rangle _{C_{0},\mathrm{ideal}}$.

For readability, we denote 
\[
\mathrm{d\left(C\right)}=\mathrm{Prob}_{old}\left(C\right)\cdot\mathrm{weight}_{new}\left(C\right)^{2}\cdot\left(\frac{\mathrm{Prob}_{new}\left(C\right)}{\mathrm{Prob}_{old}\left(C\right)}\right)^{2}~, 
\]
for $c\in\mathcal{C}$. When we use retroactive resampling, the variance of the estimator is
\begin{multline*}
\mathbb{V}\left[\mathrm{EST}_{retro}\right]=\\
\frac{1}{N_{s}N_{c}}\sum_{c\in\mathcal{C}}\mathrm{d}\left(C\right)\left[\left\langle O^{2}\right\rangle _{C,\mathrm{noisy}}-\left\langle O\right\rangle _{C,\mathrm{noisy}}^{2}\right]+\\
\frac{1}{N_{c}}\left[\sum_{c\in\mathcal{C}}\mathrm{d}\left(C\right)\left\langle O\right\rangle _{C,\mathrm{noisy}}^{2}-\left(\mathbb{E}\left[\mathrm{EST}_{retro}\right]\right)^{2}\right].
\end{multline*}

To quantify the multiplicative factor by which the variance changes due to resampling, we want to focus on the term
\[
\sum_{c\in\mathcal{C}}\mathrm{Prob}_{old}\left(C\right)\cdot\left(\frac{\mathrm{Prob}_{new}\left(C\right)}{\mathrm{Prob}_{old}\left(C\right)}\right)^{2}~.
\]
Assume, for simplicity, that we have the channel:
\[
\Lambda_{old}\left(\rho\right)=\left(1-p_{old}\right)\rho+p_{old}P\rho P^{\dagger}~,
\]
for some Pauli $P$ and some $0<p_{old}<1$. Assume that we have a
new channel:
\[
\Lambda_{new}\left(\rho\right)=\left(1-p_{new}\right)\rho+p_{new}P\rho P^{\dagger}~.
\]
Then
\begin{multline*}
\sum_{c\in\mathcal{C}}\mathrm{Prob}_{old}\left(C\right)\cdot\left(\frac{\mathrm{Prob}_{new}\left(C\right)}{\mathrm{Prob}_{old}\left(C\right)}\right)^{2}\\
=p_{old}\cdot\left(\frac{p_{new}}{p_{old}}\right)^{2}+\left(1-p_{old}\right)\cdot\left(\frac{1-p_{new}}{1-p_{old}}\right)^{2}\\
\approx e^{(p_{new}-p_{old})^{2}/p_{old}}~.
\end{multline*}
So $\sum_{c\in\mathcal{C}}\mathrm{Prob}_{old}\left(C\right)\cdot\left(\frac{\mathrm{Prob}_{new}\left(C\right)}{\mathrm{Prob}_{old}\left(C\right)}\right)^{2}>1$.

The retroactive resampling overhead can be written as a product $e^{(p_{new}-p_{old})^{2}/p_{old}}$ over all error channels in the circuit. This is problematic when $p_{old}$ is small but $p_{new}$ isn't, i.e., $p_{old}\ll p_{new}^{2}$. To avoid these cases, we can choose not to perform retroactive resampling for such parameters and add them to the bias. This gives us an overhead-bias trade-off that we can play with.
We also note that we can avoid resampling for parameters that are not statistically significantly distinct compared to the characterization error bars. 

Retroactive resampling also enables us to sample mitigation circuits from a noise model that is different from the one we characterized. This can be used to ensure ourselves against drifts and combat retroactive resampling biases due to noise-model singularities without excessive overhead.

\begin{figure}
\includegraphics[width=0.95\columnwidth]{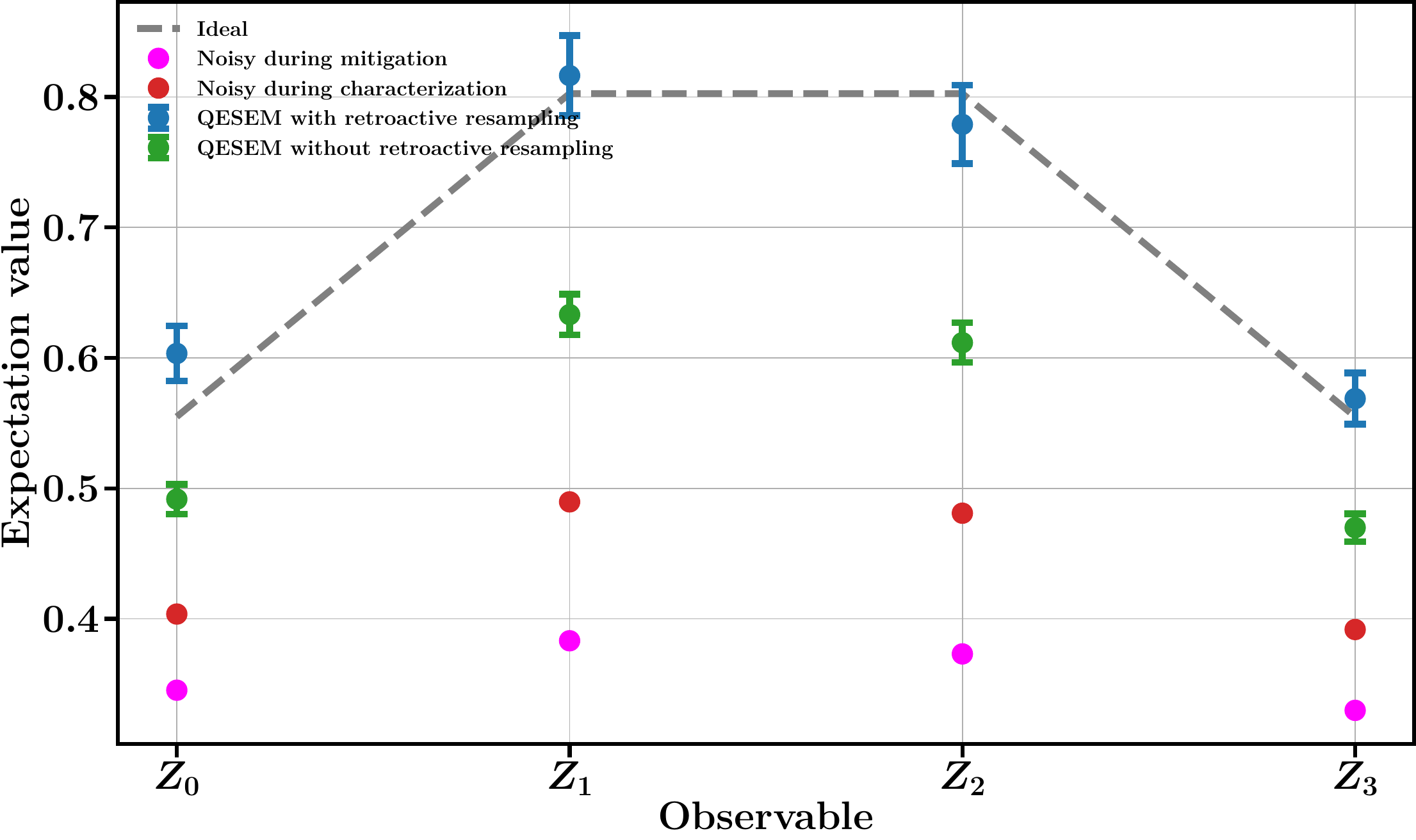}
\caption{Numerical simulation of retroactive resampling. For this simulation, an initial random noise model was generated. The mitigation circuits were sampled according to this noise model, and additionally, the noisy circuit was simulated (red points). To mimic the effect of drifts, the sampled mitigation circuits were then executed on a simulated QPU, using a new noise model with higher noise rates representing a drift in the noise. The noisy circuit with the second noise model is represented by the magenta points. The gap between the gray ideal line and the error-mitigated green points is evident. Retroactive resampling (blue points) closes this gap. }
\label{fig:resampling-simulation}

\end{figure}

In Fig.~\ref{fig:resampling-simulation}, we show a numerical simulation of retroactive resampling. In this simulation, we can see that the retroactive resampling closes the gap between the ideal expectation values and the error-mitigated expectation values. 

\subsection{Noisy expectation values from mitigation circuits}
\label{sec:Noisy-expectation-value}

We would like to identify drifts. Keeping track of the noisy expectation values (the expectation values of the desired circuit when running on a noisy device $\left\langle O\right\rangle _{C,\mathrm{noisy}}$) can identify drifts, and also can suggest how to correct their effect on mitigation, as we show in the App.~\ref{sec:Heuristic-drift-correction}. 
The noisy expectation values $\left\langle O\right\rangle _{C,\mathrm{noisy}}$ are a fairly raw quantity (compared to the mitigated expectation values), and have a smaller error bar (when compared to mitigated expectation values), and therefore are a good candidate for drift detection.
The following technique enables us to produce the noisy expectation values from the mitigation circuits that ran on the device.

In error mitigation, we want to calculate an expectation value $\left\langle O\right\rangle _{C,\mathrm{ideal}}$ of an ideal circuit $C$. When performing error mitigation, we run on the device circuits $C_{1},\ldots,C_{m}$ that are very close to the original circuit $C$, in the sense that the $C_{i}$ only differ from $C$ in local changes. In some of the circuits $C_{i}$, the local change does not affect the expectation value, i.e. $\left\langle O\right\rangle _{C}=\left\langle O\right\rangle _{C_{i}}$. 
We can check whether this is the case by checking that the local change is outside the lightcone of the observable. In other words, many circuits are effectively the same as the original circuit when calculating the observable $O$. Since these circuits are run on the noisy device, we can use them to have the noisy observable $\left\langle O\right\rangle _{C,\mathrm{noisy}}$.

\begin{figure}
\includegraphics[width=\columnwidth]{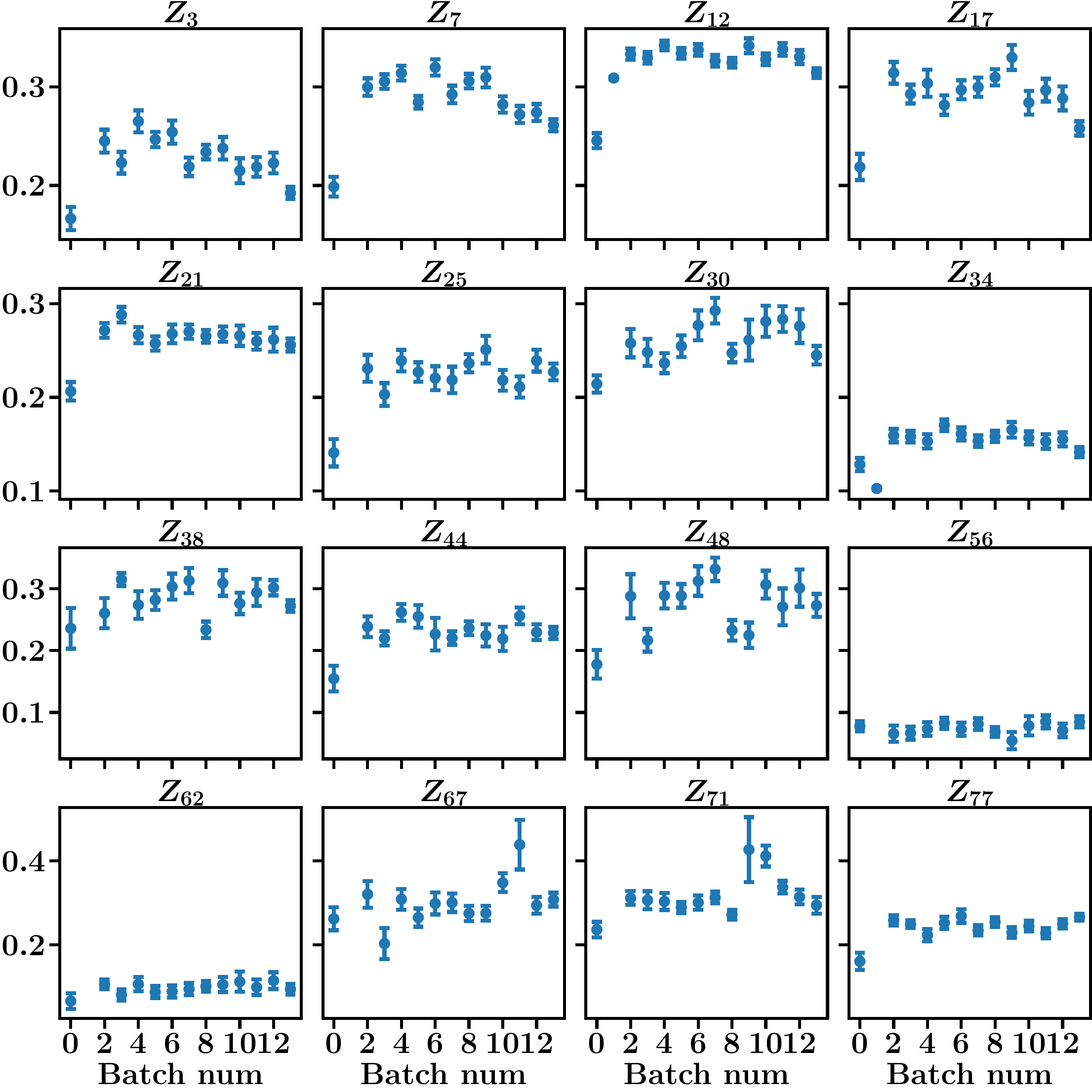}
\caption{Here we see the drifts of step 8 in the experiment discussed in Sec.~\ref{main_demo}. We chose to show only a few qubits from the experiment, but all of them experienced drift. For each qubit, we see the noisy expectation values from mitigation circuits. The x-axis represents the different batches, and the y-axis represents the noisy expectation values.
}
\label{fig:drift-step-8}
\end{figure}

In Fig.~\ref{fig:drift-step-8} we see the drifts of step 8 in the experiment discussed in Sec.~\ref{main_demo}. For each qubit shown, we see the noisy expectation values from mitigation circuits.

\subsection{Using noisy expectation values for drift mitigation}
\label{sec:Heuristic-drift-correction}

As was explained in App.~\ref{sec:Retroactive-resampling}, retroactive resampling may result in larger error bars and may also result in biased estimations for the desired expectation values. Since retroactive resampling has these undesired side effects, we examine different techniques to accommodate drifts. The following heuristic method has the advantage of being very simple to understand and to perform. 

There are heuristic error mitigation techniques, and by nature of being heuristic, they do not always work, and if they do work, their
performance is not always satisfactory. Here, we use a mathematically sound, unbiased error mitigation method, and we use a heuristic method only to account for the drift. The drifts are small relative to the average infidelity; therefore, the inaccuracies arising from the heuristic are limited to this small quantity. 

We will explain where the idea of this heuristic method comes from. If the noise on the device causes an exponential decay, we get 
\[
\left\langle O\right\rangle _{C,\mathrm{noisy}_{0}}=e^{-\lambda_{0}}\cdot\left\langle O\right\rangle _{C,\mathrm{ideal}}~,
\]
where $C$ is the target circuit, and $\lambda_{0}$ depends on the device's noise. 
In this case we can think of the mitigation as taking the noisy value $\left\langle O\right\rangle _{C,\mathrm{noisy}_{0}}$ and multiplying it by $e^{\lambda_{0}}$ the get the ideal value, i.e., $\mathrm{miti}=\left\langle O\right\rangle _{C,\mathrm{noisy}_{0}}\cdot e^{\lambda_{0}}=\left\langle O\right\rangle _{C,\mathrm{ideal}}$.
Suppose the characterization captures noise that differs from the noise during mitigation. In that case, $\left\langle O\right\rangle _{C,\mathrm{noisy}_{1}}=e^{-\lambda_{1}}\cdot\left\langle O\right\rangle _{C,\mathrm{ideal}}$ but the mitigation outputs
\begin{align}
    \mathrm{miti}=\left\langle O\right\rangle _{C,\mathrm{noisy}_{1}}\cdot e^{\lambda_{0}}~,
\end{align}
which does not reproduce the ideal $\left\langle O\right\rangle _{C,\mathrm{ideal}}$. 

We can correct the mitigation result by multiplying by $\frac{\left\langle O\right\rangle _{C,\mathrm{noisy}_{0}}}{\left\langle O\right\rangle _{C,\mathrm{noisy}_{1}}}$, because
\begin{alignat*}{1}
    \mathrm{miti\cdot}\frac{\left\langle O\right\rangle _{C,\mathrm{noisy}_{0}}}{\left\langle O\right\rangle _{C,\mathrm{noisy}_{1}}} & =\left\langle O\right\rangle _{C,\mathrm{noisy}_{1}}\cdot e^{\lambda_{0}}\cdot\frac{e^{-\lambda_{0}}\cdot\left\langle O\right\rangle _{C,\mathrm{ideal}}}{e^{-\lambda_{1}}\cdot\left\langle O\right\rangle _{C,\mathrm{ideal}}}\\
    & =\left\langle O\right\rangle _{C,\mathrm{noisy}_{1}}\cdot e^{\lambda_{1}}=\left\langle O\right\rangle _{C,\mathrm{ideal}}~.
\end{alignat*}

Here, we assumed that the noise behaves like an exponential decay. 
But we don't have to limit ourselves to exponential functions. Any extrapolation function used for ZNE (linear, polynomial, etc.) can, in principle, be used for this heuristic rescaling. As stated before, we expect the rescaling to work better than ZNE, because the drifts are small relative to the average infidelity.

There are cases in which we choose not to apply this heuristic. When the values of $\left\langle O\right\rangle _{C,\mathrm{noisy}_{0}}$ and $\left\langle O\right\rangle _{C,\mathrm{noisy}_{1}}$ are close to each other up to their error bars, i.e., statistically, they could have had the same value, we prefer not to apply this heuristic. Specifically, if we denote the estimated error of $\left\langle O\right\rangle _{C,\mathrm{noisy}_{0}}$ by $\sigma_{0}$ and the estimated error $\left\langle O\right\rangle _{C,\mathrm{noisy}_{1}}$ by $\sigma_{1}$ then we will apply the heuristic when 
\[
\left|\left\langle O\right\rangle _{C,\mathrm{noisy}_{0}}-\left\langle O\right\rangle _{C,\mathrm{noisy}_{1}}\right|>2\cdot\sqrt{\sigma_{0}^{2}+\sigma_{1}^{2}}~.
\]

There is a second condition that we check in order to decide whether to apply the heuristic. When the values $\left\langle O\right\rangle _{C,\mathrm{noisy}_{0}}$ or $\left\langle O\right\rangle _{C,\mathrm{noisy}_{1}}$ are close to zero up to their error bars, we rather not apply the heuristic. 
Specifically, we will apply the heuristic when 
\[
\left|\left\langle O\right\rangle _{C,\mathrm{noisy}_{0}}\right|\ge2\cdot\sigma_{0}\qquad\mathrm{and}\qquad\left|\left\langle O\right\rangle _{C,\mathrm{noisy}_{1}}\right|\ge2\cdot\sigma_{1}~.
\]

There is a third condition. If $\left\langle O\right\rangle _{C,\mathrm{noisy}_{0}}$ and $\left\langle O\right\rangle _{C,\mathrm{noisy}_{1}}$ have different signs, the entire idea behind this heuristic makes no sense. 
As explained in the beginning of this Section, this heuristic works perfectly in scenario in which the noise causes exponential decay, and in this case $\left\langle O\right\rangle _{C,\mathrm{noisy}_{0}}$ and $\left\langle O\right\rangle _{C,\mathrm{noisy}_{1}}$ should have the same sign. Therefore, we apply the heuristic when
\[
\mathrm{sign}\left\langle O\right\rangle _{C,\mathrm{noisy}_{0}}=\mathrm{sign}\left\langle O\right\rangle _{C,\mathrm{noisy}_{1}}~.
\]

\begin{figure}
\includegraphics[width=\columnwidth]{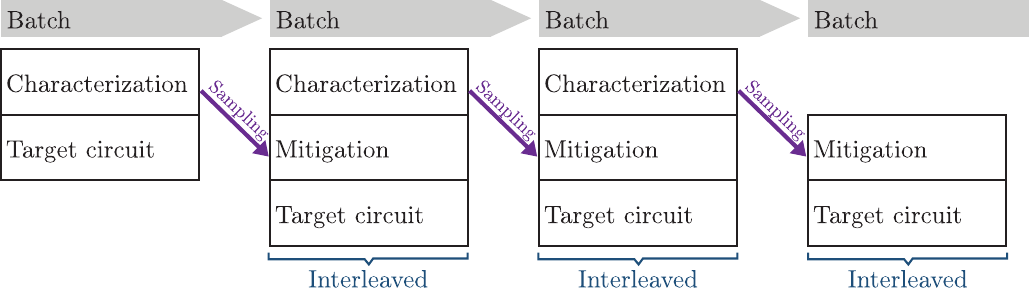}
\caption{Interleaving the target circuit in each batch allows us to use our drift rescaling method.}
\label{fig:Interleave-target-circuit}

\end{figure}

We need to explain from where we get $\left\langle O\right\rangle _{C,\mathrm{noisy}_{0}}$ and $\left\langle O\right\rangle _{C,\mathrm{noisy}_{1}}$. The most straight forwards way to get $\left\langle O\right\rangle _{C,\mathrm{noisy}_{0}}$ and $\left\langle O\right\rangle _{C,\mathrm{noisy}_{1}}$ is to interleave the target circuit $C$ in each batch (see Fig.~\ref{fig:Interleave-target-circuit}).

$\left\langle O\right\rangle _{C,\mathrm{noisy}_{1}}$ must represent the noise at the time the mitigation circuits ran on the device. We can always use the procedure in the previous section to obtain it. $\left\langle O\right\rangle _{C,\mathrm{noisy}_{0}}$ needs to be the noise at the time the characterization circuits ran on the device. 
This can always be done by mixing the original circuit when running characterization circuits. If the mitigation is run using batches as defined in App.~\ref{sec:Batches-of-characterization} then we can take $\left\langle O\right\rangle _{C,\mathrm{noisy}_{0}}$ from the previous mitigation batch (using App.~\ref{sec:Noisy-expectation-value}).
If the mitigation is run using batches, we decide whether to apply the heuristic (using the three conditions) for each batch separately. 

When the observable $O$ is a sum of Paulis: $O=\sum_{j=1}^{m}a_{j}P_{j}$, we have two options. In the first option, we calculate the mitigated expectation value using the heuristic for each Pauli separately, and then combine the results 
\[
\left\langle P\right\rangle _{\mathrm{mitigated}}=\sum_{j=1}^{m}a_{j}\left\langle P_{j}\right\rangle _{\mathrm{mitigated}}~.
\]
This enables the correction of drifts that occur locally. In the second option, we use this heuristic for the observable entire $O$ as is (without decomposing to different Paulis).

To calculate the error bar for $\left\langle O\right\rangle _{\mathrm{mitigated}}$, we use bootstrap. We resample the results for the mitigation circuits and again calculate $\left\langle O\right\rangle _{\mathrm{mitigated}}$ using the technique described above. We repeat this resampling many times and compute the standard deviation of the results, and this is our estimation of the error of the corrected mitigation value.

\subsection{Results}

\begin{figure}
\includegraphics[width=\columnwidth]{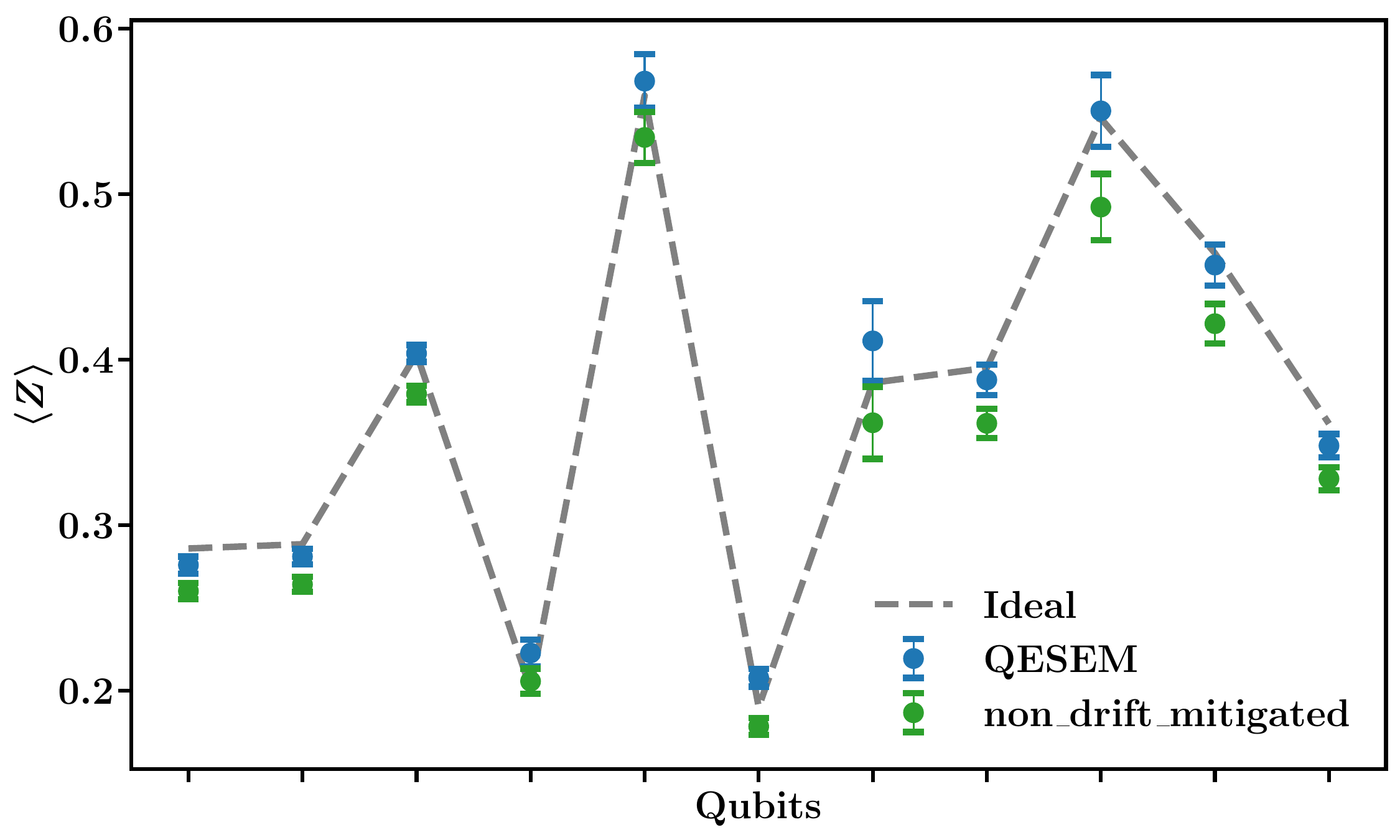}
\caption{Example of drift mitigation in the Hamiltonian simulation demonstration described in Sec.~\ref{main_demo}. The plot shows the results of step 5, restricted to qubits for which $\langle Z_i \rangle$ values obtained with and without the drift mitigation protocol are statistically inconsistent. }\label{fig:Drift-mitigation-Ising-benchmark}
\end{figure}

In Fig.~\ref{fig:Drift-mitigation-Ising-benchmark}, we present an example of drift mitigation in the Hamiltonian simulation demonstration described in Sec.~\ref{main_demo}. The data shown here correspond to the results of step 5. The observables plotted are the expectation values $\langle Z_q \rangle$, restricted to qubits $q$ for which the results with and without our drift mitigation protocol are inconsistent.
Specifically, we include only qubits whose error-bar intervals do not overlap between the two cases.
From the comparison, we see that drift mitigation improves the results: the measured expectation values are closer to the ideal values after applying the mitigation protocol.

\section{Characterization \label{Appendix: characterization}}

QP-based error mitigation methods rely on a noise model that accurately captures the implementation errors affecting the operations in a given quantum circuit. Consequently, they require an efficient and reliable characterization procedure to produce accurate results~\cite{Niroula2023, Govia2024}. Inaccuracies in the characterized noise model directly translate into mitigation bias. A characterization is considered well-suited for mitigation if the bias it introduces is small compared to the target precision while incurring only modest resource overhead.

This section outlines the characterization procedure underlying QESEM's mitigation. It improves upon standard protocols~\cite{Erhard2019, Flammia2021} by incorporating elements from self-consistent characterization~\cite{ChenChen2025} and resource reduction via noise shaping~\cite{Berg2023}, and by introducing novel components such as: (i) characterization of fractional layers, both in isolation and in combination with Clifford layers; (ii) inclusion of non-Markovian error parameters; and (iii) enhanced robustness through characterization circuits whose error propagation properties matches those of application circuits. 

\begin{figure*}
    \centering
    \includegraphics[width=\textwidth]{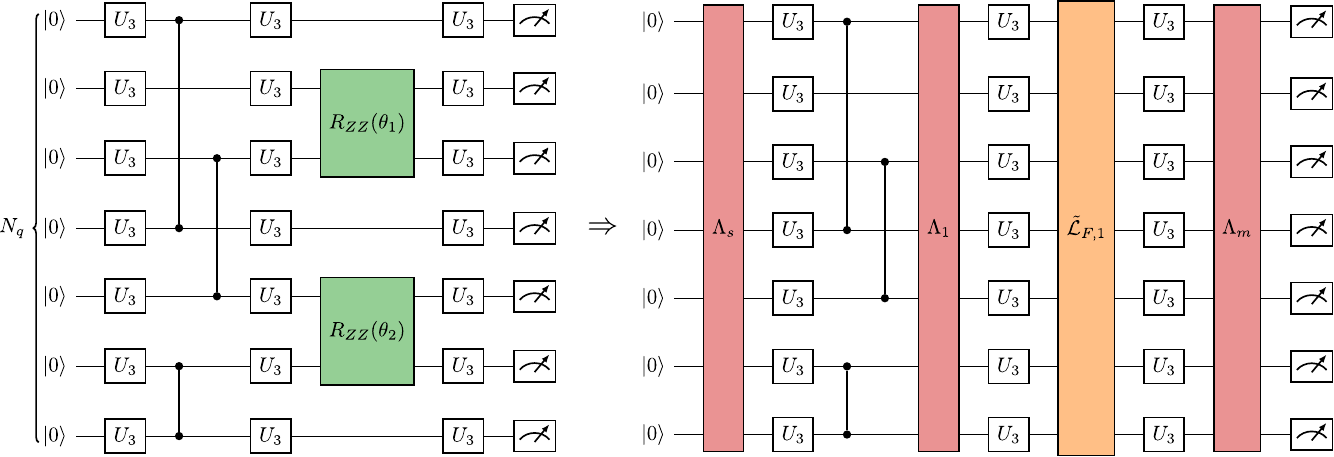}
    \caption{An example circuit with one CZ layer and one $R_{ZZ}$ layer. The ideal version appears on the left and the noisy version, as modeled by QESEM, on the right. The $U_3$ gates represent positions along the circuit where the algorithm may have SQGs. It is also the form of the executed circuits following QESEM compilation: various twirling operations and compression according to Eq.~\eqref{eq:u3_comprees}}
    \label{fig:noisy_circ}
\end{figure*}

\subsubsection{Setup}
 Consider a quantum circuit $\mathcal{C}$, acting on $N_q$ qubits, intended for execution on an unavoidably noisy quantum processor with a qubit connectivity graph $\mathcal{G}$. The circuit comprises:
\begin{enumerate}
    \item \textbf{State preparation:} all $N_q$ qubits are initialized in a known product state $\left|\rho_0\right\rrangle$, e.g., $\left| \rho_0 \right\rrangle = \left|0 \right\rangle \left\langle 0 \right|^{\otimes N_q}$.
    \item \textbf{Gate application:} A sequence of quantum gate layers, including:
    \begin{enumerate}
        \item $N_{L_C}$ unique layers of two-qubit Clifford gates $\left\{L_{C,i}\right\}_{i=1}^{N_{L_C}}$,
        \item $N_{L_{F}}$ unique layers of two-qubit fractional gates $\left\{L_{F,i}\right\}_{i=1}^{N_{L_F}}$,
        \item An arbitrary number of single-qubit-gate (SQG) layers of the form $\mathcal U = \bigotimes_{i=1}^{N_q} \mathcal U_i$
    \end{enumerate} 
    \item \textbf{Measurements:} Projective measurements of all $N_q$ qubits in the computational basis, $\{\left| E_m \right\rrangle\}_m = \{ \left| m \right\rangle \left\langle m \right|\}_{m \in \{0,1\}^{N_q}}$.
\end{enumerate}
This collection of operations defines a gate set that, in practice, is implemented imperfectly. We denote the noisy versions of these operations as:
\begin{align}
    \rho_0 \rightarrow \tilde\rho_0~, && E_m \rightarrow \tilde E_m~, && L \rightarrow \tilde{L}~.
\end{align}
 QESEM captures the noise associated with SQGs, effectively, as part of the noisy two-qubit gate layers (see details below).

\subsubsection{Noise models and circuit compilation}\label{subsubsec:noise-model-compilation}

We model noise in state preparation, measurement, and Clifford layers using quasi-local Pauli channels, i.e., 
\begin{align}
\label{eq:Pauli-channels}
    \tilde\rho_0 = \Lambda_s(\rho_0)~, \qquad \tilde E_m = \Lambda_m(E_m)~, \qquad \tilde{L} = \Lambda_{L_C}\circ L_C~,
\end{align}
where $\circ$ denotes composition. Each Pauli channel $\Lambda(\cdot)$ can be represented, equivalently, in terms of Pauli error rates $p_P$ or Pauli eigenvalues $\lambda_P$ as
\begin{align}
    \Lambda(\cdot) &= \sum_P p_P\,P\cdot P \;=\; 2^{-N_q}\!\sum_P \lambda_P\, P\, \mathrm{Tr}\!\left(P \cdot \right)~.
\end{align}
Here, $P = \otimes_j \sigma_{\mu_j}$, with $\mu_j \in \{I,X,Y,Z\}$, are $N_q$-qubit Pauli operators. Their support is $\mathrm{supp}(P) = \{j: \mu_j \neq I\}$, and the sum is restricted to $k$-local Paulis on $\mathcal{G}$\footnote{I.e., those for which the induced subgraph $\mathcal G[\mathrm{supp}(P)]$ is connected and $|\mathrm{supp}(P)|\le k$.}. Error rates and eigenvalues are related by the (Walsh–Hadamard) relations
\begin{subequations}
\begin{align}
    \label{eq:lambda_p} \lambda_{P} &= \sum_{Q} p_{Q}\,\chi(P,Q)~,\\
    p_{P} &= 4^{-N_q}\!\sum_{Q} \lambda_{Q}\,\chi(P,Q)~,
\end{align}
\end{subequations}
where $\chi(P,Q) = (-1)^{\langle P, Q \rangle} = 1$ if $[P,Q]=0$ or $-1$ otherwise. Trace preservation is ensured by $\lambda_I = 1$ (equivalently, $p_I = 1-\sum_{P\neq I}p_P$).

The eigenvalue form is convenient since $\Lambda(P)=\lambda_P\,P$ for all Pauli $P$. However, mitigation is more naturally implemented in terms of Pauli generator rates~\cite{berg2022probabilistic}: composing layers is \emph{additive} in the latter (whereas it is multiplicative in $\lambda$). To introduce a (Pauli–Lindblad/Markovian) generator, we assume the channel is \emph{strictly contractive} on the Pauli basis, i.e., $\lambda_P>0$ for all $P$. Using \eqref{eq:lambda_p}, this holds whenever
\begin{align}
\sum_{Q:\{P,Q\}=0}\!\! p_Q \;<\; \tfrac{1}{2} \qquad \forall\,P,
\end{align}
a natural condition for realistic noise near the identity (small error rates). 

Under this assumption, $\log \lambda_P$ is real for all $P$, and the Pauli channel admits a generator representation, sometimes dubbed sparse Pauli-Lindblad,
\begin{align}
\label{eq:generator_rates}
    \Lambda(\cdot) \;=\; \exp\!\Big[\sum_{P} \gamma_P\, P\cdot P\Big] \;\equiv\; e^{\mathcal{L}(\cdot)}~,
\end{align}
where the Pauli generator rates $\gamma_P$ are related to the Pauli eigenvalues via 
\begin{align}
\label{eq:lambda_gamma_relation}
    \lambda_{P} & = e^{-2\sum_{Q}\left\langle P,Q\right\rangle \gamma_{Q}}~ \\
    \gamma_{P} & = 4^{-N_q}\!\sum_{Q} \chi(P,Q)\,\log\lambda_{Q}.
\end{align}
The resulting generator rates are always \emph{real}. Trace preservation gives $\gamma_I=-\sum_{P\ne I}\gamma_P$ (equivalently, $\sum_P \gamma_P=0$).\footnote{An equivalent factorized form is $\Lambda(\cdot)=\bigcirc_{P}\!\left[\omega_P P\,\cdot\,P+(1-\omega_P)\,(\cdot)\right]$, where the commuting factors make the composition order immaterial; $\omega_P = \tfrac{1}{2}(1-e^{-2\gamma_P})$ are sometimes called Pauli conjugation weights.}

To enforce Pauli channel noise models, QESEM employs Pauli twirling~\cite{Knill2004, Wallman2016}. All circuits -- both for characterization and application -- are compiled with Clifford-layer twirling and measurement twirling. In the latter, QESEM pairs circuits with complementary $X$ gates before measurement: one circuit has $X$ gates applied to a random subset of qubits, and its pair applies $X$ gates to the complementary subset. By flipping the outcomes of the affected qubits and averaging the two circuits, measurement noise is symmetrized, exactly eliminating bias [i.e., $\Pr(1|0) \neq \Pr(0|1)$, where $\Pr(i|j)$ denotes the probability of measuring outcome $i$ when the system is in the state $j$]. Compared with standard measurement-twirling techniques~\cite{Nation2021}, this complementary pairing not only reduces the twirl-induced variance but also removes correlated measurement bias.

QESEM also performs a $U_3$ compression on all SQG layers between two-qubit gate layers. This compression rewrites the SQG into a fixed-depth structure consisting of three layers of $R_Z$ rotations interleaved with two $\sqrt{X}$ layers, i.e.
\small
\begin{align}
\label{eq:u3_comprees}
    \begin{pmatrix}
        R_Z(\phi_0) \\
        R_Z(\phi_1) \\
        \vdots \\
        R_Z(\phi_{N_q}) \\
    \end{pmatrix} 
    \begin{pmatrix}
        \sqrt{X} \\
        \sqrt{X} \\
        \vdots \\
        \sqrt{X} \\
    \end{pmatrix} 
    \begin{pmatrix}
        R_Z(\theta_0) \\
        R_Z(\theta_1) \\
        \vdots \\
        R_Z(\theta_{N_q}) \\
    \end{pmatrix} 
    \begin{pmatrix}
        \sqrt{X} \\
        \sqrt{X} \\
        \vdots \\
        \sqrt{X} \\
    \end{pmatrix} 
    \begin{pmatrix}
        R_Z(\psi_0) \\
        R_Z(\psi_1) \\
        \vdots \\
        R_Z(\psi_{N_q}) \\
    \end{pmatrix}~.
\end{align}
\normalsize
This enforces a uniform circuit structure across both characterization and application circuits, enabling any residual SQG infidelity to be effectively absorbed by adjacent two-qubit gate layers (see, e.g., the right-hand side of Fig.~\ref{fig:noisy_circ}).

Fractional gate layers cannot be twirled with the full Pauli group without altering the native parameters of $h$. Therefore, in QESEM, they are twirled only with the subgroup of Paulis that commute with $h$. This \emph{partial Pauli twirling} shapes most microscopic noise into Pauli generators acting during the layer. Certain coherent errors, such as rotation-angle miscalibration, do persist; these are eliminated by QESEM's dedicated high-accuracy calibration procedure (see App.~\ref{app:calibration} for details).

The noise model resulting from partial twirling resembles a Pauli channel but acts concurrently with the ideal gate evolution. Specifically, the noisy layer takes the form
\begin{align}
    \label{eq:during_pauli_channel}
    \tilde{L}_{F} &\equiv \exp{\left[i[h,\cdot]+\mathcal{L}\right]} = \exp{\left[i[h,\cdot] + \sum_{P} \gamma_{P} P \cdot P\right]}~,
\end{align}
where $L_F = e^{i[h,\cdot]}$ denotes the intended (ideal) layer action, generated by the Hamiltonian $h$. We refer to this model as \textit{during-layer Pauli channel}, and unlike in the Clifford case, the noise here cannot be cleanly separated from the ideal-gate unitary. Attempting such a factorization would yield a noise channel that is not strictly Pauli. 

The number of independent parameters in the noise models of Eqs.\eqref{eq:during_pauli_channel} and \eqref{eq:Pauli-channels} depends on the assumed locality of the Pauli generators and can be quite large. However, when executing twirled (and compressed) circuits on most commercially available quantum devices, we find that circuit outcomes are predicted and mitigated accurately using quasi-local Pauli channel noise models. The parameterization is configured during QESEM's adaptation to a new device. For most devices, including IBM Eagle and Heron processors, two-local Pauli generators for gate layers and single-qubit Pauli-$X$ generators for SPAM typically suffice.  

To further reduce the number of independent parameters in the noise models, QESEM performs square-root Pauli twirling (SP-twirling) to each layer. The SP-twirl group, $\mathcal{U}_{SP}$, consists of pairs of square-root Pauli SQG layers satisfying $\mathcal U_{2} = L^{-1}\mathcal U_{1}^{\dagger}L$. Similar to other twirl operations, $\mathcal U_1$, $\mathcal U_2$ operate before and after the layer $L$, i.e.,
\begin{align}
    \tilde{L} \equiv\frac{1}{dim\left\{ \mathcal{\mathcal{U}_{SP}}\right\} }\sum_{\left(\mathcal U_{1},\mathcal U_{2}\right)\in\mathcal{U}_{SP}}\mathcal U_{1} L \mathcal U_{2}~.
\end{align}
By construction, SP-twirling enforces symmetry of the noise under conjugation by the square-root Pauli elements $\mathcal U_1$, imposing symmetry constraints on the structure of the noise channel. These symmetry constraints significantly reduce the number of Pauli generators that must be learned (see also Ref.~\cite{Berg2023}). 

SP-twirling works for Clifford layer with subsequent Pauli channels and fractional layers with during-layer Pauli channels, all the same. Explicitly, for Clifford layers, the symmetrized noisy layer reads as
\begin{align}
    \tilde{L}&=\frac{1}{dim\left\{ \mathcal{\mathcal{U}_{SP}}\right\} }\sum_{\left(\mathcal U_{1},\mathcal U_{2}\right)\in\mathcal{U}_{SP}} \mathcal U_{1}e^{\mathcal{L}} L_{0}\left(L_{0}^{-1} \mathcal U_{1}^{\dagger} L_{0}\right)  \\
    \nonumber &=\left[\frac{1}{dim\left\{ \mathcal{\mathcal{U}_{SP}}\right\} }\sum_{\left(\mathcal U_{1},\mathcal U_{2}\right)\in\mathcal{U}_{SP}} \mathcal U_{1}e^{\mathcal{L}}\mathcal U_{1}^{\dagger}\right]L_{0} \equiv e^{\tilde{\mathcal{L}}} L_{0}~.
\end{align}
The symmetrized Pauli channel $e^{\tilde{\mathcal{L}}}$ is then symmetric under the action of all group elements. Namely, the following constraints are imposed
\begin{align}
\label{eq:sp_twirl_const}
    & e^{\tilde{\mathcal{L}}}=\mathcal U_{1}e^{\tilde{\mathcal{L}}} \mathcal U_{1}^{\dagger} \nonumber \\
    & \Rightarrow \tilde{\mathcal{L}}-\mathcal U_{1}\tilde{\mathcal{L}} \mathcal U_{1}^{\dagger}=0~ \nonumber  \\
    & \Rightarrow \sum_{P} \tilde{\gamma}_{P} P\cdot P -\sum_{P} \tilde{\gamma}_{P} \mathcal U_{1} P\mathcal U_1^\dagger \cdot \mathcal U P \mathcal U_{1}^{\dagger} = 0 \nonumber \\
    & \Rightarrow \sum_{P} \tilde{\gamma}_{P} P\cdot P - \sum_{Q} \tilde{\gamma}_{\mathcal U_{1}^{\dagger} Q \mathcal U_{1}} Q\cdot Q = 0 \nonumber \\
    & \Rightarrow \sum_{P} \left(\tilde{\gamma}_{P}-\tilde{\gamma}_{\mathcal U_{1}^{\dagger} P \mathcal U_{1}}\right) P\cdot P = 0 \nonumber \\
    & \Rightarrow \begin{array}{cc}
    \tilde{\gamma}_{P}=\tilde{\gamma}_{\mathcal U_{1}^{\dagger} P \mathcal U_{1}} & \forall P~,
\end{array}
\end{align}
for all $\mathcal U_1$ in $\mathcal{U}_{SP}$.

QESEM characterizes the noisy two-qubit gate layers and SPAM by executing a dedicated set of characterization circuits. These circuits are constructed to be efficiently simulable on a classical computer under the assumed noise models, enabling rapid evaluation of expected outcomes. The Pauli generator rates associated with the entire noisy gate set are then fitted to match the observed circuit outcomes. This inference is performed jointly over all layers and SPAM, yielding a consistent set of generator rates from a global fit across the full circuit collection.

To guarantee learnability, the characterization circuits must be sensitive to all independent generator rates appearing in the noisy gate set. In what follows, we describe the circuits used for Clifford two-qubit gate layers (Sec.~\ref{sec:char_cliff}) and fractional two-qubit gate layers (Sec.~\ref{sec:char_frac}). Using similar logic to Refs.~\cite{Chen2023, Chen2024}, we argue that this collection forms an amplificationally complete set: the circuits are sensitive to all learnable combinations of Pauli generator rates and amplify every term that can be amplified.

\subsection{Fractional two-qubit gate layers}
\label{sec:char_frac}
The circuits QESEM executes for the characterization of fractional two-qubit gate layers are based on Pauli-twirled \textit{composite idle germs} constructed from these layers. In particular, for each fractional layer $L_{F,i}$, a composite idle germ $g_i$ is constructed such that its noise, modeled as a Pauli channel, is determined by the layer's (during-layer Pauli channel) noise in an invertible manner. QESEM then uses the inverse mapping to directly fit the Pauli generator rates of the fractional layer to match the outcomes of characterization circuits for the identity operator.

\subsubsection{Characterization circuits for idle germs}
\begin{figure}[tb]
\centering
\subfloat[\label{fig:circs_and_germ_frac_a}]{
\includegraphics[width=0.96\columnwidth]{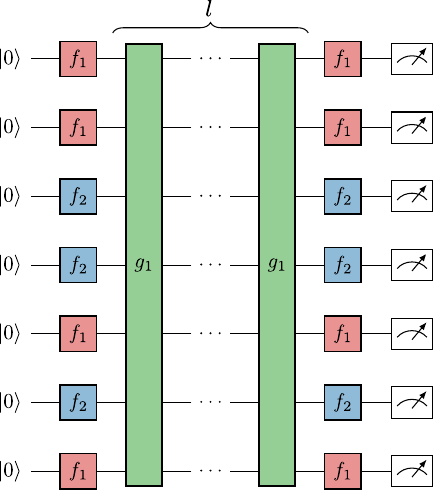}
}
\\
\subfloat[\label{fig:circs_and_germ_frac_b}]{
\includegraphics[width=0.96\columnwidth]{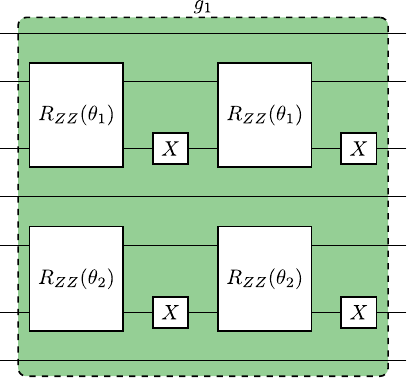}
}
\caption{
\textbf{\protect\subref{fig:circs_and_germ_frac_a}} A schematic of the characterization circuits for an idle germ $g_i$, repeated $l$ times. Here, for example, the qubit colors are red: 0, 1, 4, 6, and blue: 2, 3, 5. 
\textbf{\protect\subref{fig:circs_and_germ_frac_b}} The idling germ $g$ for an $R_{ZZ}$ layer that matches the qubit colors of \protect\subref{fig:circs_and_germ_frac_a}.
}
\end{figure}

Circuits amplifying the noise of a noisy idle $g_i$ are fairly standard and can be written concisely in the following form 
\begin{align}
\label{eq:frac_circs_l}
    C_{i,l,f} = \left\llangle E_m \right| f \left(g_i\right)^{l} f^\dagger \left| \rho_0 \right\rrangle~.
\end{align}
Here, $f$ are products of single-qubit square-root Pauli gates referred to as \textit{fiducials}. They prepare and measure Pauli eigenstates. Since $g_i$ is ideally the identity operator, its noise accumulates in this circuit with the \textit{amplification length} $l$. The circuits are visualized schematically in Fig.~\ref{fig:circs_and_germ_frac_a}.

Each fiducial configuration enables sensitivity to a particular combination of Pauli generator rates [see Eq.~\eqref{eq:lambda_gamma_relation}]. To learn all generator rates of a $k$-local Pauli channel, the fiducial layers must collectively span the eigenbases of all $k$-local Paulis. Indeed, the relation in Eq.~\eqref{eq:lambda_gamma_relation} is invertible. For $k=2$, QESEM uses fiducials of the form
\begin{equation}
     f = \prod_{q\in \text{QC1}} f_1 \prod_{q\in \text{QC2}} f_2~,
\end{equation}
where $QC1$ and $QC2$ are the two color classes of a bipartite coloring of the device's connectivity graph $\mathcal{G}$ and $f_1$, $f_2$ are square-root Pauli SQGs applied uniformly across each color class [See, for example, Fig.~\ref{fig:circs_and_germ_frac_a}]\footnote{If $\mathcal{G}$ is not bipartite, multiple such colorings are used.}. In general, full characterization of a two-local Pauli channel requires nine combinations of $(f_1, f_2)$, with $f_i \in \{I, \sqrt{Y}, \sqrt{X} \}$ (see also Ref.~\cite{Berg2023}). 

\begin{figure*}[tb]
    \centering
    \includegraphics[width=0.9\linewidth]{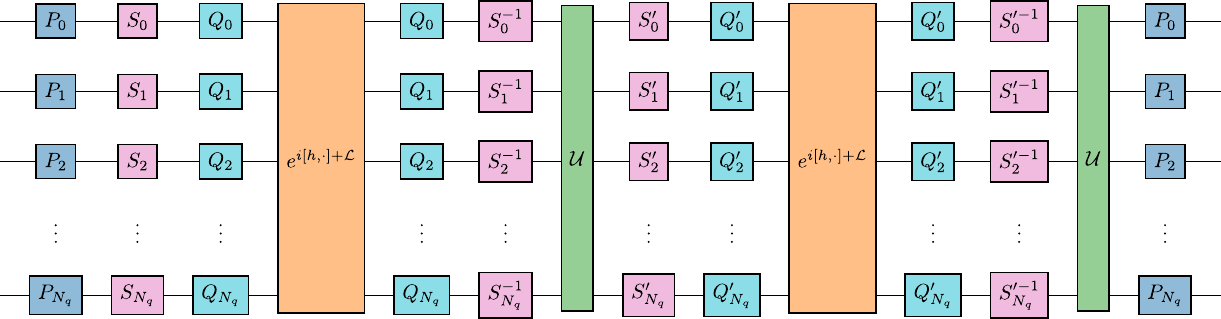}
    \caption{The various twirls of the idling germ: (i) $P_i$ are elements of the full Pauli-twirl group of the identity operator; They enclose the entire germ. (ii) $S_i$ and $S_i'$ are both elements of the SP-twirl group of the underlying fractional gate. Here we assumed they also commute with the gate, as in the case of an $R_{ZZ}$ layer and $S\in\{\sqrt{Z},I\}$. (iii) $Q_i$ and $Q_i'$ are both elements of the partial Pauli-twirl group of the underlying fractional layer. For an $R_{ZZ}$ gate, these are $Q_i \otimes  Q_j \in \{I\otimes I, I\otimes Z, X\otimes X, X\otimes Y,Y\otimes X, Y\otimes Y, Z\otimes I, Z\otimes Z\}$. Here, $\mathcal{L}$ is a general during-layer sparse Pauli-Lindblad model. $Q_i$ shape it into a during-layer Pauli-channel, $S_i$ enforce symmetries between its Pauli generator rates, and, finally, $P_i$ shape the full noise model of the germ into a Pauli channel.
    \label{fig:frac_germ_with_twirl}}
\end{figure*}

However, for many two-qubit gates, SP-twirling imposes symmetries that substantially reduce the required number of fiducial configurations. For example, SP-twirling of $R_{ZZ}$ gates, performed with $\mathcal U_1=\mathcal U_2\in\{I\otimes I, I\otimes \sqrt{Z}, \sqrt{Z}\otimes I, \sqrt{Z}\otimes \sqrt{Z}\}$, enforces invariance under conjugation by square-root $Z$ gates. This symmetry equates the rates of Pauli generators that differ only by $Y \leftrightarrow X$. As a result, all Pauli generator rates can be inferred using only $X$- and $Z$-type Pauli eigenstates, reducing the required fiducial combinations to four, with $f_i \in \{ I, \sqrt{Y} \}$. This reduction applies to all KAK-1 fractional gates, namely those that can be transpiled to an $R_{ZZ}$ gate and SQGs~\cite{KAK}. 

To gain sensitivity to all model parameters, a single amplification length is sufficient, along with $l=0$ circuits that measure SPAM errors
\begin{align}
    \label{spam_circ}
    C_0 = \left\llangle E_m \middle|  \rho_0 \right\rrangle~.
\end{align}
The optimal value of $l$ is related to the expected infidelity of the fractional layer from which the germ is constructed, and the rate of SPAM errors. However, a set of amplification lengths, $l$, can be used to gain robustness to out-of-model errors. A typical set used by QESEM consists of five characterization lengths, chosen according to a logarithmic spacing from $l=1$ to $l\sim\frac{0.25}{\widehat{IF}}$, where $\widehat{IF}$ is a prior estimate of the median infidelity. 

\subsubsection{The composite idle germ}
The germ corresponding to the layer $L_{F,i}$ reads as
\begin{align}
    g_i = L_{F,i} \mathcal{U} L_{F,i} \mathcal{U}^\dagger~,
\end{align}
where $\mathcal{U}$ is a layer of Pauli SQGs that inverts the ideal operation of the layer. For example, in the case of $R_{ZZ}$ layers -- layers of $R_{ZZ}$ gates [Eq.~\eqref{eq:during_pauli_channel} with $h=\tfrac{1}{2}\sum_{\langle i,j \rangle} \alpha_{ij} [Z_i Z_j,\cdot]$ with $Z_i Z_j$ acting only between qubits connected by gates in the layer] -- $\mathcal{U}$ can be chosen as a layer of Pauli-$X$ gates acting on one qubit of each $\langle i,j \rangle$ pair. Figure \ref{fig:circs_and_germ_frac_b} displays the germ for an example $R_{ZZ}$ layer.

\begin{figure}[!htb]
    \centering
    \includegraphics[width=0.86\columnwidth]{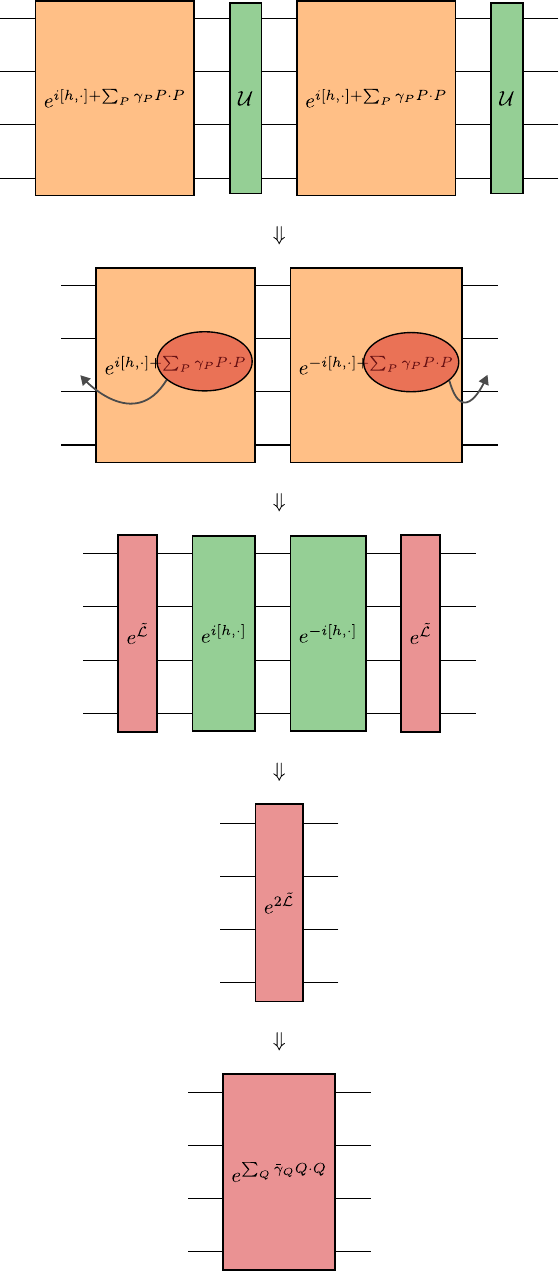}
    \caption{A visualization of the key steps described in Eq.~\eqref{integral}: At the first transition, the single qubit gates layers $\mathcal U$ flip the ideal operation of the fractional layer, but don't affect the Pauli generators. Then, at the second transition, the during-layer Pauli channels are split into before and after the ideal layers. They transform into a noise channel described by $\tilde{\mathcal{L}}$ that is not strictly a Pauli-channel. In the third step, the ideal layer and its inverse combine to form the identity. Then, the noise channels can also combine. Finally, the full twirl of the germ shapes the noise model described by $2\tilde{\mathcal{L}}$ into a Pauli channel
    \label{fig:during_to_after}}
\end{figure}

Since the ideal operation of the germ is the identity, its noisy version can be modeled as a Pauli channel, i.e., 
\begin{align}
    \tilde g_i = \Lambda_{g,i}~,
\end{align}
provided the germ is twirled using the full Pauli group. Indeed, the composite idle germs are Pauli twirled in the characterization circuits (see Fig.~\ref{fig:frac_germ_with_twirl} for details about the various twirls taking place). The relation between $\Lambda_{g_i}$ and the Pauli generator rates of the underlying during-layer Pauli channel, $L$ in Eq.\eqref{eq:during_pauli_channel}, can be obtained perturbatively in the latter using a Dyson series. Specifically,
\begin{widetext}
\begin{align}
\label{integral}
    \nonumber \Lambda_{g} =& \tilde g = \tilde{L}_{F} \mathcal{U} \tilde{L}_{F} \mathcal{U} = e^{i [h,\cdot] + \mathcal{L}} \mathcal U e^{i[h,\cdot] + \mathcal{L}} \mathcal U = e^{i [h,\cdot] + \mathcal{L}} e^{-i[h,\cdot] + \mathcal{L}} \\
    \nonumber =& \left[e^{i [h,\cdot]} + \int_0^1 dt e^{i [h,\cdot]t} \mathcal{L} e^{i [h,\cdot](1-t)} + O(\mathcal{L}^2)\right] \left[e^{-i [h,\cdot]} + \int_0^1 dt e^{-i [h,\cdot](1-t)} \mathcal{L} e^{-i [h,\cdot]t} + O(\mathcal{L}^2)\right] \\
    \nonumber =& \left[1 + \int_0^1 dt e^{i [h,\cdot]t} \mathcal{L} e^{-i [h,\cdot]t}\right] e^{i [h,\cdot]}e^{-i [h,\cdot]} \left[1 + \int_0^1 dt e^{i [h,\cdot]t} \mathcal{L} e^{-i [h,\cdot]t}\right] + O(\mathcal{L}^2) \\
    =& 1 + 2 \int_0^1 dt e^{i [h,\cdot]t} \mathcal{L} e^{-i [h,\cdot]t} + O(\mathcal{L}^2) = \exp{\left[2\int_0^1 dt e^{i[h,\cdot]t} \mathcal{L} e^{-i[h,\cdot]t} \right]} + O(\mathcal{L}^2) \equiv \mathcal I + O(\mathcal{L}^2)~,
\end{align}
\end{widetext}
where we used that $\mathcal U$ consists of Pauli operators and thus commutes with the Pauli generators in $L$. A similar use of the Dyson series was described in Ref.~\cite{Malekakhlagh2025} for synthesizing a noise model given a Lindbladian. We note that according to Eq.\eqref{integral}, $\mathcal I$ may include non-diagonal terms $\sim P\cdot P'$. These are eliminated by the full Pauli twirling of the germ and can thus be disregarded. The during-to-after transformation is visualized in Fig.~\ref{fig:during_to_after}

Plugging the explicit forms of $\mathcal{L}$, $\Lambda_g$ in terms of Pauli generators and their rates, and taking the logarithm of both sides, Eq.~\eqref{integral} takes the form $\sum_Q \bar\gamma_Q Q\cdot Q = 2 \sum_P \gamma_P w(P)$ with 
\begin{align}
    w(P) \equiv \int_0^1dt e^{i[h,\cdot]t}Pe^{-i[h,\cdot]t}\cdot e^{i[h,\cdot]t}Pe^{-i[h,\cdot]t}~.
\end{align}
It follows that the Pauli generator rates of the germ are given by
\begin{align}
\label{eq:gamma_relations}
    \bar{\gamma}_{Q}=2\sum_{P}\gamma_{P}Tr\left[Qw\left(P\right)Q\right]~,
\end{align}
where $\gamma_P$ are the Pauli generator rates of the fractional layer. For $h=0$ the integral $w(P)=P\cdot P$ and the relation in Eq.~\eqref{eq:gamma_relations} is invertible. From the continuity of $h$, it follows that the relation remains invertible in some finite region away from 0. For example, for $R_{ZZ}$ layers, the relation between the germ noise and the layer noise is invertible so long as all $\alpha_{i,j} < \frac{\pi}{2}$. 

The integral in $w(P)$ can be solved, given the specific fractional layer and locality of its during-layer Pauli channel. For $R_{ZZ}$ layers with $P$ that are two-local the solution for $w(P)$ follows:
\begin{enumerate}
\item \textbf{Commuting Pauli generators:}\\
If $\left[h,P\right]=0$ then
\begin{align}
    w\left(P\right)=P\cdot P~.
\end{align}
\item \textbf{Single non-commuting gate generator:}\\
If $\left[h,P\right]\neq0$ due to only one pair, e.g., $\left\langle m,n\right\rangle $, then
\begin{align}
\label{eq:single_non_commute}
    w\left(P\right) = &\frac{1+S_{mn}}{2} P\cdot P + \frac{1-S_{mn}}{2}P'\cdot P',
\end{align}
where $S_{mn} = \frac{\sin\left(2\alpha_{mn}\right)}{2\alpha_{mn}}$ and $P' = iZ_{m}Z_{n}P$. Non-diagonal terms $\sim \left(P\cdot P'+P'\cdot P\right)~$ are eliminated by Pauli twirling and are therefore omitted from this expression.
\item \textbf{Two non-commuting gate generators:} \\
If $\left[h,P\right]\neq0$ due to two pairs, e.g., $\left\langle m,n\right\rangle$ and $\left\langle l,p\right\rangle$, then
\small
\begin{align}
    w\left(P\right) = &\frac{1}{4}\left[1 + S_{mn} + S_{lp} + S_{mnlp,-} + S_{mnlp,+}\right] P\cdot P \\ 
    \nonumber & + \frac{1}{4} \left[1 - S_{mn} + S_{lp} -S_{mnlp,-} - S_{mnlp,+}\right] P_{1}\cdot P_{1} \\ 
    \nonumber & + \frac{1}{4}\left[1 + S_{mn} - S_{lp} -S_{mnlp,-} - S_{mnlp,+}\right] P_{2}\cdot P_{2} \\
    \nonumber & + \frac{1}{4} \left[1 - S_{mn} - S_{lp} + S_{mnlp,-} + S_{mnlp,+}\right] P_{3}\cdot P_{3}
\end{align}
\normalsize
where $S_{mnlp,\pm} = \frac{\sin\left(2\alpha_{mn} \pm 2\alpha_{lp}\right)}{4\alpha_{mn} \pm 2\alpha_{lp}}$, $P_{1}=iZ_{m}Z_{n}P$, $P_{2}=iZ_{l}Z_{p}P$, and $  P_{3}=Z_{m}Z_{n}Z_{l}Z_{p}P$. Also here, non-diagonal terms (mixing $P,P_1,P_2,P_3$) are eliminated by Pauli twirling and are therefore omitted from this expression
\end{enumerate}
Any other KAK-1 fractional layer can be transpiled to an $R_{ZZ}$ and SQG layers. Consequently, the above solution can be applied to all KAK-1 fractional gates. 

The breakdown of invertibility at the Clifford angle can be seen, for example, in the result of the single non-commuting gate generator case. Rewriting Eq.~\eqref{eq:single_non_commute} in the $(P\pm P')$ base
\begin{align}
    \nonumber w(P) =& \frac{1}{2}(P+P')\cdot (P+P') \\
    & + \frac{1}{2}S_{mn}(P-P')\cdot (P-P')~,
\end{align}
we note that $S_{mn}$, the prefactor of $(P-P')\cdot (P-P')$, vanishes at $\alpha_{mn}=\frac{\pi}{2}$. However, at the Clifford angle, the gate can be treated as a Clifford gate, i.e., twirled with the full Pauli group. Then, the combination $(P-P')$ is non-amplifiable (see Sec.~\ref{sec:char_cliff} for details).

\subsubsection{State preparation errors}
The circuit in Eq.~\eqref{spam_circ} is sensitive to the combined effects of the state preparation and measurement Pauli channels. It does not distinguish between those two types of errors. Neither do any of the circuits in Eq.~\eqref{eq:frac_circs_l}. However, when state preparation errors are non-negligible, it is pertinent to distinguish them from the measurement errors. Errors in state preparation spread within the algorithm's light cone, whereas measurement errors act locally at the end.

If the application circuit $\mathcal C$ includes Clifford layers, they can be used to construct circuits that distinguish between state preparation errors and measurement errors (see Sec.~\ref{sec:char_cliff} for details). Otherwise, QESEM estimates the state preparation noise using prior knowledge on their relation to the measurement errors, e.g., in systems employing active reset. 

In active reset, a qubit $q$ is measured and then acted on with a Pauli-$X$ gate if the measurement outcome is "1". The Pauli-$X$ gate then flips the state, assumed to be indeed $\left|1\right>$, to the desired initial state $\left|0\right>$. This measurement and conditional application of the Pauli-$X$ procedure is usually repeated a couple of times. In the limit of infinite repetitions, we solve
\begin{subequations}
\begin{align}
    p_s &= \frac{1}{2}\left[1-\sqrt{\frac{p_{1,x}-p_1}{p_{1,x}+p_1}}\right]~,\\
    p_m &= \frac{1}{2}\left[1-\sqrt{p_{1,x}^2 - p_1^2}\right]~,
\end{align}
\end{subequations}
where $p_{s/m}$ are the state/measurement bitflip probabilities and $p_1, p_{1,x}$ are the probability of measuring "1" and the probability of measuring "1" after a Pauli-$X$ gate. These can be calculated from the outcomes of the circuit in Eq.~\eqref{spam_circ} with SPAM twirling.

For a finite number of repetitions (of the active reset process), the deviation of $p_s, p_m$ from the expressions above is governed by $(2b)^{n+1}$, where $b=\frac{1}{2}\left[P_m(0|1)-P_m(1|0)\right]$ is the measurement bias. It can also be expressed in terms of $p_1, p_{1,x}$ as 
\begin{align}
    b = \tfrac{1}{2}\left[1 - p_1 - p_{1,x}\right]~.
\end{align}
It follows that one can easily estimate the quality of the infinite repetition approximation. In practice, $2b = 1 - p_1 - p_{1,x}$ is usually small enough that already three repetitions ($n=3$) suffice to get a good approximation. 

\subsubsection{Inference}
For each executed characterization circuit, QESEM processes the measurement outcomes to determine the expectation values of $Z$-basis local observables, up to, at least, the locality $k$ of the during-layer Pauli generator rates. Over all fiducial configurations, these expectation values measure all $k$-local Pauli fidelities. The Pauli fidelity obtained from measuring $\langle O\rangle$ in a given circuit $C_{i,l,f}$ [see Eq.\eqref{eq:frac_circs_l}] is expected to follow
\begin{align}
\label{eq:pauli_fidelity}
    F_{P,i} &= \lambda_{P,m} \left(\prod_l \lambda_{P,i} \right)\lambda_{P,s} = \lambda_{P,m} \lambda_{P,i}^l \lambda_{P,s} \\
    \nonumber &= \exp{\left[-2\sum_{P'}\langle P, P'\rangle \left(l\cdot\bar\gamma_{P',i}  +\gamma_{P',m}+\gamma_{P',s}\right)\right]}~,
\end{align}
where $P=fOf^\dagger$. Here, $\bar\gamma_{P', i}$ are the Pauli generator rates for the composite idle germ corresponding to the $i^{th}$ layer, and $\gamma_{P',s/m}$ are the Pauli generator rates for the SPAM Pauli channels.

The Pauli generator rates for the entire gate set $\vec \gamma$ are fitted to minimize the mean square error (MSE) cost function
\begin{align}
\label{eq:MSE}
    MSE=\left| \langle \vec O \rangle - \vec F \right|^2 = \left| \langle \vec O \rangle - e^{A\vec\gamma} \right|^2~,
\end{align}
where $\langle \vec O \rangle$ is a vector of all measured expectation values and $\vec F$ is a vector including all expected Pauli fidelities. Here, $A$ is the \textit{design matrix}, which follows from Eq.~\eqref{eq:pauli_fidelity}. QESEM performs the inference in two steps:
\begin{enumerate}
    \item \textbf{Exact solution of linearized least squares equation:}\\
    The MSE cost function can be solved analytically by taking the logarithm of the cost function and equating it to zero, i.e.,
    \begin{align}
        \log \langle \vec O \rangle  = A \vec \gamma \Rightarrow \vec\gamma = A^{-1} \log \langle \vec O \rangle~.
    \end{align}
    The invertibility of the design matrix follows from the invertibility of all $D_i$, the transformation from layer noise to germ noise. The log transformation over-weighs small $\langle O_i \rangle$~\cite{CarrollRuppertBook}. QESEM restores approximately equal emphasis to the original-scale residuals by introducing weights as
    \begin{equation}        
    \begin{aligned}
        \log\langle O_i \rangle &\rightarrow \langle O_i \rangle^{-1} \log\langle O_i \rangle~, \\
        A &\rightarrow diag(\langle O \rangle^{-1}) A~,
    \end{aligned}
    \end{equation}
    where $diag(\langle O \rangle^{-1})$ is a diagonal matrix with $\langle O \rangle^{-1}$ on its diagonal.
    \item \textbf{Gradient descent for the non-linear MSE:}\\
    Using the linearized least squares solution as the initial condition, QESEM optimizes the parameters $\vec\gamma$ further using the full non-linear MSE in Eq.~\eqref{eq:MSE}
\end{enumerate}

Here, we discussed the inference, ignoring any Clifford two-qubit gate layers that may participate in the algorithm. If any such layers are present, they require self-consistent characterization along with the fractional ones. The next section will provide a detailed treatment of Clifford layers. In particular, their contribution to the design matrix $A$ following Eq.\eqref{eq:cliff_A}. The full design matrix, including both contributions from fractional and Clifford layers (as well as SPAM), remains invertible. Consequently, the inference described here holds for the full design matrix as well.

\subsection{Clifford two-qubit gate layers}
\label{sec:char_cliff}
The set of characterization circuits executed by QESEM for Clifford two-qubit gate layers can be classified into two families:
\begin{enumerate}
    \item Circuits applying only one unique Clifford two-qubit gate layer.
    \item Circuits mixing more than one unique Clifford two-qubit gate layer.
\end{enumerate}
The former provides sensitivity to all inter-layer learnable combinations of Pauli generator rates and amplifies all intra-layer amplifiable combinations~\cite{VandenBerg2023, Chen2023}. The latter amplifies inter-layer combinations that are built from intra-layer non-amplifiable, but learnable, ones~\cite{ChenChen2025, Chen2024}. They also provide a form of \textit{spatial error spreading} that offers robustness to the characterization.

In what follows, we will refer to combinations of Pauli generator rates as self-amplifiable, amplifiable, learnable but non-amplifiable, and non-learnable, i.e., pure gauge. See Refs.~\cite{Chen2023, Chen2024} for a detailed discussion of these concepts, including formal proofs. Instead of reiterating proofs or derivations, we will focus on the example of CZ layers. This example generalizes easily to any set of Clifford layers consisting of KAK-1 two-qubit gates, as these can be transpiled to a CZ and SQG layers. QESEM's characterization is also readily applicable to higher KAK Clifford layers, e.g., $i$SWAP -- a KAK-2 gate, with some modifications. 

\subsubsection{Single-layer circuits}
For each Clifford two-qubit gate layer in the target circuit $\mathcal{C}$, QESEM executes a family of characterization circuits given by
\begin{align}
\label{eq:cliff_circs_l}
    C_1(L_{C,i},l, f) = & \left\llangle E_m \right| f_m \left(L_{C,i}\right)^{l} f_s \left| \rho_0 \right\rrangle~.
\end{align}
Here, the state fiducial $f_s$ prepares a specific Pauli eigenstate, and the measurement fiducial $f_m$ rotates the final state back to the computational basis. 

Conversely, $f_m$ rotates the measurement of Pauli observables $O$ into Pauli eigenstates. These propagate backward through the circuit and eventually rotate by $f_s$ to $Z$, which overlaps with the initial state $\left| \rho_0 \right\rrangle$. During backward propagation, the Pauli eigenstates shrink at each Pauli channel according to the corresponding Pauli eigenvalues and rotate into other Pauli eigenstates at each Clifford layer. Namely, denoting $O_m = f_m O f_m^\dagger$, the expected measured expectation value of a Pauli observable $O$ is given by
\begin{align}
    \label{eq:cliff_A}
    \langle O \rangle =& \lambda_{O,m} \prod_{n=0}^{l-1} \left(\lambda_{L_C^n O_m L_C^{n,\dagger}, i} \right)\lambda_{Z, s} \times \delta_{f_s L_C^l O_m L_C^{l,\dagger} f_s^\dagger, Z}~.
\end{align}
Here, $\delta_{i,j}$ is the Kronecker delta.

In circuits for Clifford layers whose $l^{\text{th}}$ power is the identity -- e.g., any even $l$ for a CZ layer -- $f_m$ can be taken as $f_s^\dagger$, reproducing the role of the fiducials in Eq.\eqref{eq:frac_circs_l}. The reasoning and results described in Sec.~\ref{sec:char_frac} also hold for the idling circuits considered here. In particular, four fiducial configurations (nine, if square-root Pauli twirling is not employed) are sufficient to gain sensitivity to all Pauli generator rates of the idling germ. 

For the example of a CZ layer with $l=2l'$, this idling germ is the square of the layer, its noise is amplified with an amplification length $l'$, and the four fiducials are
\begin{align}
    (f_1, f_2) \in \{ (I, I), (I, \sqrt{Y}), (\sqrt{Y}, I), (\sqrt{Y}, \sqrt{Y}) \}~.
\end{align}
The two fiducial configurations $(f_1, f_2) \in \{ (I, I), (\sqrt{Y}, \sqrt{Y}) \}$ are sufficient if idle qubits in the layer are twirled with a larger SP-twirl group. The latter is given by $\{\sqrt{X},\sqrt{Y},\sqrt{Z}\}$ instead of that induced by the layer, e.g., $\{I, \sqrt{Z}\}$ for CZ layers. In this reduced-fiducial setting, inferring a $k$-local noise model requires evaluating $k+1$-local, rather than just $k$-local, observables.

The noisy idle germ for a Clifford layer that squares to identity, e.g., CZ layer, reads as
\begin{align}
    \nonumber \tilde g &= \tilde{L}_C^2 = e^{\mathcal{L}} L_C e^{\mathcal{L}} L_C = e^{\mathcal{L}} e^{L_C \mathcal{L} L_C} \\
    &= e^{\left[\mathcal{L} + L_C \mathcal{L} L_C\right]} \equiv e^{\bar{\mathcal{L}}} = \Lambda_g~,
\end{align}
where we have used the fact that Pauli channels commute with each other. The Pauli generator rates appearing in $\bar L$ are related to those of $L$ via
\begin{align}
    \bar\gamma_P = \gamma_P + \gamma_{P'}~,
\end{align}
where $P'=L_C P$. The Paulis that commute with $L_C$, for which $P'=P$, are called self-amplifiable. The rest can only be learned by this germ in an amplified manner, only as sums. 

For $P, P'$ that have the same qubit-support, $P'$ can be rotated back to $P$ using a SQG layer $\mathcal U$. Then, $P$ will be self-amplifiable with respect to a second germ $\tilde g'=( \mathcal{U} \tilde{\mathcal{L_C}})^2$ (see Refs.~\cite{Berg2023, Chen2023} for details). The second germ is rendered moot, however, by the square-root Pauli twirling of the CZ layer. Indeed, SP-twirling imposes constraints on $\gamma_P$ that exactly set the missing combinations to zero, i.e., $\gamma_P - \gamma_{P'} = 0$.

For example, the combinations amplified by the germ $(CZ)^2$ are $\gamma_{IZ},~\gamma_{ZI},~\gamma_{ZZ}$, self-amplifiably, and 
$\gamma_{IX} + \gamma_{ZY},~    \gamma_{IY} + \gamma_{ZX},~    \gamma_{XI} + \gamma_{YZ},~    \gamma_{YI} + \gamma_{XZ},~    \gamma_{XX} + \gamma_{YY},~    \gamma_{XY} + \gamma_{YX}$ in pairs. The remaining, non-amplifiable, combinations are the differences
\begin{subequations}
\begin{align}
    &\gamma_{IX} - \gamma_{ZY},~ \\
    &\gamma_{IY} - \gamma_{ZX},~ \\
    &\gamma_{XI} - \gamma_{YZ},~ \\
    &\gamma_{YI} - \gamma_{XZ},~ \\
    &\gamma_{XX} - \gamma_{YY},~\\
    &\gamma_{XY} - \gamma_{YX}~.
\end{align}
\end{subequations}
SP-twirling a CZ is performed with $\mathcal U_1 = \mathcal U_2 \in \{I\otimes I,~I\otimes \sqrt{Z},~\sqrt{Z}\otimes I,~\sqrt{Z}\otimes \sqrt{Z}\}$. These enforce the following constraints on the Pauli channel [see Eq.~\eqref{eq:sp_twirl_const}]
\begin{subequations}
\begin{align}
    &\gamma_{IY} = \gamma_{IX}~,\\
    &\gamma_{XY} = \gamma_{YX} = \gamma_{YY} = \gamma_{XX}~,\\
    &\gamma_{YI} = \gamma_{XI}~,\\
    &\gamma_{YZ} = \gamma_{XZ}~,\\
    &\gamma_{ZY} = \gamma_{ZX}~.
\end{align}    
\end{subequations}
The second constraint eliminates exactly the last two non-amplifiable directions (setting them to zero). The rest of the constraints reduce the first four non-amplifiable directions to only two unique combinations, i.e., 
\begin{equation}
\begin{aligned}
    &\gamma_{IX} - \gamma_{ZX},~ \\
    &\gamma_{XI} - \gamma_{XZ}.
\end{aligned}
\end{equation}

These final two non-amplifiable directions of a single CZ gate correspond to pairs $(P, P')$ for which $P$ and $P'$ have different supports. They, fundamentally, cannot be learned in an amplified manner (see Ref.~\cite{Chen2023} for details). Along with the SPAM Pauli generators raters, they combine to form two learnable, but non-amplifiable, combinations and two pure gauge combinations (see Ref.~\cite{Chen2024} for details). The gauge directions are generated by single-qubit depolarizations. As such, there is exactly one gauge direction per qubit in the gate set. 

QESEM obtains sensitivity to the learnable but non-amplifiable combinations of the CZ layer via the circuits in Eq.~\eqref{eq:cliff_circs_l} with $l=1$. Each layer merits two such circuits with different fiducial configurations. Specifically, $f_m=f_s=\prod_{q\in \text{QC1}} f_1 \prod_{q\in \text{QC2}} f_2$ with $f_1=\sqrt{Y}$ and $f_2=I$ and vice versa. For the example of a layer with a single CZ gate, the expectation values of $Z \otimes I, I \otimes Z$ when the fiducial configuration is $f_m=f_s=\sqrt{Y} \otimes I, I \otimes \sqrt{Y}$, respectively, give
\begin{subequations}
\label{eq:state_smear}
\begin{align}
    \langle ZI \rangle &= \lambda_{ZI,m} \lambda_{XI} \lambda_{ZZ, s} \\
    &\nonumber = e^{-2(\gamma_{XI,m}+\gamma_{XI,s})-2(\gamma_{ZX}-\gamma_{IX})-2\gamma_{IX,s} + \text{amp. combos.}}~,\\
    \langle IZ \rangle &=\lambda_{IZ,m} \lambda_{IX} \lambda_{ZZ, s} \\
    &\nonumber = e^{-2(\gamma_{IX,m}+\gamma_{IX,s})-2(\gamma_{XZ}-\gamma_{XI})-2\gamma_{XI,s} + \text{amp. combos.}}~.
\end{align}
\end{subequations}
The SPAM combination $(\gamma_{XI,m}+\gamma_{XI,s})$ also appears in the idling circuit and in the circuit in Eq.~\eqref{spam_circ} (the latter is executed even in the absence of fractional layers), and is thus constrained to match them. The only two new directions that are otherwise unknown are
\begin{align}
    (\gamma_{ZX}-\gamma_{IX})-\gamma_{IX,s}~,\\
    (\gamma_{XZ}-\gamma_{XI})-\gamma_{XI,s}~.
\end{align}

\subsubsection{Circuits interleaving layers}
So far, we have considered isolated Clifford layers. In particular, each Pauli channel following a Clifford layer has $N_q$ non-amplifiable but learnable directions that mix with the SPAM. In the context of a full gate set, however, this counting is naive. QESEM executes circuits with germs that interleave Clifford layers to amplify inter-layer combinations and enhance the robustness of state preparation and measurement separation. The total number of non-amplifiable directions in the gate set reduces to only $N_q$.

One type of mixed-layer circuit interleaves the layers using SQG layers so that they commute. For example, for KAK-1 two-qubit gate layers, the circuits take the form 
\begin{align}
\label{eq:cliff_mixed}
    &C_2(L_{C,1},L_{C,2},l, f)  = \\
    \nonumber & \left\llangle E_m \right| f \left[\mathcal T_{CZ} (L_{C,1}) \mathcal T_{CZ}(L_{C,2})\right]^{2l} f^\dagger \left| \rho_0 \right\rrangle~,
\end{align}
where $\mathcal T_{CZ}(L)$ denotes a logical CZ layer implemented with $L$ and SQG layers. For a single CX gate, for instance, $\mathcal T_{CZ}(CNOT) = H_0 CX H_0$. In Eq.~\eqref{eq:cliff_mixed}, we set the fiducials $f_m=f_s^\dagger = f$ since the germ $g=[\mathcal T_{CZ}(L_{C1})\mathcal T_{CZ}(L_{C2})]^2$ is ideally the identity due to the CZ-layer commutativity.

The noisy germ depends on the noisy layers as
\begin{align}
    \bar{L} = \mathcal{L}_{L_2} + L_2 \mathcal{L}_{L_1} L_2 +  {L}_1  {L}_2  \mathcal{L}_{L_2} L_2  L_1 + L_1 \mathcal{L}_{L_1} L_1~.
\end{align}
For example, in a circuit mixing two CZ layers overlapping a single qubit -- while ignoring all amplified intra-layer combinations -- the germ noise is sensitive to the combinations 
\begin{subequations}
\begin{align}
    & \left(\gamma_{\cdots IXI \cdots,2} + \gamma_{\cdots ZXI\cdots ,1}\right)~, \\ 
    & \left(\tilde\gamma_{\cdots IXZ \cdots,2} + \tilde\gamma_{\cdots IXI \cdots,1}\right)~.
\end{align}
\end{subequations}
Their sum reproduces intra-layer amplified combinations, while their difference can be rewritten as a difference between two learnable but non-amplifiable intra-layer combinations:
\begin{align}
    \left(\ \gamma_{\cdots ZXI \cdots,1}- \gamma_{\cdots IXI \cdots,1}\right)-\left( \gamma_{\cdots IXZ \cdots,2}- \gamma_{\cdots IXI\cdots ,2}\right)~.
\end{align} 
This difference is amplified by the circuit in Eq.~\eqref{eq:cliff_mixed} for $l>1$. 

Another type of mixed-layer circuit interleaves the layers using SQG layers to make them non-commuting, generating nontrivial propagation of errors throughout the circuits. An important aspect of the characterization circuits described so far is that local errors remain local. In the idling circuits in Eq.~\eqref{eq:cliff_circs_l} (with $l$ a multiple of the layer order), the SPAM circuit Eq.~\eqref{spam_circ}, and even the circuits executing commuting mixed layers of Eq.~\eqref{eq:cliff_mixed}, an error on some qubit $q$ will only affect Pauli observables supported on that qubit. In particular, the measurement outcome for qubit $q$ depends on the state-preparation Pauli generators of that same qubit. Generally, application circuits do not have this unique property.

Even the circuits of Eq.~\eqref{eq:cliff_circs_l} with $l$ small enough that the germ is not idle -- e.g., $l=1$ for CZ layers -- only generate entanglement between neighboring qubits, specifically pairs $(q_1, q_2)$ corresponding to active bonds of the layer. As shown in Eq.~\eqref{eq:state_smear}, the measurement outcome of a local observable on $q_1$ can be influenced (depending on the fiducial configuration) by the state-preparation Pauli generators of $q_2$. Conversely, from an error propagation perspective, the state preparation Pauli generators on $q_2$ \text{spread} to $q_1$, becoming measurable through the outcomes of local observables on both qubits.

\begin{figure}[tb]
    \centering
    \includegraphics[width=\linewidth]{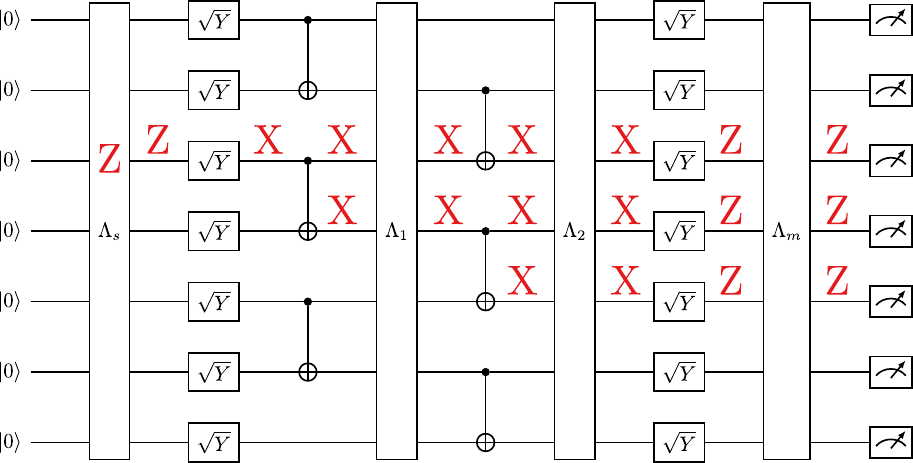}
    \caption{The propagation of a $Z$ error during the state preparation of qubit 3: the state fiducial $\sqrt{Y}$ gate rotates it to an $X$ error, which is then smeared by the CX gate onto qubit 3. This spreading continues until the layer of measurement fiducials rotates all $X$ errors back to $Z$. Consequently, the local state-preparation error of qubit 3 is measurable via local observables on qubits 3, 4, and 5.}
    \label{fig:sp}
\end{figure}

For KAK-1 two-qubit gate layers, the non-commuting mixed-layer circuits are 
\begin{align}
     C_3(L_{C,1},L_{C,2},l, f)& = \\
    \nonumber &\left\llangle E_m \right| f \left[\mathcal T_{CX}(L_{C,1})\mathcal T_{CX}(L_{C,2})\right]^{l} f^\dagger \left| \rho_0 \right\rrangle~,
\end{align}
where $\mathcal T_{CX}(L_{C})$ denotes a logical CX layer implemented with $L$ and SQG layers. Since CX layers do not commute, the Pauli generators associated with state preparation and with Clifford layers propagate through these circuits to distant qubits. Consequently, sensitivity to errors \emph{spreads} within the causal lightcone of the circuits, as in application circuits. This propagation, visualized in Fig.~\ref{fig:sp}, enhances the ability to distinguish state-preparation errors from measurement errors and improves the overall generalization capability of the characterization.

\subsubsection{Inference}
The inference of Clifford layers follows that of fractional layers. The design matrix is constructed according to Eq.~\eqref{eq:cliff_A} [see, e.g., the explicit results in Eq.~\eqref{eq:state_smear}]. Due to the gauge freedom inherent to Clifford gate sets, many Pauli generator rate configurations are equivalent for both the characterization circuits and \textit{any} application circuit. However, QP-based mitigation methods can incur vastly different circuit overheads for these equivalent configurations.  

For instance, at fixed infidelity matching that of the characterization circuits, a configuration containing negative Pauli generator rates (compensated by large positive values elsewhere) can yield the same predictions as one where the negative rates are set to zero. These negative rates are non-physical and typically stem from arbitrary gauge choices. QESEM, therefore, enforces a non-negativity constraint on all Pauli generator rates, partially fixing the gauge.

When application circuits include both fractional and Clifford two-qubit gate layers, QESEM performs a single, self-consistent inference over the full gate set. In such cases, no gauge freedom remains because the fractional layers couple to the Clifford ones via the SPAM, even without explicit mixed-layer circuits.

Finally, we note that QESEM's characterization framework -- encompassing both noise model parameterization and inference -- can also incorporate Pauli generator rates that are not tied to a specific layer or SPAM. Such parameters have proven useful for capturing non-Markovian errors on some hardware platforms, and we plan to report these results in detail in a future publication.
\begin{figure}[tb]
    \centering
    \subfloat[\label{fig:char_bench_steps}]{
    \includegraphics[width=0.96\columnwidth]{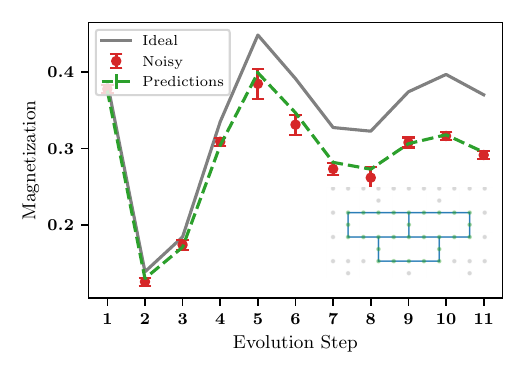}}
    \\
    \subfloat[\label{fig:char_bench_z_factors}]{
    \includegraphics[width=0.96\columnwidth]{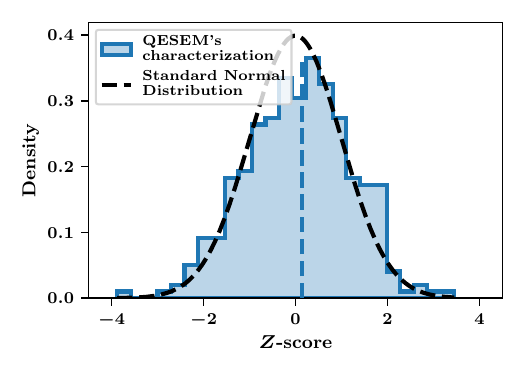}}
    \caption{QESEM accurate characterization. Prediction of noisy results for the tilted-field kicked-Ising circuits. 
    \textbf{\protect\subref{fig:char_bench_steps} Prediction of noisy magnetization values.} 
    Magnetization values for evolution steps one to eleven. Gray – ideal (noiseless); red – measured noisy results; green – predictions of the noisy results using QESEM's characterization. We observe close agreement between the noisy data and the model predictions. Inset – the ``pretzel'' geometry used for the benchmark. 
    \textbf{\protect\subref{fig:char_bench_z_factors} Prediction statistical consistency.} 
    Distribution of the $Z$-scores of all single-site $\langle Z \rangle$ observables for all measured steps (308 observables in total). The distribution closely follows a standard normal distribution, indicating accurate and unbiased characterization. The dashed line indicates the median value of 0.14.}
    \label{fig:char_benchmark}
\end{figure}

\subsection{Characterization benchmark}
We evaluate the accuracy of the characterization protocol independently of the mitigation procedure. Specifically, we compare expectation values computed from noisy circuit outcomes to those predicted by the learned noise model for the same circuits.

\subsubsection{Experiment}
For this benchmark, we use tilted-field kicked-Ising circuits with the same parameters as in Sec.~\ref{main_demo}. We restrict the geometry to a 28-qubit ``pretzel'' shape (see inset of Fig.~\ref{fig:char_bench_steps}) embedded in \texttt{ibm\_kingston}. This geometry preserves the two-dimensional structure of the circuits while reducing the number of qubits, allowing efficient computation of the corresponding model predictions.

The experiment consists of two sets of circuits (interleaved to suppress temporal drifts):
\begin{enumerate}
    \item \emph{Characterization} – The circuits used to train the model according to the QESEM characterization protocol described above. This produces a during-layer sparse Pauli-Lindblad noise model.
    \item \emph{Verification} – Circuits corresponding to the first 11 time steps of the tilted-field kicked-Ising model. Importantly, these circuits are transpiled in the same way as the characterization circuits, including twirling and SQG compression, ensuring compatibility with the learned noise model. 
\end{enumerate}

\subsubsection{Prediction}
To compute the model predictions for a given circuit, we generate an ensemble of circuits such that averaging their ideal expectation values converges to the noisy expectation value of the original circuit under the inferred noise model. Specifically, the ensemble is constructed from the original circuit by randomly inserting and replacing Pauli operators, as described in Appendix~\ref{app:QP}. Crucially, no signs or QP factors are introduced. As a result, the ensemble averaging does not constitute noise mitigation. Instead, noise is introduced probabilistically according to the inferred during-layer sparse Pauli-Lindblad noise model. 

For each verification circuit (i.e., each time step), we generated an ensemble of 1000 noise-sampled circuits\footnote{State-preparation errors were incorporated as noisy idle gates at the beginning of the circuit, while measurement errors were applied to the expectation values at the end of the simulation.}. The ideal expectation values were computed for each sampled circuit using a state-vector simulator. The predicted expectation values were obtained by averaging over these 1000 samples, and the error bars were estimated from the corresponding standard deviations.

\subsubsection{Results}
Figure \ref{fig:char_bench_steps} shows the expectation values of the magnetization operator, both measured and predicted, for evolution steps 1 through 11. We observe close agreement between the measured noisy outcomes and the predictions of the QESEM noise model. This independently demonstrates that our method accurately captures the noise using a sparse Pauli–Lindblad parametrization

In Fig.~\ref{fig:char_bench_z_factors}, we plot the distribution of $Z$-scores for all single-site $\langle Z \rangle$ expectation values (308 in total). Here, the $Z$-score is defined as
\begin{align}
Z\text{-score}=
\frac{\textrm{prediction}-\textrm{noisy}}
{\sqrt{\mathbb{V}_\textrm{noisy}+\mathbb{V}_\textrm{prediction}}},
\end{align}
where $\mathbb{V}_\textrm{noisy}$ ($\mathbb{V}_\textrm{prediction}$) denotes the variance of the measured noisy values (simulated predictions). The results are consistent with a standard normal distribution, further demonstrating the predictive accuracy of the QESEM characterization protocol.

\section{Noise-aware transpilation}
\label{app:transpilation}

To reduce the overhead of the error mitigation, QESEM is designed to optimize the transpilation of the provided circuit onto the QPU. We do this by both optimizing the mapping of algorithmic to physical qubits on the device, and by enabling parallel computation of the same circuit (for small enough circuits).
Firstly, we transpile the circuit twice; once, using the full available gate set (which includes fractional 2-qubit gates), and another time using all available single-qubit gates, but only Clifford two-qubit gates. We estimate the runtime of QESEM for both cases, and choose the option with the minimal runtime\footnote{The mitigation time can only benefit from the full gate set, but the characterization may take longer, especially if there are multiple unique fractional layers. The interplay between the two determines the chosen transpilation}.

\subsection{Qubit selection}
\label{sec:qubit-selection}
At the start of every QESEM run, the first calculation performed on the QPU is called "Device Familiarization" (DFAM) - a compact form of the characterization described in App.~\ref{Appendix: characterization}, targeted to characterize the infidelity of all the 2-qubit gates on the device in layer context. Meaning, the effects of crosstalk and large layers are taken into account, generating infidelity values that are different from those resulting from individual gate characterizations. 
Our qubit selection also takes into account the different durations of the gates, as adding a longer gate to a layer affects the infidelity of all the other gates by adding additional idle time.

QESEM's routing of the circuit onto the device's topology and qubit selection is based on the built-in methods in Qiskit. However, as subgraph isomorphism is an NP-complete problem, sampling all possible qubit mappings is not practical, and therefore, these methods are heuristic. Because of that, QESEM's optimized mapping is performed on a subset of the device that excludes the qubits with the lowest fidelity. This exclusion depends on the target circuit, as, for large enough circuits (i.e., close to utilizing the entire device), an exclusion may not be possible. However, for most circuits, this procedure is very useful and significantly improves the circuit fidelity and therefore the mitigation time.

\begin{figure} [tb]
\centering
    \includegraphics[width=0.95\columnwidth]{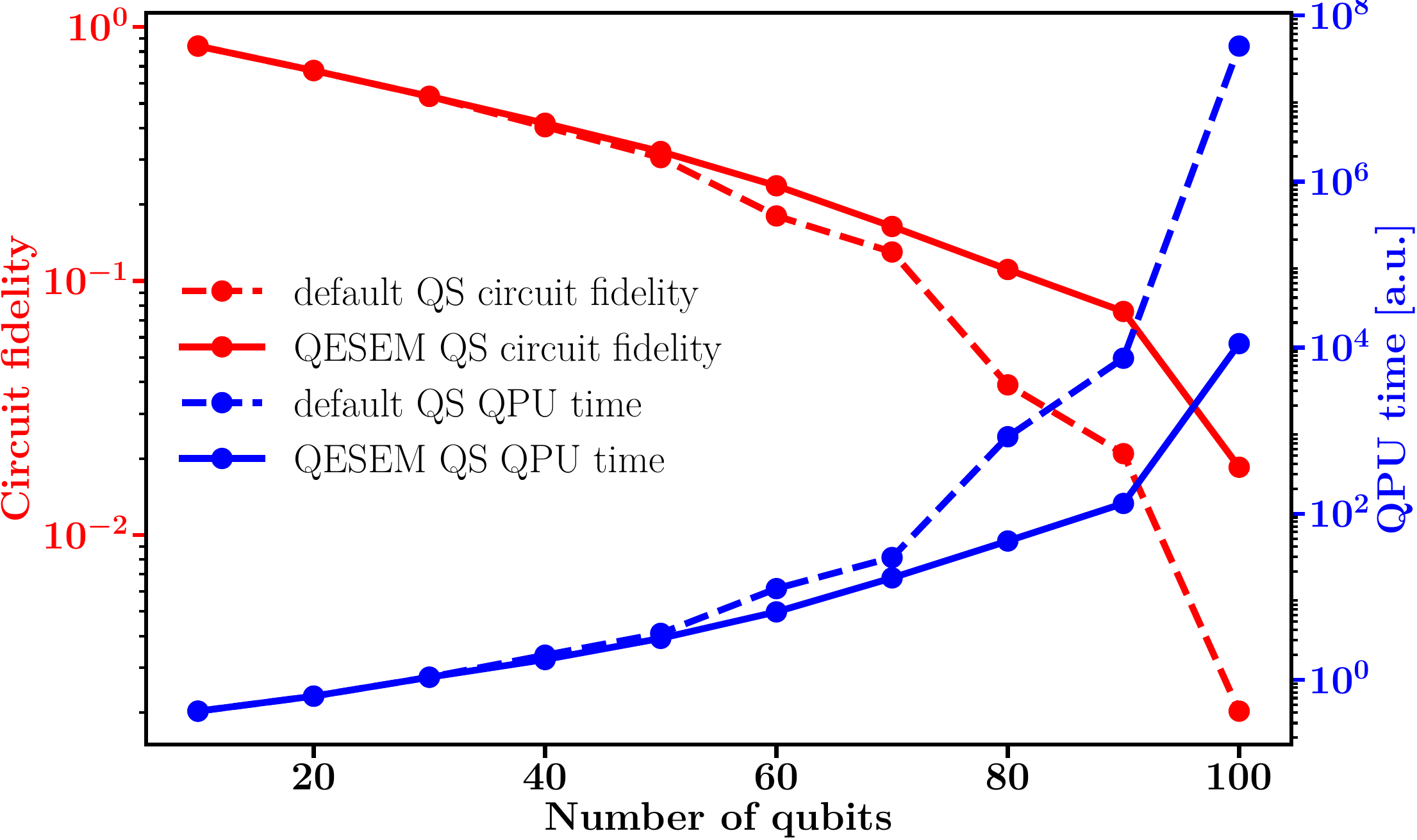}
    \caption{circuit fidelity (product of all 2-qubit gates in the circuit and the average SPAM error) and predicted mitigation time on \texttt{ibm\_kingston}, for a 1D kicked-Ising with the same parameters as the one in Sec.~\ref{main_demo}. Dashed lines denote the default qiskit qubit selection, while solid lines include QESEM's qubit exclusion. Both processes were limited to a one-minute run time, as is the default for QESEM.}
    \label{fig:qs2}
\end{figure}

Fig.~\ref{fig:qs2} shows the effects of the latter optimization. The qubit selection was performed for the backend data without DFAM, with the only difference being QESEM's qubit exclusion that enables sampling a better subset of qubit chains. Under $\sim40$ qubits, this makes no difference as the default method samples a large enough subset of possible qubit chains to select the same optimal chain. However, as the number of qubits increases, the exclusion becomes crucial. For a 100-qubit chain, our optimization decreases the required mitigation time by almost 4 orders of magnitude.

\subsection{Circuit parallelization}

For small enough circuits that can be fitted multiple times onto the device, QESEM enables parallel execution of the circuit, as was demonstrated in Sec.~\ref{vqe_demo}. The number of patches is selected using the backend error per layered gate (EPLG) data \cite{mckay2023benchmarkingquantumprocessorperformance}, to optimize the value of
\begin{equation}
    \frac{T\left(\text{EPLG}(n\times n_q)\right)}{n}~.
\end{equation}
In other words, so that we gain more from the linear acceleration factor of parallelization than we lose from increasing the infidelity of the gates due to the large layer context.
The qubit patches are then selected according to the transpilation procedure described in the previous subsection, and the final results are averaged over the different patches using inverse-variance weighting.

\section{Calibration}
\label{app:calibration}

\begin{figure}[!ht]
\centering
\subfloat[\label{fig:calibration_circuits}]{
    \includegraphics[width=0.7\columnwidth]{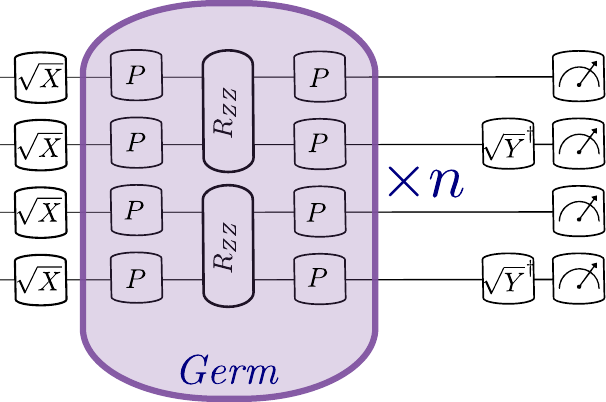}
}
\\
\subfloat[\label{fig:calibration}]{
    \includegraphics[width=0.83\columnwidth]{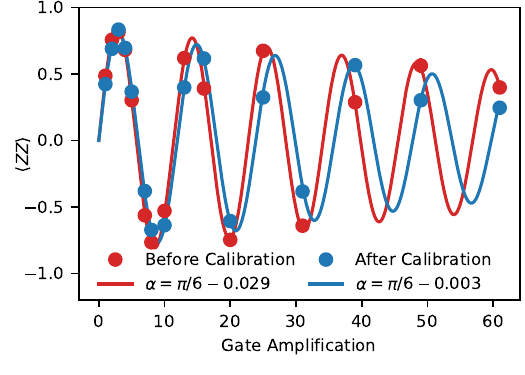}
}
\\
\subfloat[\label{fig:calibration_cdf}]{
    \includegraphics[width=0.83\columnwidth]{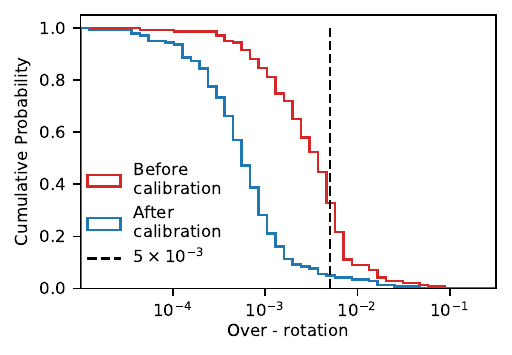}
}
\caption{Fractional gate calibration. \textbf{\protect\subref{fig:calibration_circuits} calibration circuit.} The circuit used to characterize the \(R_{ZZ}\) gate angle. The gate is repeated \(n\) times, and is sandwiched with single-qubit Pauli-twirling layers (only Pauli operators that commute with the gate). We placed initial and final fiducials according to Eq. ~\eqref{eq:fiducial_definitions}. \textbf{\protect\subref{fig:calibration} coherent characterization.} $ZZ$ expectation values as a function of the gate amplification for one of the pairs on the tested device. The Ramsey frequency \(\alpha\) is extracted from the sinusoidal fit (Eq. ~\eqref{eq:sin_decay}). Our calibration protocol reduced the OR error from \(3\times10^{-2}\) to \(3\times10^{-3}\). \textbf{\protect\subref{fig:calibration_cdf} CDF.} cumulative distribution of OR errors before and after calibration, demonstrating the significant effect of the calibration protocol.}
\end{figure}

\subsubsection{Motivation}
Most quantum hardware platforms successfully calibrate Clifford two-qubit gates (or layers composed of such gates), but the calibration of fractional (non-Clifford) gates is typically more error-prone. Moreover, while Clifford gates are fully twirled, fractional gates are only partially twirled—specifically, over-rotation (OR) errors are not twirled. This can lead to a coherent accumulation of errors in the overall circuit infidelity.

Efficient calibration is therefore especially crucial for fractional gates. To address this, the following section introduces our novel Targeted Layer Calibration proposal for coherent error characterization \cite{qedma_characterization_calibration_2023}, which combines the Ramsey experiment with refocusing techniques such as dynamical decoupling and partial twirling.

\subsubsection{Characterization of Coherent Errors}

The Ramsey experiment measures the detuning between a coherent driving field and the transition energy of a two-level quantum system (qubit). The protocol involves two $\pi/2$ fiducial pulses—one at the beginning and one at the end—that prepare the qubit in the $X$ direction and project the measurement along the $Y$ direction. During the delay between these pulses, the qubit evolves around the $Z$ axis at a rate determined by the detuning, and the accumulated phase manifests as oscillations in the $Y$ measurement.

If only a single Hamiltonian term contributes to the detuning, the resulting Ramsey signal is a clean sinusoidal oscillation. However, additional terms such as residual coupling can induce beating, degrading the signal and the fidelity of frequency estimation. 

In our Targeted Layer Calibration protocol, we repurpose the Ramsey experiment by replacing the idle delay between the two preparation and measurement fiducial pulses, $(F_p, F_m)$, with an $N$ repeated sequence of a gate-layer \emph{germ} $G$ containing the two-qubit gate under test: 
\begin{equation}
S_N = F_p\, G^N\, F_m~.
\label{eq:fidutials_and_germ}
\end{equation}

This germ $G$ is constructed to refocus all Hamiltonian terms except the one under investigation, thus preserving coherent oscillations due to the targeted term alone $P_t$.

In the following, we present two refocusing techniques employed in our protocol:

\begin{enumerate}
    \item {Concatenated Dynamical Decoupling (DD):}

    DD sequences can suppress Pauli terms that anti-commute with the applied decoupling pulses, while commuting terms are preserved. To isolate the targeted term, we construct a germ $G$ using a concatenation of multiple Pauli DD pulses $D_1, D_2, D_3, \ldots$, commuting with the targeted term $[D_i, P_t] = 0$.
    Assuming the two-qubit gate Hamiltonian contains up to two-body Pauli interactions, there are 15 nontrivial Pauli terms per qubit pair. By selecting three appropriately chosen DD directions, it is possible to refocus all undesired terms while preserving only the targeted one. 

    To isolate terms $P_t$ \emph{other than} over-rotation, the germ is:
    \[
    \begin{aligned}
    G_0 &= g \, D_1 \, g \, D_1~, \\
    G_1 &= G_0 \, D_1 \, G_0 \, D_1~, \\
    G &= G_1 \, D_2 \, G_1 \, D_3~.
    \end{aligned}
    \]
    where $D_1$ is a Pauli operator that anti-commutes with the two-qubit gate $g$, i.e., $\{D_1, g\} = 0$. $D_1$ is concatenated twice to eliminate first-order contributions from directions $P\in [D_1^\perp, g]$, where $D_1^\perp$ denotes Pauli operators that anti-commute with $D_1$. A single $D_1$ concatenation would allow such contributions to survive, and would thus violate the core principle of the DD sequence, since these residual Pauli directions commute with $D_1$.

    For OR errors, which already partially decouple themselves due to the ideal gate operation, one fewer DD layer is required:
    \[
    \begin{aligned}
    G_0 &= g \, D_1 \, g \, D_1~, \\
    G &= G_0 \, D_2 \, G_0 \, D_2~.
    \end{aligned}
    \]

    \item {Partial Twirling:}

    Partial twirling involves averaging over randomized Pauli operators $P_i$ that \emph{commute} with the targeted term $P_t$, i.e., $[P_i, P_t] = 0$.

    For OR, the germ is:
    \begin{equation}
    G(P_i) = P_i \, g \, P_i~.
    \label{eq:partial_twirling}    
    \end{equation}

    For other coherent errors, we combine partial twirling with DD:
    \[
    \begin{aligned}
    G_0 &= g \, D_1 \, g \, D_1~, \\
    G(P_i) &= P_i \, G_0 \, P_i~,
    \end{aligned}
    \]
    where $D_1$ is a Pauli operator that anti-commutes with $g$, i.e., $\{D_1, g\} = 0$.
\end{enumerate}

\paragraph{Note:} In systems where the error Hamiltonian is known or constrained—such as coupler-mediated transmon architectures (e.g., Heron, IQM) where only $Z$-axis coherent errors are present—one may simplify the refocusing sequence to avoid compensating for all Pauli directions.

The fiducial pulses are chosen to prepare and measure along Pauli directions $P_p$ and $P_m$, such that together with the targeted term $P_t$, they span an SU(2)-like algebra:
\begin{equation}
\label{eq:fiducial_definitions}
[P_i, P_j] = i \epsilon_{ijk} P_k \quad \text{for } i,j,k \in \{p,m,t\}~.
\end{equation}

The Ramsey frequency $f_t$ is extracted by fitting a decaying sinusoid of the form:
\begin{equation}
\text{fit}(n) = \exp(-[\beta + \gamma n]) \sin(f_t n)~,
\label{eq:sin_decay}
\end{equation}
where $n$ is the number of repetitions of the two-qubit gate under test in the modified Ramsey sequence, $\beta$ accounts for SPAM noise, and $\gamma$ captures decoherence. The strength of the targeted coherent term $v_t$ is then inferred from $f_t$ via a transfer function determined by the commutation relation between $P_t$ and the two-qubit gate $g$, and the rotation angle $\alpha$ of the gate:
\[
v_t =
\begin{cases}
    f_t & \text{if } [P_t, g] = 0~, \\
    f_t \dfrac{\alpha}{\sin(\alpha)} & \text{otherwise}~.
\end{cases}
\]

\subsection{Calibration}
On the Heron devices, we use the partial twirling method to accurately measure and calibrate the over-rotation error. Here, we exemplify our protocol by calibrating a \(R_{ZZ}(\pi/6)\) on all the two-qubit gate pairs of \texttt{ibm\_torino}. All the two-qubit gates in the device are divided into three layers. The Germ ~\eqref{eq:partial_twirling} (applied to each pair in the layer) is amplified to 16 different lengths (from 1 to 61) and the ZZ Pauli observable is measured with the appropriate fiducials (see Eq.~\eqref{eq:fiducial_definitions} and Fig.~\ref{fig:calibration_circuits}). In the presented experiment, we averaged 14 twirl realizations. The measured data was fitted according to Eq.~\ref {eq:sin_decay} and presented in Fig.~\ref{fig:calibration} in red (for OR measurement, i.e., ZZ error, we denote the Ramsey frequency \(f_t\) by \(\alpha\)). For the presented gate, we observed a large OR error of \(3\times10^{-2}\). The calibration is done by replacing the gate with \(R_{ZZ}(\alpha_{\mathsf{cal}})\) with
\begin{equation}    
\alpha_{\mathsf{cal}}=\frac{\alpha_{\mathsf{ideal}}^2}{\alpha_{\mathsf{meas}}}~,
\label{eq:angle_calibration}
\end{equation} 
where \(\alpha_{\mathsf{ideal}}\) is the desired (algorithmic) angle, and \(\alpha_{\mathsf{meas}}\) is the angle measured in the calibration process. 

To verify the effectiveness of the calibration protocol, we ran again the same calibration sequences, this time with the corrected angle \(R_{ZZ}(\alpha_{\mathsf{cal}})\). For the tested gate, this process reduced the OR error by an order of magnitude to \(3\times10^{-3}\) (blue markers and lines on Fig.~\ref{fig:calibration}). If we wanted to reduce the OR errors even further, we could repeat the process and refine the angle calibration according to Eq.~\eqref{eq:angle_calibration}. In Fig.~\ref{fig:calibration_cdf}, we present the cumulative distribution function (CDF) of the OR errors across all the operational two-qubit gates in the device. We can observe the outstanding improvement in the OR errors. For example, 30 percent of the original gates had an OR of \(5\times10^{-3}\) or higher, while less than 10 percent of the calibrated gates have such an error.

\section{Hamiltonian Simulation Benchmark \label{app:Ising}}

In the main text, we present the mitigation results of circuits simulating the \emph{Kicked-Ising} Hamiltonian evolution. We used a total of \(1.2\times10^5\) circuits and \(2.3\times10^7\) shots for a total of six hours and twelve minutes of QPU time to perform the mitigation. In this appendix, we present additional mitigation results acquired with QESEM, compare them to other \emph{out-of-the-box} mitigation solutions, and supply additional information regarding the methods used to calculate the ideal values.

\subsection{Single-qubit Magnetization}
\begin{figure*}[t!]
    \centering
    \includegraphics{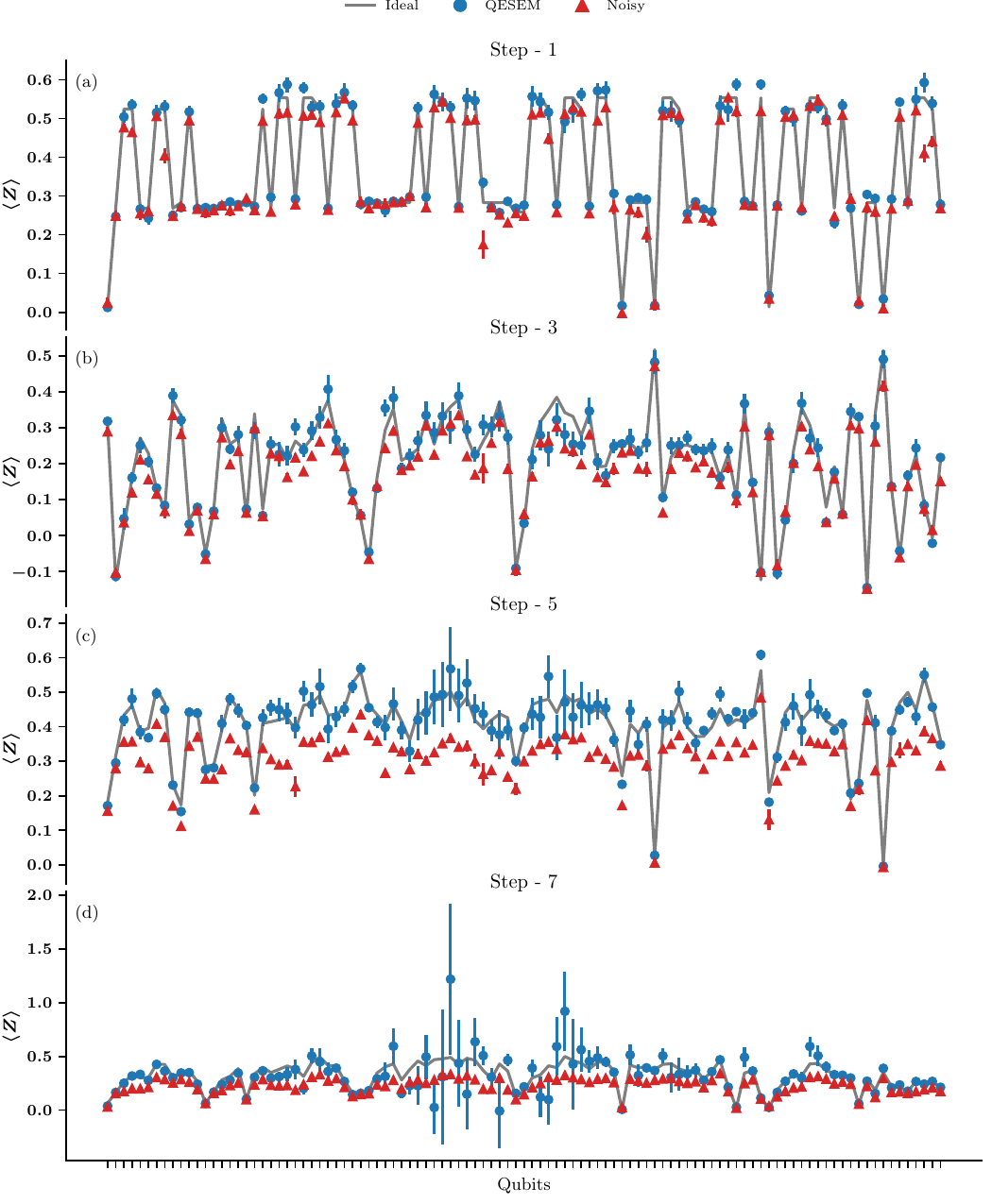} 
    \caption{QESEM mitigation per site (qubit). \textbf{(a-d)} QESEM mitigation of single-qubit magnetization for steps one, three, five, and seven. QESEM accurately mitigates the single-site magnetization. At the larger steps, some qubits have very large statistical uncertainty.}
    \label{fig:per pauli}
\end{figure*}

In Fig.~\ref{fig:ising_results} of the main text, we considered the average magnetization for the different steps. Here, in Fig.~\ref{fig:per pauli}, we present the QESEM mitigation results per site for steps one, three, five, and seven.

\subsection{Heavy-weight observables}
\label{app:heavy_weight}
\begin{figure*}[t!]
    \centering
    \includegraphics[width=\textwidth]{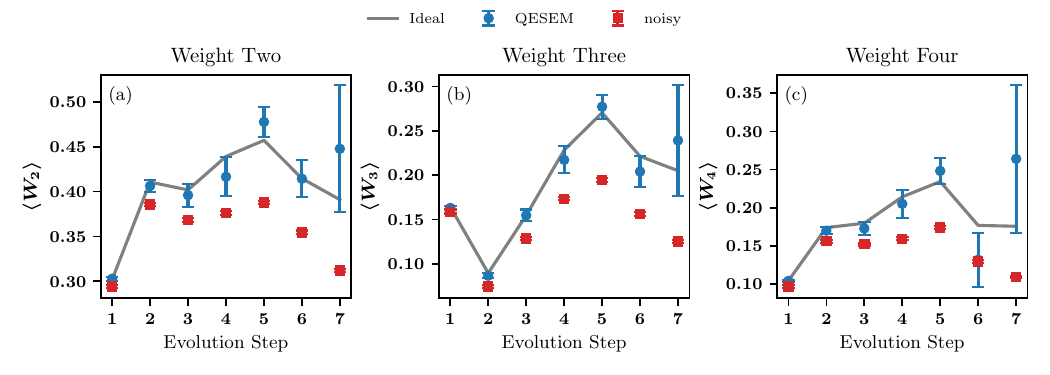} 
    \caption{QESEM mitigation for higher-weight observables. \textbf{(a-c)} QESEM mitigation of weight-two, weight-three, and weight-four observables. QESEM successfully mitigates the noisy values and reproduces the ideal values. QESEM error bars grow very large at the higher steps, as the active volumes become large.}
    \label{fig:High_weight}
\end{figure*}

In addition to the magnetization expectation values, we can consider averages of higher-weight observables. The weight-\(n\) observable is defined as
\[
\langle W_n\rangle = \frac{1}{N_n}\sum_{\langle i_0,i_1,\dots,i_{n-1}\rangle}\langle Z_{i_0}\dots Z_{i_{n-1}}\rangle,
\]
where the sum runs over all the length-\(n\) chains embedded in the device's geometry, and \(N_n\) is the number of such chains. The results are presented in Fig.~\ref{fig:High_weight} for weights two to four (the weight-one observable is the magnetization presented in the main text in Fig.~\ref{fig:ising_results}). QESEM successfully mitigates the higher-weight observables, and the ideal values are reproduced (up to statistical uncertainty).

\subsection{ZNE error mitigation}
\label{sec:zne}
\begin{figure}[tb]
    \centering
    \includegraphics{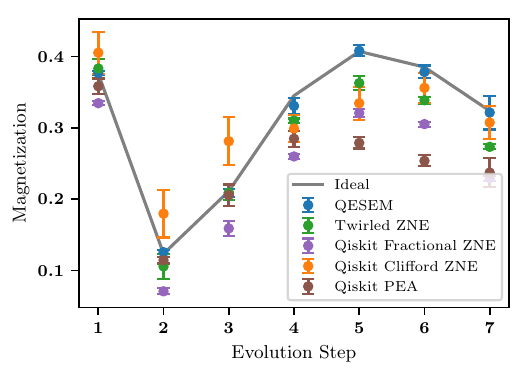} 
    \caption{Mitigation results of all mitigation methods tested, compared to QESEM. QESEM is the only method that successfully mitigated the noise and reproduced the ideal values.}
    \label{fig:ZNE}
\end{figure}

We ran the same circuits using several error mitigation solutions freely available on IBM hardware.
\begin{enumerate}
    \item \textbf{Twirled ZNE.}  
    The \(R_{ZZ}\) gates were partially Pauli-twirled (see Sec.~\ref{sec:char_frac}), and the noise was amplified locally (for each gate) via the gate-folding method. Specifically, an \(R_{ZZ}\) gate with noise level \(n\) is realized as
    \[
    R_{ZZ}^n=R_{ZZ}\left(R_{ZZ}X_0R_{ZZ}X_0\right)^{\frac{n-1}{2}},
    \]
    where \(X_0\) is a $\pi$ pulse on the first active qubit of the \(R_{ZZ}\) gate. Measurement errors were mitigated using the \emph{qiskit} Twirled Readout Error eXtinction (TREX) \cite{VanDenBerg2022}. We considered noise levels of \(n\in\left\{1,3,5\right\}\), and the expectation value of each Pauli observable was fitted to an \emph{exponent}. In order to avoid unrealistic results, we regularized the fitted Pauli observables by restricting each observable to satisfy \(-1<\langle P\rangle<1\). Similarly, we bound the variance by 1. The zero-noise levels of each Pauli observable were averaged in order to extract the zero-noise magnetization presented in the main text in Fig.~\ref{fig:ising_results}.
    
    \item \textbf{qiskit – fractional ZNE.}  
    We ran the circuits via the qiskit \emph{estimator}, with the ZNE option turned on (with default options). This performs \(\left\{1,3,5\right\}\) gate-folding noise amplification, and the zero-noise Pauli observables are extracted via exponential fitting. We then regularized the noiseless observables and averaged for the magnetization. TREX measurement error mitigation is applied as well.
    
    \item \textbf{qiskit – Clifford ZNE.}  
    Here, we compiled the same logical circuits to have only Clifford (\(CZ\)) two-qubit gates. ZNE was performed via the qiskit estimator in a similar way to the fractional ZNE.
    
    \item \textbf{qiskit – PEA.}  
    Here, we used the qiskit estimator Pauli Error Amplification (PEA) \cite{Kim2023utility} option to mitigate the noise. The gates are characterized, and the noise is amplified with Pauli insertion. Again, we used the default parameters, consisting of \(\left\{1,2,3\right\}\) noise levels. Measurement errors are mitigated with TREX. Here, due to a qiskit limitation, we ran the Clifford version of the circuits (i.e., two-qubit gates are compiled to \(CZ\)s).
\end{enumerate}
For more information on the qiskit estimator options (items 2 to 4), see \cite{ibm2025runtime}.  
In Fig.~\ref{fig:ZNE}, we plot the mitigation results of all the methods compared to QESEM. Among all the ZNE methods, the twirled-ZNE (presented in Fig.~\ref{fig:ising_results} of the main text as ZNE) performs best, but QESEM's estimates have a significantly better $Z$-score.

\subsection{PEC time estimation}
\label{app:pec}

\begin{figure}[tb]
    \centering
    \includegraphics{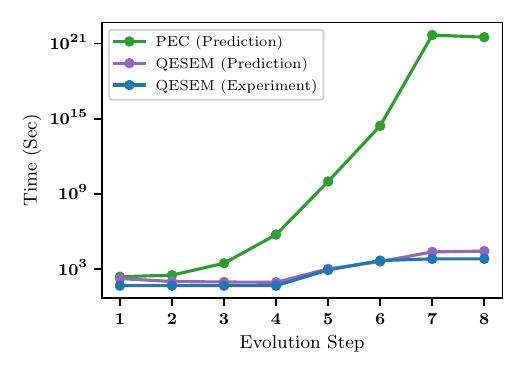}
    \caption{QPU time required for mitigation for the measured evolution steps of the kicked-Ising Hamiltonian simulation (Sec.~\ref{main_demo}). While QESEM (purple) maintains feasible mitigation times, PEC (green)times grow rapidly for deeper circuits, rendering it impractical for utility-scale circuits. Interestingly, the PEC runtime of step seven exceeds the runtime of step eight. Apparently, the effect of the slight reduction of qubit number dominates over the depth increase. Also showing the actual QPU time used in the QESEM experiment (blue), which closely follows the estimate.}
    \label{fig:PEC}
\end{figure}

In this section, we estimate the QPU runtime that of probabilistic error cancellation (PEC)\cite{Temme2017, Endo2018} would require to mitigate noise in the kicked-Ising circuits under the same conditions as in the experiment. We account for only two of the disadvantages of PEC relative to QESEM: (i) standard PEC uses Clifford entangling gates, implying so each $R_{ZZ}$ operation is compiled into two gates ($CZ$s on IBM Heron devices); and (ii) we optimistically assume that PEC mitigates only errors within the commutativity light cone (see Appendix~\ref{appendix:active-volume}), although standard implementations do not even exploit such light-cone locality.  Apart from these two assumptions, we otherwise assume optimal usage of PEC, including optimal transpilation and optimal allocation of circuits and shots.

We estimate the QESEM and PEC QPU times using the following procedure:
\begin{enumerate}
    \item We extract the average infidelity from the characterization data. For PEC, we assign twice the QESEM infidelity,
    \[
    \mathrm{IF}_{\mathrm{PEC}} = 2 \, \mathrm{IF}_{\mathrm{QESEM}},
    \]
    since in PEC the circuits are compiled into two $CZ$ gates, whereas QESEM uses a single $R_{ZZ}$ gate.

    \item For each Pauli operator $Z_i$, we calculate the active volume by counting the number of two-qubit gates inside the its light cone. For PEC we use the commutativity light cone, and for QESEM we take the light cone calculated algorithmically according to Appendix~\ref{appendix:active-volume}. We evaluate idle qubits as contributing ``half a gate'', which is approximately consistent with the performed characterization.

    \item We evaluate the effective volume per Pauli using
    \[
    \langle Z_i \rangle_{\mathrm{noisy}} =
    e^{-\mathrm{IF}\, V_{\mathrm{eff}}}
    \langle Z_i \rangle_{\mathrm{ideal}}.
    \]
    Since the effective volume is a property of the circuit, it is identical for QESEM and PEC.

    \item We use typical shot and circuit times,
    \[
    t_{\mathrm{s}} = 300~\mu\mathrm{s},
    \qquad
    t_{\mathrm{c}} = 160~\mathrm{ms}.
    \]

    \item We calculate the circuit and shot variances for each Pauli according to Eq.~\eqref{eq:phenomenological-vc-vs-pauli}:
    \begin{align}
        \mathbb{V}_{\mathrm{c}}(Z_i)
        &= \exp\!\left[\left(4V_{\mathrm{A}}(Z_i)-2V_{\mathrm{eff}}(Z_i)\right)\mathrm{IF}\right]
        \langle Z_i \rangle_{\mathrm{ideal}}^2,
        \\
        \mathbb{V}_{\mathrm{s}}(Z_i)
        &= \exp\!\left[4V_{\mathrm{A}}(Z_i)\mathrm{IF}\right]
        - \mathbb{V}_{\mathrm{c}}(Z_i).
    \end{align}

    \item To estimate the variance of the magnetization $M$, we use the variances of the single-site Pauli operators:
    \[
    \mathbb{V}_{\mathrm{c}/\mathrm{s}}(M)
    = \alpha \sum_i \mathbb{V}_{\mathrm{c}/\mathrm{s}}(Z_i).
    \]
    The prefactor $\alpha$ accounts for correlations between the $Z_i$ and is determined from the measured noisy variances of the observables:
    \[
    \alpha
    =
    \frac{\mathbb{V}_{\mathrm{noisy}}(M)}
    {\sum_i \mathbb{V}_{\mathrm{noisy}}(Z_i)}.
    \]

    \item Finally, we calculate the QPU time according to Eq.~\eqref{eq:opt_t_qpu}:
    \[
    T_{\mathrm{QPU}}
    =
    \frac{
    \left(
    \sqrt{\mathbb{V}_{\mathrm{c}}(M)\, t_{\mathrm{c}}}
    +
    \sqrt{\mathbb{V}_{\mathrm{s}}(M)\, t_{\mathrm{s}}}
    \right)^2
    }{\epsilon^2},
    \]
    where $\epsilon$ is the statistical uncertainty of the mitigated magnetization measured in the experiment.
\end{enumerate}

In Fig.~\ref{fig:PEC}, we compare the estimated mitigation times of QESEM and PEC. We find that the mitigation times required for PEC quickly become unfeasible, surpassing the age of the universe at the later evolution steps. 

\subsection{Classical simulation details\label{app: simulation}}
In order to simulate the noiseless values of the magnetization we run on the heavy-hex lattice, we use the recently introduced methods for approximate contraction and compression of tensor networks using the belief-propagation (BP) method \cite{Begusic2024, Tindall2024, Arad2021}. Specifically, we follow the approach described in \cite{Begusic2024} for Schrödinger evolution of a projected entangled pair state (PEPS). After each application of a layer of gates $U_t$, the PEPS $U_t|\psi(t)\rangle$ is compressed using \textit{lazy 2-norm belief propagation} (L2BP) such that the maximal bond dimension of all bonds is $\chi$. The expectation value of each $Z_i$ with respect to the state at time step $T$ is then computed by an approximate contraction using \textit{lazy 1-norm belief propagation} (L1BP), which uses an interpretation of the contraction
\[
\langle \psi(T) | Z_i | \psi(T) \rangle
\]
as the exponential of the Bethe free entropy.

We run the computation with a maximal bond dimension $\chi=512$ and allow a truncation error (l2 norm of truncated singular values on each bond) of up to $10^{-7}$ during the L2BP compression steps. The simulation is performed using the tensor-network Python package \textit{quimb} \cite{gray2018quimb}, including the L2BP and L1BP routines implemented in the package. While BP-based methods are only exact for tree graphs, they have been shown to provide accurate results for kicked-Ising circuits on the heavy-hex lattice in certain parameter regimes. In our example, we see good agreement between the BP simulation results and the experimental error mitigation results.

\section{VQE benchmark}
\label{app:vqe}

\begin{table}[tb]
    \centering
    \begin{tabular}{c|ccc}
        & X & Y & Z \\
        \hline
        O & 0.000 & 0.000 & 0.155 \\
        H & 0.000 & 1.193 & -0.696 \\
        H & 0.000 & -1.193 & -0.696 
    \end{tabular}
    \caption{$\mathrm{H_2O}$ stretched geometry [{\AA}] used for the VQE benchmark}
    \label{tab:h2o_geo}
\end{table}

\begin{figure} [tb]
\centering
    \includegraphics[width=0.95\columnwidth]{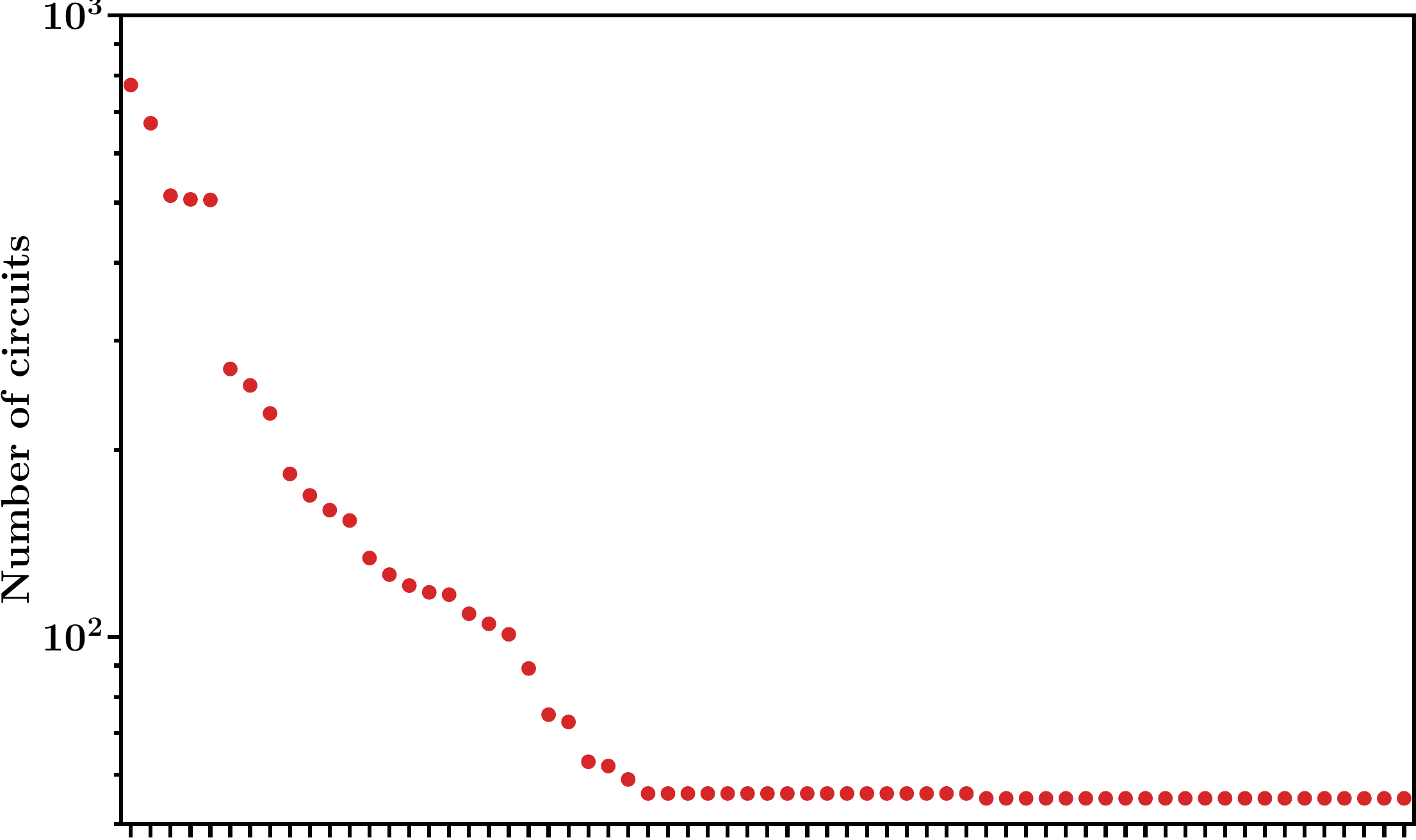}
    \caption{The optimal number of circuits per measurement base in the VQE benchmark. There are 65 measurement bases in total. \label{fig:vqe_nc_ns_bases}}
\end{figure}

The specific geometry of the water molecule for which the VQE circuit was generated is displayed in Tab.~\ref{tab:h2o_geo}. 
After QESEM's transpilation, the transpiled circuit consisted of the following  unique two-qubit gate layers
\begin{subequations}
\begin{align}
    &\left[\text{CZ}(0,1),~\text{CZ}(2,3),~\text{CZ}(4,5),~\text{CZ}(6,7)\right]~, \\
    &\left[\text{CZ}(0,1),~\text{CZ}(4,5)\right]~, \\
    &\left[\text{CZ}(0,1),~\text{CZ}(6,7)\right]~, \\
    &\left[\text{CZ}(1,2),~\text{CZ}(5,6)\right]~, \\
    &\left[\text{CZ}(2,3),~\text{CZ}(4,5)\right]~, \\
    &\left[\text{CZ}(2,3)\right]~, \\
    &\left[\text{CZ}(3,4)\right]~, \\
    \nonumber &\left[R_{ZZ}(0,1,\tfrac{\pi}{2})\right.,~R_{ZZ}(2, 3,\theta_1),~ \\
    &~~~~~~~~~~~~~~~~~~~~~~~~~~R_{ZZ}(4, 5,\tfrac{\pi}{2}),\left.~R_{ZZ}(6,7,\theta_2)\right]~, \\
    \nonumber &\left[R_{ZZ}(0,1,\theta_3)\right.,~R_{ZZ}(2, 3,\theta_3),~\\
    &~~~~~~~~~~~~~~~~~~~~~~~~~~R_{ZZ}(4, 5,\theta_4),\left.~R_{ZZ}(6,7,\theta_4))\right]~, \\
    &\left[R_{ZZ}(1,2,\tfrac{\pi}{2}),~R_{ZZ}(3, 4, \theta_5),~R_{ZZ}(5, 6, \tfrac{\pi}{2})\right]~, \\
    &\left[R_{ZZ}(2,3,~\theta_6),~R_{ZZ}(4, 5, \theta_6)\right]~, \\
    &\left[R_{ZZ}(3,4,~\theta_5)\right]~,
\end{align}
\end{subequations}
with $\theta_1 = 0.3825,~\theta_2=0.3753,~\theta_3 = 0.0114,~\theta_4=0.0856,~\theta_5=0.2607,~\theta_6=0.1582$. The circuit was applied in parallel to the qubit patches
\begin{subequations}    
\begin{align}
    1:~~&[71,~~~70,~~~69,~~~68,~~~67,~~~66,~~~65,~~~77]~,\\
    2:~~&[98,~~~111,~110,~109,~118,~129,~128,~127]~,\\
    3:~~&[140,~141,~142,~143,~136,~123,~124,~125]~,\\
    4:~~&[11,~~~12,~~~13,~~~14,~~~15,~~~19,~~~35,~~~34]~,\\
    5:~~&[53,~~~54,~~~55,~~~59,~~~75,~~~74,~~~73,~~~72]~,\\
    6:~~&[41,~~~42,~~~43,~~~44,~~~45,~~~46,~~~47,~~~48]~,\\
    7:~~&[89,~~~88,~~~87,~~~97,~~~107,~106,~105,~104]~,\\
    \label{eq:best_patch}8:~~&[114,~115,~99,~~~95,~~~94,~~~93,~~~92,~~~91]~,\\
    9:~~&[29,~~~28,~~~27,~~~17,~~~7,~~~~~6,~~~~~5,~~~~~4]~.
\end{align}
\end{subequations}
The number of sampled circuits and shots per measurement basis, as obtained by using QESEM's QPU runtime optimization (see Algs.~\ref{algo:runtime-opt} and \ref{algo:runtime-opt-meas-bases}), is shown in Fig.~\ref{fig:vqe_nc_ns_bases}.

For comparison, we prompted IBM's \textit{circuit function} with the same input circuit and observable (the Hamiltonian) used in QESEM. The \textit{optimization level} was set to 3 (maximum) and the \textit{mitigation level} to 2. At this mitigation level, ZNE is performed via gate folding together with gate twirling, measurement twirling, dynamical decoupling, and Twirled Readout Error eXtinction (TREX). 

We first set the \textit{default precision} to 0.01, below its default value of 0.015625, while otherwise using the default configuration:
\begin{enumerate}
    \item \textbf{noise factors:} 1, 3, 5.
    \item \textbf{extrapolator:} best of exponential, linear (whichever produces the smaller error bar).
\end{enumerate}
The experiment executed 320512 shots over 11.5 minutes of QPU time and yielded a large error bar, $\sim 0.1$, almost five times that of QESEM. The ideal value was contained within this interval.

\begin{figure}[tb]
\centering
    \includegraphics[width=0.95\columnwidth]{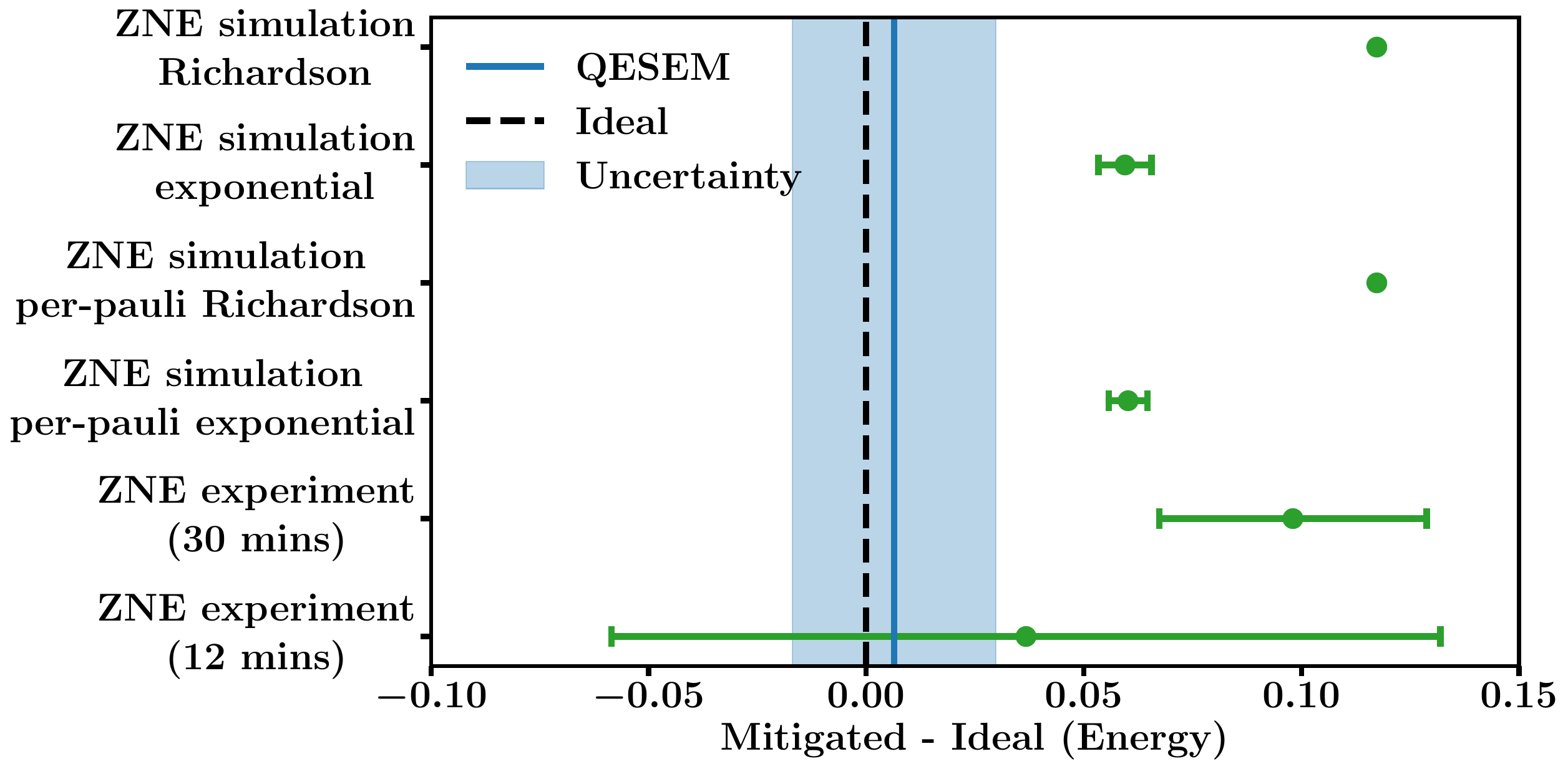}
    \caption{A comparison of QESEM and ZNE, both in experiment and simulation, shows the superiority of the former. 
    \label{fig:vqe_zne}
    }
\end{figure}

To reduce the ZNE error bar, we repeated the experiment at a higher precision, setting the target to 0.006. The total QPU time, 29 minutes and 40 seconds, was then comparable to that of QESEM. This higher-precision experiment executed 889856 shots in total and produced the result
\begin{align}
    E_{\mathrm{ZNE}} - h_I = -1.031 \pm 0.031~.
\end{align}
The error bar here is comparable to that of QESEM, albeit somewhat larger. However, the ZNE estimate is biased towards under-mitigating: the ideal energy lies nearly 3.2 standard deviations from the ZNE value.

Figure~\ref{fig:vqe_zne} shows a comparison between QESEM and the two ZNE experiments. It also shows the results from ZNE simulations we performed. As the backend noisy device, we used the QESEM characterized noise model (Pauli channels) for the \emph{best} qubit-patch (in terms of gate infidelities and SPAM error rates). We mitigated the measurement noise perfectly by setting it to zero in the backend. To amplify the noise of the layers and state preparation, we scaled their Pauli generator rates by 1, 3, and 5. 

For each amplification factor, we calculated the \emph{exact} expectation value of the Hamiltonian using the \emph{exact} density matrix following the application of the noisy VQE circuit. This procedure is akin to using infinite resources in an experiment. Various extrapolation methods were attempted (all listed in Fig.~\ref{fig:vqe_zne}); all appear biased towards under-mitigation.

\section{QESEM demonstrations on other hardware platforms}
\label{other_hardware}

\subsection{IBM Eagle}

\begin{figure*}[tb]
    \centering
    \includegraphics[width=0.95\linewidth]{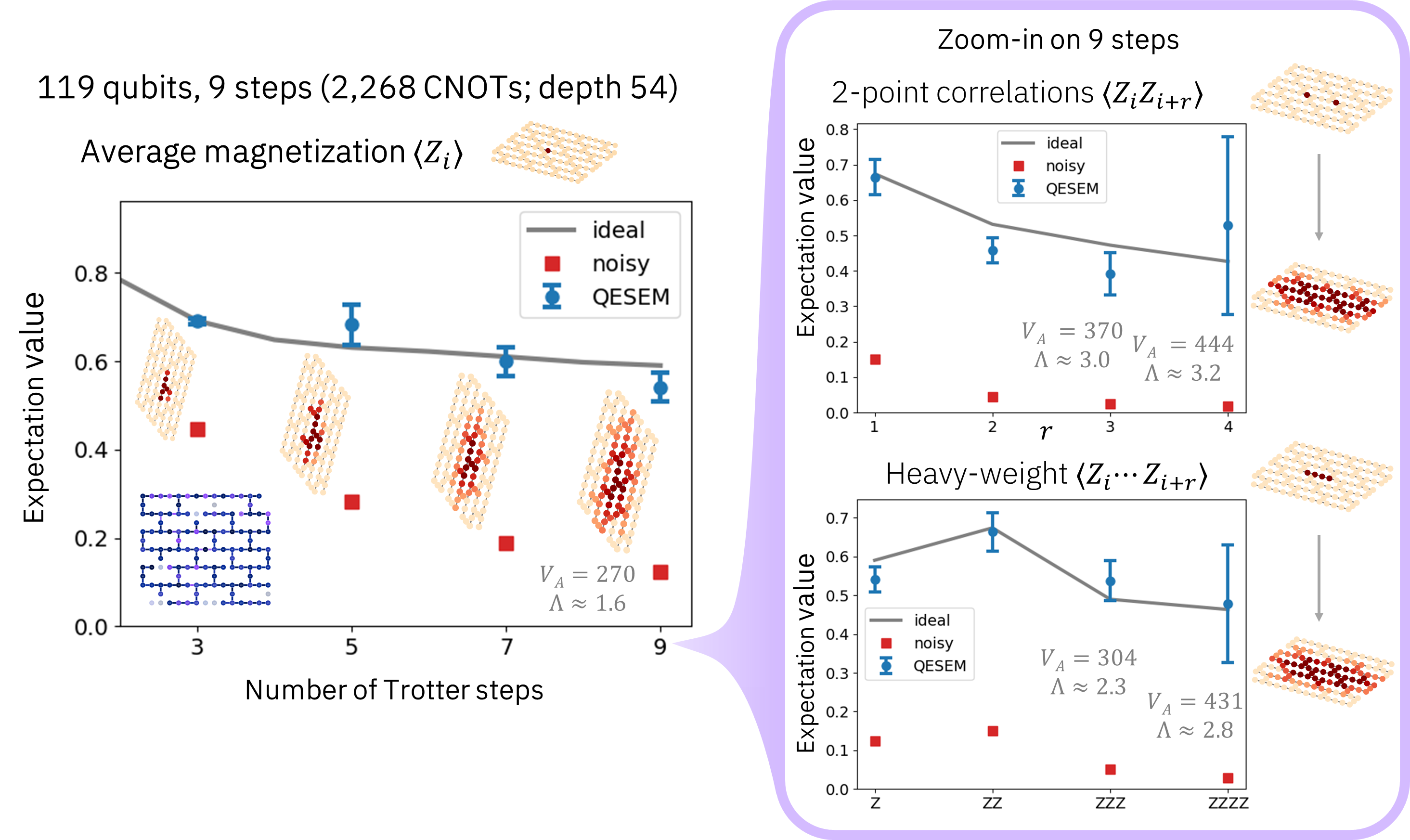}
    \caption{Utility-scale demonstration of QESEM was performed on IBM’s 127-qubit QPU \texttt{ibm\_brisbane}. The experiments involved 2D Trotter-Ising circuits with ZZ angle $\pi/6$ and X angle $\pi/8$. They were executed on the best 119 qubits of the device, selected using Qedma’s noise-aware transpilation to maximize performance (see inset within left panel).
    \textbf{Left:} The evolution of the average magnetization is shown as a function of the number of Trotter steps. The evolution of the subsystem influencing $Z_i$ for an example site is illustrated alongside the data points, demonstrating the increasing difficulty of mitigation and classical simulation.
    \textbf{Right:} A zoom-in on the ninth Trotter step, displaying two-point correlators (top) and multi-site correlators over contiguous sets of qubits (bottom). $V_A$ denotes the active volume, and $\Lambda = \log\left(\frac{O_{\text{ideal}}}{O_{\text{noisy}}}\right)$ denotes the effective total infidelity affecting the observable.}
    \label{fig:Eagle-119q}
\end{figure*}
To validate QESEM's hardware-adaptability, we show large-scale Hamiltonian simulations performed on IBM's 127-qubit fixed-frequency superconducting Eagle QPUs. While Eagle is also an IBM device, they are based on microwave-driven, echoed cross-resonance gates \cite{Sundaresan_2020}, which result in a very different noise model. Cross-resonance gates have more coherent error generators - while tunable couplers usually generate coherent errors in the $Z$ direction, cross-resonance gates can generate errors in most directions. Moreover, the noise model of the gates on Eagle devices has a greater dependence on context  (i.e., on the other gates simultaneously applied in the layer) 

QESEM mitigation on Eagle devices operates very similarly to that on Heron devices. However, Eagles don't have native fractional two-qubit gates. Instead, until recently, IBM provided pulse-level access (PLA) on Eagles. Using PLA, we were able to perform fast and precise calibration of fractional gates and add them to our gate set.
The calibration is performed by starting from an analytical estimate of the microwave amplitude, followed by a single round of correction, based on the calibration procedure described in App.~\ref{app:calibration}.
To accelerate the calibration, we developed a simplified qutrit model in which we perform the following approximation - we calculate the cross-resonance amplitude, which follows only from the coupling between the first and second excited states, diagonalize it and add it to the $\ket{1}\bra{1}$ term of the qubit cross-resonance matrix. The resulting amplitude of the ZX term of the cross-resonance gate is
\begin{align}
    \label{eq:PLA_calibration}
    & J_{ZX}= \frac{g\Omega}{2}\times\\ \nonumber
    &\sqrt{\frac{3\Omega^2+\Delta^2+(\Delta+\eta)^2-2\Delta\sqrt{2\Omega^2+(\Delta+\eta)^2}}{(\Omega^2+\Delta^2)(2\Omega^2+(\Delta+\eta)^2)}}~,
\end{align}
with $g$ and $\Delta$ being the coupling and detuning between the control and target qubit, respectively, $\Omega$ is the microwave amplitude, and $\eta$ is the anharmonicity of the control qubit (with the convention that $\eta$ is positive).
Eq.~\ref{eq:PLA_calibration} is surprisingly precise. For $\sim80\%$ of bonds on the Eagle devices, the prediction of this equation\footnote{provided the value of $g$ is extracted using the native values of the native Clifford gate, and that the always-on $ZZ$ interaction is also taken into account} is precise without any corrections. For the worst bonds, a correction of up to 5\% of the microwave amplitude may be required. This is small enough for a single iteration of the calibration procedure App.~\ref{app:calibration} to achieve the required precision.
In contrast, using the known analytical estimates from the literature \footnote{These include either ignoring the second excited state or Schrieffer–Wolff transformation using the approximation that $\Omega\ll|\Delta|,\eta$.} \cite{Magesan_2020} required significant corrections, sometimes mandating up to 5 iterations (for most cases 3 were enough) of characterization and correction to achieve a precise calibration (<0.005 over-rotation for all gates).

PLA on Eagles is also used to generate optimal "layering" of pulses with different durations and dynamical decoupling, where we make sure that the always-on $ZZ$ interaction is also decoupled, by applying the X pulses in a temporal "brick-wall" configuration.

Using all the above, we demonstrated mitigation of a 2D Trotter-Ising circuit on \texttt{ibm\_brisbane}. Fig.~\ref{fig:Eagle-119q} shows the results of the mitigation vs the noisy and ideal values. In this case, the noisy values and active volumes were calculated for a circuit transpiled using CNOT, rather than RZZ gates. The reason for this is that RZZ gates are not native to Eagle devices and are therefore not available to users running without QESEM. For the heavy-weight observables, we reached an active volume of almost 400 CNOT gates (i.e., 200 RZZ gates) while maintaining a high precision of $\sim0.05$.

\subsection{IonQ Aria}

\begin{figure*}[tb]
    \centering
    \includegraphics[width=0.9\linewidth]{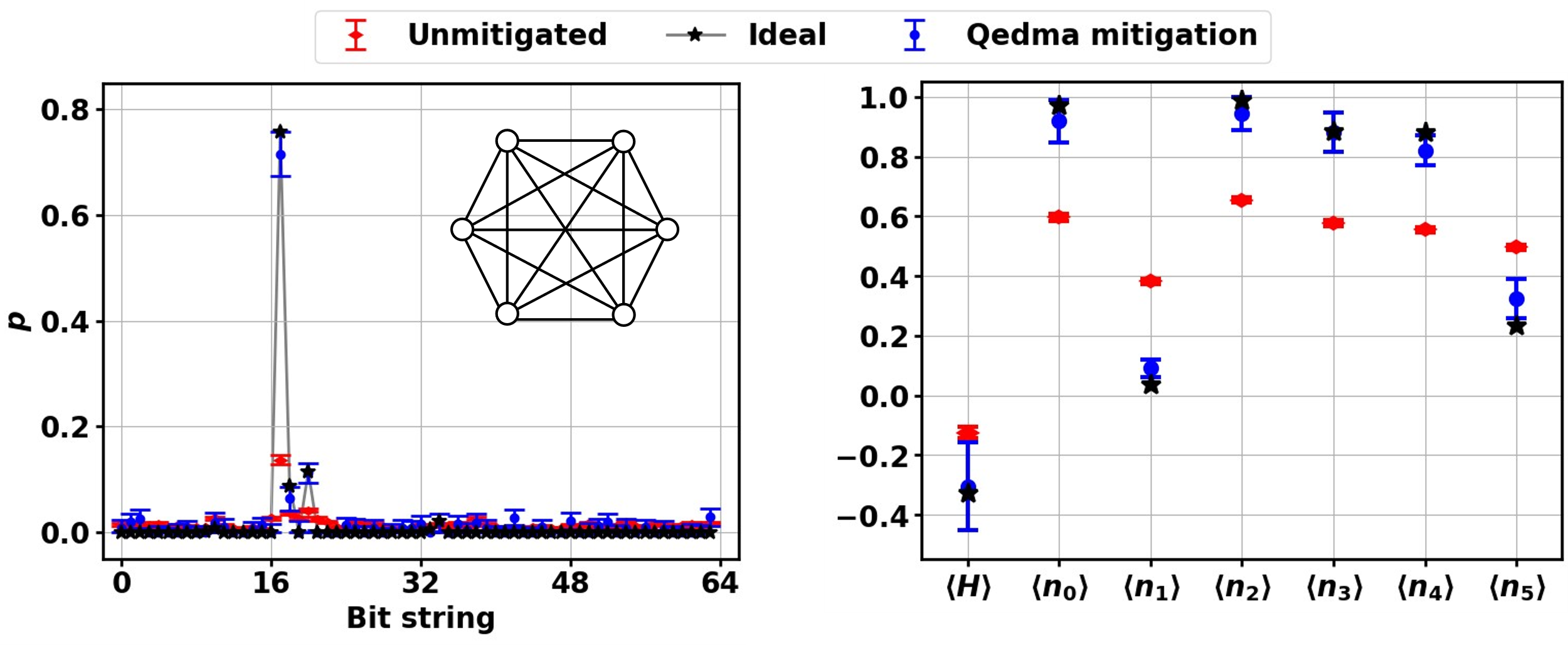}
    \caption{QESEM results for a VQE circuit simulating the NaH molecule using 6 qubits and 94 MS gates, executed on IonQ's \textit{Aria} QPU. As shown in the left panel's circuit-connectivity graph, the circuit fully leverages the device's all-to-all connectivity. The left panel demonstrates QESEM’s ability to accurately estimate the full output probability distribution of the ideal (noise-free) circuit. The right panel shows QESEM.
    }
    \label{fig:ionq-vqe-6q}
\end{figure*}

\begin{figure}[tb]
\centering
\subfloat[\label{fig:ionq_graph_12}]{
    \includegraphics[width=0.9\columnwidth]{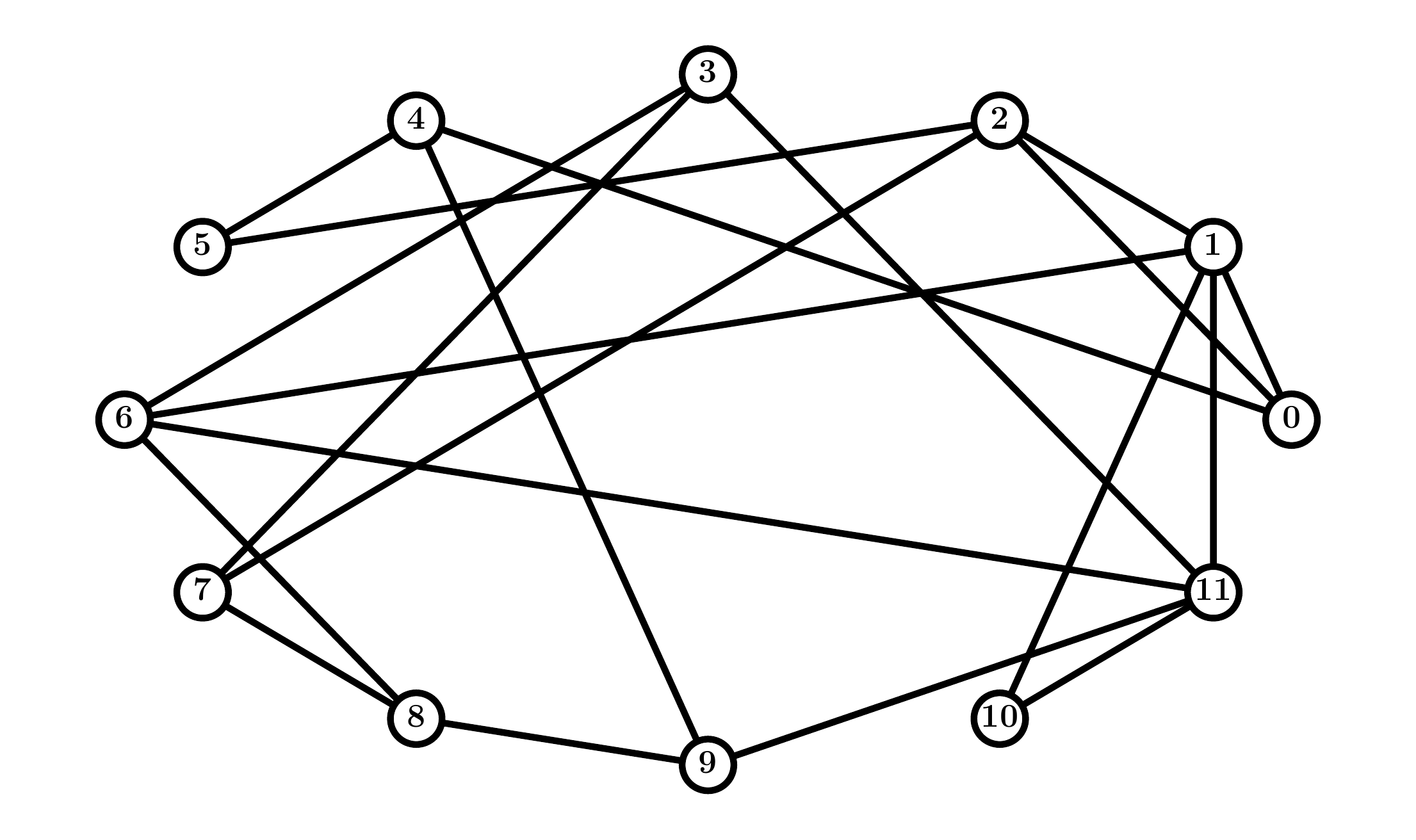}
}
\\
\subfloat[\label{fig:o2_H}]{
    \includegraphics[width=0.95\columnwidth]{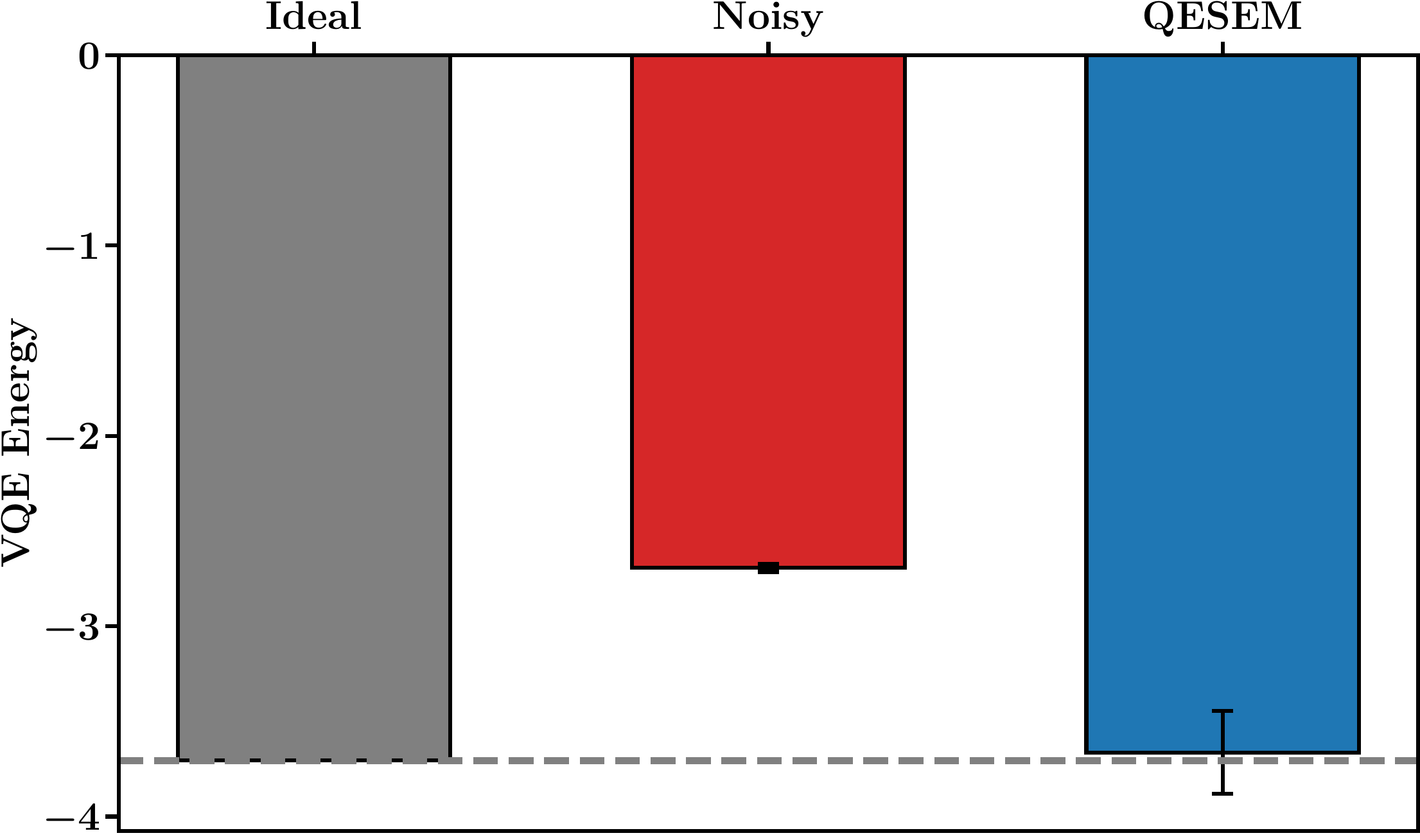}
}
\\
\subfloat[\label{fig:o2_obs}]{
    \includegraphics[width=0.95\columnwidth]{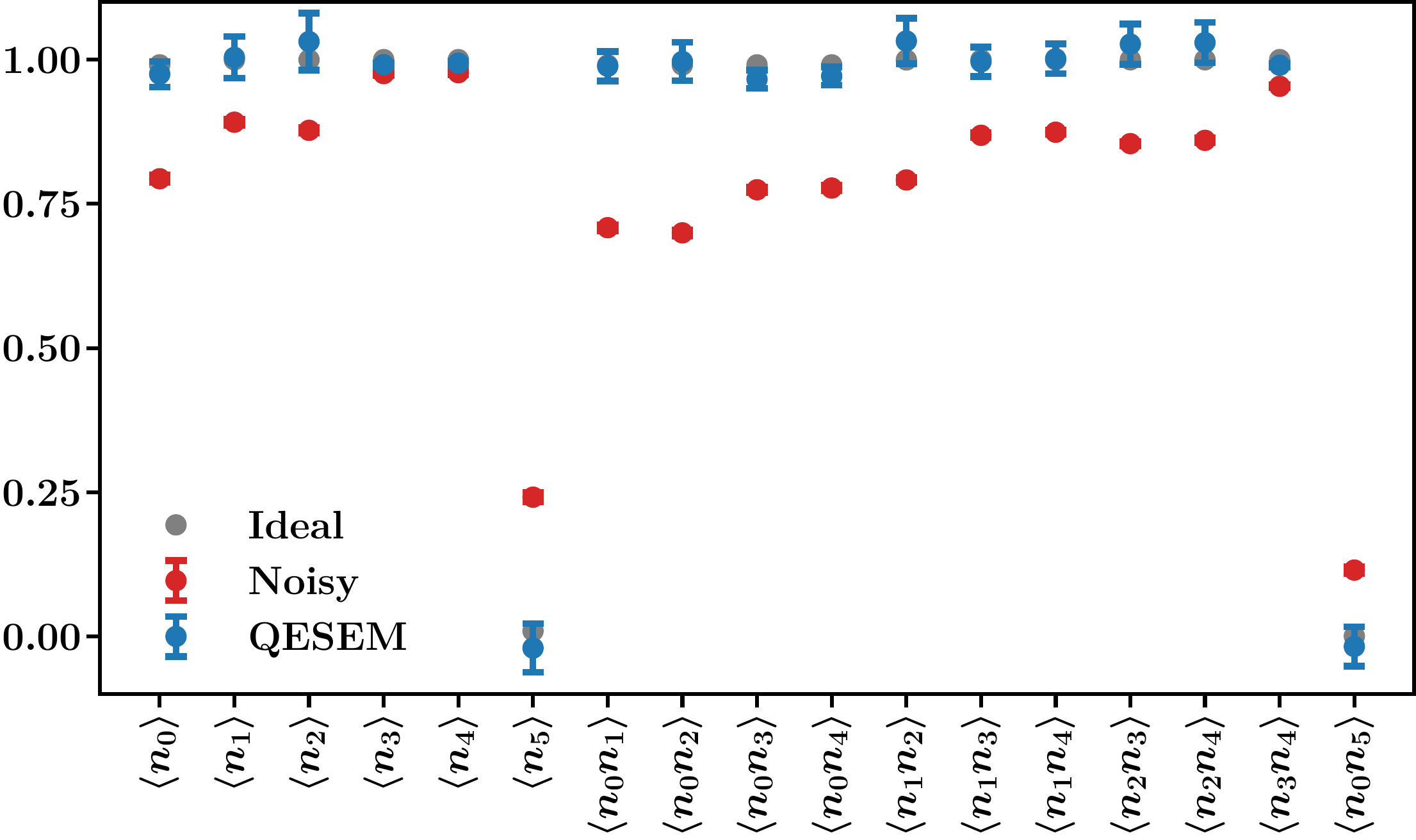}
}
\caption{QESEM results for a VQE circuit simulating the $\mathrm{O}_2$ molecule using 12 qubits and 99 MS gates, executed on IonQ's \textit{Aria} QPU. 
\textbf{\protect\subref{fig:ionq_graph_12}} The high circuit connectivity, which involves 20 unique MS gates on the 12 qubits.
\textbf{\protect\subref{fig:o2_H}} QESEM estimation of the molecular energy compared to the ideal and noisy values. 
\textbf{\protect\subref{fig:o2_obs}} QESEN estimation of some orbital occupations and two-orbital correlations within the VQE ground state compared to the ideal and noisy values.
\label{fig:ionq-vqe-12q}
}
\end{figure}

To demonstrate QESEM's hardware-agnostic capabilities, we executed several quantum circuits on IonQ’s trapped-ion QPUs. To the best of our knowledge, these experiments are the first large-scale demonstrations of error mitigation on trapped-ion hardware based on an unbiased method. They were performed via both IonQ's direct interface and Amazon Braket's cloud interface, and featured in the AWS Quantum Technologies Blog~\cite{QESEM_Braket2024}.

We simulated the $\mathrm{NaH}$ and $\mathrm{O}_2$ molecules using Variational Quantum Eigensolver (VQE) circuits implemented on IonQ’s Aria quantum processing unit (QPU). The $\mathrm{NaH}$ simulation utilized a 6-qubit circuit with 94 M\o lmer–S\o rensen (MS) gates, while the $\mathrm{O}_2$ simulation employed a 12-qubit circuit with 99 MS gates. The effects of QESEM mitigation are illustrated in Figs.~\ref{fig:ionq-vqe-6q} and \ref{fig:ionq-vqe-12q}.

\clearpage

\bibliography{bibliography.bib}

\end{document}